\documentclass[amsmath,amssymb,11pt]{article}
\usepackage{jheppub2}
\pdfoutput=1
\usepackage{graphicx,epsfig}
\usepackage{float}
\usepackage{subfig}
\usepackage{subfloat}
\usepackage[utf8]{inputenc}
\usepackage{hyperref}
\usepackage{caption}
\usepackage{color}
\usepackage{overpic}
\usepackage[dvipsnames]{xcolor}

\newcommand*{\affmark}[1][*]{\textsuperscript{#1}}

\definecolor{pink1}{rgb}{0.858, 0.188, 0.478}

\newcommand{\beq}{\begin{equation}}

\newcommand{\eeq}{\end{equation}}

\subheader{\begin{flushright}
\texttt{BRX-TH-6690}
\end{flushright}}

\title{Semi-classical  thermodynamics of quantum extremal surfaces in Jackiw-Teitelboim gravity}

\author{Juan F. Pedraza,\affmark[1,2]}
\emailAdd{j.pedraza@ucl.ac.uk}
\author{Andrew Svesko,\affmark[1]}
\emailAdd{a.svesko@ucl.ac.uk}
\author{Watse Sybesma\affmark[3]}
\emailAdd{watse@hi.is}
\author{and Manus R. Visser\affmark[4]}
\emailAdd{manus.visser@unige.ch}

\affiliation{\affmark[1]Department of Physics and Astronomy, University College London,\\
London, WC1E 6BT, United Kingdom\\
\affmark[2]Martin Fisher School of Physics, Brandeis University\\
Waltham MA 02453, USA\\
\affmark[3]Science Institute, University of Iceland\\
Dunhaga 3, 107 Reykjav\'ik, Iceland\\
\affmark[4]Department of Theoretical Physics, University of Geneva\\
24 quai Ernest-Ansermet, 1211 Gen\`{e}ve 4, Switzerland}

\abstract{Quantum extremal surfaces (QES), codimension-2 spacelike regions which extremize the generalized entropy of a gravity-matter system, play a key role in the study of the black hole information problem. The thermodynamics of QESs, however, has been largely unexplored, as a proper interpretation requires a detailed understanding of backreaction due to quantum fields. We investigate this problem in semi-classical Jackiw-Teitelboim (JT) gravity, where the spacetime is the eternal two-dimensional Anti-de Sitter  ($\text{AdS}_{2}$) black hole, Hawking radiation is described by a conformal field theory with central charge $c$, and backreaction effects may be analyzed exactly. We show the Wald entropy of the semi-classical JT theory entirely encapsulates the generalized entropy -- including time-dependent von Neumann entropy contributions -- whose extremization leads to a QES lying just outside of the black hole horizon. Consequently, the QES defines a Rindler wedge nested inside the enveloping black hole. We use covariant phase space techniques on a time-reflection symmetric slice to derive a Smarr relation and first law of nested Rindler wedge thermodynamics, regularized using local counterterms, and intrinsically including semi-classical effects.  Moreover, in the microcanonical ensemble the semi-classical first law implies the generalized entropy of the QES is stationary at fixed   energy.  Thus, the thermodynamics of the nested Rindler wedge is equivalent to the thermodynamics of the QES in the microcanonical ensemble. 
}

\begin{document}

\maketitle

\section{Introduction}
 
Classical gravity posits black holes obey a set of mechanical laws formally reminiscent of the laws of thermodynamics, where the horizon area is interpreted as a thermodynamic entropy obeying a second law. Taking this analogy seriously, Bekenstein linked horizon area $A$ with entropy $S_{\text{BH}}$ via a series of gedanken experiments  \cite{Bekenstein:1972tm,Bekenstein:1973ur}, further established by Hawking who revealed black holes emit radiation at a temperature $T_{\text H}$ proportional to their surface gravity~$\kappa$ \cite{Hawking:1974sw,Hawking:1976de}. Thus, black holes are genuine thermal systems, with a temperature and entropy, and obey a first law. Specializing to neutral, static black holes, the first law is
\beq \delta M=T_{\text H}\delta S_{\text{BH}}\;, \qquad \text{with} \qquad T_{\text{H}}=\frac{\kappa}{2\pi} \quad \text{and} \quad S_{\text{BH}} = \frac{A}{4G}\;, \label{eq:firstlawgen}\eeq
and  the ADM mass $M$ of the system is identified with the internal energy. 
When the thermal system of interest includes a black hole as well as matter in the exterior of the black hole,  the total entropy is quantified by the generalized entropy $S_{\text{gen}}$, accounting for both the classical black hole entropy $S_{\text{BH}}$ and the entropy of exterior matter $S_{\text{m}}$ \cite{Bekenstein:1974ax},
\beq S_{\text{gen}}=S_{\text{BH}}+S_{\text{m}}\;,\label{eq:genent}\eeq
such that a generalized second law $\delta S_{\text{gen}} \ge 0$ holds. Moreover, in the case of an asymptotically flat black hole surrounded by a fluid, the gravitational first law (\ref{eq:firstlawgen}) is modified by  $\delta H_{\text{m}}$, encoding classical stress-energy variations outside of the horizon \cite{Bardeen:1973gs,Iyer:1996ky,Jacobson:2018ahi}:
\beq \delta M=T_{\text H}\delta S_{\text{BH}} + \delta H_{\text{m}}\;.\label{eq:firstlawmatham}\eeq
The \emph{classical} Bekenstein-Hawking entropy formula and the subsequent thermodynamics have been derived in a variety of ways. Notably, typically one either solves the wave equations of quantum fields over the fixed classical black hole  background \cite{Hawking:1974sw,Hawking:1974rv}, or performs a Euclidean path integral analysis where the logarithm of the partition function is given by the on-shell Euclidean gravitational action \cite{Gibbons:1976ue}. Neither of these approaches addresses a fundamental conceptual problem: how does the Hawking radiation know about the (micro)-dynamics of the event horizon?\footnote{We thank Erik Curiel for bringing this to our attention.} Put another way, Hawking radiation is not standard black body radiation, \emph{i.e.}, unlike standard electromagnetic radiation of a hot iron caused by energy released from its atoms, Hawking radiation is understood as excited modes of an external quantum field, one which is generally not coupled to the background spacetime. Thus, it is unclear why the Hawking temperature should really be identified with the temperature of the black hole, and, consequently, why the first law of black hole mechanics should be identified as a first law of thermodynamics. Indeed, the first law (\ref{eq:firstlawgen}) is derived from the dynamical gravitational field equations, without any mention of quantum field theory. It is only after identifying $T_{\text H}\propto\kappa$ as the black hole temperature that the thermodynamic interpretation of the first law follows.

 A crucial ingredient missing from either derivation of black hole thermodynamics is the effect of backreaction, \emph{i.e.}, the influence quantum matter has on the classical background geometry. It is only when these semi-classical effects are included that the aforementioned conceptual puzzle can be (partially) addressed since the Hawking radiation backreacts on the geometry, and thus the temperature appearing in the semi-classical first law is naturally interpreted as the black hole temperature.

 Accounting for these backreaction effects requires solving the semi-classical gravitational field equations, \emph{e.g.}, the semi-classical Einstein equations, 
\beq G_{\mu\nu}(g)=8\pi G\langle \Psi|T_{\mu\nu}(g)|\Psi\rangle\;,\label{eq:semiclassein}\eeq
where $\langle \Psi|T_{\mu\nu}(g)|\Psi\rangle$ is the (renormalized) quantum matter stress-energy tensor, given by the expectation value of the stress-energy tensor operator in the quantum state of matter $|\Psi\rangle$, and is representative of the backreaction. These gravitational field equations are only valid in a  semi-classical regime with quantum matter fields  in a classical dynamical  spacetime. Explicitly solving   (\ref{eq:semiclassein}) is notoriously difficult as it requires one to simultaneously solve a coupled system of geometry $g_{\alpha\beta}$ and correlation functions of quantum field operators. Typically, in the case of four dimensions and higher, one studies the problem perturbatively. However, this offers limited insight, especially when backreaction effects are large.

To clarify, 
one may formally use the semi-classical Einstein equations (\ref{eq:semiclassein})  to  derive a first law which includes  semi-classical   corrections \cite{Jacobson:2018ahi}. The matter Hamiltonian variation $\delta H_{\text{m}}$ in the first law (\ref{eq:firstlawmatham}) is then replaced by the variation of the expectation value of the quantum matter Hamiltonian $\delta \langle H_{\text{m}} \rangle$. Invoking the first law of entanglement for the quantum matter fields, $\delta S_{\text{m}}=T^{-1}_{\text{H}}\delta\langle H_{\text{m}}\rangle$, the semi-classical version of the first law (\ref{eq:firstlawmatham}) becomes 
\beq
\delta M = T_{\text{H}}\delta S_{\text{gen}}\;,
\eeq
where we identified the generalized entropy \eqref{eq:genent}. Now $T_{\text{H}}$ may be interpreted as the temperature of the backreacted black hole.  However,  the identification $T_\text{H} \sim \kappa$ only holds in the semi-classical regime, and a puzzle remains whether the microscopic degrees of freedom  of the black hole in quantum gravity generate black body radiation, such that Hawking radiation is  associated to the micro-dynamics of the event horizon (see \cite{Callan:1996dv} for progress in string theory). 


Important to the thermodynamic interpretation is understanding the correct statistical ensemble one is considering, involving extremization of the thermodynamic potential of the ensemble, \emph{e.g.}, for the microcanonical ensemble  $M$ is held fixed and $S_{\text{gen}}$ is extremized. Generally, and particularly in spacetime dimensions $D\geq4$, however, the backreaction in black hole spacetimes has not yet been explicitly solved analytically. Consequently, in such contexts, one cannot generically find the extrema of the generalized entropy. 

Progress can be made in special contexts, particularly in two-dimensional dilaton gravity models such as the Callan-Harvey-Giddings-Strominger (CGHS) model \cite{Callan:1992rs} or Jackiw-Teitelboim (JT) gravity \cite{Jackiw:1984je,Teitelboim:1983ux}, since, up to minor ambiguities, the effects of the backreaction are largely fixed by the two-dimensional Polyakov action capturing the contributions of the conformal anomaly \cite{Polyakov:1981rd}. 
In fact, the semi-classical extension of both models have had success in studying the black hole information problem analytically, and address the aforementioned black hole temperature puzzle \cite{Susskind:1992gd,Russo:1992ax,Russo:1992ht,Fiola:1994ir,Fabbri:1995bz,Almheiri:2014cka} (see \cite{Fabbri:2005mw} for a pedagogical review on both models). Notably, with backreaction fully accounted for in the extended models, identifying the temperature of Hawking radiation with the black hole   temperature     is indeed valid. The Hawking temperature appears in the semi-classical first law of black hole horizons, where the classical entropy is replaced by the generalized entropy and the classical ADM energy by semi-classical energy \cite{Fiola:1994ir,Almheiri:2014cka}.

Though both (extended) models contain fully analytic black hole solutions incorporating backreaction and are capable of addressing the problem of black hole formation and evaporation, in certain respects the JT model is simpler since the background is fixed to be two-dimensional Anti-de Sitter ($\text{AdS}_{2}$) space and it is only the value of the dilaton field that tracks gravitational strength. 
For this reason, in this article we primarily focus on the JT model, however, many of the results we uncover are expected to generalize to the CGHS model (and, optimistically, beyond). 

The classical JT model can be viewed as the low-energy dynamics of a wide class of charged, near-extremal black holes and branes in higher dimensions \cite{Achucarro:1993fd,Fabbri:2000xh,Nayak:2018qej,Sachdev:2019bjn}.\footnote{The low-energy dynamics of \emph{rotating}, near-extremal black holes is also described in terms of JT gravity now coupled to extra matter fields. These extra degrees of freedom describe the deformation or ``squashing'' of the near-horizon geometry away from extremality  \cite{Castro:2018ffi,Moitra:2019bub,Castro:2021fhc}.} More recently, the gravi-dilaton theory was shown to offer a precise realization of holographic $\text{AdS}_{2}/\text{CFT}_{1}$ duality, where JT gravity arises as the low-energy gravitational dual to an integrable limit of the 1D quantum mechanical Sachdev-Ye-Kitaev (SYK) model \cite{Sachdev:1992fk,Jensen:2016pah,Maldacena:2016hyu,Maldacena:2016upp,Engelsoy:2016xyb}. It is possible to quantize classical JT gravity using canonical/covariant phase space techniques \cite{Henneaux:1985nw,NavarroSalas:1992vy,Constantinidis:2008ty,Harlow:2018tqv}, or via  a Euclidean path integral, in which the partition function of the theory is dual to some double-scaled matrix model \cite{Saad:2019lba,Stanford:2019vob,Okuyama:2019xbv,Johnson:2019eik}. 
A particularly exciting aspect of the JT model is that it offers a potential resolution of the black hole information paradox, in which the \emph{fine grained} (equivalently, von Neumann  or entanglement) entropy of the radiation may be exactly computed and follows the expected behavior for a unitary Page curve \cite{Penington:2019npb,Almheiri:2019psf,Almheiri:2019hni,Almheiri:2019yqk,Penington:2019kki,Almheiri:2019qdq,Almheiri:2020cfm}).\footnote{Similar successes were later found to hold in the CGHS model \cite{Gautason:2020tmk,Hartman:2020swn}, JT de Sitter gravity \cite{Sybesma:2020fxg,Aalsma:2021bit}, and higher-dimensional scenarios \cite{Chen:2020uac,Chen:2020hmv}. See also \cite{Geng:2021iyq} which finds the entanglement entropy characterizing communication between $\text{AdS}_{2}$ braneworld black holes follows a unitary Page curve.}

The fine grained entropy formula of radiation is inspired by the Anti-de Sitter / Conformal Field Theory (AdS/CFT) correspondence, where the entanglement entropy $S_{\text{ent}}$ of a CFT reduced to a subregion of the conformal boundary of AdS is dual to the area of an extremal bulk surface $X$ homologous to the boundary region, in the large $N$ limit of the CFT \cite{Ryu:2006bv,Hubeny:2007xt,Lewkowycz:2013nqa}. In the semi-classical approximation, the classical area term is supplemented by a bulk entanglement entropy $S^{\text{bulk}}_{\text{ent}}$ accounting for the entanglement between bulk quantum fields, including both matter and the metric, \cite{Barrella:2013wja,Faulkner:2013ana,Engelhardt:2014gca}
\beq S_{\text{ent}}(\Sigma_{X})=\underset{X}{\text{min}}\,\, \underset{X}{\text{ext}}\left[\frac{\text{Area}(X)}{4G}+S_{\text{ent}}^{\text{bulk}}(\Sigma_{X})\right]\;.\label{eq:genentQES}\eeq
Thus, the fine grained entropy $S_{\text{ent}}(\Sigma_{X})$ is given by the extremization of the generalized entropy (\ref{eq:genent}), $S_{\text{gen}}(X)$, except where now the black hole horizon $H$ has been replaced by the \emph{quantum extremal surface} (QES) $X$ \cite{Engelhardt:2014gca}, and where the matter field entropy $S_{\text{m}}$ is given by the von Neumann entropy of the bulk fields restricted to a bulk codimension-1 slice $\Sigma_{X}$ in the semi-classical limit.\footnote{A point worth clarifying is that the \emph{quantum} extremal surface is defined as the surface which extremizes the generalized entropy that includes semi-classical   corrections at \emph{all} orders in $1/G$ \cite{Engelhardt:2014gca}, while the authors of \cite{Faulkner:2013ana} only considered the leading order $1/G$ semi-classical correction, where $X$ is an ordinary extremal surface and (\ref{eq:genentQES}) is referred to as the Faulkner-Lewkowycz-Maldacena (FLM) formula.} In the limit $X$ coincides with a black hole horizon, we recover the usual generalized entropy (\ref{eq:genent}). Using either ``double holography" or a Euclidean path integral analysis invoking the replica trick, an equivalent expression for the von Neumann entropy of radiation may be derived and exhibits a \emph{dynamical} phase transition taking place near the Page time such that the state of radiation obeys unitary evolution \cite{Almheiri:2019hni,Almheiri:2019qdq}.
While understanding the precise details of this phase transition is not necessary for this article, we are highly influenced by the question of whether the Page curve follows from a Lorentzian time path integral \cite{Goto:2020wnk,Colin-Ellerin:2020mva,Colin-Ellerin:2021jev}.


The aforementioned semi-classical first law \cite{Fiola:1994ir,Almheiri:2014cka,Moitra:2019xoj} only applies to black hole horizons and not to quantum extremal surfaces. In this article we intend to derive a semi-classical first law associated with QESs in AdS$_2$. Moreover, a notable observation is the position of the QES in eternal black hole backgrounds: the QES lies \emph{outside} of the black hole horizon \cite{Almheiri:2019yqk,Gautason:2020tmk,Hartman:2020swn}, and has an associated entanglement wedge -- a ``nested"  Rindler wedge in $\text{AdS}_{2}$ -- that differs from the  AdS-Rindler  wedge of the black hole. Importantly, it is the entanglement wedge of the QES which is used to compute the quantum corrected holographic entanglement entropy, and therefore, as was the case for AdS-Rindler space \cite{Casini:2011kv,Faulkner:2013ica}, understanding the thermodynamic properties of the QES entanglement wedge is worthwhile and necessary.

Motivated by these observations, here we provide a detailed investigation of the (semi-classical) thermodynamics of surfaces that lie outside of a black hole horizon, particularly quantum extremal surfaces. More specifically, restricting to scenarios with time-reflection symmetry, we uncover a first law of thermodynamics for QESs, where the generalized entropy replaces the usual black hole entropy. When $X$ coincides with the black hole horizon, we recover a semi-classical first law of black hole thermodynamics \cite{Moitra:2019xoj}. In principle, while this study should hold for more realistic theories of gravity, we use the remarkable simplicity of the semi-classical JT model to our advantage where we can solve everything analytically. In this context the classical ``area" term is given by the value of the dilaton evaluated at $X$, while the bulk von Neumann entropy is quantified by an auxiliary field used to localize the Polyakov action, and is understood to represent conformal matter fields living in an eternal $\text{AdS}_{2}$ black hole background. We demonstrate the semi-classical Wald entropy \cite{Wald:1993nt}, with respect to the Hartle-Hawking vacuum state, \emph{exactly} reproduces the generalized entropy (\ref{eq:genentQES}), including the von Neumann contribution, and is generally time dependent. We also compute the quantum-corrected ADM mass, such that together with the entropy we derive a semi-classical Smarr relation and an associated first law.\footnote{Recently backreaction was accounted for in a specific effective theory of gravity induced via $\text{AdS}_{4}/\text{CFT}_{3}$ brane holography, leading to a quantum BTZ black hole \cite{Emparan:2020znc}, where a first law is uncovered involving the generalized entropy. This set-up is different from the one we study, however, as their derivation of the first law relies on AdS/CFT duality while ours does not. See also  \cite{Belin:2021htw} where conformal bootstrap methods are used to compute the entanglement entropy of a $\text{CFT}_{2}$ to second order in $1/c$ corrections, and, upon invoking $\text{AdS}_{3}/\text{CFT}_{2}$, is matched with $S_{\text{gen}}$ to second order corrections in $G$.}

The outline of this article is as follows. In Section \ref{sec:JTandsemiclass}, after reviewing the basics of classical JT gravity and the eternal $\text{AdS}_{2}$ black hole, we solve the semi-classical JT model in full detail. In particular, we emphasize the importance of choosing a vacuum state, defined such that its expectation value of the \emph{normal ordered} stress tensor vanishes, leading to different solutions for $\phi$ and $\chi$, the dilaton and an auxiliary field localizing the 1-loop Polyakov action, respectively. 

With the backreacted model completely solved, in Section \ref{sec:Waldentgenent} we determine the Wald entropy for the Boulware and Hartle-Hawking  vacuum, which is taken to be a thermodynamic entropy with respect to the Hartle-Hawking vacuum. Furthermore, via the presence of the $\chi$ field, we find the Wald entropy captures the full generalized entropy, that is, \textit{including} the von Neumann contribution, and is generally time dependent. Extremization of this entropy leads to a quantum extremal surface  that lies just outside the black hole horizon. 

Section \ref{sec:semithermorindler} is then devoted to the thermodynamics of   AdS-Rindler wedges  nested inside the eternal $\text{AdS}_{2}$ black hole, itself an AdS-Rindler wedge. We derive a first law associated with the nested AdS-Rindler wedges. 
Using the boost Killing vector associated with the nested Rindler wedge, and local counterterms to regulate divergences, we compute a semi-classical Smarr relation and a corresponding first law of thermodynamics of nested Rindler wedges, where the temperature  is proportional to the surface gravity of the nested  Rindler  horizon. The global Hartle-Hawking state reduced to the nested Rindler wedge is in a thermal Gibbs state at   this temperature. Allowing for  variations of the   coupling constants $(\phi_0, G, \Lambda,c)$ in the semi-classical JT action, we also derive a classical and semi-classical extended first law  where we define   different counterterm subtracted ``Killing  volumes"  conjugate to these coupling variations. 

We emphasize   the classical and semi-classical first laws hold  for \emph{any} nested Rindler wedge. However, in the microcanonical ensemble    the generalized entropy of the nested wedge is extremized, and hence  the thermodynamics of the nested wedge is equivalent to the thermodynamics of the QES.   Consequently, we argue in the microcanonical ensemble it is more natural to consider the entropy of the QES when semi-classical effects are taken into account. We summarize our findings in Section \ref{sec:conc} and discuss potential future research avenues. 

To keep this article self-contained we include a number of appendices providing computational details and formalism used in the body of the article. Appendix \ref{app:ebhcoord} summarizes the various coordinate systems of $\text{AdS}_{2}$ and how the stress tensor transforms in these coordinates. In Appendix \ref{app:embeddingformalism} we detail the construction of the boost Killing vector of the Rindler wedge inside $\text{AdS}_{2}$ using the embedding formalism. Appendix \ref{app:Waldformalism} provides a thorough account of the covariant phase space formalism for $2D$ dilaton theories of gravity, and of the construction of the ADM Hamiltonian with semi-classical corrections. In Appendix~\ref{app:extfirstlawdeets} we generalize the covariant phase formalism with boundaries to allow for coupling variations. We apply this formalism to JT gravity, and provide a detailed derivation of the extended first law. Lastly, in Appendix \ref{app:secondlaw} we provide a heuristic argument for the generalized second law.

 \section{JT gravity and semi-classical corrections} \label{sec:JTandsemiclass}
 
To set the stage we first review some basic features of eternal black hole solutions to the classical JT model. Then we analytically solve the semi-classical JT equations of motion, accounting for backreaction, paying special  attention to the choice of vacuum state. 
 
 \subsection{Classical JT and the eternal AdS$_{2}$ black hole}
 

 The classical JT action in Lorentzian signature is
\beq I_{\text{JT}}^{\text{bulk}}=\frac{1}{16\pi G}\left[\int_{M}d^{2}x\sqrt{-g}\phi_{0}R+\int_{M} d^{2}x\sqrt{-g}\phi\left(R+\frac{2}{L^{2}}\right)\right]\;.\label{eq:JTact1}\eeq
where $G$ is the two-dimensional   dimensionless  Newton's constant       which we   retain   as a book keeping device. Further $\Lambda=-1/L^{2}$ is a negative cosmological constant, where $L$ is the $\text{AdS}_{2}$ length scale characterizing the charge of the higher-dimensional black hole system,\footnote{Here we are thinking of a dimensional reduction of an asymptotically flat charged black hole, as in \emph{e.g.}, \cite{Navarro-Salas:1999zer}. For a neutral AdS black hole the $\text{AdS}_{2}$ length scale is fixed by the higher-dimensional AdS length scale (see, \emph{e.g.}, \cite{Moitra:2018jqs}).} 
$\phi$ is the dilaton arising from a standard spherical reduction of the parent theory,  and $\phi_{0}$ is a constant proportional to the extremal entropy of the higher-dimensional black hole geometry. 
To have a well-posed variational principle with  Dirichlet boundary condition, we supplement the above bulk action with a Gibbons-Hawking-York (GHY) boundary term,\footnote{Though we won't need it here,  the GHY term   can be recast as a Schwarzian theory on the boundary and is identified with the low-energy sector of the SYK action. For a review, see \emph{e.g.}, \cite{Sarosi:2017ykf,Trunin:2020vwy}.} and to cancel the divergence in the second GHY term we also subtract a boundary   counterterm, 
\beq I^{\text{bdy}}_{\text{JT}} =I^{\text{GHY}}_{\text{JT}} + I^\text{ct}_{\text{JT}}=\frac{1}{8\pi G}\left[\int_{B}dt\sqrt{-\gamma}\phi_{0}K+\int_{B}dt\sqrt{-\gamma}\phi( K - 1/L)  \right] \;,\label{eq:GHYctterms}\eeq
where $B$ is the spatial boundary of $M$, and $K$ is the trace of the extrinsic curvature of  $B$ with induced metric $\gamma_{\mu\nu}$. Due to the Gauss-Bonnet theorem the sum of the $\phi_0$ terms in the bulk \eqref{eq:JTact1} and boundary action  \eqref{eq:GHYctterms}  is proportional to the Euler character of the manifold~$M$, so these terms are purely topological in two dimensions.

\begin{figure}[t]
\begin{center}
		\begin{overpic}[width=0.60\textwidth]{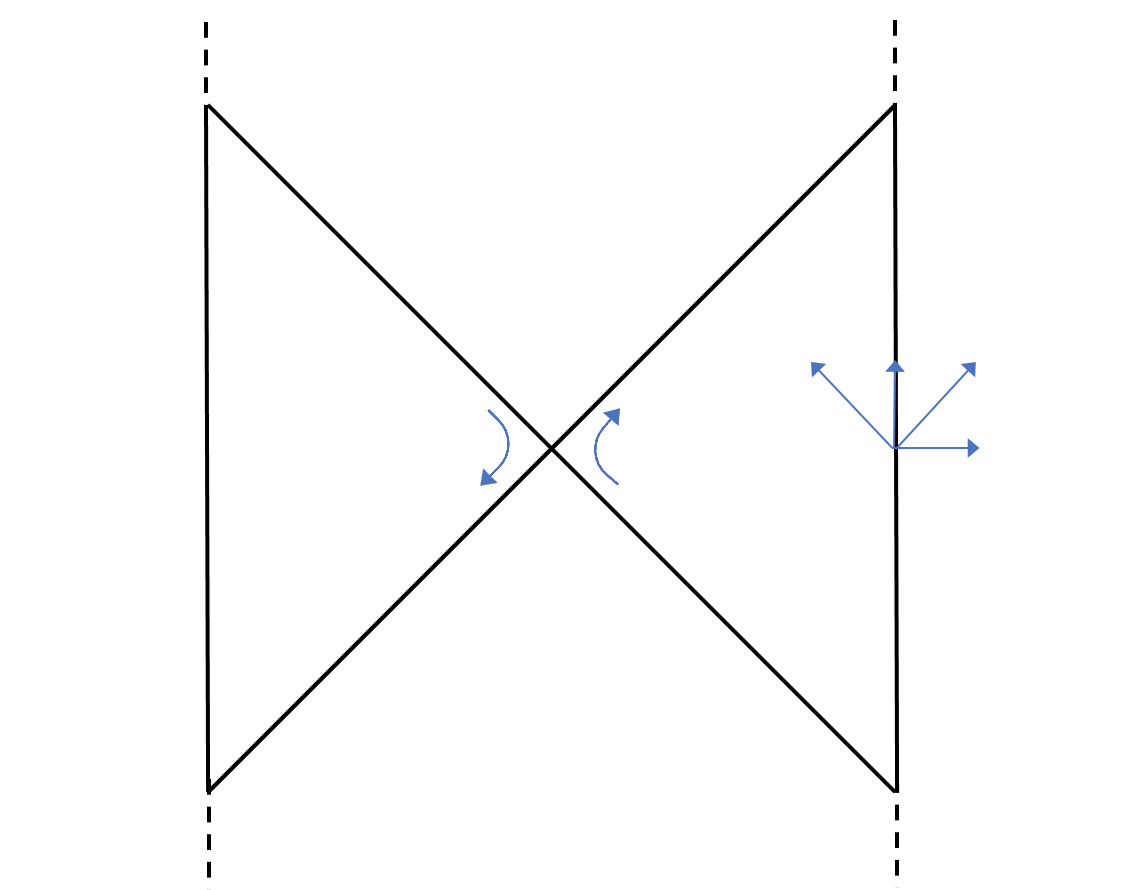}
			\put (87,38) {\footnotesize{$r_{*}$}}
			\put (84,47) {\footnotesize{$v$}}
			\put (70,47) {\footnotesize{$u$}}
			\put (79,47) {\footnotesize{$t$}}
			\put (55,49) {\rotatebox{45}{\footnotesize{$u=\infty$, $U_{\text{K}}=0$}}}
			\put (55,33) {\rotatebox{-45}{\footnotesize{$v=-\infty$, $V_{\text{K}}=0$}}}
			%
		\end{overpic}
		
	\end{center}
\centering
\caption{The relevant portion of a Penrose diagram of an eternal non-extremal AdS$_2$ black hole is presented. The diagonal lines denote the bifurcate Killing horizons and the vertical lines denote the conformal boundary. The dashed edges indicate that this is only a portion of the full Penrose diagram. The curved arrows indicate the directions of the Killing flow.}
\label{fig:coords}
\end{figure}
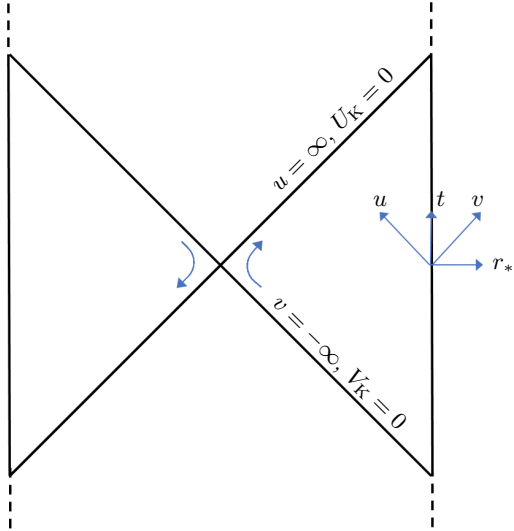

The gravitational and dilaton equations of motion are, respectively, 
\beq T_{\mu\nu}^{\phi}=0\;,\quad \text{with} \quad T_{\mu\nu}^{\phi} \equiv -\frac{2}{\sqrt{-g}}\frac{\delta I_{\text{JT}}}{\delta g^{\mu\nu}} =-\frac{1}{8\pi G}\left(g_{\mu\nu}\Box-\nabla_{\mu}\nabla_{\nu}-\frac{1}{L^{2}}g_{\mu\nu}\right)\phi\;,\label{eq:graveom}\eeq
\beq R+\frac{2}{L^{2}}=0\;.\label{eq:dileom}\eeq
The dilaton equation of motion \eqref{eq:dileom} fixes  the background geometry to be purely $\text{AdS}_{2}$.  In static Schwarzschild-like coordinates the $\text{AdS}_{2}$ metric takes the form
\beq ds^{2}=-N^{2}dt^{2}+N^{-2}dr^{2}\;,\quad N^{2}=\frac{r^{2}}{L^{2}}-\mu\; .\label{eq:ads2bhstat}\eeq
This describes an eternal AdS$_2$ black hole, which is equivalent to  AdS$_2$-Rindler space, as shown in Appendix \ref{app:ebhcoord}. Further, we take the dilaton solution to the gravitational equation of motion  \eqref{eq:graveom}  to be (see \emph{e.g.}, \cite{Witten:2020ert})
\beq   \phi(r)=\frac{r}{L}\phi_{r}\;. \label{eq:dilaton1}\eeq
The horizon of the black hole is located at  $r_{H}=L\sqrt{\mu}$, where $L$ is the $\text{AdS}$ curvature scale and  $\mu\geq0$ is a dimensionless mass parameter proportional to the ADM mass (c.f. \cite{Moitra:2019xoj})
\beq M_{\phi_r}=\frac{1}{16\pi G}\frac{\mu\phi_{r}}{L}\;.\label{eq:ADMmass}\eeq
The conformal boundary is at $r\to\infty$, where $\phi_{r}>0$ is the boundary value of $\phi$ such that at a cutoff $r=\frac{1}{\epsilon}$ near the boundary ($\epsilon=0$), $\phi\to\frac{\phi_{r}}{\epsilon L}$. In the conventions of \cite{Moitra:2019xoj}, the boundary value is given by $\phi_{r}=1/(\mathcal{J} L )$, where $\mathcal{J}$ is an energy scale.\footnote{Note the dilaton $\phi$ and the constant $\phi_{r}$ are dimensionless.}
Further, at the horizon the dilaton is equal to $\phi_H = \sqrt{\mu} \phi_r.$ 
Throughout this article we will express the $\text{AdS}_{2}$ black hole in static  and Kruskal coordinates. We refer to Appendix \ref{app:ebhcoord} for the definitions of various $\text{AdS}_{2}$  coordinate systems and the   transformations relating them. In Figure \ref{fig:coords} we depict  the Penrose diagram  of an  AdS$_2$ black hole   to keep track of   the different coordinates.

Upon Wick rotating the Lorentzian time $t$ coordinate $t\to-i\tau$ and identifying $\tau\sim \tau+\beta$ to remove the conical singularity of the resulting Euclidean cigar, the eternal black hole (\ref{eq:ads2bhstat}) is in thermal equilibrium at a  temperature $T_{\text H}$  inversely proportional to the period $\beta$ of the Euclidean time circle,
\begin{equation} 
T_{\text H}=\frac{N'(r_{H})}{4\pi}=\frac{\sqrt{\mu}}{2\pi L}\;.\label{eq:Hawktemp}
\end{equation}
  We observe extremality, $T_{\text H}=0$, occurs in the massless limit $\mu=0$.

The associated ``Bekenstein-Hawking'' entropy of the black hole is well known and given by 
\beq S_{\text{BH}}=\frac{\phi_{0}}{4G}+\frac{\phi_H}{4G}=S_{\phi_0}+S_{\phi_r}\;,\label{eq:entbh1}\eeq
where $S_{\phi_0}=\frac{\phi_{0}}{4G}$ is the entropy of the extremal black hole  while $S_{\phi_r} =\frac{\phi_{r}\sqrt{\mu}}{4G}$ is the non-extremal contribution to the black hole entropy. Consequently, the thermodynamic variables $M_{\phi_r}$ (\ref{eq:ADMmass}), $T_{\text H}$ (\ref{eq:Hawktemp}), and $S_{\phi_r}$ (\ref{eq:entbh1}), obey a Smarr relation and corresponding first law \cite{Grumiller:2007ju}
\beq 2 M_{\phi_r}= T_{\text H} S_{\phi_r}\;,\qquad \delta M_{\phi_r}=T_{\text H} \delta S_{\phi_r}\;, \label{eq:classicaljtsmarrfirst}\eeq
where only $\mu$ is varied in the first law and $\phi_0$, $\phi_r$ and $G$ are kept fixed. In Eq. \eqref{eq:smarrJTv2}  we present a different form of the Smarr relation which does not have a factor of two on the left-hand side and which contains a new  term proportional to the cosmological constant. Further, in Eq. \eqref{eq:extendedfirstlawasentfirstlaw} we give the extended first law for the eternal black hole   allowing for variations of the   coupling constants. 

A chief goal of this article is to uncover the quantum corrected first law and Smarr relation. To do this we  discuss the semi-classical JT model, where we emphasize the importance of the choice of vacuum state.



  \subsection{Backreaction and vacuum  states}
 

One of the novel and exciting features of JT gravity, similar to the CGHS model \cite{Callan:1992rs}, is that 1-loop quantum effects may be incorporated and continue to yield fully analytic solutions, allowing for a complete study of backreaction. These semi-classical effects are entirely captured by a (bulk) non-local Polyakov action \cite{Polyakov:1981rd} and its associated Gibbons-Hawking-York boundary term, which take the following form in Lorentzian signature \cite{Almheiri:2014cka}: 
\beq I_\chi = I^{\text{Poly}}_{\chi}+I_{\chi}^{\text{GHY}}=-\frac{c}{24\pi}\int_M d^{2}x\sqrt{-g}\left[(\nabla\chi)^{2}+\chi R-\frac{\lambda}{L^{2}}\right]-\frac{c}{12\pi}\int_{B}dt\sqrt{-\gamma}\chi K\;.\label{eq:semiclassactionconts}\eeq
Here $\chi$ is a local auxiliary field such that the non-local Polyakov contribution appears local. 
The Polyakov term is motivated from the conformal anomaly associated with a classical CFT having central charge $c$.
The equation of motion for the auxiliary field $\chi$ is simply
\beq 2\Box\chi=R\;.\label{eq:chieom}\eeq
Upon substituting the solution for $\chi$ into the Polyakov action one recovers the standard non-local expression.\footnote{Specifically, substituting the formal solution $\chi =\frac{1}{2}\int d^{2}y\sqrt{-g(y)}G(x,y)R(y)$, where $G(x,y)$ is the Green's function for the D'Alembertian  $\Box$ operator, into the local form of the Polyakov action~(\ref{eq:semiclassactionconts}), and ignoring boundary terms,  one recovers the non-local  form,  $I_{\text{Poly}}=-\frac{c}{96 \pi}\int d^{2}x\sqrt{-g(x)}\int d^{2}y\sqrt{-g(y)}R(x)G(x,y)R(y)\;.$}
Lastly, note we have included a cosmological constant term $\lambda/L^{2}$ which is always allowed at the quantum level due to an ambiguity in the conformal anomaly \cite{Fabbri:2005mw}. In the majority of our results we follow the standard convention and set $\lambda=0$, however, in the following analysis we demonstrate a special choice of $\lambda$ removes all semi-classical corrections to the dilaton in the Hartle-Hawking vacuum state.

The $\chi$ field models $c$ massless scalar fields (reflecting a CFT of central charge $c$), which are taken to represent the Hawking radiation of an evaporating black hole   (\emph{e.g.}, \cite{Nayak:2018qej}). The classical limit thus corresponds to $c\to0$.\footnote{More precisely, here $\hbar$ has been set to unity. To recover factors of $\hbar$, one simply replaces $c\to c\hbar$.} The semi-classical approximation is only valid in the regime
\beq \label{eq:semicl}
\phi_{0}/G\gg\phi_H/G = \sqrt{\mu} \phi_r /G \gg c\gg1\;.
\eeq
 This follows from the following reasoning. Recall the classical JT model arises from a   spherical reduction of a near-extremal black hole. The extremal  black hole  entropy is  $S_{\phi_0}\sim\phi_{0}/G$, and  deviation from extremality is captured by the entropy $S_{\phi_r}\sim\phi_H/G$, such that near extremality implies $\phi_{0}\gg\phi_H$, with $\phi_{0}/G\gg 1$. The classical solution \eqref{eq:dilaton1} for the dilaton at the horizon is $\phi_H =(r_H /L) \phi_r =  \sqrt{\mu} \phi_r$. To keep curvatures small from the higher-dimensional perspective, one works with large black holes, \emph{i.e.}, $r_H \gg L$ or $\sqrt{\mu} \gg 1$. The semi-classical approximation is understood as quantizing $c$ conformal fields, leaving the background geometry classical. This is allowed since we are ignoring stringy-like ghost corrections to the dilaton or metric, provided $c\gg1$. However, for $c$ to amount to a \emph{correction} to the classical result, we impose $\phi_H /G\gg c$. Combining these approximations we find the regime of validity \eqref{eq:semicl}.

Combining the semi-classical action (\ref{eq:semiclassactionconts}) with the classical JT model (\ref{eq:JTact1}) augments the classical gravitational equation  of motion (\ref{eq:graveom}) with backreaction fully characterized by $\langle T^{\chi}_{\mu\nu}\rangle\equiv-\frac{2}{\sqrt{-g}}\frac{\delta I_{\chi}}{\delta g^{\mu\nu}}$, such that the semi-classical gravitational equation  of motion is
\beq T_{\mu\nu}^{\phi}+\langle T^{\chi}_{\mu\nu}\rangle=0\;,\label{eq:semiclassgraveom}\eeq
with 
\beq \langle T^{\chi}_{\mu\nu}\rangle=\frac{c}{12\pi}\left[(g_{\mu\nu}\Box-\nabla_{\mu}\nabla_{\nu})\chi+(\nabla_{\mu}\chi)(\nabla_{\nu}\chi)-\frac{1}{2}g_{\mu\nu}(\nabla\chi)^{2}+\frac{\lambda}{2L^{2}}g_{\mu\nu}\right]\;.\label{eq:stresstenchi}\eeq
Here $\langle T^{\chi}_{\mu\nu}\rangle$ denotes the stress tensor expectation value with respect to some unspecified quantum state $| \Psi \rangle$. In any coordinate frame $\langle T^{\chi}_{\mu\nu}\rangle$ yields a conformal anomaly:
\beq g^{\mu\nu}\langle T^{\chi}_{\mu\nu}\rangle=\frac{c}{24\pi}\left(R+\frac{2\lambda}{L^{2}}\right)\;,\eeq
where we used the equation of motion for $\chi$ (\ref{eq:chieom}). Using the on-shell relation $R=-\frac{2}{L^{2}}$, we note the conformal anomaly is eliminated by selecting $\lambda=1$, \emph{i.e.}, $g^{\mu\nu}\langle T^{\chi}_{\mu\nu} \rangle=0$. It is worth emphasizing, moreover, since $\chi$ does not directly couple to the dilaton $\phi$, the system continues to admit eternal black hole solutions of the same type as in the classical JT model, but with the full backreaction taken into account. This is not to say the dilaton will not receive semi-classical quantum corrections; indeed it will generically, however, the solution for $\phi$ will depend on the vacuum state and value of $\lambda$. As we will see momentarily, the choice $\lambda=1$ eliminates all semi-classical corrections to $\phi$ in the Hartle-Hawking state.

Below we will primarily work in conformal gauge, where any two-dimensional spacetime is conformally flat in some set of light cone coordinates $(y^{+},y^{-})$,
\beq ds^{2}=-e^{2\rho}dy^{+}dy^{-}\;,\label{eq:confgauge2}\eeq
with a conformal factor $e^{2\rho(y^{+},y^{-})}$. In conformal gauge,   taking into account backreaction (\ref{eq:stresstenchi}), the equations of motion for the dilaton $\phi$ (\ref{eq:dileom}), auxiliary field $\chi$ (\ref{eq:chieom}) and metric (\ref{eq:graveom}) become, respectively, 
\beq 4\partial_{+}\partial_{-}\rho+\frac{e^{2\rho}}{L^{2}}=0\;,\eeq
\beq \partial_{+}\partial_{-}(\chi+\rho)=0\;,\label{eq:chieomconfgauge}\eeq
\beq -\partial_{\pm}^{2}\phi+2(\partial_{\pm}\rho)(\partial_{\pm}\phi) =-e^{2\rho}\partial_{\pm}(e^{-2\rho}\partial_{\pm}\phi)=8\pi G\langle T^{\chi}_{\pm\pm}\rangle\;,\label{eq:graveomsjt2}\eeq
\beq 2\partial_{\pm}\partial_{\mp}\phi+\frac{1}{L^{2}}e^{2\rho}\phi=16\pi G\langle T^{\chi}_{\pm\mp}\rangle\;,\label{eq:graveomsjt1}\eeq
where we  employed the conformal gauge identities \eqref{eq:usefulid}, and the components of the stress tensor for $\chi$ (\ref{eq:stresstenchi}) are 
\beq \langle T^{\chi}_{\pm\mp}\rangle=\frac{c}{12\pi}\partial_{+}\partial_{-}\chi-\frac{c}{48\pi}\frac{\lambda}{L^{2}}e^{2\rho}\;,\label{eq:pmmptchi}\eeq
\beq \langle T^{\chi}_{\pm\pm}\rangle=\frac{c}{12\pi}[-\partial_{\pm}^{2}\chi+2(\partial_{\pm}\rho)(\partial_{\pm}\chi)+(\partial_{\pm}\chi)^{2}]\;.\label{eq:pmpmtchi}\eeq
Here $\langle T^{\chi}_{\pm\pm}\rangle$ is shorthand for $\langle \Psi|T^{\chi}_{y^{\pm}y^{\pm}}|\Psi\rangle$.
The equation of motion for $\chi$ (\ref{eq:chieomconfgauge}) is particularly easy to solve, leading to
\beq\label{eq:solconfgauge} 
    \chi
    =
    -\rho+\xi\;, 
    \quad 
    \text{with} \quad 
    \xi(y^{+},y^{-})
    =
    \xi_{+}(y^{+})
    +
    \xi_{-}(y^{-}) \;, 
\eeq
where $\xi$ solves the wave equation $\Box\xi=0=\partial_{+}\partial_{-}\xi$.
Substituting the $\chi$ solution into the components of the stress tensor (\ref{eq:pmmptchi}) and (\ref{eq:pmpmtchi}) leads to
\beq \langle T^{\chi}_{\pm\mp}\rangle=-\frac{c}{48\pi}\left(4\partial_{+}\partial_{-}\rho+\frac{\lambda}{L^{2}}e^{2\rho}\right)=\frac{c}{48\pi L^{2}}e^{2\rho}(1-\lambda)\;, \label{eq:tensorchi12} \eeq
\beq \langle T^{\chi}_{\pm\pm}\rangle=\frac{c}{12\pi}[\partial_{\pm}^{2}\rho
-(\partial_{\pm}\rho)^{2}+(\partial_{\pm}\xi)^{2}-\partial_{\pm}^{2}\xi]=-\frac{c}{12\pi}[(\partial_{\pm}\rho)^{2}-\partial^{2}_{\pm}\rho]-\frac{c}{12\pi}t_{\pm}(y^{\pm})\;, \label{eq:tensorchi2}\eeq
where we have defined 
\beq t_{+}(y^{+})\equiv\partial^{2}_{+}\xi_{+}-(\partial_{+}\xi_{+})^{2}\;,\quad  t_{-}(y^{-})\equiv\partial_{-}^{2}\xi_{-}-(\partial_{-}\xi_{-})^{2}\;.\label{eq:tpmfuncgen}\eeq
The functions $t_{\pm}(y^{\pm})$ characterize the  quantum  matter   state and are related to the expectation value of the \emph{normal-ordered} stress-tensor, as we will now review, including some additional new insights.\footnote{Equivalently, $t_{\pm}$ can be found from the conformal anomaly and the continuity equation $\nabla^{\mu}\langle T^{\chi}_{\mu\nu}\rangle=0$, where the $t_{\pm}$ arise as  ``constants" of integration \cite{Almheiri:2014cka}.}

\subsubsection{Normal-ordered stress tensors and vacuum states}

In what follows it will be useful to know how to transform the components of $\langle T^{\chi}_{\mu\nu}\rangle$ in various coordinate systems in conformal gauge. This will be crucial in our interpretation of a choice of vacuum state. First note that since $\rho$ is a component of the metric it transforms as such. Specifically, under a conformal transformation from $(y^{\pm})$ to another set of conformal coordinates $(x^{\pm})$ one has (see, \emph{e.g.}, Eq. (5.68) of \cite{Fabbri:2005mw})
\beq \rho(y^{+},y^{-})\to\rho(x^{+},x^{-})=\rho(y^{+},y^{-})+\frac{1}{2}\log\left(\frac{dy^{+}}{dx^{+}}\frac{dy^{-}}{dx^{-}}\right)\;.\label{eq:rhotranslaw}\eeq
Since $\chi=-\rho+\xi$ is a scalar field, but $\rho$ transforms as a tensor, one finds the functions $\xi_{\pm}$ must transform under a conformal transformation as
\beq \xi_{\pm}(y^{\pm})\to\xi_{\pm}(x^{\pm})=\xi_{\pm}(x^{\pm}(y^{\pm}))+\frac{1}{2}\log\frac{dy^{\pm}}{dx^{\pm}}\;.\label{eq:transformxi}\eeq
This implies   the  functions $t_{\pm}(y^{\pm})$ (\ref{eq:tpmfuncgen}) transform as
\beq t_{\pm}(x^{\pm})=\left(\frac{dy^{\pm}}{dx^{\pm}}\right)^{2}t_{\pm}(y^{\pm})+\frac{1}{2}\{y^{\pm},x^{\pm}\}\;,\label{eq:tpmtrans}\eeq
where $\{y^{\pm},x^{\pm}\}$ is the Schwarzian derivative of $y^{\pm}$ with respect to $x^{\pm}$:
\beq \{y^{\pm},x^{\pm}\}\equiv\frac{(y^{\pm})'''}{(y^{\pm})'}-\frac{3}{2}\left(\frac{(y^{\pm})''}{(y^{\pm})'}\right)^{2}\;, \qquad \text{with} \qquad (y^{\pm})'\equiv\frac{dy^{\pm}}{dx^{\pm}}\;.\eeq
 In fact, $t_{\pm}$ turns out to be proportional to the expectation value of the \emph{normal ordered} stress tensor \cite{Fabbri:2005mw}  
\beq \langle \Psi | :T^{\chi}_{\pm\pm}:| \Psi \rangle = -\frac{c}{12\pi}t_{\pm}(y^{\pm})\;,\label{eq:normordgen}\eeq
where we reiterate that $|\Psi\rangle$ is some unspecified quantum state.
By normal ordering we mean  
$
:T^{\chi}_{\pm\pm} (y^\pm):\;\equiv T_{\pm\pm}^{\chi}(y^\pm)-\langle0_{y}|T^{\chi}_{\pm\pm}(y^\pm)|0_{y}\rangle\;,
$
with $|0_{y}\rangle$   the vacuum state with respect to the $(y^{\pm})$ coordinate system, \emph{i.e.}, the vacuum state defined with respect to the positive frequency modes in coordinates $y^{\pm}$, ${a}_{y}|0_{y}\rangle=0$. 

Crucially, from the definition of normal ordering we have that \emph{the vacuum state $|0_{y}\rangle$ is the state such that the expectation value of the normal ordered energy-momentum tensor vanishes},
\beq\langle 0_y| :T_{\pm\pm}^{\chi}(y^{\pm}):|0_y \rangle=0\;,\quad \Leftrightarrow \quad t_{\pm}(y^{\pm})=0\;.\label{eq:gendefvacstate}\eeq
Moreover, from equation \eqref{eq:tpmtrans} we see the  normal-ordered stress tensor obeys an   anomalous transformation law under a conformal transformation  $y^{\pm}\to x^{\pm}(y^{\pm})$
\beq :T^{\chi}_{\pm\pm}(x^{\pm}):\;=\left(\frac{dy^{\pm}}{dx^{\pm}}\right)^{2}\,:T^{\chi}_{\pm\pm}(y^{\pm}):-\frac{c}{24\pi}\{y^{\pm},x^{\pm}\}\;.\label{eq:transnormord}\eeq
Notice that taking the expectation value with respect to $|0_y \rangle$ on both sides yields
\beq \langle 0_{y}|:T^{\chi}_{\pm\pm}(x^{\pm}):|0_{y}\rangle=-\frac{c}{24\pi}\{y^{\pm},x^{\pm}\}\;.\label{eq:changeinvacstnov2}\eeq 
The stress-energy tensor itself follows a standard coordinate transformation law
\beq \langle \Psi | T^{\chi}_{\pm\pm}(x^{\pm})|\Psi \rangle=\left(\frac{dy^{\pm}}{dx^{\pm}}\right)^{2}\langle \Psi | T^{\chi}_{\pm\pm}(y^{\pm})|\Psi\rangle\;,\label{eq:coordtranstmunu}\eeq
because the expectation value of the stress tensor is a rank 2 tensor. This transformation is consistent with the anomalous transformation law for the normal-ordered stress tensor, since combining \eqref{eq:tensorchi2} and  \eqref{eq:normordgen} with   \eqref{eq:transnormord}  yields \eqref{eq:coordtranstmunu} above (following a similar computation as in Appendix~A of \cite{Moitra:2019xoj}).  However, under a change of vacuum the stress tensor expectation value does transform in an anomalous way  
\beq \langle 0_{y}|T^{\chi}_{\pm\pm}(x^{\pm})|0_{y}\rangle=\langle 0_{x}|T^{\chi}_{\pm\pm}(x^{\pm})|0_{x}\rangle-\frac{c}{24\pi}\{y^{\pm},x^{\pm}\}\;,\label{eq:changeinvacst}\eeq
which follows from combining \eqref{eq:tensorchi2}, \eqref{eq:normordgen}, \eqref{eq:gendefvacstate} and \eqref{eq:changeinvacstnov2}.
 





As noted earlier, the metric and dilaton are unaffected by the introduction of the auxiliary field $\chi$ and the semi-classical cosmological constant term, and thus we still have an eternal black hole solution. However, as we have just seen, the stress-energy tensor expectation value takes different forms depending on the vacuum state one is in. The solutions for $\chi$ and $\phi$, found by solving the semi-classical equations of motion with $\langle T^{\chi}_{\mu\nu}\rangle$, are hence state dependent. 
Let us thus write down the components of $\langle T_{\mu\nu}^{\chi}\rangle$ for the black hole solution in (static)  advanced and retarded time coordinates $(v,u)$ (\ref{eq:ads2confnull}), and Kruskal coordinates $(V_{\text K},U_{\text K})$ (\ref{eq:scalekruskalcoord}) and find how the stress tensors transform under a change in coordinates. Below we summarize the results and  we leave the detailed calculations for the interested reader in Appendix \ref{subsecapp:stress}. 

First note the eternal black hole metric in   advanced and retarded time coordinates $( v,u)$ is given by
\beq ds^{2}=-e^{2\rho(v,u)}dudv\;,\quad e^{2\rho}=\frac{\mu}{\sinh^{2}[\frac{\sqrt{\mu}}{2L}(v-u)]}\;,\label{eq:staticcoorduv}\eeq
while the line element in Kruskal coordinates is\footnote{Here we  work with the scaled coordinates (\ref{eq:scalekruskalcoord}).} $(U_{\text K},U_{\text K})$, 
\beq ds^{2}=-e^{2\rho(V_{\text K},U_{\text K})}dU_{\text K}dV_{\text K}\;,\quad e^{2\rho}=\frac{4\mu}{(1+\frac{\mu}{L^{2}}U_{\text K}V_{\text K})^{2}}\;.\label{eq:Kruscoordsys}\eeq
The conformal factors $\rho$ are related in the two coordinate systems via the transformation rule (\ref{eq:rhotranslaw}), from which
\beq \rho(v,u)
=\rho(V_{\text K},U_{\text K})+\frac{\sqrt{\mu}}{2 L}(v-u)=
\rho(V_{\text K},U_{\text K})+\frac{1}{2}\log\left(-\frac{\mu}{L^{2}}U_{\text K}V_{\text K}\right)\;,\eeq
where we used $(x^{+},x^{-})=(v,u)$ and $(y^{+},y^{-})=(V_{\text K},U_{\text K})$.

Relation (\ref{eq:transformxi}) allows us to relate the function $\xi$ in the different coordinate systems
\beq 
\begin{split}
&\xi_{u}=\xi_{U_{\text K}}-\frac{\sqrt{\mu}}{2 L}u=\xi_{U_{\text K}}+\frac{1}{2}\log\left(-\frac{\sqrt{\mu}}{L}U_{\text K}\right)\;,\\
&\xi_{v}=\xi_{V_{\text K}}+\frac{\sqrt{\mu}}{2L}v=\xi_{V_{\text K}}+\frac{1}{2}\log\left(\frac{\sqrt{\mu}}{L}V_{\text K}\right)\;,
\end{split}
\label{eq:coordtransxiuv}\eeq
such that 
\beq \xi(v,u)=\xi(V_{\text K},U_{\text K})+\frac{1}{2}\log\left(-\frac{\mu}{L^{2}}U_{\text K}V_{\text K}\right)\;.\eeq
Consequently, the transformation rule for the normal-ordered stress tensors (\ref{eq:transnormord}) yields
\beq 
\begin{split}
:T^{\chi}_{vv}:&=4\pi^{2}T^{2}_{H} V_{\text K}^{2}:T^{\chi}_{V_{\text K}V_{\text K}}:+\frac{c\pi}{12}T^{2}_{H}\;,
\end{split}
\label{eq:normordTchiv}\eeq
\beq 
\begin{split}
:T^{\chi}_{uu}:&=4\pi^{2}T^{2}_{}U_{\text K}^{2}:T^{\chi}_{U_{\text K}U_{\text K}}:+\frac{c\pi}{12}T^{2}_{H}\;,
\end{split}
\label{eq:normordTchiu}\eeq
where we used $(\frac{\sqrt{\mu}}{L})^{2}=4\pi^{2}T_{\text H}^{2}$ with $T_{\text H}$ the Hawking temperature \eqref{eq:Hawktemp}. Upon taking the expectation value, notice for special states the stress tensor will be purely thermal. We will return to this point shortly.

Lastly, using the normal ordered relation (\ref{eq:normordgen})
together with the following relations between $t_{V_{\text K}}$ and $t_{v}$, and $t_{U_{\text K}}$ and $t_{u}$:
\beq t_{v}=\left(\frac{\sqrt{\mu}}{L}V_{\text K}\right)^{2}t_{V_{\text K}}-\frac{\mu}{4 L^{2}}\;,\quad t_{u}=\left(\frac{\sqrt{\mu}}{L}U_{\text K}\right)^{2}t_{U_{\text K}}-\frac{\mu}{4 L^{2}}\;,\eeq
we find the diagonal stress-energy tensor expectation values in static and Kruskal coordinates are, respectively,
\beq \langle \Psi| T^{\chi}_{uu}|\Psi\rangle=-\frac{c\mu}{12\pi L^{2}}U_{\text K}^{2}t_{U_{\text K}}\;,\quad \langle \Psi|T^{\chi}_{vv}|\Psi\rangle=-\frac{c\mu}{12\pi L^{2}}V_{\text K}^{2}t_{V_{\text K}}\;,\eeq
\beq \langle\Psi| T^{\chi}_{U_{\text K}U_{\text K}}|\Psi\rangle=-\frac{c}{48\pi}\frac{1}{U_{\text K}^{2}}-\frac{c L^{2}}{12\pi\mu}\frac{t_{u}}{U_{\text K}^{2}}\;,\quad \langle \Psi| T^{\chi}_{V_{\text K}V_{\text K}}|\Psi\rangle=-\frac{c}{48\pi}\frac{1}{V_{\text K}^{2}}-\frac{c L^{2}}{12\pi\mu}\frac{t_{v}}{V_{\text K}^{2}}\;.\eeq
Observe that, depending on the coordinate frame, $\langle \Psi|T^{\chi}_{\mu\nu}|\Psi\rangle$ may diverge in the limit one approaches the horizon ($u\to\infty, v\to-\infty$ or $U_{\text K}=V_{\text K}=0$). This feature is in fact dependent on the choice of vacuum state as we will now describe. 


\subsubsection*{Choice of vacuum: Boulware \emph{vs.} Hartle-Hawking}

Let us now specify the vacuum states considered in this paper.  Famously there are two vacuum states of interest for $\text{AdS}_{2}$ black holes \cite{Spradlin:1999bn}: (i) the \emph{Boulware state} $|\text{B}\rangle$ \cite{Boulware:1974dm} and  (ii) the \emph{Hartle-Hawking state} $|\text{HH}\rangle$ \cite{Hartle:1976tp}. The Boulware vacuum is the state for the eternal black hole defined by positive frequency modes in static coordinates $(v,u)$.  Equivalently, as in  \eqref{eq:gendefvacstate},  the expectation value of the $uu$ and $vv$ components of the normal-ordered stress tensor vanishes in the Boulware state
\beq \langle \text{B}|:T^{\chi}_{uu}:|\text{B}\rangle=\langle \text{B}|:T^{\chi}_{vv}:|\text{B}\rangle=0\;\;\Leftrightarrow \;\;t_{u}=t_{v}=0\;.\label{eq:boulvacdef}\eeq
Consequently, from (\ref{eq:normordTchiv}) and (\ref{eq:normordTchiu}), the expectation value of the $V_{\text K}V_{\text K}$ and $U_{\text K}U_{\text K}$ components of the normal-ordered stress tensor are
\beq \langle \text{B}|:T^{\chi}_{V_{\text K}V_{\text K}}:|\text{B}\rangle=-\frac{c}{48\pi}\frac{1}{V_{\text K}^{2}}\;,\quad \langle \text{B}|:T^{\chi}_{U_{\text K}U_{\text K}}:|\text{B}\rangle=-\frac{c}{48\pi}\frac{1}{U_{\text K}^{2}}\;.\eeq
Equivalently, 
\beq t_{U_{\text K}}=\frac{1}{4U_{\text K}^{2}}\;,\quad t_{V_{\text K}}=\frac{1}{4V_{\text K}^{2}}\;.\eeq
Moreover, the expectation value of the stress tensor becomes
\beq 
\begin{split}
&\langle \text{B}|T_{uu}^{\chi}|\text{B}\rangle=\langle \text{B}|T^{\chi}_{vv}|\text{B}\rangle=-\frac{c\mu}{48\pi L^{2}}\;,\\
&\langle \text{B}|T^{\chi}_{U_{\text K}U_{\text K}}|\text{B}\rangle=-\frac{c}{48\pi U_{\text K}^{2}}\;,\quad \langle \text{B}|T^{\chi}_{V_{\text K}V_{\text K}}|\text{B}\rangle=-\frac{c}{48\pi V_{\text K}^{2}}\;.
\end{split}
\label{eq:expvalTuuB}\eeq
Thus, in Kruskal coordinates, the renormalized stress tensor, normal ordered or not, diverges at the horizon, as is common for the Boulware state. The expectation value of the stress tensor in static conformal coordinates, moreover, gives rise to a negative `energy' and is interpreted as a Casimir energy in the presence of a boundary at the black hole horizon \cite{Spradlin:1999bn}. 

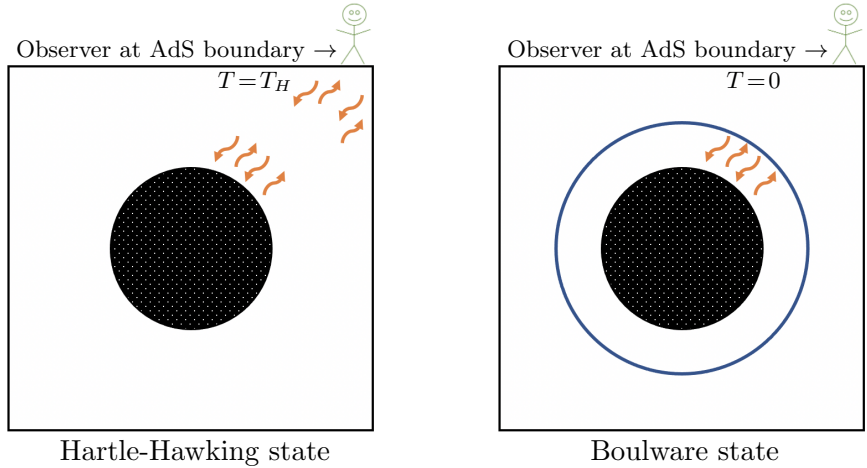
\begin{figure}[t]
\begin{center}
		\begin{overpic}[width=0.75\textwidth]{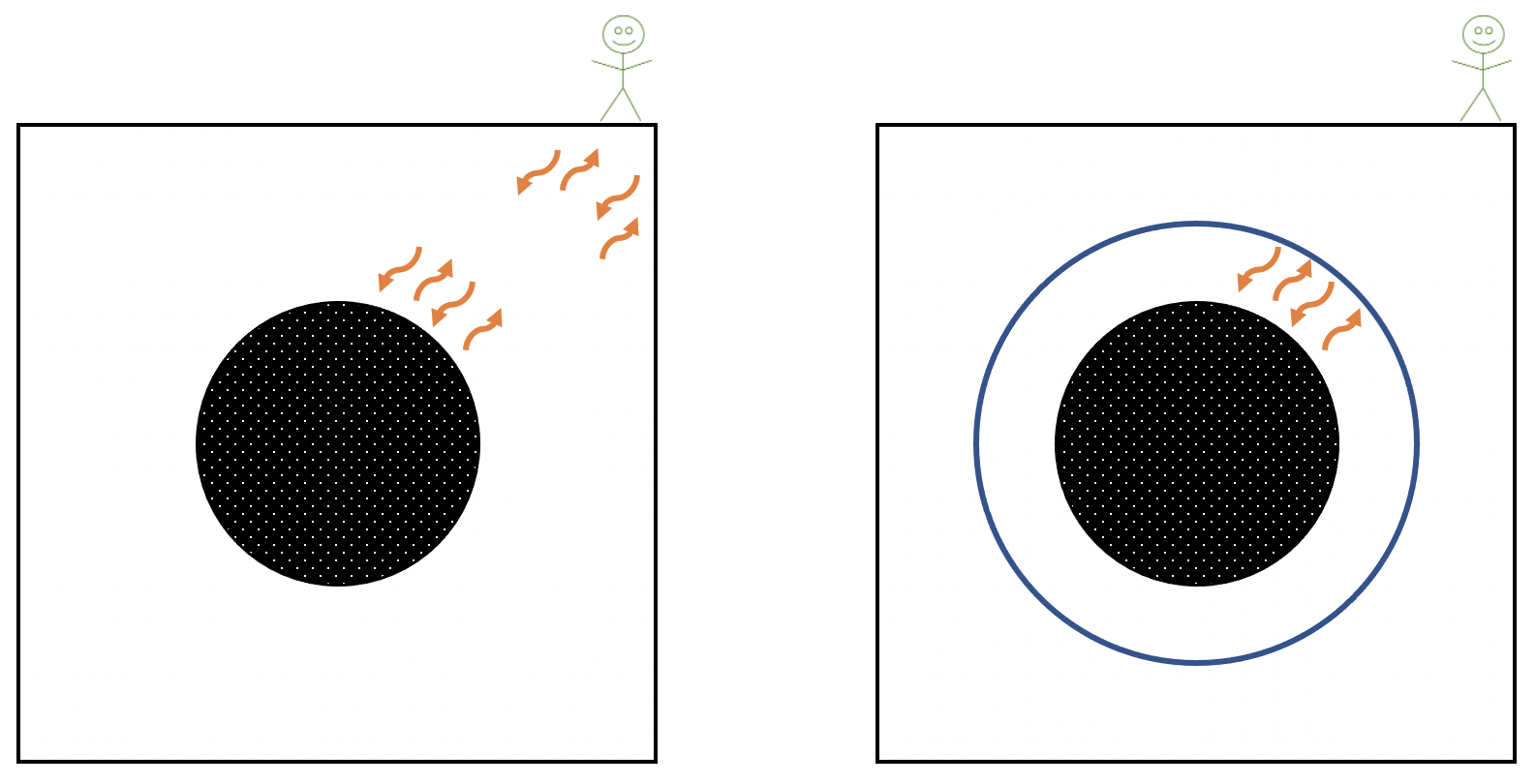}
		    \put (2,44) {\footnotesize{Observer at AdS boundary $\rightarrow$}}
			\put (25,40.5) {\footnotesize{$T\!=\!T_{H}$}}
			\put (7,-2) {Hartle-Hawking state}
		    \put (58,44) {\footnotesize{Observer at AdS boundary $\rightarrow$}}
			\put (83,40.5) {\footnotesize{$T\!=\!0$}}
			\put (67.5,-2) {Boulware state}
		\end{overpic}
		
	\end{center}
\centering
\caption{The black circles denote a black hole and the wavy arrows represent Hawking radiation. In the Hartle-Hawking state (left Figure) a stationary observer at the AdS boundary measures $T=T_{H}$, whereas a stationary observer at the boundary measures $T=0$ in the Boulware state (right Figure). In the Hartle-Hawking state the black hole is in thermal equilibrium with all of its surroundings whereas the Boulware state can be interpreted as the black hole being in thermal equilibrium only with a membrane wrapped tight around the event horizon.}
\label{fig:starrynight}
\end{figure}

Alternatively, the \emph{Hartle-Hawking} vacuum state $|\text{HH}\rangle$ is defined by positive frequency modes in Kruskal coordinates $(V_{\text K},U_{\text K})$, such that the expectation value of the $U_{\text K}U_{\text K}$ and $V_{\text K}V_{\text K}$ components of the normal-ordered stress tensor vanishes:
\beq \langle \text{HH}|:T^{\chi}_{U_{\text K}U_{\text K}}:|\text{HH}\rangle=\langle \text{HH}|:T^{\chi}_{V_{\text K}V_{\text K}}:|\text{HH}\rangle=0\;\;\Leftrightarrow\;\; t_{U_{\text K}}=t_{V_{\text K}}=0\;.\label{eq:defHHvac}\eeq
Correspondingly, we have
\beq t_{u}=t_{v}=-\frac{\mu}{4L^{2}}\;.\eeq
The expectation value of $uu$ and $vv$ components of the normal-ordered stress-tensor then becomes
\beq \langle \text{HH}|:T^{\chi}_{uu}|:\text{HH}\rangle=\langle \text{HH}:|T^{\chi}_{vv}:|\text{HH}\rangle=\frac{c\pi}{12}T_{\text H}^{2}\;.\label{eq:expvalHHuuvv}\eeq
Lastly, the expectation value of the diagonal components of the stress-tensor in $(u,v)$ and $(U,V)$ coordinates vanishes
\beq \langle \text{HH}|T^{\chi}_{U_{\text K}U_{\text K}}|\text{HH}\rangle=\langle \text{HH}|T^{\chi}_{V_{\text K}V_{\text K}}|\text{HH}\rangle=\langle \text{HH}|T^{\chi}_{uu}|\text{HH}\rangle=\langle \text{HH}|T^{\chi}_{vv}|\text{HH}\rangle=0\;.\eeq
Thus, with respect to a static observer in $(v,u)$ coordinates, the  Hartle-Hawking vacuum state is interpreted as the state with the energy density of a thermal bath of particles at Hawking temperature $T_{\text H}$. This is completely analogous to an accelerating (Rindler) observer seeing the Minkowski vacuum as being populated with particles in thermal equilibrium \cite{Fulling:1972md,Davies:1974th,Unruh:1976db}; indeed, the $\text{AdS}_{2}$ eternal black hole   is simply AdS$_2$-Rindler space. Therefore, the Hartle-Hawking state allows us to identify the black hole as a thermal system with a temperature, thermodynamic energy and entropy, unlike the Boulware vacuum. Nonetheless, below we will solve the semi-classical JT equations of motion with respect to both vacuum states. In Figure~\ref{fig:starrynight} we provide an overview of the physical picture we advocate.

As a final remark, note that had we performed the same analysis in   Poincar\'e coordinates \eqref{eq:poincconfcoord} $(V,U)$, we would have found the Poincar\'e vacuum state, \emph{i.e.}, the state defined with respect to  positive frequency modes in Poincar\'e coordinates, is equivalent to the Hartle-Hawking state $|\text{HH}\rangle$ \cite{Spradlin:1999bn}, a feature that persists in higher dimensions \cite{Danielsson:1998wt}.

\subsubsection{Backreacted solutions}

We now have all the necessary ingredients to solve for the dilaton $\phi$ and the auxiliary field $\chi$ in the semi-classical JT model, taking into account the   backreaction of Hawking radiation. We will look to solve the dilaton equation in static coordinates $(t,r_{\ast})$, where we assume $\phi$ is static, \emph{i.e.}, $\phi=\phi(r_{\ast})$, however we will allow $\chi$ to be potentially time dependent. As we will see below, these assumptions are self-consistent, \emph{i.e.}, they lead to valid and physically sensible solutions to the equations of motion. Moreover, the time dependence of $\chi$ will be crucial in the next section; in particular, it will allow us to interpret the Wald entropy as the (time-dependent) generalized entropy.

Using  $t=\frac{1}{2}(v+u)$ and the tortoise coordinate  $r_{\ast}=\frac{1}{2}(v-u)$, and the relations
\beq \partial_{v}=\frac{1}{2}(\partial_{t}+\partial_{r_{\ast}})\;,\;\; \partial_{u}=\frac{1}{2}(\partial_{t}-\partial_{r_{\ast}})\;,\;\; \partial_{u}\partial_{v}=\frac{1}{4}(\partial^{2}_{t}-\partial^{2}_{r_{\ast}})\;,\eeq
 the gravitational equation (\ref{eq:graveomsjt1}) becomes:
\beq 
\begin{split}
&2\partial_{u}\partial_{v}\phi+\frac{e^{2\rho}\phi}{L^{2}}=16\pi G\langle T^{\chi}_{uv}\rangle
~\Rightarrow~\partial_{r_{\ast}}^{2}\phi=\frac{2\mu}{L^{2}}\frac{1}{\sinh^{2}\left(\frac{\sqrt{\mu}}{L}r_{\ast}\right)}\left[\phi+(\lambda-1)\frac{cG}{3}\right]\;,
\end{split}
\label{eq:firstgraveomuv}\eeq
where we assumed $\partial_{t}\phi=0$. Meanwhile, the two equations in (\ref{eq:graveomsjt2})
become
\beq \partial_{r_{\ast}}\left[\sinh^{2}\left(\frac{\sqrt{\mu}}{L}r_{\ast}\right) \partial_{r_{\ast}}\phi \right]= \frac{8 c G}{3}\left(\frac{ \mu}{4 L^{2}}+ t_{v}(v)\right)\sinh^{2}\left(\frac{\sqrt{\mu}}{L }r_{\ast}\right)\;,\label{eq:scaleomsjt1}\eeq
\beq \partial_{r_{\ast}}\left[\sinh^{2}\left(\frac{\sqrt{\mu}}{L}r_{\ast}\right) \partial_{r_{\ast}}\phi \right]=\frac{8 cG}{3}\left(\frac{ \mu}{4 L^{2}}+ t_{u}(u)\right)\sinh^{2}\left(\frac{\sqrt{\mu}}{L}r_{\ast}\right)\;.\label{eq:scaleomsjt2}\eeq
In order for the left-hand side to be solely a function of $r_{\ast}$, and since $t_{v}(v)$ ($t_{u}(u)$) is solely a function of $v$ ($u$), we see that $t_{u}(u)\equiv \tau=t_{v}(v)$  must be a constant \cite{Almheiri:2014cka}. 

For $t_{v}=t_{u}=\tau$, the most general solution to the system of differential equations is
\beq \phi=-\phi_{r}\sqrt{\mu}\text{coth}\left(\frac{\sqrt{\mu}}{L}r_{\ast}\right)+\frac{Gc}{3}\left(1+\frac{4\tau L^{2}}{\mu}\right)\left(1+\frac{\sqrt{\mu}}{L}r_{\ast}\right)\text{coth}\left(\frac{\sqrt{\mu}r_{\ast}}{L}\right)-\frac{Gc}{3}\left(\lambda+\frac{4\tau L^{2}}{\mu}\right)\;.\label{eq:dilgensol}\eeq
In the limit $c\to0$ we recover the classical solution for the dilaton.

To proceed, we must specify the choice of vacuum state, thereby fixing the value of $\tau$. We thus evaluate the general solution (\ref{eq:dilgensol}), as well as the possible solutions for the state dependent $\chi$ in the Boulware and Hartle-Hawking vacuum states separately. 


\subsection*{Boulware vacuum}

The Boulware vacuum $|\text{B}\rangle$ is defined by (\ref{eq:boulvacdef}), $t_{u}=t_{v}=\tau=0$,
such that the dilaton (\ref{eq:dilgensol}) simplifies to \cite{Cadoni:1994uf,Fabbri:2005mw}
\beq \phi=-\phi_{r}\sqrt{\mu}\text{coth}\left(\frac{\sqrt{\mu}}{L}r_{\ast}\right)+\frac{Gc}{3}\left(1+\frac{\sqrt{\mu}}{L}r_{\ast}\right)\text{coth}\left(\frac{\sqrt{\mu}r_{\ast}}{L}\right)-\frac{\lambda Gc}{3}\;.\label{eq:dilsolbol}\eeq
We see $\phi$   diverges as $\phi(r_{\ast})\sim r_{\ast}$ as one approaches the horizon, $r_{\ast}\to-\infty$.

Meanwhile, from   (\ref{eq:tpmfuncgen}) we have two possible solutions for $\xi_{u}$ and $\xi_{v}$:
\begin{equation}\begin{aligned}
\xi_{u}^{(1)}&=c_{u}\,,&
 \xi_{v}^{(1)}&=c_{v}\,,\\
\xi^{(2)}_{u}&=c'_{u}-\log\left(\pm\frac{\sqrt{\mu}}{L}u+K_{u}\right)\,,&
\xi^{(2)}_{v}&=c'_{v}-\log\left(\pm\frac{\sqrt{\mu}}{L}v+K_{v}\right)\,,
\end{aligned}
\label{eq:solnxibouluv}
\end{equation}
where $c_{u},c_{v}, c_{u}',c_{v}',K_{u}$, and $K_{v}$ are constants, to be fixed by physical boundary conditions. We will see momentarily these constants can be arranged in such a way that the Wald entropy captures the full generalized entropy. Note that $\xi(v,u)=\xi_{u}+\xi_{v}$ only has one time-independent solution, namely, $\xi^{(1)}=\xi^{(1)}_{u}+\xi^{(1)}_{v}$.  Using \eqref{eq:solconfgauge} $\chi=-\rho+\xi$, and the fact, as a component of the metric, $\rho$ is always time independent, we find  a time-independent solution $\chi^{(1)}$ upon substitution of $\xi^{(1)}$ 
\beq
\chi^{(1)}=\frac{1}{2}\log\left[\frac{1}{\mu}\sinh^{2}\left(\frac{\sqrt{\mu}r_{\ast}}{L}\right)\right]+C\;,
\label{eq:chi1bol}\eeq
where $C \equiv c_{u}+c_{v}$. A time-dependent solution for $\chi$ also exists, via substitution of the choice $\xi^{(2)}=\xi^{(2)}_{u}+\xi^{(2)}_{v}$, namely,
\beq 
\begin{split}
\chi^{(2)}&=\frac{1}{2}\log\left[\frac{1}{\mu}\sinh^{2}\left(\frac{\sqrt{\mu}r_{\ast}}{L}\right)\right]+C'-\log\left(\pm\frac{\sqrt{\mu}}{L}u+K_{u}\right)-\log\left(\pm\frac{\sqrt{\mu}}{L}v+K_{v}\right)\;,
\end{split}
\label{eq:chitimedepB}\eeq
with $C'\equiv c_{u}'+c_{v}'$. This solution will prove useful when we compare the Wald entropy to the von Neumann entropy.

From transformations (\ref{eq:coordtransxiuv}), we can express the two solutions for $\xi$ in Kruskal coordinates.  Considering only the time-indepenent solution $\xi^{(1)}$  yields
\beq \xi^{(1)}_{U_{\text K}}=c_{u}-\frac{1}{2}\log\left(-\frac{\sqrt{\mu}}{L}U_{\text K}\right)\;,\quad \xi^{(1)}_{V_{\text K}}=c_{v}-\frac{1}{2}\log\left(\frac{\sqrt{\mu}}{L}V_{\text K}\right)\;.\eeq
The corresponding $\chi$ field is then
\beq 
    \chi^{(1)}
    =
    -\frac{1}{2}\log
    \left[
        \frac{4\mu}{
        \left(
            1
            +
            \frac{\mu U_{\text K}V_{\text K}}{
            L^{2}}
        \right)^{2}}
    \right]
    -
    \frac{1}{2}\log
    \left(
        -\frac{\mu}{L^{2}}
        U_{\text K}V_{\text K}
    \right)+C\;.
\eeq
To summarize, the eternal $\text{AdS}_{2}$ black hole, characterized by $\rho$, remains a solution of the semi-classical equations of motion, for which, with respect to the Boulware vacuum, the full backreaction is encoded in the dilaton $\phi$ (\ref{eq:dilsolbol}) and the auxiliary field $\chi$ (\ref{eq:chi1bol}) or \eqref{eq:chitimedepB}. Notice $\phi$ is the only field explicitly modified from its classical counterpart, such that in the classical limit $c\to0$, where physically we can imagine turning off the influence of $c$~background conformal fields, we recover the classical JT result. The semi-classical effect, moreover, is non-trivial as in the limit we approach the horizon the solution diverges.


\subsection*{Hartle-Hawking vacuum}

The Hartle-Hawking vacuum $|\text{HH}\rangle$ is defined in (\ref{eq:defHHvac}), \emph{i.e.}, $t_{U_{\text K}}=t_{V_{\text K}}=0$  or  $t_{u}=t_{v}=-\frac{\mu}{4 L^{2}}$.
In this case the dilaton (\ref{eq:dilgensol}) is only shifted by a constant with respect to the classical solution \cite{Fabbri:2005mw,Almheiri:2014cka}
\beq \phi(r_{\ast})=-\phi_{r}\sqrt{\mu}\text{coth}\left(\frac{\sqrt{\mu}}{L}r_{\ast}\right)+(1-\lambda)\frac{Gc}{3}=\frac{r\phi_{r}}{L}+(1-\lambda)\frac{Gc}{3}\;,\label{eq:dilasolnHH}\eeq
where we recall $r$ is the standard radial Schwarzschild coordinate. Thus, in the Hartle-Hawking vacuum the dilaton is trivially modified by backreaction. Moreover, selecting the (Polyakov) cosmological constant to be $\lambda=1$ entirely eliminates the semi-classical modification to $\phi$  \cite{Fabbri:2005mw}. Almheiri and Polchinski work with the $\lambda=0$ solution   \cite{Almheiri:2014cka}. Note that they  consider only the Hartle-Hawking state  and discard the Boulware state.

In the Hartle-Hawking vacuum there are three solutions for $\xi$:
\begin{align}
\xi^{(3)}_{u}&=c_{u}+\frac{\sqrt{\mu}}{2L}u\;,&\quad \xi^{(3)}_{v}&=c_{v}-\frac{\sqrt{\mu}}{2L}v\;, \nonumber\\
\xi^{(4)}_{u}&=c_{u}-\frac{\sqrt{\mu}}{2L}u\;,&\quad \xi^{(4)}_{v}&=c_{v}+\frac{\sqrt{\mu}}{2L}v\;,\label{eq:xisolnsHHuv}\\
\xi^{(5)}_{u}&=c'_{u}-\log\left[2\cosh\left(\mp\frac{\sqrt{\mu}}{2L}u+K_{u}\right)\right]\;,&\quad \xi_{v}^{(5)}&=c'_{v}-\log\left[2\cosh\left(\pm\frac{\sqrt{\mu}}{2L}v+K_{v}\right)\right]\;, \nonumber 
\end{align}
where there is an inherent sign ambiguity on the linear term differing the $\xi^{(3)}$ solutions from the $\xi^{(4)}$ solutions.\footnote{Here we denote $\xi^{(3)}=\xi^{(3)}_{u}+\xi^{(3)}_{v}$. Technically, mixed solutions  such as $\xi = \xi^{(3)}_{u} + \xi^{(4)}_{v}$  are also allowed, but we will not consider them in this paper.} Note that  both $\xi^{(3)}$ and $\xi^{(4)}$ solutions lead to a time-independent $\xi$, while the solution $\xi^{(5)}$ is generically time dependent. Furthermore, in the horizon limit  the   solution $\xi^{(5)}$ diverges in the same way as   $\xi^{(3)}$ and $\xi^{(4)}$, 
\beq \lim_{r_{\ast}\to-\infty}\xi_{u}^{(5)}=c'_{u}-K_{u}\mp\frac{\sqrt{\mu}}{2L}u+...\;,\quad \lim_{r_{\ast}\to-\infty}\xi_{v}^{(5)}=c'_{v}-K_{v}\pm\frac{\sqrt{\mu}}{2L}v+...\;,\eeq
where we identified $c_{u}=c'_{u}-K_{u}$ and $c_{v}=c'_{v}-K_{v}$.  

Using transformation (\ref{eq:coordtransxiuv}), the above three solutions read in Kruskal coordinates, respectively,
\beq
\begin{aligned}
\xi^{(3)}_{U_{\text K}}
&=c_{u}-\log\left(-\frac{\sqrt{\mu}}{L}U_{\text K}\right)\;,&\quad \xi^{(3)}_{V_{\text K}}
&=c_{v}-\log\left(\frac{\sqrt{\mu}}{L}V_{\text K}\right)\;,\\
\xi^{(4)}_{U_{\text K}}&=c_{u}\;,&\quad \xi^{(4)}_{V_{\text K}}&=c_{v}\;,\\
\xi^{(5)}_{U_{\text K}}&=k_{U_{\text K}}-\log\left[-\frac{\sqrt{\mu}}{L}U_{\text K}+K_{U_{\text K}}\right]\;,&\quad \xi^{(5)}_{V_{\text K}}&=k_{V_{\text K}}-\log\left[\frac{\sqrt{\mu}}{L}V_{\text K}+K_{V_{\text K}}\right]\;,
\end{aligned}
\eeq
where we combined $c'_{u}\mp K_{u}\equiv k_{U_{\text K}}$, $c'_{v}\mp K_{v}\equiv k_{V_{\text K}}$, and identified $e^{\mp 2K_{u}}\equiv K_{U_{\text K}}$ and $e^{\mp 2K_{v}}\equiv K_{V_{\text K}}$.   When $K_{U_{\text K}}=K_{V_{\text K}}=0$, the time-dependent solution $\xi^{(5)}$ reduces to the third, time-independent solution, while for $K_{U_{\text K}}=K_{V_{\text K}}=1$ we have $\xi^{(5)} \to\xi^{(4)}$ in the horizon limit  $U_{\text{K}}, V_{\text{K}}\to0$. 

With the $\xi$ solutions  in hand, we find three associated $\chi=-\rho+\xi$ fields, namely,
\beq 
\chi^{(3)} =\frac{1}{2}\log\left[\frac{1}{\mu}\sinh^{2}\left(\frac{\sqrt{\mu}r_{\ast}}{L}\right)\right]-\frac{\sqrt{\mu}r_{\ast}}{L}+C =-\frac{1}{2}\log\left[\frac{4\mu}{\left(1+\frac{\mu U_{\text K}V_{\text K}}{L^{2}}\right)^{2}}\right] - \log \left ( -\frac{\mu}{L^2}U_{\text{K}}V_{\text{K}}\right)+C\;,
\eeq
\beq 
\begin{split}
\chi^{(4)}&=\frac{1}{2}\log\left[\frac{1}{\mu}\sinh^{2}\left(\frac{\sqrt{\mu}r_{\ast}}{L}\right)\right]+\frac{\sqrt{\mu}r_{\ast}}{L}+C=-\frac{1}{2}\log\left[\frac{4\mu}{\left(1+\frac{\mu U_{\text K}V_{\text K}}{L^{2}}\right)^{2}}\right] +C\;,
\end{split}
\label{eq:chi4sol}
\eeq
\begin{align}
\chi^{(5)}&=\frac{1}{2}\log\left[\frac{1}{\mu}\sinh^{2}\left(\frac{\sqrt{\mu}r_{\ast}}{L}\right)\right]-\frac{\sqrt{\mu}r_{\ast}}{L}-\log\left[\left(1+K_{U_{\text K}}e^{\sqrt{\mu} u/L}\right)\left(1+K_{V_{\text K}}e^{-\sqrt{\mu}v/L}\right)\right]+k\;, \nonumber \\
&=-\frac{1}{2}\log\left[\frac{4\mu}{\left(1+\frac{\mu U_{\text K}V_{\text K}}{L^{2}}\right)^{2}}\right]-\log\left[\left(-\frac{\sqrt{\mu}U_{\text K}}{L}+K_{U_{\text K}}\right)\left(\frac{\sqrt{\mu}V_{\text K}}{L}+K_{V_{\text K}}\right)\right]+k\;, \label{eq:chi5}
\end{align}
where in the fifth solution we defined $k\equiv k_{U_{\text K}}+k_{V_{\text K}}$. Observe that for $K_{U_{\text K}}=K_{V_{\text K}}=1$ the solutions $\chi^{(4)}$ and $\chi^{(5)}$ approach the same constant on the horizon, while for $K_{U_{\text K}}=K_{V_{\text K}}=0$ the field $\chi^{(5)}$ reduces to the static solution $\chi^{(3)}$, which blows up at the black hole horizon. In \cite{Almheiri:2014cka} Almheiri and Polchinski took only solution $\chi^{(4)}$     into account, because they required $\chi$ to be constant on the horizon, but they neglected the fifth solution which is also constant on the horizon for some   $K_{U_{\text{K}}}$ and $K_{V_{\text K}}$. 
Thus, different choices of integration constants $K_{U_{\text K}},K_{V_{\text K}}$ lead to potentially dramatically different physics. As we show below, there exists yet another choice of integration constants for the fifth solution, such that the Wald entropy is equal to the complete generalized entropy, including time dependence.

\section{Wald entropy is generalized entropy} \label{sec:Waldentgenent}

Having exactly solved the problem of backreaction in an eternal $\text{AdS}_{2}$ black hole background, here we begin to study semi-classical corrections to the thermodynamics, starting with the entropy. Semi-classical corrections to the \emph{black hole} entropy and associated first law of thermodynamics were accounted for in \cite{Almheiri:2014cka,Moitra:2019xoj}, where it was shown that the generalized entropy appears in the first law \cite{Almheiri:2014cka} and obeys a generalized second law \cite{Moitra:2019xoj}.  In contrast, here we consider the entropy of \emph{quantum extremal surfaces} (QES) \cite{Engelhardt:2014gca}, defined as the surface extremizing the generalized entropy, which is generally not the black hole horizon. Moreover, we demonstrate the Wald entropy of the semi-classical JT model entirely captures the generalized entropy, and, for certain physical scenarios, includes time dependence. We won't dwell on the time dependence, as we are primarily interested in systems in thermal equilibrium (such that we focus about a point of time-reflection symmetry), however, we believe this observation may be useful in studying the black hole information problem, specifically the dynamics of entanglement wedge  islands. Indeed, a time-dependent von Neumann entropy only makes sense when a bath is attached at the boundary where the radiation is deposited. We will have more to say about this connection in Section \ref{sec:conc}. 

There are two points worth emphasizing here. First, the eternal $\text{AdS}_{2}$ black hole is an AdS-Rindler wedge whose thermodynamics has been extensively studied  before in higher dimensions $D>2$. In particular, the Bekenstein-Hawking entropy of  higher-dimensional AdS-Rindler space, a.k.a. the massless ``topological" black hole, may be identified with the entanglement entropy of a holographic CFT in the vacuum reduced to a boundary subregion of the pure AdS spacetime \cite{Casini:2011kv}. As we will show, the quantum extremal surface  lies outside of the eternal black hole horizon, and is itself the bifurcation point of the Killing horizon of  a nested Rindler wedge; thus the QES also defines a topological black hole. Second, in   $D>2$  the thermodynamics of a QES corresponds to (i) semi-classical ``entanglement" thermodynamics when the region of interest is a subset of the full AdS boundary, or (ii) semi-classical black hole thermodynamics, when the region of interest is the full boundary. In $D=2$  there is no   distinction between (i) and (ii), since there are no subregions to the $(0+1)$-dimensional boundary, so this distinction is not relevant for the present article. 


 \subsection{Wald entropy}
 
 For diffeomorphism invariant gravity theories other than general relativity, the entropy for stationary black holes is  quantified by the Wald entropy functional $S_{\text{Wald}}$, or equivalently the Noether charge associated with the Killing vector field generating the bifurcate horizon of the black hole   normalized to have unit surface gravity $\kappa=1$  \cite{Wald:1993nt,Iyer:1994ys},
\beq S_{\text{Wald}}=-2\pi\int_{H}dA\frac{\partial \mathcal{L}}{\partial R^{\mu\nu\rho\sigma}}\epsilon_{\mu\nu}\epsilon_{\rho\sigma}\;,\eeq
where $\epsilon_{\mu\nu}$ is the binormal satisfying $\epsilon_{\mu\nu}\epsilon^{\mu\nu}=-2$, $dA$ the infinitesimal area element of the codimension-2 Killing horizon $H$, and $\mathcal{L}$ is the Lagrangian density defining the theory. For two-dimensional theories  the integral is replaced with evaluating the integrand at the horizon.  The Wald entropy can be associated to the AdS$_2$ black hole horizon $H$, since this is a bifurcate Killing horizon. In fact, any point outside the black hole horizon in two dimensions  may be viewed as a bifurcation point of the Killing horizon of a nested Rindler wedge, and hence the Wald entropy may also be evaluated on these points, as we will do later on.\footnote{For extremal surfaces   in higher dimensions, which are not bifurcation surfaces of Killing horizons, the holographic entanglement entropy functional   is not equal to  the Wald entropy for generic higher curvature theories of gravity, but is modified with extrinsic curvature terms  \cite{Hung:2011xb,Dong:2013qoa,Camps:2013zua}.}


Black hole entropy generically receives quantum corrections when backreaction is taken into account. In the present case the semi-classical effects are encoded in the dilaton $\phi$ and auxiliary field $\chi$. Using
\beq \frac{\partial \mathcal{L}}{\partial R^{\mu\nu\rho\sigma}}=\frac{\partial(\mathcal{L}_{\text{JT}}+\mathcal{L}_{\chi})}{\partial R^{\mu\nu\rho\sigma}}=\left(\frac{\phi_{0}+\phi}{32\pi G}-\frac{c}{48\pi}\chi\right)(g^{\mu\rho}g^{\nu\sigma}-g^{\mu\sigma}g^{\nu\rho})\;,\eeq
the Wald entropy for the semi-classical JT model   is
\beq S_{\text{Wald}}=\frac{1}{4G}(\phi_{0}+\phi)|_{H}-\frac{c}{6}\chi|_{H}\;,\label{eq:waldbackreactgen}\eeq
with $\chi|_{H}=(-\rho+\xi)|_{H}$.  The precise form of $\phi$ and $\chi$ depends on the choice of vacuum state. The Hartle-Hawking vacuum is a thermal density matrix when restricted to the exterior region of the black hole  \cite{Fulling:1972md,Davies:1974th,Unruh:1976db}  
\beq
    \rho_{\text{HH}} = \text{tr}_{\text {ext}} | \text{HH} \rangle \langle \text{HH}|=\frac{1}{Z} e^{- H / T_{\text{H}}}\;,
\eeq where $H$ is the Hamiltonian generating   evolution with respect to the Schwarzschild time~$t$. Hence the Wald entropy for the Hartle-Hawking vacuum is a thermodynamic entropy. However,  the Boulware vacuum is not thermal when restricted to the exterior of the black hole, and therefore it is dubious to assign a thermodynamic interpretation to the Wald entropy for the Boulware state. Nonetheless, for completeness below  we will compute the Wald entropy for both   vacuum states, finding that not all solutions lead to sensible values of entropy.

\subsection*{Boulware vacuum}

In the Boulware vacuum state, where the solution of the dilaton takes the form (\ref{eq:dilsolbol}), we have, before evaluating on the horizon, 
\beq S_{\text{Wald}}=S_{\phi_0}+\frac{1}{4G}\left[-\phi_{r}\sqrt{\mu}\text{coth}\left(\frac{\sqrt{\mu}}{L}r_{\ast}\right)+\frac{Gc}{3}\left(1+\frac{\sqrt{\mu}}{L}r_{\ast}\right)\text{coth}\left(\frac{\sqrt{\mu}r_{\ast}}{L}\right)-\frac{\lambda Gc}{3}\right]-\frac{c}{6}\chi\;.\label{eq:waldentboul1}\eeq
Then, considering the time-independent solution $\chi^{(1)}$ (\ref{eq:chi1bol}) and taking the limit $r_{\ast}\to-\infty$, we find the black hole entropy including semi-classical backreaction effects is given by
\beq S^{(1)}_{\text{Wald}}=S_{\phi_0}+S_{\phi_r}-(\lambda+1)\frac{c}{12}+\frac{c}{6}\log(2\sqrt{\mu})-\frac{c}{6}C+\frac{c}{12}\frac{\sqrt{\mu}}{L}r_{\ast}|_{r_{\ast}\to-\infty}\;,\eeq
where $S_{\phi_0}+S_{\phi_r}=\frac{\phi_{0}}{4G}+\frac{\phi_{r}\sqrt{\mu}}{4G}$ is the classical entropy of the black hole.  With the additional semi-classical corrections, we see the entropy is diverging like $r_{\ast}$ to negative infinity when evaluated at the black hole horizon. This divergent result might be related to the negative Casimir energy   \eqref{eq:expvalTuuB} for the Boulware vacuum and to the fact that the Wald entropy cannot be interpreted as a thermodynamic entropy in this case. 

\subsection*{Hartle-Hawking vacuum}

In the Hartle-Hawking vacuum state the dilaton solution is given by (\ref{eq:dilasolnHH}), such that 
\beq S_{\text{Wald}}=S_{\phi_0}+\frac{1}{4G}\left[-\phi_{r}\sqrt{\mu}\text{coth}\left(\frac{\sqrt{\mu}}{L}r_{\ast}\right)+(1-\lambda)\frac{Gc}{3}\right]-\frac{c}{6}\chi\;.\label{eq:WaldentHHgen}\eeq
From the solutions for $\xi$ in (\ref{eq:xisolnsHHuv}), we have in the horizon limit $r_{\ast}\to-\infty$ the entropy corresponding to $\chi^{(3)}$ diverges to negative infinity, while the entropy coorresponding to $\chi^{(4)}$ 
\beq S^{(4)}_{\text{Wald}}=S_{\phi_0}+S_{\phi_r}+(1-\lambda)\frac{c}{12}+\frac{c}{6}\log(2\sqrt{\mu})-\frac{c}{6}C\; \label{eq:waldentropypolchinski} \eeq
is constant on the horizon, since the divergences in $-\rho(r_*)$ and $\xi^{(4)}(r_*)$ cancel each other. When $\lambda=0$ and $C=0$, this is precisely the entropy found in \cite{Almheiri:2014cka},\footnote{The apparent factor of two mismatch in the logarithm -- they find a term $\frac{c}{6} \log (4 \sqrt{\mu})$ -- is explained by a different convention for the mass parameter. By replacing $\sqrt{\mu} \to 2 \sqrt{\mu}$, $\phi_0 \to 1$, $\phi_r \to 1/2$ in our expression for $S^{(4)}_{\text{Wald}}$ it matches with equation (5.27) in \cite{Almheiri:2014cka}.} which can be interpreted as a generalized entropy, with $\chi$ capturing the von Neumann entropy of conformal matter fields. Further, in \cite{Almheiri:2014cka} the generalized entropy was   shown to obey a  first law, if $\mu$ or equivalently $T_{\text{H}}$ is being varied,
\beq
dM=T_{\text H}dS^{(4)}_{\text{Wald}}\;. \label{eq:semiclassicalfirst} \eeq  Here, $M$ is the semi-classically corrected ADM energy, which we rederive in Appendix \ref{app:energies} from the renormalized boundary stress tensor,  
\beq
M = M_{\phi_r} + M_c = \frac{1}{16 \pi G} \frac{\mu \phi_r}{L} + \frac{c \sqrt{\mu}}{12 \pi L}\;. \label{eq:masschi4full}
\eeq
Note that the semi-classical first law \eqref{eq:semiclassicalfirst} reduces to the classical result \eqref{eq:classicaljtsmarrfirst} for $c\to 0$.

Lastly, substituting in $\chi^{(5)}$  leads to a time-dependent Wald entropy (before evaluating on the black hole horizon $U_{\text{K}},V_{\text{K}}\to 0$)
\beq
\begin{aligned}
 S^{(5)}_{\text{Wald}}&=S_{\phi_0}+S_{\phi_r}+\frac{c}{12}(1-\lambda)+\frac{c}{12}\log\left[\frac{4\mu}{\left(1+\frac{\mu U_{\text K}V_{\text K}}{L^{2}}\right)^{2}}\right]_{U_{\text K},V_{\text K}\to0}\\
&+\frac{c}{12}\log\left[\left(-\frac{\sqrt{\mu}}{L}U_{\text K}+K_{U_{\text K}}\right)^{2}\left(\frac{\sqrt{\mu}}{L}V_{\text K}+K_{V_{\text K}}\right)^{2}\right]\biggr|_{U_{\text K},V_{\text K}\to0}-\frac{c}{6}k\;.
\end{aligned}
\eeq
As we will now show, a specific choice of constants $K_{U_{\text K}}$ and $K_{V_{\text K}}$ allows us to identify the time-dependent Wald entropy with the full time-dependent generalized entropy, and, moreover,  an extremization procedure leads to a QES other than the black hole  horizon.

 \subsection{von Neumann entropy}
 
Ordinarily, when considering quantum fields outside of a stationary black hole horizon, the Wald entropy   represents the UV divergent contribution of the entanglement entropy between field degrees of freedom inside and outside the horizon \cite{Bombelli:1986rw}, and   
 is separate from any additional matter von Neumann entropy. Together, the gravitational Wald entropy   and matter von Neumann entropy form the generalized entropy $S_{\text{gen}}$ attributed to the black hole. An important lesson garnered from gravitational thermodynamics, however, is that entropy can be assigned to surfaces other than black hole horizons. Of particular interest are extremal surfaces \cite{Ryu:2006bv,Hubeny:2007xt}, wherein the context of AdS/CFT, the gravitational entropy 
 is identified with the entanglement entropy of a dual holographic CFT. When quantum corrections are included, the holographic entropy formula is modified by a von Neumann entropy term of bulk quantum fields \cite{Faulkner:2013ana,Engelhardt:2014gca}, such that the holographic entanglement entropy is interpreted as the generalized entropy. In this context, the generalized entropy is attributed to a quantum extremal surface $X$, the surface which extremizes $S_{\text{gen}}$, where the classical  gravitational entropy $S_{\text{Wald}}$ captures only a portion of the full generalized entropy.
 
 Working directly with the semi-classical JT model, and without invoking AdS/CFT, here we prove the semi-classical Wald entropy computed above is exactly equivalent to the generalized entropy, where the dilaton gives the gravitational entropy and the semi-classical correction associated to the $\chi$ field accounts for the matter von Neumann contribution, 
 \beq
 S_{\text{Wald}} =  S_{\text{Wald}}^\phi +  S_{\text{Wald}}^\chi=   \frac{1}{4G}(\phi_0 + \phi) - \frac{c}{6} = S_{\text{BH}} + S_{\text{vN}}=S_{\text{gen}}\;.
 \eeq
It remains to be shown that the Wald entropy associated to the time-dependent $\chi$ field is equal to the time-dependent von Neumann entropy, which we will do in this section. Then, in the next section, upon extremizing the Wald/generalized entropy, we uncover quantum extremal surfaces which lie just outside the bifurcate horizon of the black hole.

 Since the auxiliary $\chi$ field is a stand in for a collection of external conformal matter fields, recall the von Neumann entropy of a $2D$ CFT with central charge $c$ in vacuum over an interval $[(x_{1},y_{1}),(x_{2},y_{2})]$ in a curved background in conformal gauge $ds^{2}=-e^{2\rho(x,y)}dx dy$ is (see \emph{e.g.}, \cite{Fiola:1994ir})
 \beq 
 \begin{split}
 S_{\text{vN}}&=\frac{c}{6}\log\left[\frac{1}{\delta_{1}\delta_{2}}(x_{2}-x_{1})(y_{2}-y_{1})e^{\rho(x_{1},y_{1})}e^{\rho(x_{2},y_{2})}\right]\\
 &=\frac{c}{6}(\rho(x_{1},y_{1})+\rho(x_{2},y_{2}))+\frac{c}{6}\log\left[\frac{1}{\delta_{1}\delta_{2}}(x_{2}-x_{1})(y_{2}-y_{1})\right]\;.
 \end{split}
 \label{eq:vnentgen}\eeq
 Here $\delta_{1,2}$ are independent UV regulators located at the endpoints of the interval. Since the von Neumann entropy depends on the vacuum state, and should be evaluated in coordinates defining the vacuum, it will be different for the Boulware versus  the Hartle-Hawking vacuum.
 
 Let us now show the semi-classical contribution encased in $\chi$  has the general structure of the von Neumann entropy (\ref{eq:vnentgen}),  
\beq S_{\text{Wald}}^{\chi}=-\frac{c}{6}\chi=S_{\text{vN}}\;.\label{eq:Ssemiclassent}\eeq
It  has been noted previously  in \cite{Almheiri:2014cka,Moitra:2019xoj} that the (static) entanglement entropy with one endpoint at the black hole horizon is captured by a static choice of $\chi$, given by our equation~\eqref{eq:chi4sol}. Here we instead recognize the integration constants appearing in $\chi$  give us the freedom to match $S_{\text{Wald}}=S_{\text{gen}}$, including the full time dependence of the entanglement entropy. For example, with respect to the \emph{Hartle-Hawking vacuum}, we may identify constants in $\chi^{(5)}$ (\ref{eq:chi5})
\beq 
\begin{split}
&k=-\frac{1}{2}\log\left[\frac{4\mu}{\left(1+\frac{\mu U^{B}_{\text K}V^{B}_{\text K}}{L^{2}}\right)^{2}}\right]-\frac{1}{2}\log\left[\frac{1}{(\sqrt{\mu}\delta/L)^{2}(\sqrt{\mu}\delta_{B}/L)^{2}}\right]\;,\\
&K_{U_{\text K}}=\frac{\sqrt{\mu}}{L}U^{B}_{\text K}\;,\quad K_{V_{\text K}}=-\frac{\sqrt{\mu}}{L}V_{\text K}^{B}\;,
\end{split}
\label{eq:constidsHH}\eeq
such that the von Neumann entropy $S^{\text{HH}}_{\text{vN}}$ in Kruskal coordinates $(y,x)=(V_{\text K},U_{\text K})$ is
\beq 
\begin{split}
S_{\text{vN}}^{\text{HH}}&=-\frac{c}{6}\chi^{(5)}=\frac{c}{12}\log\left[\frac{4\mu}{\left(1+\frac{\mu U_{\text K}V_{\text K}}{L^{2}}\right)^{2}}\right]+\frac{c}{12}\log\left[\frac{4\mu}{\left(1+\frac{\mu U^{B}_{\text K}V^{B}_{\text K}}{L^{2}}\right)^{2}}\right]\\
&+\frac{c}{12}\log\left[\frac{1}{\delta^{2}\delta_{B}^{2}}(U_{\text K}^B-U_{\text K})^{2}(V_{\text K}^B-V_{\text K})^{2}\right]\;.
    \end{split}
\label{eq:vnentHH}\eeq
The two endpoints in Kruskal coordinates are taken to be, respectively, $(y_1,x_1)=(V_{\text K},U_{\text K})$ and $(y_2,x_2)=(V_{\text K}^{B},U_{\text K}^{B})$, where $B$ is some generic endpoint, and $\delta$ and $\delta_{B}$ are the UV regulators located at the two  endpoints. 
The von Neumann entropy above is generically time dependent, but it becomes  independent of the times $t$ and $t_B$ at the two endpoints  when
\beq
\frac{U_{\text{K}}}{V_{\text{K}}}=\frac{U_{\text{K}}^B}{V_{\text{K}}^B} \qquad \text{or} \qquad t = t_B\;, \label{eq:extrem1}
\eeq
since then the argument of the logarithm in \eqref{eq:vnentHH} only depends on the tortoise coordinates: $(U_{\text K}^B-U_{\text K})(V_{\text K}^B-V_{\text K})= \frac{L^2}{\mu} \left [- e^{2 \sqrt{\mu} r_{*,B} / L}+ 2 e^{\sqrt{\mu} (r_{*,B} + r_*)/ L}- e^{2 \sqrt{\mu} r_* / L} \right]$.
Although the entanglement entropy is static in this case, it nonetheless leads to a QES just outside the black hole  horizon, after extremizing the generalized entropy with respect to $r_*$. In fact,  the relation \eqref{eq:extrem1} is one of the conditions for the existence of the QES, which follows from extremizing the generalized entropy with respect to $U_{\text{K}}$ and $V_{\text{K}}$, see equation \eqref{eq:cond2kruskal} below. 

There are two singular cases of \eqref{eq:extrem1}  where a non-trivial QES fails to exist though. The first is when $U^{B}_{\text K}=V^{B}_{\text K}=0$,   such that the point $B$ is located at the black hole horizon, in which case the extremum of the generalized entropy lies at the black hole horizon. Second, for   $U_{\text{K}}= V_{\text K}=0$, such that the first endpoint is located at the black hole horizon,   the surface extremizing the generalized entropy is again  the black hole horizon.  The Wald entropy $S_{\text{Wald}}^{(4)}$ \eqref{eq:waldentropypolchinski} associated to the field $\chi^{(4)}$ is an example of the second case (where in addition  we have $U_{\text{K}}^B= - L / \sqrt{\mu} =-V_{\text{K}}^B$)  which is the reason why a non-trivial QES was not found before, in \emph{e.g.} \cite{Almheiri:2014cka,Moitra:2019xoj}, from extremizing the Wald entropy.

It is worth emphasizing the semi-classical contribution to the Wald entropy, for the right choice for the integration constants in $\chi$, yields the entanglement entropy. This demonstrates the Wald formalism is fully capable of attaining the logarithmic term, in contrast to what was argued in the context of the flat space RST model \cite{Myers:1994sg}; in fact, the same line of reasoning used above shows the generalized entropy in the RST model is fully characterized by the semi-classical Wald entropy.

Note we can also match the Wald entropy with the von Neumann entropy in the \emph{Boulware vacuum}. To do so we identify the integration constants appearing in the time-dependent solution $\chi^{(2)}$ in (\ref{eq:chitimedepB}) as
\beq
\begin{split}
&C'=-\frac{1}{2}\log\left[\frac{1}{\mu}\sinh^{2}\left(\frac{\sqrt{\mu}}{2L}(v_{B}-u_{B})\right)\right]-\frac{1}{2}\log\left[\frac{1}{(\sqrt{\mu}\delta/L)^{2}(\sqrt{\mu}\delta_{B}/L)^{2}}\right]\;,\\
&K_{u}=-\frac{\sqrt{\mu} u_{B}}{L}\;,\quad K_{v}=\frac{\sqrt{\mu}v_{B}}{L}\;,
\end{split}
\label{eq:constidsboul}\eeq
such that the von Neumann entropy $S_{\text{vN}}$ in static conformal coordinates $(v,u)$ takes the form
\beq
\begin{split}
S_{\text{vN}}^{\text{B}}&=\frac{c}{12}\log\left[\frac{1}{\mu}\sinh^{2}\left(\frac{\sqrt{\mu}}{2L}(v-u)\right)\right]+\frac{c}{12}\log\left[\frac{1}{\mu}\sinh^{2}\left(\frac{\sqrt{\mu}}{2L}(v_{B}-u_{B})\right)\right]\\
&+\frac{c}{12}\log\left[\frac{1}{\delta^{2}\delta_{B}^{2}}(v_B-v)^{2}(u_B-u)^{2}\right]\;.
\end{split}
\label{eq:vnentBoulw}\eeq
In a moment we will show this von Neumann contribution leads to a QES, extremizing the generalized entropy in Boulware vacuum, which also lives outside   the black hole horizon.



 
\subsection*{Fixing constants of integration}
 Thus, generically we can arrange the integration constants in the solutions for $\chi$ such that the semi-classical Wald entropy is exactly equal to the generalized entropy, including time dependence. 
So far, however, we matched these constants in an \emph{ad hoc} manner. 
Here we present a derivation justifying our choice of integration constants. We do this in two steps. First, we generically solve the equation of motion for  $\chi$ imposing a Dirichlet boundary condition  in flat space, and then we perform a Weyl transformation back to an arbitrary curved spacetime, leading to a general solution for $\chi$ in  conformal coordinates. Next we place this generic derivation of $\chi$ into the context of semi-classical JT gravity by providing an interpretation of the endpoint where the Dirichlet boundary condition is imposed. For this derivation we adopt general conformal coordinates $y^{\pm}$, i.e.  $ds^{2}=-e^{2\rho(y^{+},y^{-})}dy^{+}dy^{-}$, for which we are agnostic about whether they are, \emph{e.g.} Kruskal or static conformal coordinates. We will show that these coordinates are specified by having vanishing normal-ordered fluxes, \emph{i.e.}, $t_+(y^{+})=t_-( y^{-})=0$.


Recall the equation of motion for $\chi$ in \eqref{eq:chieom}, $2\Box \chi=R$, which defines the solution of $\chi$ in an arbitrary curved spacetime. We expect $S_{\text{vN}}=-\frac{c}{6}\chi$,    where the general solution of $\chi$ in conformal gauge is   $\chi=-\rho+\xi$, and for  
 the Hartle-Hawking vacuum the time-dependent $\chi$ should be  given by \eqref{eq:chi5}.
Before analyzing the solution of the $\chi$ equation of motion in curved spacetime, let us first perform a Weyl transformation and solve (\ref{eq:chieom}) in flat space. 
The Weyl transformation effectively amounts to setting $R=0$, or  equivalently  $\rho=0$.
The solution for $\chi$ in flat space thus becomes
\begin{equation}
	\chi_{\text{flat}}
	=
	\xi
	\,,
\end{equation}
where it is important to recall that in a general curved spacetime $\chi$ transforms as a scalar whereas $\xi$ does not.

The boundary condition we impose on this solution   is a Dirichlet boundary condition  at the endpoint $(y^{+}_{B},y^{-}_{B})$. When the inverse Weyl transformation is performed, we take the endpoint $B$ to be close to the conformal boundary of AdS, but it does not matter \emph{where}  it is located near the boundary.   Explicitly, the boundary condition is
\begin{equation}\label{eq:lim1}
	\lim_{(y^{+},y^{-})~\to~(y^{+}_{B},y^{+}_{B})}
	\xi
	=
	C
	\,,
\end{equation}
where $C$ is some real constant that does not depend on $(y^{+}, y^{-})$. 
Typically in the literature \cite{Almheiri:2014cka,Moitra:2019xoj} a boundary condition is imposed on  $\chi$ in curved space, instead of on $\xi$ in flat space, where specifically the flux of $\chi$ is required to go to zero at the boundary.  It will turn out that we can also put the same requirement on $\chi$, but it is important for what follows that we require \eqref{eq:lim1}   for $\xi$ as well.

Since we are solving a harmonic equation $\Box \xi=0$ with a Dirichlet boundary condition, without any loss of generality we know that
\begin{equation}
	\xi(y^{+},y^{-},y^{+}_{B},y^{-}_{B})
	=
	\xi(y^{+}_{B}-y^{+},y^{-}_{B}-y^{-})
	\,.
\end{equation}
This forces the solution for $\chi_{\text{flat}}$ to be of the form
\begin{equation}
	\chi_{\text{flat}}
	=
	\xi
	=
    -\frac{1}{2}
	\log
	\left[
		\frac{1}{\tilde{\delta}^{4}}(y^{+}_{B}-y^{+} )^{2}(y^{-}_{B}-y^{-})^{2}
	\right]
	\,,
\end{equation} 
where $\tilde{\delta}$ is some arbitrary length scale which regulates the limit \eqref{eq:lim1}. Using definition \eqref{eq:tpmfuncgen} we see that this solution  yields   $t_\pm(y^{\pm})=0$  \eqref{eq:gendefvacstate}, so the Dirichlet boundary condition selects the vacuum state with respect to the ($y^\pm$) coordinates. We recognize $\chi_{\text{flat}}$ can be written as a two-point correlation function of primary operators $\partial \chi$
\begin{equation}
	\chi_{\text{flat}}
	=
	\log
	\left[
		\hat{\delta}^{2}
			\langle
				\partial \chi(y^{+},y^{-})
				\partial \chi(y^{+}_{B},y^{-}_{B})
			\rangle_{\text{flat}}
	\right]
	\,,
\end{equation} 
where $\hat{\delta}$ is another arbitrary length scale. 
An elementary result from conformal field theory instructs us how the two-point correlation function transforms under a Weyl transformation.\footnote{Let $\mathcal{O}$ be primary operators with weight $\Delta_{\mathcal{O}}$, then under Weyl transformations we can relate correlation functions as $\langle\mathcal{O}(x)\mathcal{O}(y)\rangle_{\Omega^{2}g_{\mu\nu}}
=
\Omega(x)^{-\Delta_{\mathcal{O}}}
\Omega(y)^{-\Delta_{\mathcal{O}}}
\langle\mathcal{O}(x)\mathcal{O}(y)\rangle_{g_{\mu\nu}}$. In the case considered in this work $\Delta_{\mathcal{O}}=1$.} Undoing the Weyl transformation, we obtain the full curved spacetime expression for $\chi$, 
\begin{equation}\label{eq:neumann3}
	\chi
	=
    -\frac{1}{2}
	\log
	\left.
	\left[
		\frac{e^{2\rho}e^{2\rho_{B}}}{\tilde{\delta}^{4}}(y^{+}_{B}-y^{+})^{2}(y^{-}_{B}-y^{-})^{2}
	\right]
	\right|_{t_\pm  =0}
	\,,
\end{equation}
which coincides with the regulated von Neumann entropy \eqref{eq:vnentHH} up to a constant in the HH vacuum or with \eqref{eq:vnentBoulw} in the Boulware vacuum. We in fact reproduced the general von Neumann entropy formula in 2D CFT for a single interval in curved space,  previously obtained in \cite{Fiola:1994ir} from field theory methods and in    \cite{Almheiri:2019hni,Gautason:2020tmk}  using holographic methods.
Notice that even if $e^{2\rho_{B}}$ takes a singular value somewhere at the boundary, we can always regulate this divergence such that $\chi$ goes to zero on the boundary. As a last remark we point out that from the holographic perspective the cut-off $\tilde{\delta}$ takes care of the regularization of the endpoints and as such is able to generate the conformal factors that appear in the logarithm through an appropriate change of coordinates of the holographic boundary coordinate, see e.g. \cite{Almheiri:2019hni}.
\subsection*{von Neumann  contribution to generalized entropy in general $2D$ gravity}
The argument we presented above to derive the von Neumann entropy from the $\chi$ fields was rather general. 
Let us now argue how general its conclusions are.
Any two-dimensional theory of gravity with a minimally coupled CFT with large central charge $c$ gives a conformal anomaly at the semi-classical level, captured by the Polyakov term. The Polyakov term \eqref{eq:semiclassactionconts} can always be localized by introducing $\chi$ fields whose equation of motion is given by~\eqref{eq:chieom}. Furthermore, the Wald entropy
receives a functional contribution from $\chi$, independent of the value of the dilaton or metric.

In general one has to make a choice for what the point  $(y^{+}_{B},y^{-}_{B})$ denotes. We generically require the product $y^{+}_{B}y^{-}_{B}$ to define a timelike surface on which the metric is constant, as is the case for, \emph{e.g.}, the conformal boundary in AdS or near $\mathcal{I}^{+}$ in Minkowski spacetime. Note the interpretation of this timelike surface depends on the context of the problem of interest.
For example, the case of $2D$ de Sitter gravity is special as there the asymptotic boundary   $\mathcal{I}^{+}$ is spacelike, instead of timelike or null. Nevertheless one can, from an algebra\"{i}c point of view, perform the same manipulations on a timelike surface close to the poles of de Sitter.

Once the meaning of the endpoint $(y^{+}_{B},y^{-}_{B})$ has been established one can employ the Dirichlet condition  above and, aided by the fact  that   any two-dimensional spacetime is Weyl equivalent to flat spacetime, 
we generically arrive at the solution (\ref{eq:neumann3})
for $\chi$. Thus, we find $S_{\text{Wald}}^\chi= -\frac{c}{6}\chi = S_{\text{vN}}$ for any $2D$ gravity theory coupled to a CFT. 
Two remarks worth mentioning are, first, the above derivation follows through for different vacua as well, modifying only the coordinates $(y^+,y^-)$ with respect to which the vacuum state is defined. 
Second, as we only made assumptions about the existence of a timelike surface on which the metric is constant,  the argument presented here holds for static and non-static backgrounds alike.
 \subsection{Quantum extremal surfaces}
Having verified the semi-classical Wald entropy is equivalent to the generalized entropy,  let us now look for the surfaces which extremize the entropy. These surfaces are known as quantum extremal surfaces (QES) \cite{Engelhardt:2014gca}, and are generalizations of the Ryu-Takayanagi surfaces necessary to compute quantum corrected holographic entanglement entropy. We will not need this interpretation below, however, we are certainly motivated by it. Rather, the point we aim to emphasize here is that,   when including the backreaction of  the Hawking radiation of a stationary black hole in thermal equilibrium, it is natural to associate thermodynamics with a QES, such that the bifurcate Killing horizon can be viewed as the classical limit of the QES. 

Since the form of the Wald entropy is dependent on the vacuum state, so too will the QES depend on the choice of vacuum. Let us thus extremize the Wald entropy in both Hartle-Hawking and Boulware vacua and look for the QESs in either case. 

\subsection*{Hartle-Hawking QES}

Consider the time-dependent generalized entropy $S^{\text{HH}}_{\text{gen}}$ given by (\ref{eq:WaldentHHgen}), where the semi-classical contribution is the von Neumann entropy $S_{\text{vN}}^{\text{HH}}$ in (\ref{eq:vnentHH}), and where we set $\lambda=0$. Specifically, 
\beq S^{\text{HH}}_{\text{gen}}=S_{\phi_0}+\frac{1}{4G}\left[\phi_{r}\sqrt{\mu}\left(\frac{1-\frac{\mu}{L^{2}}U_{\text K}V_{\text K}}{1+\frac{\mu}{L^{2}}U_{\text K}V_{\text K}}\right)+\frac{Gc}{3}\right]+S_{\text{vN}}^{\text{HH}}\;.\label{eq:waldentHH2}\eeq
We extremize with respect to the Kruskal coordinates $U_{\text K}$ and $V_{\text K}$ of the first endpoint, while keeping the second endpoint $B$ fixed, 
\beq \partial_{U_{\text K}}S_{\text{gen}}^{\text{HH}}=\partial_{V_{\text K}}S_{\text{gen}}^{\text{HH}}=0\;.\eeq
We find 
\beq
\begin{split}
 \partial_{V_{\text K}}S^{\text{HH}}_{\text{gen}}&=\frac{3L^{3}\mu^{3/2}U_{\text K}(V^{B}_{\text K}-V_{\text K})+Gc\frac{L}{\phi_{r}}  (L^{2}+\mu U_{\text K}V_{\text K})(L^{2}+\mu U_{\text K}V^{B}_{\text K})}{6G\frac{L}{\phi_{r}}(L^{2}+\mu U_{\text K}V_{\text K})^{2}(V_{\text K}-V^{B}_{\text K})}=0\;,
\end{split}
\eeq
and similarly for $\partial_{U_{\text K}}S^{\text{HH}}_{\text{gen}}=0$. 

Subtracting the two extremization conditions we uncover the following relation:
\beq 0=\partial_{U_{\text K}}S_{\text{gen}}^{\text{HH}}-\partial_{V_{\text K}}S_{\text{gen}}^{\text{HH}}\quad \Leftrightarrow \quad  \frac{U_{\text K}}{V_{\text K}}=\frac{U^{B}_{\text K}}{V^{B}_{\text K}}\;.\label{eq:cond2kruskal}\eeq
Meanwhile, adding the two extremization conditions leads to 
\beq 
\begin{split}
0&=2\epsilon+\frac{6\mu}{L^{2}}\left(1-\frac{V_{\text K}}{V^{B}_{\text K}}\right)V_{\text K}U^{B}_{\text K}+\frac{2\epsilon\mu}{L^{2}}\left[1+\frac{V_{\text K}}{U^{B}_{\text K}}+\frac{\mu}{L^{2}}\frac{(V_{\text K})^{2}U^{B}_{\text K}}{V^{B}_{\text K}}\right]V_{\text K}U^{B}_{\text K}\;.
\end{split}
\eeq
where we used (\ref{eq:cond2kruskal}) and introduced the small parameter, 
\beq\epsilon\equiv \frac{Gc}{\sqrt{\mu}\phi_{r}}\ll 1\;, \eeq which follows from the semi-classical regime of validity \eqref{eq:semicl}. 
The effect of the cubic term can be neglected (it changes the location of the QES at subleading order in $\epsilon$) and we can instead solve for $V_{\text K}$ with the simplified quadratic equation, such that for small $\epsilon$ we have the following two solutions for $V_{\text K}$:
\beq V_{\text K}^{(1)}\approx-\frac{L^{2}}{3\mu}\frac{\epsilon}{U^{B}_{\text K}}+O(\epsilon^{2})\;,\quad V_{\text K}^{(2)}\approx V^{B}_{\text K}+\frac{(L^{2}+2\mu U^{B}_{\text K}V^{B}_{\text K})}{3\mu U^{B}_{\text K}}+O(\epsilon^{2})\;.\eeq
The second solution is  deemed unphysical, since it does not coincide with the black hole horizon in the classical limit $c \to 0$. In fact, classically it coincides with the anchor point $B$, and semi-classically it lies inward and we find that the generalized entropy is parametrically larger than the entropy of the first solution. Thus, we neglect the second solution,  leaving us with a single QES located at
\beq  V_{\text K}^{\text{QES}}\approx-\frac{L^{2}}{3\mu}\frac{\epsilon}{U^{B}_{\text K}}\;,\quad U_{\text K}^{\text{QES}}\approx-\frac{L^{2}}{3\mu}\frac{\epsilon}{V^{B}_{\text K}}\;.\label{eq:QESHHloc}\eeq
In terms of Schwarzschild coordinates, the QES has the radial position 
\beq r_{\text{QES}}=L\sqrt{\mu}\left(\frac{1-\frac{\mu}{L^{2}}U_{\text K}V_{\text K}}{1+\frac{\mu}{L^{2}}U_{\text K}V_{\text K}}\right)\approx r_{H}\left(1+\frac{2\epsilon^{2}}{9}e^{-\frac{2\sqrt{\mu}}{L}r_{\ast,B}}\right)\;,\label{eq:QESlocHH}\eeq
where $r_{\ast,B}=\frac{L}{2\sqrt{\mu}}\log\left(-\frac{\mu}{L^{2}}U^{B}_{\text K}V^{B}_{\text K}\right)$ is the tortoise coordinate of the second endpoint near the conformal boundary. Importantly, in the limit we approach the conformal boundary, $r_{\ast,B}\to0$ or $-\frac{\mu}{L^{2}}U_{\text K}^{B}V_{\text K}^{B}=1$,    the QES still exists. 

Substituting the QES (\ref{eq:QESHHloc}) back into the generalized entropy (\ref{eq:waldentHH2}) we find:
\begin{align}
  S_{\text{gen}}\big |_{\text{QES}}&=S_{\phi_0}+S_{\phi_r}+\frac{c}{12}+\frac{c}{6}\log(2\sqrt{\mu})+\frac{c}{6}\log\left(-\frac{\mu U^{B}_{\text K}V^{B}_{\text K}}{L^{2}}\right)+\frac{c}{12}\log\left[\frac{4\mu}{\left(1+\frac{\mu}{L^{2}}U_{\text K}^{B}V_{\text K}^{B}\right)^{2}}\right]\nonumber \\
  &+\frac{cL^{2}\epsilon}{9\mu U^{B}_{\text K}V_{\text K}^{B}}+\mathcal{O}(\epsilon^{2})\;.
\end{align}
Thus, to leading order in $\epsilon$, the generalized entropy of the QES is simply that of the black hole. Critically, however, since $U_{\text K}<0$ and $V_{\text K}>0$, the $\epsilon$ correction is negative; indeed in the conformal boundary limit  $-\frac{\mu}{L^{2}}U_{\text K}^{B}V_{\text K}^{B}=1$, where the divergence in the generalized entropy is regulated by a cutoff, the above becomes
\beq S_{\text{gen}}\big|_{\text{QES}}\overset{r_{*,B}\to0}{=}S_{\text{gen}}\big|_{H}-\frac{c\epsilon}{9}+\frac{\sqrt{\mu}\phi_{r}}{18G}\epsilon^{2}+\mathcal{O}(\epsilon^{3})\;,\eeq
where we have included the $\mathcal{O}(\epsilon^{2})$ for further clarity. Thence, the generalized entropy evaluated at the QES   is \emph{less} than the entropy evaluated on the eternal black hole horizon, 
\beq S_{\text{gen}}\big|_{\text{QES}}< S_{\text{gen}}\big|_{H}\;.\eeq
A few comments are in order. First, notice in the classical limit $\epsilon\to0$ we observe the QES is located at the bifurcate Killing horizon. This is consistent with the fact the classical Wald entropy is extremized by the black hole horizon, while semi-classical effects alter the position of the extremal surface. In fact, in the case of the static background considered here, we see the QES lies just \emph{outside} of the black hole  horizon. Specifically, from (\ref{eq:QESlocHH}), the QES is approximately located at a distance $r_H \left(\frac{G  c }{\sqrt{\mu} \phi_r} \right)^2 \ll r_H$ above the horizon.   This suggests, in the context of an eternal black hole background at least, the QES may be viewed as a stretched horizon or membrane \cite{Price:1986yy}, which are well known to obey a first and second law of thermodynamics. We emphasize, however, the stretched horizon of a black hole is located outside the horizon even when one does not take into account semi-classical effects (whereas the QES is located at the event horizon classically).  Moreover, in dynamical scenarios such as an evaporating black hole, the QES is found to be located inside of the horizon \cite{Penington:2019npb}. Thus, the QES is generally different from a stretched horizon. 

Next, note the condition (\ref{eq:cond2kruskal}) attained from extremality, yields the Schwarzschild time~$t$ is fixed to the time at $B$, $t_{B}$, 
\beq t=t_{B}\;.\eeq
Equivalently, in static null coordinates we find $(u-u_{B})=-(v-v_{B})$. As we mentioned below~\eqref{eq:extrem1}, substituting this time symmetry back into  the time-dependent solution for $\chi$~(\ref{eq:chi5}) we find the logarithmic contribution -- where the time dependence resides -- solely becomes a function of $(u-u_{B})$, such that time dependence drops out. Thus, restricted to a time slice $t=t_{B}$, the generalized entropy becomes time independent. Nonetheless, as can be explicitly verified, had we implemented (\ref{eq:cond2kruskal}) into $S_{\text{gen}}$ initially and performed the same extremization procedure as above we would have recovered the same position of the QES, further indicating the QES arises from semi-classical effects and not time dependence of the von Neumann entropy. This observation is also relevant for the interpretation of the generalized entropy as a thermodynamic entropy, a fact we use in the next section.

It is worth comparing our set-up and extremization of the generalized entropy to that in~\cite{Moitra:2019xoj}. The set-up in \cite{Moitra:2019xoj} distinguishes between two scenarios: (i) the semi-classical JT model, with $\chi$ representing $c$ scalar fields, with the same form of the generalized entropy as in (\ref{eq:waldbackreactgen}), defined at the horizon, however, where $\chi$ is vanishing  at the asymptotic boundary of a collapsing $\text{AdS}_{2}$ black hole (not connected to a thermal bath, such that outgoing radiation bounces off the boundary, falling back into the black hole), and (ii) working directly with $c$ scalar fields $\psi_{i}$ obeying Dirichlet boundary conditions, which has another generalized entropy $S^{\psi}_{\text{gen}}$, given by the classical contribution of the black hole and the von Neumann entropy of the fields $\psi_{i}$ outside the horizon. Crucially, contrary to our set-up,  where we find a QES outside of the black hole horizon and $S_{\text{gen}}|_{H}>S_{\text{gen}}|_{\text{QES}}$, the authors of \cite{Moitra:2019xoj} find $S^{\psi}_{\text{gen}}$ is extremized for points other than the horizon, however, discard such solutions since they claim $S_{\text{gen}}^{\psi}$ is larger at these points than at the bifurcate horizon, $S^{\psi}_{\text{gen}}|_{H}<S^{\psi}_{\text{gen}}|_{\text{QES}}$. This may be an artifact of working with the $\chi$ system, which is the $c\to\infty$ limit of the $\psi_{i}$ system, or perhaps a consequence of working with different boundary conditions for $\chi$. 

Lastly, as a passing comment, note we found a QES outside of the black hole horizon in the eternal $\text{AdS}_{2}$ background, akin to the entanglement wedge islands exterior to the horizon in static background systems in \cite{Almheiri:2019yqk}, used to resolve a particular form of the information paradox by deriving a unitary Page curve. It is thus tempting to interpret our findings above as likewise locating islands outside of the horizon. However, in \cite{Almheiri:2019yqk} a thermal bath is joined at the conformal boundary, such that the Hawking radiation disappears into the bath and the $\text{AdS}_{2}$ black hole evaporates, which is contrary to our set-up. Nonetheless, the similarities are striking and we will revisit this topic in the discussion.

\subsection*{Boulware QES}

Before moving on to the next section where we study the thermodynamics of the QES, let us point out the generalized entropy in the Boulware state is also extremized by a quantum extremal surface.  In the Boulware vacuum, the generalized Wald entropy (\ref{eq:waldentboul1}) with $\lambda=0$ is, 
\beq S^{\text B}_{\text{gen}}=S_{\phi_0}+\frac{1}{4G}\left[-\phi_{r}\sqrt{\mu}\text{coth}\left(\frac{\sqrt{\mu}}{2\ell}(v-u)\right)+\frac{Gc}{3}\left(1+\frac{\sqrt{\mu}}{2L}(v-u)\right)\text{coth}\left(\frac{\sqrt{\mu}(v-u)}{2L}\right)\right]+S^{\text B}_{\text{vN}}\;,\label{eq:waldentboul2}\eeq
with $S^{\text B}_{\text{vN}}$ equal to (\ref{eq:vnentBoulw}). We now extremize (\ref{eq:waldentboul2}) with respect to both $v$ and $u$,
\beq \partial_{v}S^{\text B}_{\text{gen}}=\partial_{u}S^{\text B}_{\text{gen}}=0\;.\eeq
The rest of the analysis is nearly identical as before. Adding the derivatives we again find $(u-u_{B})=-(v-v_{B})$, while,  upon substituting in this time symmetry condition, the difference yields,
\beq 
\begin{split}
0= 1-\frac{\epsilon}{3}&-\frac{\epsilon}{6} \frac{\sqrt{\mu}}{ L}(-2 u + v_B + u_B)   +\frac{\epsilon}{2}\sinh\left[\frac{\sqrt{\mu}}{L}(-2u+v_{B}+u_{B})\right]\\
&-\frac{4 \epsilon}{3} \frac{  L \sqrt{\mu}  }{ u-u_B } \sinh^2 \left[ \frac{\sqrt{\mu}}{2L}(-2u + v_B+ u_B) \right]\;.
\end{split}
\eeq
Consequently, the position of the QES in advanced and retarded coordinates $( v,u)$ is 
\beq
\begin{split}
&v_\text{QES}
\approx\frac{1}{2}(u_{B}+v_{B})+\frac{L}{\sqrt{\mu}}\log\left(\frac{\sqrt{\epsilon}}{2}\right) \;,\\
&u_\text{QES}
\approx\frac{1}{2}(u_{B}+v_{B})-\frac{L}{\sqrt{\mu}}\log\left(\frac{\sqrt{\epsilon}}{2}\right)\;.
\end{split}
\eeq 
Equivalently, in Kruskal coordinates 
\beq \frac{\sqrt{\mu}}{L}V_{\text K}^\text{QES}\approx\frac{1}{2}\left(-\frac{V^{B}_{\text K}}{U^{B}_{\text K}}\right)^{1/2}\sqrt{\epsilon}\;,\quad   \frac{\sqrt{\mu}}{L}U_{\text K}^\text{QES}\approx\frac{1}{2}\left(-\frac{U^{B}_{\text K}}{V^{B}_{\text K}}\right)^{1/2}\sqrt{\epsilon} \;,\eeq
or in Schwarzschild coordinates, 
\beq r_{\text{QES}}\approx r_{H}\left(1+\frac{\epsilon}{2}\right)\;.\eeq
Thus, as in the Hartle-Hawking vacuum, in the classical limit $\epsilon\to0$ we find $r_{\text{QES}}\to r_{H}$, but generally $r_{\text{QES}}$ lies outside of the black hole horizon. Moreover, if we evaluate the generalized entropy at the QES we find it leads to a value smaller than the generalized entropy when evaluated on the bifurcate horizon, however, both are formally negatively divergent on the horizon. The overall point here is that while the Boulware vacuum does not describe a system in thermal equilibrium, the generalized entropy in the Boulware vacuum is nonetheless extremized by a QES outside of the black hole horizon.

\subsection*{A potential causality paradox and the quantum focusing conjecture} 

As we will describe below in detail, a QES lying outside of the black hole horizon defines an AdS-Rindler region nested inside the eternal black hole, where the bifurcate horizon of the nested region coincides with the QES (see Figure \ref{fig:nestedwedges}). This construction leads to a naive causality paradox \cite{Almheiri:2019yqk}: it appears a signal from the island region associated to the QES may reach the physical $\text{AdS}_{2}$ boundary, which is in conflict with the quantum focusing conjecture (QFC) \cite{Bousso:2015mna}.  In two-dimensional JT gravity
the QFC takes the form
\beq \frac{d^{2}S_{\text{gen}}}{d\lambda^{2}}\leq 0\quad \Leftrightarrow\quad \nabla_{+}^{2}\left(\frac{\phi}{4G}+S_{\text{vN}}\right)\leq0\;,\label{eq:QFCv1}\eeq
where $\lambda$ is the affine parameter along a null geodesic $N$ extending from the QES $X$ outwards toward the future, and $\nabla_{+}$ is a covariant derivative with respect to a metric $ds^{2}=e^{2\rho}dy^{+}dy^{-}$. Moreover, here $S_{\text{vN}}$ is the entropy of matter fields to the right of the null geodesic $N$. As explained in \cite{Almheiri:2019yqk}, a stronger inequality \cite{Wall:2011kb} holds for a $\text{CFT}_{2}$ confined to a single interval:
\beq \frac{1}{4G}\nabla^{2}_{+}\phi+\nabla_{+}^{2}S_{\text{vN}}+\frac{6}{c}(\nabla_{+}S_{\text{vN}})^{2}\leq0\;.\label{eq:QFCv2}\eeq
Likewise, it is straightforward to check this QFC inequality holds for the generalized entropy (\ref{eq:waldentHH2}) with $S_{\text{vN}}$ given in (\ref{eq:vnentHH}), with $y^+ =V_{\text{K}}$ and $y^-=U_{\text{K}}$. In fact, using $\nabla_{+}^{2}f=\partial_{+}^{2}f-2(\partial_{+}\rho)(\partial_{+}f)$ for an arbitrary scalar function $f$, we see the inequality is saturated since
\beq \nabla_{+}^{2}\phi=0\;,\quad \nabla_{+}^{2}S_{\text{vN}}+\frac{6}{c}(\nabla_{+}S_{\text{vN}})^{2}=0\;.\eeq
  Consequently, the QFC (\ref{eq:QFCv1}) is satisfied since $\nabla_{+}^{2}S_{\text{vN}}\leq0$,  and hence
the aforementioned apparent causality paradox is resolved. That the QFC holds also implies $S_{\text{gen}}$ obeys a second law \cite{Bousso:2015mna}, for which we provide a heuristic derivation of in Appendix \ref{app:secondlaw}. 

A potentially confusing point is that for spacetimes with reflecting boundary conditions at infinity, as in $\text{AdS}_{2}$, a QES must lie \emph{inside} the horizon \cite{Engelhardt:2014gca}.\footnote{We thank Aron Wall for discussions about this point.} The reason for this is as follows: Suppose $X$ lies outside of the horizon, the null geodesic $N$ eventually reaches the $\text{AdS}_{2}$ boundary. By the QFC (\ref{eq:QFCv1}) and since $\nabla_{+}S_{\text{gen}}=0$ at the QES, $S_{\text{gen}}$ decreases monotonically along $N$, $\nabla_{+}S_{\text{gen}}\leq0$. However, since the dilaton $\phi$ diverges at the boundary, so too does $S_{\text{gen}}$, leading to a contradiction and thus a QES cannot lie outside of the horizon. Naively, this fact is in tension with our finding of a QES lying outside of the eternal black hole horizon. A similar confusion arises in \cite{Almheiri:2019yqk}, where upon decoupling the two sides of $\text{AdS}_{2}$ from Minkowski $\mathbb{R}^{1,1}$ bath regions, it seems the location of the QES leads to an inconsistency with the QFC. The problem is following the null geodesic $N$ all of the way to the boundary. This was avoided in \cite{Almheiri:2019yqk} by considering causal diamonds whose future tip is displaced away from the boundary. Similarly, consistency with our above QFC calculation follows if we place a cut-off surface just away from the conformal boundary $r_{\ast,B}\neq0$, as  for instance in equation \eqref{eq:QESlocHH},  such that the nested AdS-Rindler region is the domain of dependence of the spatial slice extending between the bifurcation point (the QES) and the cutoff surface. The QES is then behind the horizon of a nested AdS-Rindler region which is anchored at the boundary, consistent with the QFC. Essentially, the QES uncovered above thus lies behind the horizon of a nested AdS-Rindler region, but outside the classical  eternal black hole horizon.

 \section{Semi-classical thermodynamics of  AdS$_2$-Rindler space} \label{sec:semithermorindler}
 
 Above we showed the Wald entropy $S_{\text{Wald}}$ associated with the semi-classical JT model is exactly equivalent to the generalized entropy, including the von Neumann entropy contribution associated with a CFT with central charge $c$ outside of the eternal black hole. 
Moreover, we found a quantum extremal surface lying just outside of the black hole horizon which extremizes the Wald entropy and, crucially, leads to a lower value of the entropy than when $S_{\text{Wald}}$ is evaluated on the bifurcate horizon of the black hole. Thus, we may attribute a thermodynamic entropy to the QES, suggesting the QES may also obey a first law. 

The thermodynamics of the QES on display may be seen as a consequence of the fact the entanglement wedge of the QES is a Rindler wedge.\footnote{Given a ``bulk" asymptotically  AdS spacetime with boundary subregion $A$, the \emph{entanglement wedge} $\mathcal{E}_{A}$ \cite{Czech:2012bh,Wall:2012uf,Headrick:2014cta} is the domain of dependence of any (achronal) codimension-1 bulk spatial surface with boundary $A\cup\Gamma_{A}$, where $\Gamma_{A}$ is the extremal codimension-2 bulk region homologous to $A$, \emph{i.e.}, the HRT surface \cite{Hubeny:2007xt}. The entanglement wedge is generally larger than the \emph{causal wedge} \cite{Hubeny:2012wa}, the intersection of bulk causal future and past boundary domain of dependence of $A$, however, the two wedges sometimes coincide, \emph{e.g.}, for spherical boundary subregions $A$ in vacuum AdS, in which case $\mathcal{E}_{A}$ is equal to the AdS-Rindler wedge.}  First, we recall the $\text{AdS}_{2}$ black hole background  is simply $\text{AdS}_{2}$-Rindler space, such that the black hole thermodynamics is a consequence of the thermal behavior of AdS-Rindler space, namely, that the density matrix of the Hartle-Hawking vacuum state reduced to the Rindler wedge is a thermal Gibbs state. It is this realization which leads to the first law of black hole thermodynamics, where the temperature is given by the Hawking temperature, the energy by the ADM mass, and the entropy by the Wald entropy. By extension, since the entanglement wedge associated with the QES is also a Rindler wedge,  the corresponding thermal character  leads to a first law of thermodynamics of quantum extremal surfaces, where the entropy is given by the generalized entropy, \emph{i.e.}, the semi-classically corrected Wald entropy, and the asymptotic energy is identified with a semi-classical modification of the ADM mass (we will comment on the temperature momentarily). Therefore, the first law of QESs is understood as the natural semi-classical generalization of the first law of black hole thermodynamics. In this section we derive this semi-classical first law of QES thermodynamics. 

An important feature to point out is that, since the QES lies outside of the black hole, the QES (right) Rindler wedge is nested inside of the (right) Rindler wedge associated with the black hole, where, in the limit the QES and bifurcate horizon coincide, $c\to0$, the Rindler wedges are identified. Thus, we are in fact interested in studying the thermal behavior of the nested Rindler wedge. Crucially, the global Hartle-Hawking state restricted to the nested Rindler wedge is a again   a thermal Gibbs state. In fact, we will show below the static observers in the eternal black hole and in the nested Rindler wedge both see the same Hartle-Hawking vacuum   as a thermal state, but at different temperatures. Indeed, the surface gravity of the nested Rindler horizon is the parameter which naturally appears in the first law of the nested wedge, and is identified as the temperature of the entanglement wedge of the QES when in equilibrium.

Before moving on to the derivation of the semi-classical first law, it is worth mentioning prior important studies of AdS-Rindler thermodynamics. In the seminal work \cite{Casini:2011kv}, it was shown the entanglement entropy of a CFT vacuum state in flat space $\mathbb{R}^{d}$ reduced to a ball of radius $R$, via a conformal mapping, is equal to the thermal entropy of a Gibbs state in the hyperbolic cylinder $\mathbb{R}\times\mathbb{H}^{d-1}$, where the temperature is inversely proportional to the radius of the ball. When the CFT is holographic, the thermal CFT state is then dual to a massless hyperbolically sliced black hole, where the thermal entropy of the Gibbs state is identified with the Bekenstein-Hawking entropy of the black hole at the same temperature. For holographic CFTs dual to higher curvature theories of gravity, the Bekenstein-Hawking entropy is replaced by the Wald entropy.\footnote{In fact, the statement holds for holographic CFTs dual to \emph{arbitrary} higher derivative theories of gravity in higher and lower dimensions, including JT gravity \cite{Rosso:2020zkk}.} The massless hyperbolic black hole, of course, is simply $(d+1)$-dimensional AdS-Rindler space, such that the Rindler wedge is identified with the entanglement wedge of the ball on the boundary flat space. Thus, via the Casini-Huerta-Myers map, the vacuum entanglement entropy of a holographic CFT reduced to a ball is equal to the Wald entropy of a AdS-Rindler eternal black hole. Furthermore, the first law of entanglement entropy \cite{Blanco:2013joa,Wong:2013gua} for ball-shaped regions in the vacuum boundary CFT is holographically dual to the first law of the AdS-Rindler horizon \cite{Faulkner:2013ica}.

 The first law we derive below may be interpreted as an exact semi-classical extension of the    first law of the AdS-Rindler wedge, however, there are some conceptual differences with   previous works. First, our work is not based on the AdS/CFT correspondence. Indeed, we work directly in a $(1+1)$-dimensional AdS-Rindler space, for which the dual one-dimensional CFT is obscure. Therefore we do not invoke the existence of a  holographic CFT.
 Second, due to the semi-classical effects of backreaction, we uncover a first law associated with a Rindler wedge nested inside the AdS-Rindler black hole. We will return to discuss the similarities between our semi-classical extension and the work of \cite{Casini:2011kv,Faulkner:2013ica,Blanco:2013joa} in the discussion.

 \subsection{Rindler wedge inside a Rindler wedge}
 
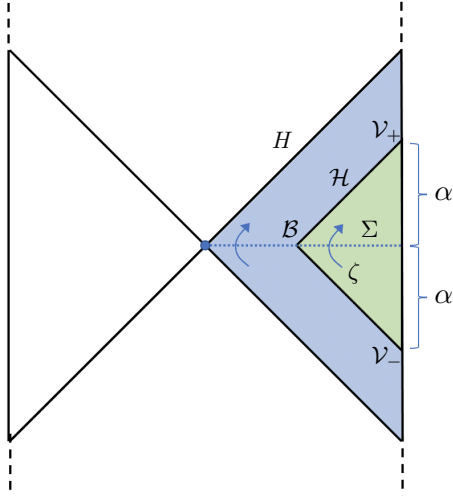
\begin{figure}[t]
\begin{center}
		\begin{overpic}[width=0.45\textwidth]{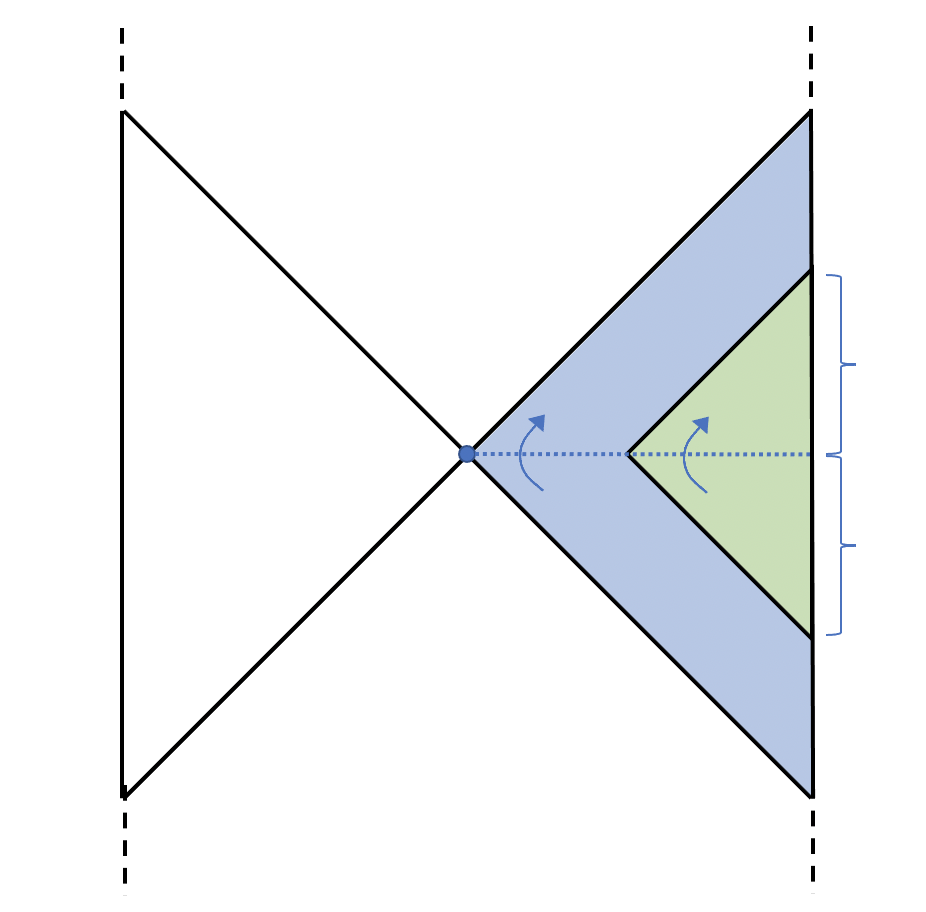}
			\put (82,70) {\footnotesize{$\mathcal{V}_{+}$}}
			\put (82,27) {\footnotesize{$\mathcal{V}_{-}$}}
			\put (80,51) {\footnotesize{$\Sigma$}}
			\put (65,51) {\footnotesize{$\mathcal{B}$}}
			\put (74,61) {\footnotesize{$\mathcal{H}$}}
			\put (63,68) {\footnotesize{$\emph{H}$}}
		    \put (77.5,43) {\footnotesize{$\zeta$}}
			\put (94,58) {$\alpha$}
			\put (94,38.5) {$\alpha$}
		\end{overpic}
		
	\end{center}
\centering
\caption{Nested Rindler wedges. The larger  AdS-Rindler wedge (shaded in blue) envelopes a smaller  Rindler wedge (shaded in green). The bifurcation points of the larger and smaller (right) Rindler patches generally do not coincide, but they do in the limit when the boundary time interval goes to infinity, $\alpha\to\infty$. In this limit the boost Killing vector $\zeta$ of the nested  Rindler  wedge becomes proportional to the time-translation Killing vector $\partial_{t}$ of  the   black hole metric. We have illustrated a nested Rindler wedge whose extremal slice $\Sigma$ is located at $t=0$, but we also consider nested wedges centered at $t=t_0 \neq0.$}
\label{fig:nestedwedges}
\end{figure}

 Since the quantum extremal surface lies outside of the AdS$_{2}$ black hole horizon, we are motivated to study the thermodynamics associated with a Rindler wedge nested inside the eternal $\text{AdS}_{2}$ black hole, whose exterior is itself a Rindler wedge (see Figure \ref{fig:nestedwedges}). Here we summarize the main aspects of the geometric set-up,  more details may be found in Appendix~\ref{app:embeddingformalism}.

One way to see that the eternal $\text{AdS}_{2}$ black hole is AdS-Rindler space is by the   coordinate transformation \eqref{eq:adsrindlercoordtr}
\beq  \kappa \sigma=\sqrt{\mu}t/L\;, \quad \varrho=\frac{L}{\sqrt{\mu}}\sqrt{\frac{r^{2}}{L^{2}}-\mu}\;,\eeq
such that the line element \eqref{eq:ads2bhstat} becomes
\beq ds^{2}=-\kappa^2 \varrho^{2} d\sigma^{2}+\left(\frac{\varrho^{2}}{L^{2}}+1\right)^{-1}d\varrho^{2}\;.\eeq
Here $\kappa$ is the surface gravity of the AdS-Rindler horizon associated to the boost Killing vector~$\partial_\sigma$. However, the surface gravity depends on the normalization of the boost Killing vector, which   is not unique since the boost generator cannot be    normalized  at spatial infinity in (AdS-)Rindler space, like the stationary Killing vector  of asymptotially flat black holes.   The radial coordinate $\varrho$ lies in the range $\varrho\in[0,\infty)$, with the horizon located at $\varrho=0$ and the asymptotic boundary at $\varrho=\infty$.  Lines of constant $\varrho$ parametrize the worldlines of   Rindler observers with proper acceleration $a = \sqrt{1/\varrho^2 + 1/L^2}$. 

Note we need not work with coordinates $(\sigma,\varrho)$ explicitly since the Schwarzschild coordinates also cover AdS-Rindler space. Indeed, the proper Rindler time-translation Killing vector $\partial_{\sigma}$ is proportional to the Schwarzschild time-translation Killing vector $\partial_{t}$; lines of constant $t$ likewise represent the worldlines of uniformly accelerating observers, such that the $\sigma=0$ slice coincides with the $t=0$ slice. Consequently, we often refrain from working with coordinates $(\sigma,\varrho)$, and instead opt for the advanced and retarded null coordinates $(v,u)$. 

Below we let $(\bar{t},\bar{r})$ denote the Schwarzschild coordinates adapted to the nested Rindler region with associated advanced and retarded time coordinates $(\bar{v},\bar{u})$, defined in a completely equivalent way as for the larger Rindler wedge. The nested AdS-Rindler wedge is the domain of dependence of the time slice $\Sigma$ bounded by the bifurcation point $\mathcal B$ (defined   in Eq. \eqref{eq:bifpoint}) and the conformal boundary at $r\to\infty$, which has vanishing extrinsic curvature and is hence extremal. We focus on the ``right" AdS-Rindler wedge, neglecting the ``left" AdS-Rindler wedges, and we consider AdS-Rindler patches whose extremal slice $\Sigma$ is not located on the $t=0$ time slice, but rather on an arbitrary $t_{0}\neq0$ slice.  


\subsection*{Boost Killing vector of the nested Rindler wedge}

The generator of proper Rindler time $\bar{t}$ translations is the boost Killing vector of the nested Rindler wedge. We are interested in how this Killing vector is expressed in terms of the coordinates of the enveloping black hole. This problem is addressed in Appendix  \ref{app:embeddingformalism} using the  embedding formalism, where $\text{AdS}_{2}$ is embedded as a hyperboloid in a $(-,-,+)$ signature Minkowski space. The boost Killing vector  $\zeta=\partial_{\bar{t}}$  takes the following form in the   advanced and retarded time coordinates $(v,u)$ of the $\text{AdS}_{2}$ black hole
\begin{equation}
    \zeta 
    = 
    \frac{L\kappa/\sqrt{\mu}}
    {\sinh (\sqrt{\mu} \alpha / L)}  \Big[ (\cosh(\sqrt{\mu} \alpha / L)  -  \cosh (\sqrt{\mu} (v-t_0) / L))\partial_v+   ( \cosh(\sqrt{\mu} \alpha / L) -  \cosh(\sqrt{\mu} (u-t_0) /L)) \partial_u     \Big]\;.
\label{eq:boostzetanull} 
\end{equation}
Equivalently, in Schwarzschild coordinates $(t,r)$  we have 
\begin{equation}
\begin{split}
 \zeta = \frac{\kappa}{\sinh(\sqrt{\mu} \alpha / L)} &\Bigg[ \frac{L}{\sqrt{\mu}} \left ( \cosh\left(\frac{\sqrt{\mu}}{L} \alpha \right)- \frac{r / L}{\sqrt{\frac{r^2}{ L^2} - \mu}} \cosh \left(\frac{\sqrt{\mu}}{L} (t-t_0)\right)\right) \partial_t \\
&+ L \sqrt{\frac{r^2}{ L^2} - \mu} \sinh\left(\frac{\sqrt{\mu}}{L} (t-t_0)\right) \partial_r  \Bigg]\;.
 \end{split}
\label{eq:boostzetaschwarrz}\end{equation}
The   parameter $\alpha$ may be   interpreted as a boundary time interval. This can be seen via the following coordinate relationship between enveloping black hole coordinates $(v,u)$ and the nested Rindler wedge coordinates $(\bar{v},\bar{u})$:
\beq
  \frac{\sqrt{\mu}}{L}   u = \log \left [ \frac{e^{\sqrt{\mu}t_0/L} + e^{\kappa \bar u +   \sqrt{\mu}(t_0 + \alpha)/L}}{ e^{\sqrt{\mu} \alpha / L}+e^{\kappa\bar u}} \right]\;,\quad  \frac{\sqrt{\mu}}{L}v = \log \left [ \frac{e^{\sqrt{\mu}t_0 /L} + e^{\kappa \bar v    +\sqrt{\mu}(t_0+ \alpha) /L}}{ e^{\sqrt{\mu} \alpha / L}+e^{\kappa \bar v}} \right]\;,
\label{eq:transuvtobaruv}
\eeq
where $(\bar{v},\bar{u})\to(v,u)$ in the limit $\alpha\to\infty$ (upon identifying $\kappa = \sqrt{\mu}/L$). In the nested coordinates the boundary vertices, where the nested AdS-Rindler wedge intersects the global conformal AdS boundary, are located  at $\mathcal V_\pm = \{\bar u = \bar v =\pm \infty\}$, with the plus sign for the future vertex and the minus sign for the past vertex. Inserting these points in the coordinate transformation above we find that the vertices are   at  \beq\mathcal V_\pm = \{t=t_0 \pm\alpha, r=\infty\}\eeq in Schwarzschild coordinates, hence $\alpha$ is the boundary time interval between $\Sigma$ and $\mathcal V_\pm$.  In the limit $\alpha\to\infty$ the boundary time interval is identified with the entire AdS  boundary such that the nested Rindler wedge coincides with the exterior of the black hole. Moreover, in this limit we see the Killing vector~$\zeta$ (\ref{eq:boostzetaschwarrz}) is proportional to the generator of Schwarzschild time translations, 
\beq \zeta \big|_{\alpha\to\infty}\to\frac{\kappa L}{\sqrt{\mu}}\partial_{t}\;.\eeq
The past and future AdS-Rindler Killing horizons are located at the null surfaces where $\zeta$ becomes null, $\zeta^{2}\big |_{\mathcal H}=0$, which  in Schwarzschild coordinates   are given by
 \begin{equation}
 \mathcal H = \left\{   \cosh \left ( \frac{\sqrt{\mu}}{L}(t  - t_0 \pm\alpha)\right) = \frac{r / L}{\sqrt{r^2 / L^2 - \mu}} \right \}\;,
 \end{equation}
 where the plus sign corresponds to the past horizon and the minus sign to the future horizon.
 From the expression \eqref{eq:boostzetanull} for $\zeta$  in  terms of null coordinates we see the future horizon of the right Rindler wedge is at   $ u = t_0 + \alpha$ and the past horizon is at $v = t_0 - \alpha $. 
 
 Further, the boost  Killing vector $\zeta$ vanishes at   the future and past vertices  $\mathcal V_\pm$  and at the bifurcation point $\mathcal{B}$, the intersection of the future and past AdS-Rindler horizons, which is characterized in Schwarzschild coordinates   by 
 \begin{equation}
 \mathcal{B}=\left\{t =t_0,\; r = L \sqrt{\mu} \coth(\sqrt{\mu} \alpha / L) \right\}\;.
 \label{eq:bifpoint}\end{equation}
 The boundary intersection points $\mathcal V_\pm$ and the bifurcation surface $\mathcal B$ are thus fixed points of the flow generated by $\zeta$. Observe that in the $\alpha\to\infty$ limit the bifurcation surface $\mathcal{B}$ coincides with the bifurcation point of the black hole event horizon.  

A few more comments are in order. First, since $\zeta$ is the sum of two future null vectors in the interior of the (right) Rindler wedge, it is timelike and future directed. Moreover, $\zeta$ also acts outside the right Rindler wedge, and is in fact null on  four different Killing horizons $\{u=t_0\pm \alpha,v=t_0\pm \alpha\}$. It is also worth remarking,   that in the form (\ref{eq:boostzetanull}) $\zeta$ is identical to the \emph{conformal} Killing vector preserving a spherical causal diamond in a   maximally symmetric spacetime   \cite{Jacobson:2018ahi}. Thirdly, note the normalization of $\zeta$ is chosen such that the (positive) surface gravity, defined via $\nabla_{\mu}\zeta^{2}   = -2\kappa\zeta_{\mu}$    on the   (future) Killing horizon,  is equal to $\kappa$, which we keep arbitrary in the main body of the paper (in Appendix \ref{app:embeddingformalism} we set $\kappa=1$). 


\subsection*{Some useful derivative identities}

There are some useful observations we should remark on before we move to derive the  Smarr relation and first law associated with the nested Rindler wedge. First, in order to derive the first law we will need to know the Lie derivative of the dilaton $\phi$ with respect to the boost Killing vector $\zeta$. Explicitly, since $\phi$ is time-independent in Schwarzschild coordinates,
 \begin{equation}
  \mathcal L_\zeta \phi = \zeta^{r}\partial_{r}\phi =\frac{\kappa\phi_{r}}{\sinh (\sqrt{\mu}\alpha/L)} \sqrt{\frac{r^2}{L^2} - \mu} \sinh \left(\frac{\sqrt{\mu}}{L}(t-t_{0})\right)\quad \Rightarrow \quad \mathcal L_\zeta \phi \big |_\Sigma =0\;,
 \label{eq:Liedervphi}\end{equation}
where $\Sigma$ is defined as the time slice $t =t_0$. Thus, the boost Killing vector is an instantaneous symmetry of the dilaton at $\Sigma$, which will prove crucial momentarily. 

 
 In what follows we   also need to know the covariant derivative of the Lie derivative of $\phi$,  
 \begin{equation}
 \nabla_\nu \left (  \mathcal L_\zeta \phi \right)  \big |_\Sigma = -u_\nu \frac{\kappa \phi_r \sqrt{\mu}}{L\sinh(\sqrt{\mu}\alpha/L)} =-  u_\nu \frac{\kappa \phi_r }{L}  \sqrt{\frac{r_{\mathcal{B}}^2}{L^{2}} - \mu}\;, \label{eq:computationcovariantliephi}
 \end{equation}
where we used the value of $\alpha$ in terms of the bifurcation point (\ref{eq:bifpoint}), and identified $u^\nu$ as the future-pointing unit normal to $\Sigma$:
\begin{equation}
    u^\nu = \frac{1}{\sqrt{r^2 / L^2 - \mu}}  \partial_t^\nu\;, \qquad u_\nu =- \sqrt{\frac{r^2}{ L^2 }- \mu} \delta_\nu^t\;.
\label{eq:futurepointingnormal}\end{equation}
Given the \emph{proper length} $\ell$ of $\Sigma$ between the bifurcation point $\mathcal B$ of the nested Rindler horizon and the asymptotic boundary  
\beq
\ell = \int_{r_{\mathcal B} }^{ \infty} \frac{dr}{\sqrt{r^2/ L^2 - \mu}} =\lambda( \infty) - \lambda(r_{\mathcal B}) ,
\,\, \text{with} \,\,\,\lambda(r ) =  L \, \text{arctanh} \left ( \frac{r/L}{\sqrt{r^2/L^2 - \mu}} \right) ,
\eeq
we  can define the derivative of the dilaton at $\mathcal B$ with respect to the proper length $\ell$ by
\beq \phi'_{\mathcal{B}}\equiv\frac{\partial\phi_{\mathcal B}}{\partial\ell} = -\frac{\partial \phi_{\mathcal B}}{\partial\lambda(r_{\mathcal B})} =  - \frac{\phi_r }{L}  \sqrt{\frac{r_{\mathcal{B}}^2}{L^{2}} - \mu}\;,\label{eq:geodesiclength}\eeq
where in the last equality we used $\phi_{\mathcal B}= \frac{\phi_r}{L} r_{\mathcal B}$. 
Comparing with \eqref{eq:computationcovariantliephi} we thus find 
\begin{equation}
    u^\nu \nabla_\nu (\mathcal L_\zeta \phi) \big |_\Sigma=- \kappa \phi_{\mathcal{B}}'\;.
\label{eq:geodderiv}\end{equation}
Importantly, note that $\phi'_{\mathcal{B}}$ is \emph{constant} on $\Sigma$, a crucial property used to prove   the existence of a geometric form of the first law for AdS-Rindler space in (classical) JT gravity.\footnote{A similar property is necessary to prove the geometric form of the first law of spherical causal diamonds in higher-dimensional maximally symmetric spacetimes \cite{Jacobson:2015hqa,Jacobson:2018ahi}.}

Moreover, since they will be useful momentarily, it is straightforward to show
\beq \mathcal{L}_{\zeta}\chi\Big|_{\Sigma}=0\;,\label{eq:Liederivchi}\eeq
\beq 
    \nabla_{\nu}(\mathcal{L}_{\zeta}\chi)\Big|_{\Sigma}
    =
    -\frac{\kappa\sqrt{\mu}}{L}\frac{\tanh(\sqrt{\mu}\alpha/2L)}{ \frac{r}{L}-\sqrt{\frac{r^{2}}{L^{2}}-\mu} }u_{\nu}
    =
    -\frac{\kappa}{L}
    \frac{ \frac{
                r_{\mathcal{B}}
                }
                {L}
                -\sqrt{\frac{r_{\mathcal{B}}^{2}}{L^{2}}-\mu}}{\frac{r}{L}-\sqrt{\frac{r^{2}}{L^{2}}-\mu}}u_{\nu}\;.
\label{eq:covdliechiu}\eeq
The first equation holds for   the time-independent as well as   time-dependent solutions for $\chi$, whereas the second equation is specific to the time-\emph{dependent} solution $\chi^{(5)}$ (\ref{eq:chi5}), with constants (\ref{eq:constidsHH}) and where $t_{0}=t_{B}$. 
We observe $u^{\nu}\nabla_{\nu}(\mathcal{L}_{\zeta}\chi)\Big|_{\Sigma}$
is not constant on $\Sigma$. This result will play a role in the derivation of the semi-classical first law. 


  \subsection{Classical Smarr formula and   first law}
  
  Here we derive the \emph{classical} Smarr relation and associated first law for the nested AdS$_{2}$-Rindler wedge described above, leaving its semi-classical extension for the next section. Our derivation relies on the covariant phase space formalism \cite{Crnkovic:1986ex,Lee:1990nz,Wald:1993nt,Iyer:1994ys,Wald:1999wa}, particularly the methodology of   Wald   used to study the first law of black hole thermodynamics. In Appendix~\ref{app:Waldformalism} we briefly review Wald's Noether charge formalism and apply it to generic two-dimensional dilaton-gravity theories. We specialize to classical and semi-classical JT gravity in the next two sections. For previous applications of the covariant phase formalism to classical CGHS and JT models, see, \emph{e.g.}, \cite{NavarroSalas:1992vy,NavarroSalas:1994gs,Iyer:1994ys,Kummer:1995qv,Harlow:2019yfa}.
  

  \subsubsection{Smarr relation} \label{sec:classicalsmarr}
  
The Smarr identity \cite{Smarr:1972kt} is a fundamental relation between thermodynamic variables of a stationary black hole.  For example, in the case of a $D>3$ dimensional   AdS black hole of mass $M$ and horizon area $A$, 
the Smarr relation is  \cite{Kastor:2009wy}
  \beq \frac{D-3}{D-2}M=\frac{\kappa}{8\pi G}A-\frac{2\Lambda}{(D-2)8\pi G} \Theta
  \;,\label{eq:SmarrgenBH}\eeq
  where $\Theta\equiv  8\pi G\left(\frac{\partial M}{\partial\Lambda}\right)_A$ is the quantity conjugate to the cosmological constant $\Lambda$. In the literature (minus) $\Theta$ is better known as the ``thermodynamic volume'' since $-\Lambda/8\pi G$ is interpreted as a bulk pressure. Naively, from expression \eqref{eq:SmarrgenBH} it appears a Smarr relation  is ill defined in $D=2$ spacetime dimensions. Nonetheless, a Smarr identity for $(1+1)$-dimensional AdS black holes does exist, see for instance \eqref{eq:classicaljtsmarrfirst}, but it has to be derived separately for $2D$ dilaton gravity (see also \cite{Grumiller:2014oha,Frassino:2015oca}). 

  Here we derive the Smarr identity for the $(1+1)$-dimensional nested Rindler wedge by equating the Noether   current $j_{\zeta}$ associated with diffeomorphism symmetry generated by the boost Killing vector $\zeta$ to the exterior derivative of the associated Noether charge $Q_{\zeta}$, and integrating over the Cauchy slice $\Sigma$ at $t=t_{0}$. A similar generic  method  was used in the context of Lovelock black holes \cite{Liberati:2015xcp},   spherical causal diamonds in (A)dS \cite{Jacobson:2018ahi}, and for   higher curvature gravity theories coupled to  scalar and vector fields \cite{Ortin:2021ade}, all of which are based on Wald's Noether charge formalism \cite{Wald:1993nt}.  We mostly follow the notation of \cite{Jacobson:2018ahi}.
  
We recall the Noether current 1-form $j_{\zeta}$   is defined as  
  \beq j_{\zeta}\equiv \theta(\psi,\mathcal{L}_{\zeta}\psi)-\zeta\cdot L\;,\label{eq:noethercurrentgen} \eeq
  where $\theta$ is the symplectic potential 1-form, $\psi$ denotes a collection of dynamical fields (namely, the metric $g_{\mu\nu}$ and dilaton $\phi$), $L$ is the Lagrangian 2-form, whose general field variation is $\delta L=E^{\psi}\delta\psi+d\theta(\psi,\delta\psi)$, and $\delta_\zeta \psi=\mathcal{L}_{\zeta}\psi$ is the field variation induced by the flow of vector field~$\zeta$. For diffeomorphism covariant theories, for which $\delta_\zeta L = \mathcal L_\zeta L$,   the Noether current is closed for all $\zeta$ when the equations of motion hold, $E_{\psi}=0$, and hence $j_{\zeta}$ can be cast as an exact form, 
  \beq j_{\zeta}=dQ_{\zeta}\;, \label{eq:jdQ}\eeq
  where $Q_{\zeta}$ is the Noether charge 0-form. 
  
  The Smarr relation follows from the integral version of (\ref{eq:jdQ}), 
  \beq \int_{\mathcal{R}}j_{\zeta}=\int_{\mathcal{R}} dQ_{\zeta}=\oint_{\partial\mathcal{R}}Q_{\zeta}\;,\label{eq:intSmarrgen}\eeq
  where, generically, $\mathcal{R}$ is a codimension-one submanifold with boundary $\partial \mathcal{R}$. The second equality follows from Stokes' theorem. For a black hole with a bifurcate Killing horizon, $\zeta$ is taken to be the  horizon generating Killing   field, and $\mathcal{R}$ the hypersurface extending from the bifurcation surface to spatial infinity. Using $\mathcal{L}_{\zeta} g_{\mu \nu}=0$   one arrives at the Smarr relation~(\ref{eq:SmarrgenBH}) for stationary   black holes \cite{Jacobson:2018ahi}. In our case, where $\zeta$ is the boost Killing vector associated with the nested Rindler wedge, we have in addition $\mathcal{L}_{\zeta}\phi=0$  (\ref{eq:Liedervphi}) only along the  $t=t_{0}$ 1-dimensional hypersurface $\Sigma$ extending from the bifurcatation point $\mathcal{B}$ to the asymptotic  boundary. Thus, 
\beq \int_{\Sigma}j_{\zeta}=\oint_{\infty} Q_{\zeta}-\oint_{\mathcal{B}}Q_{\zeta}\;,\label{eq:Smarrintformgenus}\eeq
where the orientation of the Noether charge integral on $\mathcal B$ is chosen to be outward, towards spatial infinity. The Noether charge integral over either location in two dimensions can be  replaced by evaluating $Q_{\zeta}$ at either point $\oint_{\infty}Q_{\zeta}=\lim_{r\to\infty}Q_{\zeta}$ and $\oint_{\mathcal{B}}Q_{\zeta}=\lim_{r\to r_{\mathcal{B}}}Q_{\zeta}$.

Let's first focus on the right-hand side of (\ref{eq:Smarrintformgenus}). As derived in Appendix \ref{app:Waldformalism}, the Noether charge    (\ref{eq:Noethercharge}) for JT gravity is
\beq Q_{\zeta}^{\text{JT}}=-\frac{1}{16\pi G}\epsilon_{\mu\nu}\left[(\phi+\phi_{0})\nabla^{\mu}\zeta^{\nu}+2\zeta^{\mu}\nabla^{\nu}(\phi+\phi_{0})\right]\;,\label{eq:QnoetherJT}\eeq
where $\epsilon_{\mu\nu}$ is the binormal volume form for $\partial\Sigma$, which we may write in terms of the timelike and spacelike unit normals, $u_{\mu}$ and $n_{\mu}$, respectively, such that $\epsilon_{\mu\nu}=-2u_{[\mu}n_{\nu]}$ (where we used $\epsilon_{\partial \Sigma }=1$ in two dimensions).
At the bifurcation point $\mathcal{B}$  the boost Killing vector vanishes, $\zeta|_{\mathcal{B}}=0$, and $\nabla_{\mu}\zeta_{\nu}|_{\mathcal{B}}=\kappa \epsilon_{\mu\nu}$, where $\kappa$ is the surface gravity. Since $\epsilon_{\mu\nu}\epsilon^{\mu\nu}=-2$, we find
\beq \oint_{\mathcal{B}}Q_{\zeta}^{\text{JT}}=\frac{\kappa}{8\pi G}(\phi_{0}+\phi_{\mathcal{B}})\;,\label{eq:intNoetherchargeJT}\eeq
where $\phi_{\mathcal{B}}$ is the value of the dilaton   at the bifurcation  point. In the limit the nested Rindler horizon coincides with the black hole horizon, $\mathcal{H}\to H$, we recover the ``area" term. That is, had we been working with general relativity, this contribution gives  $\kappa A_{H}/8\pi G$, such that one interprets $\phi_{0}+\phi_{H}$ as the horizon ``area" of the higher-dimensional black hole from which JT gravity is reduced. 
Meanwhile, the Noether charge evaluated at the  intersection of the spatial boundary $B$ and the extremal slice $\Sigma$, given by the point $\{t=t_0,r=r_B \}$ in Schwarzschild coordinates, is
\beq
Q_{\zeta}^{\text{JT}}\Big |_{ \{t_0,r_B\}  }=\frac{\kappa\phi_{0}\coth(\sqrt{\mu}\alpha/L)}{8\pi GL\sqrt{\mu}}r_B-\frac{\kappa\phi_{0}}{8\pi G \sqrt{\mu}}\frac{\sqrt{r_B^{2}/L^2-\mu  }}{\sinh(\sqrt{\mu}\alpha/L)}+\frac{\kappa\sqrt{\mu}}{8\pi G}\coth(\sqrt{\mu}\alpha/L)\phi_{r}\;,
\label{eq:QzetaJTclass}
\eeq
which  diverges as $r_B\to\infty$. Thus, the first term on the right-hand side (\ref{eq:Smarrintformgenus}) is
\beq \oint_{\infty} Q^{\text{JT}}_{\zeta} =\lim_{r_B\to\infty}  \frac{\kappa\phi_{0} \tanh(\sqrt{\mu}\alpha/2L)}{8\pi GL\sqrt{\mu}} r_B + \frac{\kappa \sqrt{\mu} \phi_r \coth(\sqrt{\mu}\alpha/L)}{8 \pi G}  \;.\label{eq:RHSclassJTsmarr}\eeq
We will return to the divergent contribution momentarily.

We now evaluate the left-hand side of (\ref{eq:Smarrintformgenus}), where we need the form of the Noether current $j_{\zeta}$. We can either work directly with the current  (\ref{eq:Noethercurrent})  specific to JT gravity, or,  use general the form (\ref{eq:noethercurrentgen}) together with the symplectic potential (\ref{eq:sympotgen})   and the Lagrangian  
\beq L_{\text{JT}}=\frac{\epsilon}{16\pi G}\left[(\phi_{0}+\phi)R+\frac{2}{L^{2}}\phi\right]\;, \label{eq:lagrform}\eeq
where $\epsilon$ is the volume form for the full spacetime manifold. In the current context it is simpler to work with the latter since $\zeta$ is a Killing field, such that $\mathcal{L}_{\zeta}g_{\mu\nu} =\nabla_{\nu}\mathcal{L}_{\zeta}g_{\mu\nu} =0$, and the symplectic potential $\theta (\psi, \delta \psi)$ (\ref{eq:sympotgen}) for JT gravity is linear in $\delta g_{\alpha \beta}$ and $\nabla_\nu \delta g_{\alpha \beta}$, hence
\beq \theta_{\text{JT}}(\psi,\mathcal{L}_{\zeta}\psi)=0\;.\eeq
The Noether current (\ref{eq:noethercurrentgen}) is therefore 
\beq j^{\text{JT}}_{\zeta}=-\zeta \cdot L_{\text{JT}}=\frac{\phi_{0}}{8\pi GL^{2}}(\zeta\cdot \epsilon)\;,\eeq
where the Lagrangian is taken on-shell in the last equality by inserting  $R=-\frac{2}{L^{2}}$.
Consequently, the left-hand side of the relation (\ref{eq:Smarrintformgenus}) is 
\beq \int_{\Sigma}j^{\text{JT}}_{\zeta}=\frac{\phi_{0}}{8\pi GL^{2}}\int_{\Sigma}|\zeta|d\ell\;,\eeq
where we used $\int_{\Sigma}\zeta\cdot\epsilon=\int_{\Sigma}|\zeta|d\ell$, with $d\ell$ being the infinitesimal proper length   $d\ell=\frac{dr}{\sqrt{r^{2}/L^{2} -\mu}}$, and $|\zeta| \equiv \sqrt{- \zeta \cdot \zeta}$ is the norm of the Killing vector $\zeta$. It is straightforward to show this integral is divergent as $r\to\infty$. 

To remove the undesired divergences in the integrals of $j^{\text{JT}}_{\zeta}$ and $Q^{\text{JT}}_{\zeta}|_{\infty}$, one can either introduce local (boundary) counterterms \cite{Balasubramanian:1999re,Emparan:1999pm}, or regulate with respect to the extremal ($\mu=0$) background. The latter method is a standard regularization technique in the Smarr formula for AdS black holes \cite{Kastor:2009wy}, however, here we use the local counterterm method, which was employed in \cite{Visser:2021eqk} to derive a finite Smarr formula.   Specifically, as in \eqref{eq:GHYctterms} we add a   boundary term to the JT action, such that the total action becomes     
\beq I_{\text{JT}} = \int_M L_{\text{JT}} - \oint_{B} b_{\text{JT}}\, ,  \qquad \text{where} \qquad b_{\text{JT}}=-\frac{\epsilon_B}{8\pi G}\left[(\phi_{0}+\phi)K-\frac{\phi}{L}\right]\; \label{eq:bdryformandct}\eeq
is the boundary Lagrangian 1-form on the timelike boundary $B$ of $M$, which contains both the Gibbons-Hawking boundary term (the first term proportional to $K$) and a local counterterm (the second term). Here, $\epsilon_{B}$ is the volume form on $B$, and   $K$ is the trace of the extrinsic curvature of $B$, which restricted to $t=t_0$ in Schwarzschild coordinates takes the form 
\beq K=\gamma^{\mu\nu}\nabla_{\mu}n_{\nu} \Big |_{\Sigma }=\frac{\frac{r_B}{L^2} \cosh ( \sqrt{\mu } \alpha  /L )-\frac{1}{L}\sqrt{r_B^2/L^2-\mu   }}{   \sqrt{\left(r_B^2/L^2-\mu   \right) \cosh ^2( \sqrt{\mu } \alpha  /L )-2 \frac{r_B}{L} \sqrt{r_B^2/L^2-\mu  } \cosh ( \sqrt{\mu } \alpha  /L )+r_B^2/L^2}}
   \;,\eeq
with $n^{\nu}$   the unit normal vector (\ref{eq:normalnvec}) to  boundary $B$. Importantly, vector $n^{\mu}$ is not necessarily  proportional to $\partial^{\mu}_{r}$ everywhere, since by definition it is orthogonal to the vector field $\zeta^\mu$,   leading to the complex expression for $K$ above. In the  $\alpha\to\infty$ limit the trace of the extrinsic curvature simplifies to $K=\frac{r_B/L^2}{ \sqrt{r_B^{2}/L^2-\mu}}$ and the normal to $n^{\mu}= \sqrt{r^2_B / L^2 - \mu} \partial^{\mu}_{r}$.   

Next, we regulate the divergences in the on-shell integral identity \eqref{eq:Smarrintformgenus} by subtracting on both sides of the equation  (the integral of) the 0-form $\zeta \cdot b_{\text{JT}}$ at spatial infinity. This term arises as follows. Subtracting an exact form from the Lagrangian, $L \to L - db$, is known to modify    the symplectic potential, $\theta \to \theta - \delta b $,  the Noether current, $j_\zeta \to j_\zeta - d(\zeta \cdot b)$, and the Noether charge, $Q_\zeta \to Q_\zeta - \zeta \cdot b$ \cite{Iyer:1994ys}. In the present case,   we     add  a boundary term to the action only at the spatial boundary $B$, and we take the limit  $B \to \infty$ to the asymptotic timelike boundary.  This regulates the Noether current and charge       at infinity, and not at the inner boundary~$\mathcal B$ of $\Sigma$,\footnote{At the bifurcation point $\mathcal B$ we find that $\zeta \cdot b  $ is nonzero, since $\zeta \to 0$ while $K \to \infty$ at $\mathcal B$, and is given by $\zeta \cdot b_{\text{JT}} |_{\mathcal B}=- \frac{\kappa}{8\pi G} (\phi_0 + \phi_{\mathcal B} )= - \oint_{\mathcal B} Q_\zeta^{\text{JT}}$. We could have added it to the regularized integral identity, since it yields the same Smarr formula, but we think it is more natural to only add boundary terms at infinity.} such that the regularized integral identity becomes:
\beq \int_{\Sigma}j_{\zeta}-\oint_\infty\zeta\cdot b =\oint_{\infty}(Q_{\zeta}-\zeta\cdot b)-\oint_{\mathcal{B}}Q_{\zeta}\;. \label{eq:regulatedintegralidentity}\eeq
Specifically,
\beq
\zeta \cdot b_{\text{JT}} \big |_\infty =- \lim_{r_B \to \infty} \frac{\kappa \phi_0 \tanh (\sqrt{\mu}\alpha/2L)}{8 \pi G L \sqrt{\mu}} r_B - \frac{\kappa \sqrt{\mu} \phi_r}{16 \pi G} \coth(\sqrt{\mu}\alpha/2L)\;,
\eeq
where we used $\zeta \cdot \epsilon_B \big |_{\Sigma} =- \zeta^\mu u_\mu$. Adding this to $Q_{\zeta}^{\text{JT}}$ yields a finite total Noether charge at infinity and hence the regulated version of (\ref{eq:RHSclassJTsmarr}) becomes
\beq
\oint_{\infty} (Q_{\zeta}^{\text{JT}}-\zeta\cdot b_{\text{JT}} )
=\frac{\kappa \sqrt{\mu} \phi_r}{16 \pi G}\tanh (\sqrt{\mu}\alpha/2L) = E_\zeta^{\phi_r}\;,
\label{eq:Qzetatot}\eeq
where $E_\zeta^{\phi_r}$ is the asymptotic energy   (\ref{eq:asympenergyclassJTapp}), which in the limit $\alpha \to \infty$ reduces to  the classical ADM energy (\ref{eq:ADMmass}) of the black hole, $\lim_{\alpha\to\infty}E_{\zeta}^{\phi_r}=M_{\phi_r}$.  

Meanwhile, the regulated Noether current  is now
\beq \int_{\Sigma} j_{\zeta}^{\text{JT}}-\oint_\infty \zeta\cdot b_{\text{JT}} \equiv\frac{\phi_{0}}{8\pi GL^{2}}\Theta^G_{\zeta}\;,\label{eq:Noethercurrentjzeta}\eeq
where we have introduced the counterterm subtracted ``$G$-Killing volume" $\Theta^G_{\zeta}$, formally defined by
\beq \Theta^G_{\zeta}\equiv  \int_{\Sigma}\zeta\cdot\epsilon -\frac{8\pi GL^{2}}{\phi_{0}} \oint_{\infty} \zeta \cdot b_{\text{JT}}\;. \label{eq:thermovoldef}\eeq
We refer to $\Theta^G_{\zeta}$ as a Killing volume since the first integral is the   proper volume of $ \Sigma$ locally weighted by the norm of the boost Killing vector.   It is analogous to the background subtracted ``Killing volume'' $\Theta$, a.k.a. ``thermodynamic volume'', in   ordinary AdS black hole mechanics  \cite{Jacobson:2018ahi,Kastor:2009wy}. However, we will see in Section \ref{sec:extendedfirst} that $\Theta^G_\zeta$ is   not the conjugate quantity to the cosmological constant in the extended first law, as is usual with the thermodynamic volume, but rather   to Newton's constant. That is why we    have added a superscript $G$ on $\Theta^G_\zeta$.   Explicitly, we find
\beq \Theta^G_{\zeta}=-\frac{\kappa L^{2}\sqrt{\mu}\phi_{r}}{2\phi_{0}}\coth(\sqrt{\mu}\alpha/2L)-\kappa L^{2}\;.
\label{eq:thermovolJT}\eeq
 Therefore, combining  (\ref{eq:Qzetatot}) and (\ref{eq:Noethercurrentjzeta}) we arrive to the classical Smarr relation for nested Rindler wedges:
 \beq E_{\zeta}^{\phi_r}=\frac{\kappa}{8\pi G}(\phi_{0}+\phi_{\mathcal{B}})-\frac{\phi_{0}\Lambda}{8\pi G}\Theta^G_{\zeta}\;.\label{eq:smarrJTv2}\eeq
 Inserting  expression \eqref{eq:thermovolJT} for the Killing volume  we obtain a different form of the Smarr law
 \beq E_{\zeta}^{\phi_r}+\frac{\kappa\sqrt{\mu}\phi_{r}}{16\pi G}\coth(\sqrt{\mu}\alpha/2L)=\frac{\kappa}{8\pi G}\phi_{\mathcal{B}}\;.\label{eq:SmarrJT}\eeq
From this form of the Smarr   law,  in  the limit $\alpha\to\infty$   where $r_{\mathcal{B}}\to r_{H}$, and identifying $\kappa=\sqrt{\mu}/L$,  we recover  the original Smarr relation (\ref{eq:classicaljtsmarrfirst})  for the eternal $\text{AdS}_{2}$ black hole. In particular the left-hand side of \eqref{eq:SmarrJT} becomes equal to $2M_{\phi_r}$ in the $\alpha \to \infty $ limit.  Notably, the Smarr relation is not of the standard form in higher dimensions (\ref{eq:SmarrgenBH}), and is not identical to the $(1+1)$-dimensional Smarr relation   in Eq. (4.7) of \cite{Frassino:2015oca}, which differs by a factor of two from our equation \eqref{eq:smarrJTv2}. The difference between the $2D$ Smarr formula  in \cite{Frassino:2015oca} and our Smarr relation    is that in the former one rescales the $D$-dimensional Newton's constant $G$ by $(D-2)$ so as to have a well defined $D\to2$ limit. Here we made no such rescaling.

  
  \subsubsection{First law of nested  AdS-Rindler wedge mechanics}
  
The Smarr relation and the first law of black hole mechanics can be derived from each other for stationary black holes in general relativity. In fact, using Euler's scaling theorem for homogeneous functions,  Smarr \cite{Smarr:1972kt} originally used the first law of black holes to derive his relation,   identifying the ADM mass as the homogeneous function. Here we derive the first law for the nested AdS-Rindler wedge,  which is an extension  of the first law \eqref{eq:classicaljtsmarrfirst} for the eternal AdS$_2$ black hole.   As it turns out to be difficult to derive this first law from the Smarr relation~\eqref{eq:smarrJTv2},  we follow Wald's method of deriving the first law of   black holes \cite{Wald:1993nt} by varying the integral identity (\ref{eq:intSmarrgen}). The variations under consideration are arbitrary variations of the dynamical fields $\psi$ (the metric and dilaton) to nearby solutions, whilst keeping the Killing   field $\zeta$, and the extremal slice $\Sigma$ of the unperturbed     wedge  fixed.

On-shell, the variation of the Noether current 1-form (\ref{eq:noethercurrentgen}) is  formally given by the  fundamental variational identity   \cite{Wald:1993nt,Iyer:1994ys}
\beq \delta j_{\zeta}=\omega(\psi,\delta\psi,\mathcal{L}_{\zeta}\psi)+d(\zeta\cdot\theta(\psi,\delta\psi))\;,\eeq
with $\omega(\psi,\delta_{1}\psi,\delta_{2}\psi)\equiv\delta_{1}\theta(\psi,\delta_{2}\psi)-\delta_{2}\theta(\psi,\delta_{1}\psi)$   the symplectic current 1-form. This follows from the variation of the Lagrangian 2-form, $\delta L = E^\psi \delta \psi + d \theta (\psi, \delta \psi)$, and from   Cartan's magic formula applied to the symplectic potential, $\mathcal L_\zeta \theta = d(\zeta \cdot \theta) + \zeta \cdot d \theta.$ Substituting $\delta j_{\zeta}$ into the variation of the integral identity (\ref{eq:intSmarrgen}) we obtain
 \begin{equation}
     \int_\Sigma \omega (\psi, \delta \psi, \mathcal L_\zeta \psi) = \oint_{\partial \Sigma} [\delta Q_\zeta - \zeta \cdot \theta (\psi, \delta \psi)]\;.
\label{eq:variationalidJT1} \end{equation}
where we have taken the hypersurface $\mathcal R$ to be the extremal slice $  \Sigma$. By Hamilton's equations the left-hand side is equal to the variation of the Hamiltonian $H_\zeta$ generating evolution along the flow of $\zeta$, 
\beq
\delta H_\zeta = \int_\Sigma \omega (\psi, \delta \psi, \mathcal L_\zeta \psi) \,,  \label{eq:hameqns}
\eeq
hence we find an integral variational identity stating the Hamiltonian variation is a boundary charge on-shell
\beq
\delta H_\zeta = \oint_{\partial \Sigma} [\delta Q_\zeta - \zeta \cdot \theta (\psi, \delta \psi)]\;.
\eeq
 We first evaluate the right-hand side. As before, $\partial\Sigma$ consists of two points, the bifurcation point $\mathcal{B}$ and a point at the asymptotic boundary $B\to\infty$. At spatial infinity on $\Sigma$   the boost Killing vector (\ref{eq:boostzetaschwarrz})  is proportional to the Schwarzschild time-translation Killing vector, \emph{i.e.}, $\lim_{r \to \infty,t \to t_0} \zeta= \mathcal A \partial_{t}$.
 As reviewed in Appendix \ref{app:Waldformalism}, the variation of the form $\delta Q_\zeta - \zeta \cdot \theta$ evaluated at spatial infinity is equal to the  asymptotic energy $\delta E_{\zeta}^{\phi_r}$.\footnote{Following \cite{Harlow:2019yfa}, in Appendix \ref{app:Waldformalism} we subtract the combination $\delta C(\mathcal L_\zeta \psi) - \mathcal L_\zeta C(\delta \psi)$, evaluated at the spatial boundary $B$, from both sides of   equation \eqref{eq:variationalidJT1}, where the $0$-form $C$ is defined   via the restriction of the symplectic potential to   $B$, $\theta |_B = \delta b + dC$, and Dirichlet boundary conditions are imposed. This results in a different definition of the asymptotic energy variation, since $[\delta Q_\zeta - \zeta \cdot \theta(\psi, \delta \psi) - \delta C(\psi, \mathcal L_\zeta \psi) + \mathcal L_\zeta C(\psi, \delta \psi) ] \big |_B = \delta [Q_\zeta - \zeta \cdot b - C(\psi, \mathcal L_\zeta \psi)]$ where we used  Cartan's magic formula and the restriction of $\theta $ on the asymptotic spatial boundary. With weak Dirichlet conditions, $\delta \phi |_B =0={\gamma_\alpha}^{\mu} {\gamma_\beta}^{\nu} \delta  \gamma_{\mu \nu} |_B$, the $C$ term contributes non-trivially to the energy, but with stronger boundary conditions, $\delta \phi |_B =0= \delta \gamma_{\mu \nu} |_B$, the $C$ term vanishes identically. In the main body of the paper we take the latter perspective, and in the Appendix we allow for   weaker conditions. \label{fn:bdrycond}}
 Explicitly, referring to (\ref{eq:asympenergyclassJTapp}) we find that the asymptotic energy is equal to  the product of the normalization $\mathcal A$ and the ADM mass associated to the  time-translation generator $\partial_t$, and its variation is hence
 \begin{equation}
  \delta E_\zeta =  \oint_\infty [\delta Q_\zeta^{\text{JT}} - \zeta \cdot \theta_{\text{JT}} (\psi, \delta \psi)] = \delta \oint_\infty (Q_\zeta^{\text{JT}} - \zeta \cdot b_{\text{JT}} ) =\delta (\mathcal{A}  M_{\phi_{r}})\;, \label{eq:energyvariationmain}
 \end{equation}
 where $\mathcal{A}\equiv (\kappa L/ \sqrt{\mu}) \tanh ( \sqrt{\mu}\alpha/2L)$ and $M_{\phi_{r}}$ is the classical   mass \eqref{eq:ADMmass} of the static black hole. In the last equality we inserted equation \eqref{eq:Qzetatot}. Notice when $\mathcal{A}=1$, \emph{i.e.}, when $\alpha\to\infty$ and $\kappa = \sqrt{\mu }/L$, we recover the expected variational result for black holes \cite{Iyer:1994ys,Papadimitriou:2005ii}. Furthermore, if the form $\delta Q_\zeta - \zeta \cdot \theta$ on  the right-hand side of the variational identity is evaluated at the bifurcation point $\mathcal{B}$ (where $\zeta=0$, and $\kappa$ is constant) we have from the Noether charge (\ref{eq:intNoetherchargeJT}), 
 \beq \oint_{\mathcal{B}} [\delta Q^{\text{JT}}_\zeta - \zeta \cdot \theta_{\text{JT}} (\psi, \delta \psi)]=\delta\oint_{\mathcal{B}} Q^{\text{JT}}_{\zeta}=\frac{\kappa}{8\pi G}\delta \phi_{\mathcal{B}} \;,\eeq
 where the orientation of the integral is outward, towards the AdS boundary. If we allow for variations of the couplings $\phi_0$ and $G$, then the Noether charge variation at $\mathcal B$ would be equal to $\frac{\kappa}{2\pi}\delta(\frac{\phi_{0}+\phi_{\mathcal{B}}}{G}) $.
Thus, 
\beq \oint_{\partial \Sigma} [\delta Q_\zeta^{\text{JT}} - \zeta \cdot \theta_{\text{JT}} (\psi, \delta \psi)]=\delta E_{\zeta}^{\phi_r}-\frac{\kappa}{8\pi G}\delta \phi_{\mathcal B}\;.\label{eq:RHSfirstlawJT}\eeq
Let us now turn to the left-hand side of (\ref{eq:variationalidJT1}), where we must evaluate the symplectic current $\omega$ on the Lie derivative of the fields along the Killing vector $\zeta$. In general relativity the symplectic current $\omega (g, \delta g, \mathcal L_\zeta g)$  is linear in $\mathcal L_\zeta g$ and hence vanishes when $\zeta$ is a Killing vector, whereas in JT gravity, due to the nonminimal coupling of the dilaton, the current $\omega (g, \delta g, \mathcal L_\zeta g)$  is  generically nonzero for a Killing vector $\zeta$ since the dilaton is not everywhere invariant under the flow of $\zeta$. The full symplectic current   (\ref{eq:sympcurrentgen}), specialized to JT gravity, is
\begin{align}
	\omega_{\text{JT}}&(\psi,\delta_{1}\psi,\delta_{2}\psi)  = \frac{\epsilon_\mu }{16\pi G}\Big [ (\phi_{0}+\phi)S^{\mu \alpha \beta \nu \rho \sigma}\delta_1 g_{\rho \sigma} \nabla_\nu \delta_2 g_{\alpha \beta} + \frac{1}{2} g^{\mu \beta} g^{\alpha \nu} g^{\rho \sigma} \delta_1 g_{\rho \sigma} \delta_2 g_{\alpha\beta} \nabla_\nu (\phi_{0}+\phi) \nonumber \\
	&+ (g^{\mu \beta}g^{\alpha\nu} - g^{\mu \nu} g^{\alpha \beta}) [\delta_1( \phi_{0}+\phi)  \nabla_\nu \delta_2 g_{\alpha \beta} - \delta_1 (\nabla_\nu (\phi_{0}+\phi))\delta_2 g_{\alpha \beta}]- [1 \leftrightarrow 2] \Big ]\;, \label{eq:sympcurrentJT} 
\end{align}
where the tensor $S^{\mu \alpha \beta \nu \rho \sigma}$ is given by \eqref{eq:tensorS}.  Identifying $\delta_{1}\equiv\delta$ and $\delta_{2}\equiv\mathcal{L}_{\zeta}$ and using $\mathcal{L}_{\zeta}g_{\mu \nu} =\nabla_{\nu}\mathcal{L}_{\zeta}g_{\alpha\beta} =0$ and $\mathcal{L}_{\zeta}\phi|_{\Sigma}=0$ (\ref{eq:Liedervphi}), we find the   expression above dramatically reduces when evaluated on the Lie derivative along $\zeta$ and restricted to~$\Sigma$:
\beq 
\begin{split}
\omega_{\text{JT}}(\psi,\delta\psi,\mathcal{L}_{\zeta}\psi) \Big |_{\Sigma} 
 =\frac{\epsilon_{\mu}u^{\mu}}{16\pi G}(u^{\nu}h^{\alpha\beta}-u^{\beta}h^{\alpha\nu})(\nabla_{\nu}\mathcal{L}_{\zeta}\phi)\delta g_{\alpha\beta}  
 =-\frac{\kappa\phi'_{\mathcal{B}}}{16\pi G}h^{\alpha\beta}\delta g_{\alpha\beta}(u\cdot \epsilon)  \;.
\end{split}
\label{eq:sympcurrentsteps}\eeq
 Here we decomposed the metric on $\Sigma$, $g_{\mu\nu}=-u_{\mu}u_{\nu}+h_{\mu\nu}$, where $u^\nu$ is the future-pointing unit normal to $\Sigma$   and $h_{\mu \nu}$ is the induced metric. In the second equality we used    
  identity (\ref{eq:geodderiv}) together with $\epsilon_{\mu}|_{\Sigma}=-u_{\mu}(u\cdot\epsilon)$. Finally, we note $(u\cdot \epsilon)h^{\alpha\beta}\delta g_{\alpha\beta}= 2\delta(u\cdot\epsilon)= 2\delta(d \ell)$, where $d\ell$ is the infinitesimal proper length.   Since both $\phi'_{\mathcal{B}}$ in \eqref{eq:geodesiclength} and $\kappa$ are constant over~$\Sigma$, we may pull them out of the integral such that, 
\beq \delta H_{\zeta}^{\phi_r}\equiv\int_{\Sigma}\omega_{\text{JT}}(\psi,\delta\psi,\mathcal{L}_{\zeta} \psi) = -\frac{\kappa\phi'_{\mathcal{B}}}{8\pi G}\delta\ell\;.\label{eq:LHSfirstlawJT}\eeq
Notice this term arises since $u^\nu \nabla_\nu \mathcal L_\zeta \phi $, given by  (\ref{eq:geodderiv}), does \emph{not} vanish at $\Sigma$. Thus, combining (\ref{eq:RHSfirstlawJT}) and (\ref{eq:LHSfirstlawJT}), we arrive to the classical mechanical first law of the nested Rindler wedge,
  \beq \delta E_{\zeta}^{\phi_r}=\frac{\kappa}{8\pi G} \left(\delta \phi_{\mathcal{B}}- \phi'_{\mathcal{B}} \delta\ell \right)\;.\label{eq:firstlawJTclass}\eeq
As a reminder to the reader we recall the notation used here: $E^{\phi_r}_\zeta$ is the asymptotic energy associated to the boost Killing vector $\zeta$ of the nested AdS-Rindler patch, $\kappa$ is the surface gravity, $\phi_{\mathcal B}$ is the value of the dilaton at the bifurcation point $\mathcal B$ of the nested AdS-Rindler Killing horizon, $\phi'_{\mathcal B}$ is the derivative of the dilaton at $\mathcal B$ along   $\ell$, and $\ell$ is the proper ``volume'' of the extremal slice $\Sigma$ with endpoints at the bifurcation point and spatial infinity.

Several comments are in order. First, notice in the limit $\alpha\to\infty$ we have $\phi'_{\mathcal{B}}=0$ and hence we recover the usual first law \eqref{eq:classicaljtsmarrfirst} for  black holes in JT gravity. More generally, the proper length variation contribution is non-zero and is   analogous to the proper volume variation of a spherical region $\Sigma$ in the first law of (A)dS causal diamonds \cite{Jacobson:2018ahi}, where, however, $\zeta$ is the conformal Killing vector preserving the diamond. It would be interesting to understand this analogy in more detail. We further emphasize the $\delta\ell$ contribution is absent in the higher-dimensional first law of AdS-Rindler space in general relativity (c.f. \cite{Faulkner:2013ica}). 

Importantly, as a consistency check, we   verify   the first law   holds if  all variations are induced by changing the  proper length $\ell$ of $\Sigma$. The asymptotic energy stays the same if the proper length varies, $\delta_{\ell} E_\zeta=0$, hence the left-hand side of the first law vanishes, while the  dilaton variation becomes $\delta_\ell \phi_{\mathcal B} = \frac{\partial \phi_{\mathcal B}}{\partial \ell}\delta \ell = \phi'_{\mathcal B} \delta \ell$, thus the right-hand side is also zero.
This explains why the coefficient of   $\delta \ell$   is given by $ -\phi'_{\mathcal B}.$  

Further, a first law for JT gravity was previously derived in Eq. (5.7) of \cite{Callebaut:2018xfu}, where the AdS-Rindler wedge was extended to a causal diamond, and was shown to encode the dynamics of the kinematic space of a two-dimensional CFT. Notably,   the variation of the asymptotic energy is absent, since there is no asymptotic infinity, whereas a variation of the matter Hamiltonian is included, identical to the term $\delta H_\zeta^{\text{m}}$ in \eqref{eq:firstlawwithmatter} below. Moreover, the variation of the metric is taken to be zero, such that there is no variation of the proper length as in our first law.

There are also a number of ways in which the first law (\ref{eq:firstlawJTclass}) may be generalized or extended. Firstly, one may consider adding in classical matter contributions,  where now the variation of the Hamiltonian \eqref{eq:hameqns} includes a contribution coming from the matter Hamiltonian $\delta H^{\text{m}}_{\zeta}$, characterized by an additional matter energy-momentum tensor $T_{\mu\nu}^{\text{m}}$ which vanishes in the AdS background, and can be cast as $\delta H^{\text{m}}_{\zeta}=-\int_{\Sigma}\delta( {T_{\mu}}^\nu)^{\text{m}} \zeta^{\mu}\epsilon_{\nu}$ \cite{Iyer:1996ky,Jacobson:2018ahi}. The classical first law with matter Hamiltonian variation reads
 \beq \delta E_{\zeta}^{\phi_r}=\frac{\kappa}{8\pi G} \left (\delta \phi_{\mathcal{B}} -\phi'_{\mathcal B} \delta \ell \right) + \delta H_\zeta^{\text{m}}\;. \label{eq:firstlawwithmatter}
\eeq
Secondly, we can consider variations of coupling constants and other parameters, \emph{e.g.}, $\phi_{0} ,G,$ and $\Lambda$, thereby leading to an extended first law. We  discuss this generalization below.


\subsubsection{An extended first law}
 \label{sec:extendedfirst}

Above we kept all couplings and parameters   fixed. It is often natural to include variations with respect to the couplings $\lambda_{i}$ of the theory, which can enrich the interpretation of the first law.  For example, when the cosmological constant $\Lambda$ and Newton's constant $G$ are allowed to vary, the first law for static, neutral AdS black holes  is extended to   \cite{Kastor:2009wy,Kastor:2010gq}
\beq \delta M=\frac{\kappa}{8\pi G}\delta A - \frac{M}{G} \delta G +\frac{\Theta}{8\pi G}\delta\Lambda\;, \label{eq:extendedfirstlawadsbh}\eeq
where the quantity $\Theta$ conjugate to $\Lambda$ is the same as the one in  the Smarr equation \eqref{eq:SmarrgenBH}. Standardly, $\Lambda$ is interpreted as a pressure, such that (minus) $\Theta$ plays the role of ``thermodynamic volume'',  which leads to an extended first law of black hole thermodynamics,  defining the study of \emph{extended black hole thermodynamics} or \emph{black hole chemistry}.  Including a pressure-volume work contribution enriches the phase space structure of black holes (for a review see~\cite{Kubiznak:2016qmn}). 

In the context of the AdS/CFT correspondence, gravitational coupling variations in the AdS bulk are dual to    central charge variations of the boundary CFT \cite{Kastor:2009wy,Johnson:2014yja,Dolan:2014cja,Kastor:2014dra,Karch:2015rpa,Visser:2021eqk}.   For example,  according to the  AdS/CFT dictionary, the (generalized) central charge $a^{\ast}_{d}$ in a CFT, defined as the universal coefficient of the vacuum entanglement entropy for ball-shaped regions, is dual to $ L^{D-2}/G$ for Einstein gravity in asymptotically AdS spacetimes \cite{Myers:2010tj}. Recall the CFT vacuum state reduced to a ball on   flat space is dual to  AdS-Rindler space, such that the first law of entanglement is dual to the first law of AdS-Rindler space  \cite{Casini:2011kv,Faulkner:2013ica}. AdS-Rindler space is identical to the massless static hyperbolic AdS black hole,   so the Smarr formula \eqref{eq:SmarrgenBH} reduces to $(D-2) \kappa A = 2 \Theta \Lambda$ and the extended first law \eqref{eq:extendedfirstlawadsbh} can be written as  \cite{Caceres:2016xjz,Sarkar:2020yjs}
\beq
\delta M = \frac{\kappa}{8\pi  }\delta \left ( \frac{A}{G} \right) - \frac{\kappa A}{8\pi G} \left[  (D-2)\frac{\delta L}{L} - \frac{\delta G}{G} \right] \;,
\eeq
where we used $\delta \Lambda / \Lambda = - 2 \delta L / L.$
 Crucially, the combination of variations between brackets on the right-hand side is dual to the variation of the (generalized) central charge $a^{\ast}_{d} $ of the CFT. Therefore, the extended bulk first law of the massless hyperbolic AdS black hole can be  identified with the extended first law of entanglement \cite{Kastor:2014dra}
 \beq
 \frac{2\pi}{\kappa} \delta\langle H_{\text{ball}}\rangle= \delta S_{\text{ent}} - \frac{S_{\text{ent}}}{a^{\ast}_{d}}\delta a^{\ast}_{d} \;, \label{eq:extendedfirstlawent}\eeq
where $S_{\text{ent}}$ is the vacuum entanglement entropy of the CFT across the ball, dual to $A/4G$ due to the Ryu-Takayanagi formula \cite{Ryu:2006bv}, and $  H_{\text{ball}}$ is the modular Hamiltonian defining the CFT vacuum state reduced to the ball, and the variation of its expectation value  $\delta \langle H_{\text{ball}} \rangle $ is identified with $\delta M $ in the bulk. If the generalized central charge  is kept fixed, $\delta a^{\ast}_{d}=0$, then  Eq. \eqref{eq:extendedfirstlawent} reduces to the standard first law of entanglement  $\delta\langle H_{\text{ball}}\rangle=  \delta S_{\text{ent}}$, where  the surface gravity is   conveniently normalized as $\kappa=2\pi$  \cite{Blanco:2013joa,Wong:2013gua}. The extended first law of entanglement was verified to hold for arbitrary coupling variations and specific higher derivative theories of gravity in \cite{Caceres:2016xjz} and further generalized to arbitrary bulk theories in high and low dimensions in \cite{Rosso:2020zkk}, including an extended first law for JT gravity (though, notably, the $\delta\ell$ variation is absent; see also Appendix G of \cite{Svesko:2020yxo} for a derivation using covariant phase techniques).

Before moving on to the extended first law of the nested Rindler wedge, it is worth briefly reviewing the derivation of the extended first law in the covariant phase space formalism \cite{Urano:2009xn,Caceres:2016xjz}. Moreover, we refine this derivation here by adding appropriate boundary terms to the action. In Appendix \ref{app:extfirstlawdeets} we present more details of the derivation of the extended first law, where we also allow for a nonzero $C(\psi, \delta \psi)$   which appears in the covariant phase space formalism with boundaries \cite{Harlow:2019yfa}. 

Consider the bulk off-shell Lagrangian $d$-form $L = \mathcal L_{\text{bulk}}(\psi,\lambda_{i})\epsilon$, which  depends on fields $\psi$ and couplings $\lambda_{i}$, and whose   variation is given by
\beq \delta L =E_{\psi}\delta\psi+  E_L^{\lambda_{i}}\delta\lambda_{i}+d\theta(\psi,\delta\psi)\;,\eeq
where $E_L^{\lambda_{i}}\equiv\frac{\partial\mathcal{L}_{\text{bulk}}}{\partial\lambda_{i}}\epsilon$  and summation over $i$ is understood. Allowing for such coupling variations modifies the on-shell integral identity (\ref{eq:variationalidJT1}) as follows  \cite{Urano:2009xn,Caceres:2016xjz} 
 \beq
     \int_\Sigma \omega (\psi, \delta \psi, \mathcal L_\zeta \psi) = \oint_{\partial \Sigma} [\delta Q_\zeta - \zeta \cdot \theta (\psi, \delta \psi)] + \int_{\Sigma}\zeta\cdot E_L^{\lambda_{i}}\delta\lambda_{i}\;,
\label{eq:variationalidJText} \eeq
where the `$\delta$' is a general place holder for the variation with respect to both the  couplings and dynamical fields $\psi$. When $\delta$ is a variation solely with respect to the   dynamical fields, we attain the standard first law of black hole mechanics (without coupling variations). 

Next, we specialize to the geometric setup   where $\Sigma$ is a Cauchy slice with inner boundary at the bifurcation surface $\mathcal B$ of a bifurcate Killing horizon and an outer boundary at spatial infinity. At the outer boundary $B \to \infty$ the symplectic potential is a total variation (if we set $C(\psi , \delta \psi )=0$, see footnote \ref{fn:bdrycond}) \cite{Wald:1993nt,Iyer:1994ys}
    \beq 
\theta \big |_{B\to\infty} = \delta_\psi b = \delta b - E_b^{\lambda_i} \delta \lambda_i\;,
  \eeq 
   where $\delta_\psi b$ is the variation of $b$ only with respect to the dynamical fields, and $E_b^{\lambda_i} \equiv \frac{\partial   \mathcal L_{\text{bdy}}}{\partial \lambda_i}  \epsilon_{B}$, with $\mathcal{L}_{\text{bdy}}$ being the off-shell boundary Lagrangian density, such that $b =\mathcal{L}_{\text{bdy}} \epsilon_B$. We also include local counterterms in the 1-form $b$ to regulate any divergences, so $b = b_{\text{GHY}} + b_{\text{ct}}$. Combined, the variational identity (\ref{eq:variationalidJText}) becomes
    \beq 
  \int_\Sigma \omega (\psi, \delta \psi, \mathcal L_\zeta \psi) =\delta \oint_\infty \left ( Q_\zeta - \zeta\cdot b\right ) - \delta \oint_{\mathcal B} Q_\zeta + \int_{\Sigma}\zeta\cdot E^{\lambda_{i}}_L \delta\lambda_{i} + \oint_\infty \zeta \cdot E^{\lambda_i}_b \delta \lambda_{i}\;.
 \label{eq:variationalidJTextv2}\eeq 
This identity can be seen as a generalization of the extended Iyer-Wald formalism described in \cite{Urano:2009xn,Caceres:2016xjz} to the case of spacetimes with boundaries. 

In the context of classical JT gravity, the relevant coupling constants defining the theory are $\lambda_{i}=\{\phi_{0},G,\Lambda\}$. Leaving the computations to Appendix \ref{app:extfirstlawdeets}, we obtain the following result for the extended first law of the nested AdS-Rindler wedge  
\beq \delta E_{\zeta}^{\phi_r}= \frac{\kappa}{8\pi} \delta \left( \frac{\phi_0 + \phi_{\mathcal B}}{G} \right)-\frac{\kappa \phi'_{\mathcal B}}{8\pi G} \delta \ell +  \frac{\Theta^{\phi_0}_\zeta}{8\pi GL^2} \delta \phi_0  - \frac{\phi_0 \Theta^G_{\zeta}}{8 \pi G L^2}\frac{\delta G}{G} - \frac{\Theta^{\Lambda}_{\zeta}}{8 \pi G L^2}\frac{\delta \Lambda}{\Lambda}\;,
\label{eq:extfirstlawclassJT}\eeq
where  we have introduced a counterterm subtracted ``$\phi_0$-Killing volume"
\beq
\Theta^{\phi_0}_\zeta \equiv  \int_\Sigma \zeta \cdot \epsilon -L^2\oint_\infty K \zeta \cdot \epsilon_{B} \label{eq:TVphi0v1}\; ,
\eeq
and a counterterm subtracted ``$\Lambda$-Killing volume"  
\beq
  \Theta^\Lambda_{\zeta}\equiv\int_\Sigma \phi \zeta \cdot \epsilon - \frac{L}{2} \oint_\infty \phi \zeta \cdot \epsilon_{B}\;.\label{eq:dilatonTVv1}\eeq
 Notice the dilaton arises inside the   integrals  above    because of its non-minimal coupling to the metric in the JT Lagrangian. Although the definition includes a ``counterterm'' the result for $\Theta^\Lambda_\zeta$ is generally divergent and is given by \eqref{eq:dilatonKillvolexp}, where we have introduced a cutoff.
 
 We can perform some consistency checks of the extended first law \eqref{eq:extfirstlawclassJT} by considering specific variations. Firstly, the variations with respect to $\phi_{0}$ cancel between the first and the third term on  the right-hand side, 
 \beq
 \frac{\kappa}{8\pi G} \delta \phi_0 +  \frac{\Theta^{\phi_0}_\zeta}{8\pi GL^2} \delta \phi_0   =0 \qquad \longrightarrow \qquad  \Theta_\zeta^{\phi_0} = - \kappa L^2\;. \label{eq:phi0nontrivialkilling}
 \eeq
 We have verified the last identity explicitly using the definition of the $\phi_0$-Killing volume. Secondly, the variations of Newton's constant cancel between  the left and right-hand side, 
 \beq
 \delta_G E_{\zeta}^{\phi_r} = -\frac{\kappa}{8 \pi G} (\phi_0 + \phi_{\mathcal B})\frac{\delta G}{G}  - \frac{\phi_0 \Theta_\zeta^G}{8\pi G L^2}   \frac{\delta G}{G} \quad \longrightarrow \quad   E_{\zeta}^{\phi_r}   =  \frac{\kappa}{8 \pi G} (\phi_0 + \phi_{\mathcal B})   + \frac{\phi_0   \Theta_\zeta^G}{8\pi G L^2 }    \;,  \label{eq:originalsmarrnestedwedge}
 \eeq
where we used $E_\zeta^{\phi_r}\sim 1/G$. This  identity is precisely  the classical Smarr law (\ref{eq:smarrJTv2}). Thirdly, verifying the extended first law if the variations are induced by changing the cosmological constant (or AdS length) is a bit more subtle, since  the $\delta L$ variation coming from the $\Lambda$-Killing volume is divergent as $r_B\to\infty$. However, this divergence is precisely cancelled by the $r_B\to\infty$ divergence appearing in the $\delta_{L}\ell$ term, see (\ref{eq:omegaLvar}). In fact, equating the variation of $E_{\zeta}^{\phi_r}$ with respect to $L$ to the  sum of these two separately divergent $\delta L$ variations   yields yet another nontrivial relation: 
\beq 
\delta_L E_{\zeta}^{\phi_r}= -\frac{\kappa\phi'_{\mathcal{B}} L}{8\pi G} \delta_L \ell +\frac{\Theta_{\zeta}^{\Lambda}}{4\pi GL^{2}} \frac{\delta L}{L}  \qquad \longrightarrow \qquad E_{\zeta}^{\phi_r}= \frac{\kappa\phi'_{\mathcal{B}} L}{8\pi G}\frac{\partial \ell}{\partial L}-\frac{\Theta_{\zeta}^{\Lambda}}{4\pi GL^{2}}    \;. \label{eq:newsmarrlawforlambda}
\eeq
This  follows from observing $\delta_L \ell = \frac{\partial \ell}{\partial L} \delta L$ and  $\delta_L E_{\zeta}^{\phi_r} = - E_\zeta^{\phi_r} \frac{\delta L}{L}$, since from \eqref{eq:asympenergyclassJTapp} we see $E^{\phi_r}_\zeta \sim \kappa$ and by dimensional analysis we have $\kappa\sim1/L$. We check this identity more explicitly in  \eqref{eq:checkofdeltaLvariation}. Note that it  also involves the asymptotic energy, just like the Smarr relation, but not the entropy of the bifurcation point, in contrast to the standard Smarr relation. 
Thus, the extended first law has uncovered   three nontrivial relations for the three   Killing volumes, one of which \eqref{eq:originalsmarrnestedwedge} is the Smarr relation. 

We emphasize that here, unlike in the case of pure gravity theories, we find three generally different ``Killing volumes'', $\Theta^{G}_{\zeta}$, $\Theta^{\phi_{0}}_{\zeta}$, and  $\Theta^{\Lambda}_{\zeta}$, defined in (\ref{eq:thermovoldef}), (\ref{eq:TVphi0v1}), and \eqref{eq:dilatonTVv1}, which appear in the extended first law as conjugate variables to Newton's constant $G$, $\phi_{0}$, and the cosmological constant $\Lambda$, respectively. None of them separately have  the interpretation of a thermodynamic volume, since they are not conjugate to the bulk pressure $P = - \Lambda / 8\pi G$ (note we are also considering variations of $G$ here). The reason we have three Killing volumes is a consequence of the fact we are studying a non-minimally coupled dilaton-gravity theory with three different coupling constants, since if we set $\phi=1$ the volume integrals in the definitions of  $\Theta^{\phi_0}_\zeta$, $\Theta^G_\zeta$ and $\Theta^\Lambda_\zeta$  match (neglecting the   boundary integrals at infinity).

Furthermore, let us discuss the extended first law for the eternal $\text{AdS}_{2}$  black hole. In the $\alpha\to\infty$ limit, where $\phi'_{\mathcal{B}}=0$ such that the   $\delta\ell$ term in the first law vanishes, we recover the classical extended first law of the black hole  
\beq \delta M_{\phi_r}= \frac{\kappa}{8\pi} \delta \left( \frac{\phi_0 + \phi_H}{G} \right) +  \frac{\Theta_{\zeta,\alpha \to \infty}^{\phi_0}}{8\pi G L^2} \delta \phi_0  - \frac{\phi_0 \Theta^G_{\zeta, \alpha \to \infty}}{8 \pi G L^2}\frac{\delta G}{G} - \frac{\Theta^{\Lambda}_{\zeta, \alpha \to \infty}}{8 \pi G L^2}\frac{\delta \Lambda}{\Lambda} \;, \label{eq:blackholefirstlawcouplings1}
\eeq
  where the Killing volumes associated with the eternal black hole are (setting $\kappa=\sqrt{\mu}/L$)
  \beq
 \Theta^{\phi_0}_{\zeta}|_{\alpha \to \infty} = - \sqrt{\mu} L\;, \quad \Theta^{G}_{\zeta}|_{\alpha\to\infty}=-\frac{\mu L\phi_{r}}{2\phi_{0}}-\sqrt{\mu}L  \;, \quad \text{and}\quad \Theta^{\Lambda}_{\zeta}|_{\alpha\to\infty}=-\frac{\mu L\phi_{r}}{4} \;.
 \label{eq:volliimts} \eeq
 In this limit we observe the volume $\Theta^{G}_{\zeta}$ is a combination of the volumes $\Theta^{\Lambda}_{\zeta}$ and $\Theta^{\phi_{0}}_{\zeta}$, namely,  $\Theta^{G}_{\zeta}= 2\Theta^{\Lambda}_{\zeta}/\phi_{0}+\Theta^{\phi_{0}}_{\zeta}$. Moreover, explicitly, in terms of the variables of the black hole, we have the extended first law
\beq
  \delta M_{\phi_r} = \frac{\sqrt{\mu}}{8\pi L} \delta \left( \frac{\phi_0 + \phi_H}{G} \right) - \frac{\sqrt{\mu}}{8 \pi G L} \delta \phi_0 + \left ( \frac{\mu\phi_{r}}{16 \pi GL} + \frac{\sqrt{\mu}\phi_0}{8\pi GL} \right) \frac{\delta G}{G} -\frac{\mu \phi_r}{16 \pi G L} \frac{\delta L}{L}\;. \label{eq:blackholefirstlawcouplings2}
  \eeq
Simplifying the expression,  and  in addition allowing for variations of $\phi_r$, yields
  \beq \delta ( M_{\phi_{r}} L) =\frac{\sqrt{\mu}}{8\pi  } \delta \left( \frac{  \phi_H}{G} \right)  -M_{\phi_{r}} L  \frac{\delta (\phi_r/G)}{\phi_r/ G} \;.\eeq
Interestingly, in terms of the dimensionless mass $\bar M_{\phi_r} = M_{\phi_r} L$, the  dimensionless inverse Hawking temperature $\bar \beta_{\text{H}} =2 \pi / \sqrt{\mu}$, the  entropy $S_{\phi_r} = \sqrt{\mu} \phi_r / 4G$ above extremality, and the   gravitational coupling $a = \phi_r / G$, the black hole extended first law can be written as\footnote{Alternatively, we could include the variation of the entropy $S_{\phi_0} = \phi_0 / 4G$ of the extremal black hole into the extended first law, but then we have to subtract off a term proportional to the variation of   $\phi_0/G$  
\beq
 \bar \beta_{\text H}   \delta \bar M_{\phi_r} = \delta( S_{\phi_0}+ S_{\phi_r}) - \frac{S_{\phi_0}}{a_{\phi_0}} \delta a_{\phi_0} - \frac{1}{2}  \frac{   S_{\phi_r} }{a_{\phi_r}} \delta a_{\phi_r} \;, \label{eq:extendedfirstlawasentfirstlaw3}
\eeq
where now there are two couplings $a_{\phi_0} = \phi_0/G$ and $a_{\phi_r}=\phi_r/ G.$ Note  the variation of the black hole entropy with respect to the couplings is $\delta_{a_i} ( S_{\phi_0}+ S_{\phi_r})=\frac{S_{\phi_0}}{a_{\phi_0}} \delta a_{\phi_0} + \frac{S_{\phi_r}}{a_{\phi_r}} \delta a_{\phi_r}.$
} 
\beq
 \bar \beta_{\text H}   \delta \bar M_{\phi_r} = \delta S_{\phi_r}- \frac{1}{2}  \frac{   S_{\phi_r} }{a} \delta a \;. \label{eq:extendedfirstlawasentfirstlaw}
\eeq
This is a generalization  of the standard first law \eqref{eq:classicaljtsmarrfirst} to nonzero coupling variations.  It is very similar to the extended first law of entanglement \eqref{eq:extendedfirstlawent}, dual to the extended first law for   higher-dimensional AdS-Rindler space, except for the factor one half in the $\delta a$ term. This factor does not appear in the extended first law of entanglement since the modular Hamiltonian vanishes in the CFT vacuum, so $\delta_{a^{\ast}_{d}} \langle H_{\text{ball}} \rangle    =0$, and hence the left-hand side of \eqref{eq:extendedfirstlawent} vanishes identically. The right-hand side of \eqref{eq:extendedfirstlawent} also vanishes identically because the vacuum CFT entanglement entropy of a ball   in flat space is proportional to the central charge, $S_{\text{ent}} \sim a^{\ast}_{d}$, hence $\delta_{a^{\ast}_{d}} S_{\text{ent}}= \frac{S_{\text{ent}}}{a^{\ast}_{d}}\delta  a^{\ast}_{d}$. On the other hand, in the AdS$_2$-Rindler extended first law \eqref{eq:extendedfirstlawasentfirstlaw} the energy variation induced by changing the coupling $a$ is nonzero, since $\delta_a \bar M_{\phi_r} = \frac{\bar M_{\phi_r}}{a} \delta a $, and the entropy variation takes the same form as the entanglement entropy variation, $\delta_a S_{\phi_r} = \frac{S_{\phi_r}}{a} \delta a.$ Inserting these variations into \eqref{eq:extendedfirstlawasentfirstlaw} yields   a true identity due to the classical Smarr law \eqref{eq:classicaljtsmarrfirst} $M_{\phi_r} =\frac{1}{2} T_{\text{H}}S_{\phi_r}$. 

To summarize, the coupling variation of the entropy for JT gravity is still of the form $\delta_a S_{\phi_r} = \frac{S_{\phi_r}}{a} \delta a$,  with $a = \phi_r / G$, just like in the first law of entanglement (in contrast to the findings of \cite{Rosso:2020zkk}). However, the coefficient of the $\delta a$ term in the extended   first law for AdS$_2$ black holes differs from the coefficient of the $\delta a^{\ast}_{d}$ term in the extended first law of entanglement.  Further, the coupling $a^{\ast}_{d}$ in the extended first law of entanglement has the interpretation of a generalized central charge in $d \ge 2$ dimensional CFTs, so in analogy it would be interesting to find a microscopic interpretation of the JT gravitational coupling $a $ in the dual double-scaled random matrix theory \cite{Saad:2019lba} or in the SYK model \cite{Maldacena:2016hyu}.



Lastly, it is worth recalling JT gravity arises from a spherical reduction of a charged higher-dimensional black hole, where $\phi$  represents the transversal area of the sphere and $\Lambda$ is a function of  the charge $q$ and higher-dimensional AdS length. Therefore, in a certain sense, the extended first law can be seen as the dimensionally reduced extended first law of a charged static black hole. For example, in the case of a spherical reduction of a four-dimensional AdS-RN black hole of charge $q$ and AdS length $L_{4}$, the AdS length of the $2D$ black hole is given by $L=q^{3/2}L_{4}$ \cite{Navarro-Salas:1999zer}. Consequently, the variation with respect to $L$ encodes the variation of the black hole charge and higher-dimensional AdS length, $\frac{\delta L}{L}=\frac{3}{2}\frac{\delta q}{q}+\frac{\delta L_{4}}{L_{4}}$. It would be interesting to study the relation between our extended first law for JT gravity and the extended first law of a  higher-dimensional charged black hole in more detail.

 \subsection{Semi-classical Smarr formula and first law}
 
 Having derived a classical Smarr relation and first law for the nested Rindler wedge, here we aim to determine their semi-classical extension. Mathematically this amounts to accounting for the quantum corrections to the dilaton $\phi$ and including the auxiliary field $\chi$.

\subsubsection{Smarr relation}

Let us first  derive the Smarr relation including semi-classical backreaction effects. The general procedure used to derive the classical Smarr law   (\ref{eq:SmarrJT}) is the same, where now we need to compute the Noether current and charge associated to the Polyakov action and to  the 1-loop quantum corrections to dilaton $\phi$. As we will see shortly, the Smarr relation is generalized accordingly, where   the classical entropy is replaced by the generalized entropy, and a semi-classically corrected asymptotic energy.

We again make use of the regulated integral identity (\ref{eq:regulatedintegralidentity}). First, we consider the right-hand side of the identity. Working in the Hartle-Hawking state, the semi-classical correction to the dilaton is only   a constant (\ref{eq:dilasolnHH}), $\phi_c= \frac{Gc}{3}$ (where we have set $\lambda=0$). Thus here we primarily focus on the effect of the auxiliary field $\chi$. Using the generic expression for the Noether charge  (\ref{eq:Noethercharge}), the contribution from the $\chi$ field is
\beq Q^{\chi}_{\zeta}=\frac{c}{24\pi}\epsilon_{\mu\nu}\left[\chi\nabla^{\mu}\zeta^{\nu}+2\zeta^{\mu}\nabla^{\nu}\chi\right]\;.\label{eq;noetherchargechi}\eeq
Evaluating this at the bifurcation surface $\mathcal{B}$ leads to 
\beq \oint_{\mathcal{B}}Q_{\zeta}^{\chi}=-\frac{\kappa}{2\pi}\frac{c\chi_{\mathcal{B}}}{6}\;,\label{eq:QchiB}\eeq
which we recognize as the  Wald entropy associated to the $\chi$ field (times the temperature). Note we did not need to specify the precise form of $\chi$ such that this expression even holds   in the Boulware vacuum. We ignore the Boulware vacuum solution, however, as it does not describe a system in thermal equilibrium.

Next we  evaluate $\oint_\infty Q^{ c}_{\zeta}$,
where $Q^{ c}_{\zeta}$ is the total Noether charge due to the semi-classical corrections in $\phi$ and $\chi$, $Q_{\zeta}^c=Q_{\zeta}^{\phi_c}+Q_{\zeta}^{\chi}$. In addition we subtract the term $\zeta\cdot b_{c}=\zeta \cdot {b_{\phi_c}}+\zeta \cdot b_\chi$ at infinity, where $b_{\phi_c}$  is the local counterterm associated with $\phi_{c}$ (see (\ref{eq:bdryformandct}) with $\phi$ replaced by $\phi_{c}=\frac{Gc}{3}$  and $\phi_{0}=0$), and $b_{\chi}$ is the    Polyakov boundary counterterm 1-form  \cite{Almheiri:2014cka}
\beq b_{\chi}=\frac{c}{12\pi}\chi K\epsilon_{B}+\frac{c}{24\pi L}\epsilon_{B}\;.\label{eq:bdryctPoly}\eeq
Working in the   Hartle-Hawking state and using the static solution $\chi^{(4)}$ (\ref{eq:chi4sol}) we find
\beq \oint_\infty \left (  Q^{c}_{\zeta}-\zeta\cdot b_c \right) =\frac{c\kappa}{12\pi}\tanh(\sqrt{\mu}\alpha/2L)=E_{\zeta}^{c}\;,\label{eq:Ezetacv2}\eeq
 where $E_{\zeta}^{c}$ is the semi-classical correction to the asymptotic energy in (\ref{eq:asymptotenSJT}). In the limit $\alpha\to\infty$, with $\kappa=\sqrt{\mu}/L$ we recover the semi-classical correction to the eternal black hole mass   $M_{c}=\frac{c\sqrt{\mu}}{12\pi L}$  \cite{Almheiri:2014cka,Moitra:2019xoj}.

Consider now the left-hand side of the identity (\ref{eq:regulatedintegralidentity}).  For the semi-classical correction to the dilaton, we need to compute
\beq \int_{\Sigma}j^{\phi_{c}}_{\zeta}- \oint_\infty \zeta \cdot b_{\phi_c}=\int_{\Sigma}[\theta_{\text{JT}}(\phi_{c},\mathcal{L}_{\zeta}\phi_{c})-\zeta\cdot L_{\phi_{c}}]- \oint_\infty \zeta \cdot b_{\phi_c}\;,\eeq
where $b_{\phi_{c}}$ is the boundary counterterm Lagrangian 1-form (\ref{eq:bdryformandct}) with $\phi_{0}=0$ and $\phi\to\phi_{c}$, and similarly for $L_{\phi_{c}}$. Since $L_{\phi_{c}}=0$ on-shell, and $\theta_{\text{JT}}(\phi_{c},\mathcal{L}_{\zeta}\phi_{c})=0$, and $\zeta \cdot b_{\phi_c}|_\infty=0$ we find
\beq \int_{\Sigma}j^{\phi_{c}}_\zeta -\oint_\infty \zeta\cdot b_{\phi_{c}} =0\;.\eeq
 Meanwhile,   the contribution from $\chi$ to the left-hand side is
\beq \int_{\Sigma}j^{\chi}_{\zeta}- \oint_\infty \zeta \cdot b_{\chi}=\int_{\Sigma}[\theta_\chi(\psi,\mathcal{L}_{\zeta}\psi)-\zeta\cdot L_{\chi}]- \oint_\infty \zeta \cdot b_{\chi}\;.\eeq
From (\ref{eq:sympotgen}) and using $\mathcal{L}_{\zeta}\chi|_{\Sigma}=0$  (\ref{eq:Liederivchi}), we see the symplectic potential for $\chi$ will not contribute:
$ \theta_\chi (\psi,\mathcal{L}_{\zeta}\psi)|_{\Sigma} =0$. 
 Further, with the Polyakov Lagrangian 2-form  (setting $\lambda =0$ in \eqref{eq:semiclassactionconts})
\beq L_{\chi}=-\frac{c}{24\pi}\epsilon[\chi R+(\nabla\chi)^{2}]\;,\eeq
and the boundary 1-form (\ref{eq:bdryctPoly}), it is straightforward to show\footnote{Here for convenience we use the static $\chi^{(4)}$ solution, however, similar expressions hold for the generally time-dependent $\chi^{(5)}$ solution.}
\beq 
\begin{split}
 \int_{\Sigma}j^{\chi}_{\zeta}- \oint_\infty \zeta \cdot b_{\chi}&=-\frac{c}{12\pi L^{2}}\Theta_{\zeta}^{c}\;,
\end{split}
\label{eq:jzetachisjt}\eeq
where we have introduced the  counterterm subtracted ``semi-classical $c$-Killing volume" $\Theta_{\zeta}^{c}$ analogous to (\ref{eq:thermovoldef}),   on shell given by
\beq \Theta^{c}_{\zeta}\equiv\int_{\Sigma}\left(\chi-\frac{L^{2}}{2}(\nabla\chi)^{2}\right)\zeta\cdot\epsilon+\oint_{\infty}\left(L^{2}\chi K+\frac{L}{2}\right)\zeta\cdot \epsilon_{B}\;.\label{semiclassicalvolumeccc}\eeq
Or, more explicitly, for the time-independent solution $\chi^{(4)}$ (\ref{eq:chi4sol}) it is equal to 
\beq
\Theta_{\zeta}^{c}=\frac{\kappa L^{2}}{\sinh(\sqrt{\mu}\alpha/L)}+\frac{\kappa L^{2}}{2}\left[1+2( \sqrt{\mu}\alpha/L-\coth(\sqrt{\mu}\alpha/L))-\log\left(\frac{1}{\mu}  \sinh^{2}(\sqrt{\mu}\alpha/L) \right)\right]\;.
\label{eq:semiclassicalTV}\eeq
Combining (\ref{eq:QchiB}), (\ref{eq:Ezetacv2}), and (\ref{eq:jzetachisjt}) together with the classical Smarr relation (\ref{eq:smarrJTv2}), we arrive at the semi-classical Smarr law for nested AdS-Rindler wedges 
\beq 
\begin{split}
E_{\zeta}^{\phi_r} +E_{\zeta}^{c}&=\frac{\kappa}{2\pi}\left( \frac{1}{4G}(\phi_{0}+\phi_{\mathcal{B}})-\frac{c}{6}\chi_{\mathcal{B}}\right)-\frac{\phi_{0}\Lambda}{8\pi G}\Theta^G_{\zeta}+\frac{c\Lambda}{12\pi}\Theta_{\zeta}^{c}\;.
\end{split}
\label{eq:semiJTsmarr}\eeq
As we will argue momentarily, the first term on the right-hand side is recognized to be $T S_{\text{gen}}$. In the limit $\alpha\to\infty$, we uncover the semi-classical Smarr relation for the $\text{AdS}_{2}$ black hole (setting $\kappa = \sqrt{\mu}/L$)
\beq
M_{\phi_r} + M_c = \frac{\sqrt{\mu}}{2\pi L}\left( \frac{1}{4G}(\phi_{0}+\phi_{H})-\frac{c}{6}\chi_{H}\right)-\frac{\phi_{0}\Lambda}{8\pi G}\Theta^G_{\zeta, \alpha \to \infty}+\frac{c\Lambda}{12\pi}\Theta_{\zeta, \alpha \to \infty}^{c}\;.
\eeq
For the time-independent $\chi^{(4)}$  the $\alpha\to \infty$ limit of the semi-classical Killing volume is
\beq
\Theta_{\zeta, \alpha \to \infty}^c = -\frac{\sqrt{\mu} L}{2}\left ( 1 - 2 \log (2 \sqrt{\mu})\right)\;.
\label{eq:cvollimit}\eeq
The semi-classical Smarr law for the black hole can thus be simplified to
\beq
2M_{\phi_r} +  M_c =\frac{\sqrt{\mu}}{2\pi L} \left ( S_{\phi_r} + S_c \right) +  \frac{c \Lambda}{12 \pi} \Theta^c_{\zeta, \alpha \to \infty}\;,
\eeq
 with $S_{\phi_r} = \frac{\phi_r \sqrt{\mu}}{4G} $ and $S_c =\frac{c}{12} +   \frac{c}{6}\log(2\sqrt{\mu}) .$ This is the semi-classical extension of the standard Smarr formula \eqref{eq:classicaljtsmarrfirst} for the eternal black hole.

\subsubsection{First law of nested backreacted  AdS-Rindler wedge mechanics}

Now we derive the first law of the nested Rindler wedge, taking into account the full backreaction. The steps are morally the same as deriving the classical first law (\ref{eq:firstlawJTclass}), where we make use of the integral variational identity (\ref{eq:variationalidJT1}). Since  in the Hartle-Hawking state the semi-classical correction to the dilaton is a constant, it trivially modifies the classical first law. We therefore mainly discuss the $\chi$ contribution.

We first focus on evaluating the right-hand side of (\ref{eq:variationalidJT1}). As in the classical case, using the Noether charge at spatial infinity \eqref{eq:Ezetacv2} and at the bifurcation point (\ref{eq:QchiB}) we find 
\beq  \oint_{\partial \Sigma} [\delta Q^{\text{c}}_\zeta - \zeta \cdot \theta (\psi, \delta \psi)]=\delta E^{c}_{\zeta}-\frac{\kappa}{2\pi}\delta\left(\frac{\phi_c}{4G} -\frac{c}{6}\chi_{\mathcal{B}}\right) \;,\label{eq:RHSvaridSJT}\eeq
where the second term is the variation of  the semi-classical contribution to the Wald entropy.

Moving on, the quantum correction $\phi_c$ to the dilaton  does not modify the left-hand side of (\ref{eq:variationalidJT1}) since  it is a constant. Thus, the only addition comes from the auxiliary field $\chi$. The symplectic current 1-form (\ref{eq:sympcurrentgen}) with respect to the Polyakov action is
\beq \omega_{\chi}(\psi,\delta\psi, \mathcal{L}_{\zeta}\psi)\Big|_{\Sigma}=-\frac{c}{24\pi}\epsilon_{\mu}\left[(g^{\mu\beta}g^{\alpha\nu}-g^{\mu\nu}g^{\alpha\beta})\nabla_{\nu}(\mathcal{L}_{\zeta}\chi)\delta g_{\alpha\beta}-2\nabla^{\mu}(\mathcal{L}_{\zeta}\chi)\delta\chi\right] \;, \label{eq:omegabeginningchi} \eeq
where we used $\mathcal L_\zeta g_{\mu \nu}=0$ and $\mathcal{L}_{\zeta}\chi|_\Sigma=0$. The same algebraic steps leading to (\ref{eq:sympcurrentsteps}) yield
\beq 
\begin{split}
\omega_{\chi}(\psi,\delta \psi, \mathcal{L}_{\zeta}\psi)\Big|_{\Sigma}&=
\frac{c}{24\pi}(h^{\alpha\beta}\delta g_{\alpha\beta}-2\delta\chi)(u^{\nu}\nabla_{\nu}\mathcal{L}_{\zeta}\chi)d\ell\\
&=\frac{c\kappa\sqrt{\mu}}{24\pi L}\tanh(\sqrt{\mu}\alpha/2L)\frac{(h^{\alpha\beta}\delta g_{\alpha\beta}-2\delta\chi)}{ \frac{r}{L}-\sqrt{\frac{r^{2}}{L^{2}}-\mu} }d\ell\;,
\end{split} \label{eq:symplecticcurrentchi1}
\eeq
where   we used the derivative identity (\ref{eq:covdliechiu}). Since $u^\nu \nabla_\nu \mathcal L_\zeta \chi$ is not constant on $\Sigma$ we cannot explicitly  integrate the symplectic current   over $\Sigma.$ Therefore, we write the integral over the symplectic current  \eqref{eq:symplecticcurrentchi1} formally, via Hamilton's equations, as the Hamiltonian variation associated to the $\chi$ field
 \beq
  \delta H_\zeta^\chi = \int_\Sigma \omega_{\chi}(\psi, \delta \psi, \mathcal L_\zeta \psi) \label{eq:hamvarchifield}\,.
 \eeq 
Thus, altogether, combining (\ref{eq:RHSvaridSJT}) and \eqref{eq:hamvarchifield} with the classical first law (\ref{eq:firstlawJTclass}), the semi-classical first law is
\beq \delta (E_{\zeta}^{\phi_r}
+ E_{\zeta}^{c}) = \frac{\kappa}{2\pi} \delta \left ( \frac{  \phi_{\mathcal B}}{4 G} - \frac{c\chi_{\mathcal B}}{6}  \right)  - \frac{\kappa \phi'_{\mathcal B}}{8 \pi G} \delta \ell  + \delta H^\chi_\zeta\;.
\label{eq:SJTfirstlawv1}\eeq
  Importantly, note the first term on the right-hand side  is simply $T\delta S_{\text{gen}}$, evaluated at the bifurcation point of the nested Rindler wedge $\mathcal{B}$ opposed to the black hole horion. In the next section we will see in the microcanonical ensemble the semi-classical first law  is equivalent to extremizing the generalized entropy, such that the semi-classical first law of nested Rindler wedges becomes the semi-classical first law of quantum extremal surfaces.

\subsection*{Extended semi-classical first law}

We conclude this section by briefly commenting on extending the semi-classical first law to include coupling variations, where we now also consider variations of the central charge~$c$. The central charge appears both in the Polyakov action as well as in the semi-classical correction to the dilaton. 
As detailed in Appendix \ref{app:extfirstlawdeets}, the semi-classical extended first law is
\beq 
\begin{split}
&\delta (E_{\zeta}^{\phi_r}+E_{\zeta}^{c})= \frac{\kappa}{8\pi} \delta \left( \frac{\phi_0 + \phi_{\mathcal B}}{G} -\frac{c\chi_{\mathcal{B}}}{6}\right)-\frac{\kappa \phi'_{\mathcal B}}{8\pi G} \delta \ell +\delta H^{\chi}_{\zeta}\\
&+\frac{\Theta^{\phi_0}_\zeta}{8\pi GL^2} \delta \phi_0  - \frac{\phi_0 \Theta^G_{\zeta}}{8 \pi G L^2}\frac{\delta G}{G} - \frac{\Theta^{\Lambda}_{\zeta}+\Theta^{\Lambda,c}_{\zeta}}{8 \pi G L^2}\frac{\delta \Lambda}{\Lambda}-\frac{c\Theta^{c}_{\zeta}}{12\pi L^{2}}\frac{\delta c}{c}+\frac{c\Theta^{B}_\zeta}{48\pi L}\frac{\delta \Lambda}{\Lambda}\;,
\end{split}
\label{eq:semiextfirstlawclassJT}\eeq
where we have introduced the  counterterm subtracted ``semi-classical $\Lambda$-Killing volume" 
\beq \Theta_{\zeta}^{\Lambda,c}=\phi_{c}\left(\int_\Sigma \zeta \cdot \epsilon -\frac{L}{2}\oint_\infty \zeta \cdot \epsilon_{B}\right)  \;,\eeq
where $\phi_{c}$ comes out of the integrals because it is  constant and $\Theta^{B}_{\zeta}\equiv\oint_{\infty}\zeta \cdot \epsilon_{B}$ is the Killing volume of $B$. Both $\Theta^{\Lambda,c}_{\zeta}$ \eqref{eq:semidilatonKV} and $\Theta^{B}_{\zeta}$ are formally divergent, however, it can be checked these divergences precisely cancel, see \eqref{eq:LvarEphiEc}. Using the $\alpha\to\infty$ limit of the Killing volumes (\ref{eq:volliimts}) and (\ref{eq:cvollimit}),  we attain
\beq
\begin{split}
 & \delta (M_{\phi_r}+M_{c}) = \frac{\sqrt{\mu}}{2\pi L} \delta \left( \frac{\phi_0 + \phi_H}{4G} -\frac{c\chi_{H}}{6}\right) - \frac{\sqrt{\mu}}{8 \pi G L} \delta \phi_0 + \left ( \frac{\mu\phi_{r}}{16 \pi GL} + \frac{\sqrt{\mu}\phi_0}{8\pi GL} \right) \frac{\delta G}{G} \\
  &-(M_{\phi_{r}}+M_{c})\frac{\delta L}{L}+\left(  M_{c}-\frac{c\sqrt{\mu}}{24\pi L}-\frac{c\sqrt{\mu}}{12\pi L}\log(2 \sqrt{\mu})\right)\frac{\delta c}{c}\;, \label{eq:semiextblackholefirstlaw}
 \end{split}
 \eeq
where we note the $\delta H^{\chi}_{\zeta}$ contribution has dropped out in this limit. This is the semi-classical generalization of the extended first law \eqref{eq:blackholefirstlawcouplings2}.  We have inserted the time-independent solution $\chi^{(4)}$ here, for which the semi-classical mass $M_c$ is given by \eqref{eq:quasilocasymenHH}. 
 The semi-classical extended first law for the eternal black hole can be simplified as
 \begin{equation}
\delta ( M L ) = \frac{\sqrt{\mu}}{2\pi} \delta \left ( \frac{\phi_H}{4 G} - \frac{c \chi_H}{6}\right) - M_{\phi_r} L \frac{\delta (\phi_r / G)}{\phi_r /G} + \left ( M_c L - \frac{c\sqrt{\mu}}{24\pi }-\frac{c\sqrt{\mu}}{12\pi  }\log(2 \sqrt{\mu}) \right) \frac{\delta c}{c}\;,
\end{equation}
where $M= M_{\phi_r} + M_c,$ or more compactly as,
\begin{equation}
\bar \beta_{\text{H}} \delta \bar  M = \delta (S_{\phi_r} + S_c) -(S_{\phi_r} - \beta_{\text{H}}M_{\phi_r}) \frac{\delta a}{a}    - (S_c - \beta_{\text{H}}M_c) \frac{\delta c}{c}\;,
\end{equation}
where $\bar{M}\equiv M L$ and $S_ c$ is the semi-classical contribution to the Wald entropy (\ref{eq:Ssemiclassent}). This form of the extended first law is the semi-classical generalization of \eqref{eq:extendedfirstlawasentfirstlaw}.

\subsection{Thermodynamic interpretation: canonical and microcanonical ensembles}

Now we are going to provide a thermodynamic interpretation to the classical and semi-classical first laws of the nested Rindler wedges.  We keep all the coupling constants $\{ \phi_0, G, \Lambda, c\}$ fixed in this section. Specifically, in the classical limit, we make the following identifications with the temperature $T$ and entropy $S_{\text{BH}}$
\beq T= \frac{\kappa}{2\pi}\;,\quad S_{\text{\text{BH}}}=\frac{\phi_{0}+\phi_{\mathcal B}}{4G}\;.\label{eq:identificationtempentr}\eeq
The classical  first law (\ref{eq:firstlawJTclass}) of nested Rindler horizons  can thus be written as
\beq
\delta E_\zeta^{\phi_r} = T \delta S_{\text{BH}} + \delta H^{\phi_r}_\zeta\;, \label{eq:classicalthermofirstlaw}
\eeq
where $\delta E^{\phi_r}_\zeta$ is the variation of the classical asymptotic energy and $\delta H^{\phi_{r}}_\zeta$ the variation of the JT Hamiltonian   generating time evolution along the flow of $\zeta.$ The asymptotic energy clearly represents   the internal energy of the system, whereas the thermodynamic interpretation of $\delta H^{\phi_{r}}_\zeta$ is less obvious. It is a gravitational energy variation which appears on the same side of the first law as $T \delta S_{\text{BH}}$, and hence might be interpreted as a gravitational work term. We recall the Hamiltonian variation \eqref{eq:LHSfirstlawJT} is   negative, $\delta H^{\phi_{r}}_{\zeta}=-\frac{\kappa\phi'_{\mathcal{B}}}{8\pi G}\delta\ell$, thereby it thus contributes negatively to the asymptotic energy variation, just like the work term $\delta W$ in the first law of thermodynamics, $dE = \delta Q - \delta W$, contributes negatively to the change in internal energy, suggesting the identification $\delta W=-\delta H_\zeta^{\phi_r}$.

In the semi-classical case the temperature identification remains the same, while the entropy is replaced by the generalized entropy of the bifurcation point $\mathcal B$
\beq 
S_{\text{gen}}=\frac{1}{4G} (\phi_{0}+\phi_{\mathcal{B}})-\frac{c}{6}\chi_{\mathcal{B}}\;.
\eeq
The semi-classical first law (\ref{eq:SJTfirstlawv1}) thus becomes
\beq
\delta \left(E_\zeta^{\phi_r} + E^c_\zeta \right) = T \delta S_{\text{gen}} + \delta H^{\phi_r}_\zeta + \delta H^\chi_\zeta \;, \label{eq:semiclassicalthermofirstlaw}
\eeq
where we note $\delta H^{\phi_r}_\zeta = \delta H^{\phi}_\zeta$, since $\delta H^{\phi_{c}}_{\zeta}=0$. Here $\delta E^c_\zeta $ is the   semi-classical contribution to the asymptotic energy variation, and $\delta H^\chi_\zeta$ is the Hamiltonian variation associated to the $\chi$ field in the Polyakov action. The role of the asymptotic energy variation and the Hamiltonian variation in the thermodynamic first law is the same as in the classical case.

The above identification \eqref{eq:identificationtempentr} for the temperature follows from the fact an asymptotic observer in the nested AdS-Rindler wedge sees the Hartle-Hawking state as a thermal state with  temperature~$T$, as we will now show.

\subsection*{A comment on vacuum states in adapted coordinates}

Solving the fully backreacted semi-classical JT model required us to a pick a specific vacuum state of the quantum matter. Since we are interested in studying thermodynamics, we solved the system with respect to the Hartle-Hawking state, a thermal state according to a static observer in $(v,u)$ coordinates such that the black hole radiates at the Hawking temperature $T_{\text H}=\frac{\sqrt{\mu}}{2\pi L}$, with $\sqrt{\mu}/L$ the surface gravity evaluated at the future Killing horizon. The QES naturally led us to consider a nested AdS-Rindler wedge characterized by the boost Killing vector $\zeta$, where the Killing horizon has surface gravity $\kappa$. To justifiably attribute thermodynamics to this system, we are interested to know what a static observer in adapted coordinates $(\bar{v},\bar{u})$ would detect outside of the QES in the global Hartle-Hawking state.  

 The global Hartle-Hawking state of the eternal black hole is thermal with respect to the static coordinates $(v,u)$, cf. Eq. \eqref{eq:expvalHHuuvv}, 
\beq \langle \text{HH}|:T^{\chi}_{uu}:|\text{HH}\rangle=\langle \text{HH}|:T^{\chi}_{vv}:|\text{HH}\rangle=\frac{c\pi}{12}T_{\text H}^2\;.\label{eq:HHexpuubar2}\eeq
To write  down a thermodynamic first law for the nested Rindler wedge, accounting for effects of backreaction, as done above, requires us to know the form of the solutions $\phi$ and $\chi$, which depend on the choice of vacuum state. We choose the global Hartle-Hawking state, since this is known to be a thermal state in the exterior of the black hole, but we now show  it is also thermal in the  nested Rindler wedge. To this end, we compute the expectation value of the normal-ordered stress tensor in the nested coordinates $(\bar v, \bar u )$ in the Hartle-Hawking state.   Recall the transformation rule (\ref{eq:transnormord}), such that
\beq :T^{\chi}_{\bar u \bar u}:=\left(\frac{d {u}}{d \bar u}\right)^{2}:T^{\chi}_{uu}:-\frac{c}{24\pi}\{ {u},\bar u\}\;,\eeq
and similarly for $\bar v\bar v$ and $vv$ components. Taking the expectation value of both sides with respect to the Hartle-Hawking vacuum, and using (\ref{eq:HHexpuubar2}) together with the coordinate relation (\ref{eq:transuvtobaruv}), we obtain 
\beq \langle \text{HH}|:T^{\chi}_{\bar u  \bar u }:|\text{HH}\rangle= \left(\frac{d u }{d \bar u}\right)^{2}\frac{c\pi}{12}T_{\text H}^{2} -\frac{c}{24\pi}\{ u, \bar u\}=\frac{c\pi}{12}\left(\frac{\kappa}{2\pi}\right)^{2}\;,\label{eq:HHTmunuallkap}\eeq
\emph{for all} $\kappa$ and $t_{0}\neq0$. In other words, asymptotic    $(\bar{v},\bar{u})$ observers   see the Hartle-Hawking state as a thermal state at the   temperature $T=\kappa / 2\pi$, where the surface gravity is associated to the nested AdS-Rindler horizon. This surface gravity,  and  hence this temperature,      appears  in both the classical and semi-classical  thermodynamic  first laws.

To be clear, the Hartle-Hawking state $|\text{HH}\rangle$ above is the global Hartle-Hawking state of the enveloping black hole, \emph{i.e.}, the vacuum state with respect to global Kruskal coordinates  ($V_{\text{K}}, U_{\text{K}}$). Naturally, one may wonder whether the Hartle-Hawking state defined with respect to the Kruskal coordinates adapted to the nested wedge $(\bar{V}_{\text K},\bar{ U}_{\text K})$ also appears thermal and coincides with the global Hartle-Hawking state. To check the latter, one may compute the expectation value of the normal ordered $T^{\chi}_{U_{\text{K}}U_{\text{K}}}$ with respect to the ``adapted" Hartle-Hawking state $|\overline{\text{HH}}\rangle$ using the relation (\ref{eq:changeinvacstnov2}). A straightforward calculation yields 
\beq \langle \overline{\text{HH}}|:T^{\chi}_{U_{K}U_{K}}:|\overline{\text{HH}}\rangle=-\frac{c}{24\pi}\{\bar{U}_{\text{K}},U_{\text{K}}\}=0\;,\eeq
where $\bar{U}_{\text{K}}=-\frac{1}{\kappa}e^{-\kappa\bar{u}}$. Therefore, the ``adapted" and global Hartle-Hawking states are one and the same: $|\overline{\text{HH}}\rangle=|\text{HH}\rangle$!  The Hartle-Hawking vacuum can thus  be written as a thermofield double state between left and right AdS-Rindler wedges of the eternal black hole or the left and right nested Rindler wedges, where, however, the associated temperature depends on the location where the spacetime is cut in half. Correspondingly, the reduced density matrix is thermal in the exterior region, where the temperature is dependent on the Rindler wedge one has access to. Note that we are not saying the vacuum state $|0_{u,v}\rangle$ with respect to $(v,u)$ coordinates is equivalent to the vacuum state $|0_{\bar{u},\bar{v}}\rangle$ in $(\bar{v},\bar{u})$ coordinates, \emph{i.e.}, the Boulware vacuum states of the different wedges are generally different, $\langle 0_{\bar{u}}|:T^{\chi}_{uu}:|0_{\bar{u}}\rangle\neq0$, as can be shown from a calculation similar to the above. Only in the $\alpha \to\infty$ limit, when the nested Rindler wedge coincides with the eternal black hole, do the Boulware vacua match.


It is worth comparing the above comment to the recent findings of \cite{Lochan:2021pio}, which considers an infinite family of nested Rindler wedges in Minkowski spacetime. The Minkowski (inertial) vacuum appears thermal with respect to each nested Rindler wedge, at a temperature proportional to the Unruh temperature of the corresponding wedge. 
The above calculation (\ref{eq:HHTmunuallkap}) is a realization of the observation made in \cite{Lochan:2021pio} for a single nested AdS-Rindler wedge, and a similar result is expected for subsequent nested AdS-Rindler wedges.

\subsubsection*{Generalized second law}

There is additional evidence suggesting we should interpret $S_{\text{BH}}$ and $S_{\text{gen}}$ as  thermodynamic entropies: both quantities, under particular assumptions, obey a second law. That is, upon throwing some matter into the system, the total entropy of the system monotonically increases along the future Killing horizon defining the nested wedge, such that, semi-classically, one has the generalized second law, $\frac{dS_{\text{gen}}}{d\lambda}\geq0$, where $\lambda$ is an affine parameter along a null generator of the future  Killing horizon associated with the boost Killing vector $\zeta$. We provide a heuristic derivation of the generalized second law in Appendix \ref{app:secondlaw}. Thus, akin to black holes, we note the nested Rindler wedge obeys a generalized second law of thermodynamics.

\subsubsection*{Equilibrium conditions and thermodynamic ensembles}

In standard thermodynamics, the stationarity of  free energy at fixed temperature follows from the first law of thermodynamics, $dE = \delta Q - \delta W$, and  Clausius' relation, $\delta Q = T dS$. Indeed, assuming $\delta W=0$ for the moment, the first law and Clausius' relation  yield $d E=Td S$,\footnote{Below we colloquially call this the thermodynamic ``first law'', which is standard terminology in black hole thermodynamics literature.} with $E=E(S)$. This implies the Helmholtz free energy $F\equiv E-TS$,  with $F=F(T)$, is stationary at fixed temperature  $d F|_{T}=-Sd T=0$. Thus, the stationarity of the free energy arises from the first law, from which the free energy or thermodynamic potential actually may be identified.
 The same line of reasoning has been applied to the first law of black hole mechanics, where the Helmholtz free energy of   a  static, neutral  black hole system corresponds to $F=M-T_{\text H}S_{\text{BH}}$, such that the first law implies $F$ is stationary at a fixed Hawking temperature $T_{\text H}$.

From the classical thermodynamic first law (\ref{eq:classicalthermofirstlaw}), the (Helmholtz) free energy $F_{\zeta}^{\text{cl}}$  is identified as 
\beq F_{\zeta}^{\text{cl}}=E_{\zeta}^{\phi_r}-TS_{\text{BH}}-H^{\phi_r}_{\zeta}\;.\eeq
Note that the absolute value of $H^{\phi_r}_\zeta$ appears here instead of its variation.\footnote{In the covariant phase space formalism, the variation of the Hamiltonian $\delta H_{\zeta}$ is a well-defined quantity, however, $H_{\zeta}$ is not precisely determined. Typically one imposes ``integrability conditions" (see, \emph{e.g.}, Eq. (80) of \cite{Iyer:1994ys}) such that $H_{\zeta}$ is ambiguous up to an additive constant. Here, following \cite{Harlow:2019yfa}, the integrability follows from the boundary condition $\theta|_{B}=\delta b+dC$, a consequence of  requiring the variational principle is well posed.} Upon invoking the first law it  is straightforward to verify the free energy is stationary at fixed temperature
\beq \delta F_{\zeta}^{\text{cl}}\big|_{T}=-S_{\text{BH}}\delta T=0\;.\eeq
In the limit $\alpha\to\infty$, where the nested Rindler wedge becomes the full exterior of the eternal AdS black hole and $\phi'_{\mathcal{B}}=0$, we recover the standard result for black holes. The free energy of the eternal black hole in our notation is $ F=M_{\phi_r} - T_{\text{H}} S_{\text{BH}}.$ 

In the semi-classical regime, the quantum corrected (Helmholtz) free energy and its stationarity condition are given by
\beq F_{\zeta}^{\text{semi-cl}}=E_{\zeta}^{\phi_r}+E_{\zeta}^{c}-TS_{\text{gen}}-H^{\phi_r}_{\zeta}-H_{\zeta}^{\chi}\;,\quad \delta F_{\zeta}^{\text{semi-cl}}\big|_{T}=-S_{\text{gen}}\delta T=0\;,\eeq
where the stationarity condition follows from the semi-classical first law \eqref{eq:semiclassicalthermofirstlaw}. This establishes the nested AdS-Rindler wedge is an equilibrium state, even when semi-classical corrections are taken into account.

The variable dependence and stationarity condition for the free energy also define  the statistical ensemble the system is characterized by. The \emph{canonical ensemble} is defined to be the ensemble at fixed temperature, \emph{i.e.}, a system in thermal equilibrium, such that the Helmholtz free energy is stationary in equilibrium in this ensemble. Moreover, thermodynamic stability requires the Helmholtz free energy be minimized at equilibrium in the canonical ensemble with respect to any unconstrained internal variables of the system. 
By analogy,  the free energy $F^{\text{cl}}_{\zeta}$ and its semi-classical counterpart define a canonical ensemble at fixed temperature~$T$.

 Different ensembles may be transformed into one another via an appropriate Legendre transform of a particular thermodynamic potential. Indeed, in standard thermodynamic parlance, the Helmholtz free energy $F(T)$ is simply a Legendre transform of the internal energy $E$ of the system with respect to the entropy $S$, $F=E-TS$. Likewise, the microcanonical entropy $S(E)$ defining the microcanonical ensemble is equivalent to the (negative) Legendre transform of $\beta F$  with respect to $\beta=T^{-1}$, $S(E)=-(\beta F-E\beta)$.\footnote{This may also be derived at the level of the partition functions, where the canonical partition function $Z(\beta)$ is equal to the Laplace transform of the microcanonical partition function $\Omega(E)$, with $S(E)=\log\Omega(E)$.}  Furthermore, in standard thermodynamics, the stationarity of free energy at fixed temperature in the canonical ensemble is equivalent to the stationarity of entropy at fixed energy in the microcanonical ensemble. This is because $dF |_T = dE - T dS $ and $dS |_E = dS - \beta dE$, so $dF |_T = - T dS |_E$, and hence $dF|_T=0$ is equivalent to $dS|_E=0.$
 
 Transforming between different ensembles also holds for self-gravitating systems, including  black holes \cite{Brown:1992bq}. In this context, the  microcanonical description is specified by fixing the energy surface density as boundary data, while the canonical ensemble fixes the surface temperature. Changing the boundary data thus corresponds to a Legendre transform. Therefore, as in the case of black holes in general relativity \cite{Brown:1992bq}, or more general theories of gravity \cite{Iyer:1995kg}, a suitable  Legendre transformation of the free energy $F_{\zeta}^\text{cl}$ casts the classical entropy $S_{\text{BH}}$ as the microcanonical entropy, describing the nested Rindler wedge in the \emph{microcanonical ensemble}. Specifically, 
 \beq S_{\text{BH}}=- \beta F^{\text{cl}}_{\zeta}+\beta(E_{\zeta}^{\phi_r}-H_{\zeta}^{\phi_r}) \;.\eeq
 Likewise, transforming $F^{\text{semi-cl}}_{\zeta}$ allows us to describe the microcanonical ensemble for semi-classical JT gravity, where the generalized entropy $S_{\text{gen}}$ is equal to the microcanonical entropy 
 \beq S_{\text{gen}}=- \beta F^{\text{semi-cl}}_{\zeta} +\beta (E_{\zeta}^{\phi_r} + E_\zeta^{c} -H_{\zeta}^{\phi_r}-H_{\zeta}^{\chi})  \;.\eeq
 It now follows   from the semi-classical first law (\ref{eq:semiclassicalthermofirstlaw}) that the generalized entropy is stationary at fixed asymptotic energy $E_{\zeta}\equiv E_{\zeta}^{\phi_r} + E_\zeta^{c}$ and Hamiltonian $H_{\zeta}\equiv H_{\zeta}^{\phi_r}+H_{\zeta}^{\chi}$:
 \beq
\delta S_{\text{gen}} \big |_{E_\zeta, H_\zeta} = 0 \;. \label{eq:stationarityentropygen}
\eeq
This is the equilibrium condition in the microcanonical ensemble for semi-classical JT gravity in the nested AdS-Rindler wedges. 

  To summarize, in the semi-classical regime the two thermodynamic ensembles and corresponding stationarity conditions are defined by
 \begin{align}
 & \text{canonical ensemble:} \qquad \qquad \, \, \, \text{fixed} \, T ,  \qquad \qquad  \delta F^{\text{semi-cl}}_{\zeta}=0 \;,\\
& \text{microcanonical ensemble:} \qquad \text{fixed}\, E_\zeta,H_{\zeta}   , \qquad \delta S_{\text{gen}} =0 \;.
 \end{align}
Crucially, since quantum extremal surfaces extremize the generalized entropy,   the thermodynamics of nested AdS-Rindler wedges reduces to the thermodynamics of quantum extremal surfaces in the microcanonical ensemble, when backreaction effects are taken into account. 
 
 It is worth emphasizing that the classical and semi-classical first laws, (\ref{eq:classicalthermofirstlaw}) and (\ref{eq:semiclassicalthermofirstlaw}), respectively, hold for \emph{any} nested AdS-Rindler wedge. The above discussion on ensembles, however, leads to a fundamental insight into semi-classical thermodynamics of gravitating systems: \emph{in the microcanonical ensemble it is more natural to consider the entropy of quantum extremal surfaces than black hole horizons when semi-classical effects are included}. This is because it is the QES, and not the black hole horizon, which extremizes the generalized entropy.   In this way, in the microcanonical ensemble  the semi-classical first law may be genuinely regarded as a first law of thermodynamics of quantum extremal surfaces. 
 
 Finally, the stationarity of generalized entropy   at fixed energy $E_\zeta$ and $H_\zeta$ for nested AdS-Rindler wedges 
 is strikingly similar to Jacobson's ``entanglement equilibrium" proposal \cite{Jacobson:2015hqa} for small causal diamonds: the generalized (entanglement) entropy (a.k.a. $S_{\text{gen}}$) of causal diamonds is stationary in a maximally symmetric vacuum at fixed   proper volume. This   equilibrium condition was derived in    \cite{Jacobson:2018ahi} from the semi-classical first law of causal diamonds in maximally symmetric spacetimes, applied to small diamonds. 
 The first law  of causal diamonds contains   the volume variation of the maximal slice, which is proportional to the variation of the gravitational ``York time'' Hamiltonian. In order to obtain the stationarity of generalized entropy in this setting, the volume or, equivalently, the York time Hamiltonian  must be held fixed. This suggests the entanglement equilibrium condition should be interpreted in the microcanonical ensemble, as recently acknowledged in~\cite{Jacobson:2021talk}.



 \section{Conclusion} \label{sec:conc}
 
 In this article we considered the full backreaction due to Hawking radiation in an eternal $\text{AdS}_{2}$ black hole using the semi-classical JT model. Importantly, we showed the time-dependent auxiliary field $\chi$, which localizes the Polyakov action modeling the Hawking radiation, leads to a time-dependent Wald entropy which is equal to the generalized entropy under appropriate identifications. This time-dependent solution emphasized here appears to have been overlooked, however, we believe it will prove useful  in studying the Hawking information puzzle in certain contexts, as the entropy associated with $\chi$ is precisely identified with a (time-dependent) von Neumann entropy.

Extremizing the (time-dependent) generalized entropy along the time slice $t=t_{B}$ led to a crucial insight: the appearance of a quantum extremal surface that lies just outside of the black hole horizon, consistent with previous investigations on the black hole information paradox in eternal backgrounds, \emph{e.g.}, \cite{Almheiri:2019yqk,Gautason:2020tmk,Hartman:2020swn}. The QES arises purely due to semi-classical effects, such that in the classical limit the QES coincides with the bifurcation surface of the eternal black hole. Notably, we find the QES is always present and, moreover, the generalized  entropy evaluated at the QES is smaller than  $S_{\text{gen}}$ evaluated at the black hole horizon, contrary to what was observed previously \cite{Moitra:2019xoj}. Since the QES lies outside of the horizon, such that the QES describes an AdS-Rindler wedge nested inside the eternal black hole, we derived the classical and semi-classical Smarr relations and associated first laws of the nested Rindler wedge. Actually, we   established first laws for  generic  nested AdS$_{2}$-Rindler wedges, for which the bifurcation point is at an arbitrary location outside the black hole.  Notably absent in higher dimensions, we showed the first laws naturally incorporate  the variation of the proper length, which is absent in the limit the nested wedge concurs with eternal AdS black hole. From the first law we uncovered the semi-classical free energy in the canonical ensemble, and, via a Legendre transform, identified the microcanonical entropy as the generalized entropy. This leads to the important observation that it is more natural in the microcanonical ensemble to assign thermodynamics to quantum extremal surfaces over black hole horizons when semi-classical effects are included, as it is the QES which extremizes the generalized entropy.  


There are several exciting prospects and applications of this work. Firstly, while here we focused on JT gravity in AdS, our methods may be appropriately applied to JT gravity in de Sitter space (see, \emph{e.g.}, \cite{Sybesma:2020fxg}), and the semi-classical RST \cite{Russo:1992ax} model of $2D$ dilaton-gravity in flat space, for which \cite{Gautason:2020tmk,Hartman:2020swn} will be of use. Indeed, a chief technical result of this article is the covariant phase formalism for a wide class of $2D$ dilaton-gravity theories, which will   prove useful in deriving the Smarr relation and first law in the RST    model. Second, we extended the first laws derived here by allowing for coupling variations. As with higher-dimensional black holes, these new terms enrich the phase space of the black hole thermodynamics and it would be interesting to study this in more detail.

 A principal question in the context of gravitational thermodynamics is which surface does the entropy increase with respect to (c.f. \cite{Wall:2009wm,Wall:2011kb} for a thorough examination of this question). For example, in the case of JT gravity, the classical Wald entropy of the black hole monotonically increases along the black hole event horizon as well as the apparent horizon, however, when the semi-classical correction is included the generalized entropy of the black hole only monotonically increases along the event horizon and not along the apparent horizon\footnote{More precisely, one considers the future holographic screen, \emph{i.e.}, the codimension-1 spacelike hypersurface foliated by marginally trapped surfaces whose area, given by the dilaton in the $2D$ context, is the apparent horizon and obeys the classical focusing conjecture. When semi-classical corrections are included, the generalized entropy $S_{\text{gen}}$ does not increase monotonically along the apparent horizon. Alternatively, the generalized entropy does increase along a future Q-screen \cite{Bousso:2015mna,Bousso:2015wca}, the analog of the future holographic screen where the classical area is replaced by the generalized entropy and obeys a quantum focusing conjecture.} \cite{Moitra:2019xoj}. In Appendix \ref{app:secondlaw} we provided a heuristic argument suggesting $S_{\text{gen}}$ obeys a second law along the future Killing horizon of the nested Rindler wedge. For the sake of consistency with our thermodynamic interpretation, it would be worthwhile to have a more detailed proof of the generalized second law initiated here.

As noted before  the first law of entanglement entropy is holographically equivalent to the first law of (AdS-Rindler) black hole thermodynamics \cite{Blanco:2013joa}. The relation between the first laws of entanglement and AdS-Rindler thermodynamics was used in \cite{Faulkner:2013ica} to further show the first law of holographic entanglement encodes the linearized Einstein's equations. Similarly, the semi-classical first law of AdS-Rindler space, where  bulk quantum corrections  are included dual to $1/N$ corrections in the CFT, is equivalent to the (linearized) semi-classical Einstein's equations \cite{Swingle:2014uza}. It would be interesting to see whether we can use the semi-classical first law uncovered here as an input and derive gravitational equations of motion for semi-classical JT gravity, following an analysis similar to \cite{Callebaut:2018xfu}, where one invokes the proposal of entanglement equilibrium \cite{Jacobson:2015hqa}.

A compelling feature of JT gravity (and its supersymmetric extensions), defined as a Euclidean path integral over manifolds of constant negative curvature, is that it may be completely captured by a double-scaled random matrix model \cite{Saad:2019lba,Stanford:2019vob}, including all non-perturbative effects \cite{Okuyama:2019xbv,Johnson:2019eik}. Thus, JT gravity provides an example of an exactly solvable model of $2D$ quantum gravity, whose thermodynamics is only now being investigated (c.f. \cite{Engelhardt:2020qpv, Johnson:2020mwi, Okuyama:2021pkf, Janssen:2021mek}) using matrix model techniques. Most recently, it was shown the matrix models provide an explicit and detailed description of the underlying microscopic degrees of freedom \cite{Johnson:2021zuo}, allowing for a deeper understanding of the quantum properties of black holes. It would be interesting to see how the semi-classical thermodynamics studied here incorporates this microscopic physics.

A particularly tantalizing application of this work is its ability to potentially address the information paradox in eternal backgrounds using entanglement wedge islands. As briefly described in the introduction, the fine grained gravitational entropy is given by the generalized entropy, where the QES represents an extremum of this entropy. Invoking either double holography, or the replica trick, the fine grained radiation entropy in an evaporating $2D$ black hole background may be computed, which can be used to show the radiation entropy follows a unitary Page curve, where the entropy undergoes a dynamical phase transition near the Page time. At times before the Page time, the entropy  follows the Hawking curve, eventually turning over at the Page time where von Neumann entropy   includes the degrees of freedom of an entanglement wedge island, the (typically) disconnected portion to the codimension-1 entanglement wedge of the radiation, bounded by the QES. In the case of eternal black holes, an information paradox still exists, and the island region remains outside of the black hole horizon \cite{Almheiri:2019yqk,Gautason:2020tmk,Hartman:2020swn}. 

The derivation involving the replica trick \cite{Almheiri:2019qdq,Penington:2019kki,Goto:2020wnk}  is specifically compelling as it does not require AdS/CFT holography, and argues the phase transition may be understood as an exchange in dominant saddle points in a Euclidean path integral, where at times after the Page time, the Euclidean path integral used in the replica trick is dominated by a ``replica wormhole'' saddle. Understanding this phase transition precisely in Lorentzian time remains an open question, though for recent work focused on addressing this problem, see \cite{Colin-Ellerin:2020mva,Colin-Ellerin:2021jev}. Given that we uncovered a time-dependent generalized entropy, and that the entropy may be identified as the thermodynamic potential in a microcanonical ensemble, it would be very interesting to apply this observation to the question at hand. 
Moreover, at least in eternal backgrounds, we would then be able to attribute the thermodynamics of quantum extremal surfaces to the associated islands. Studying this exciting avenue was recently explored \cite{Pedraza:2021ssc}. For dynamical settings, such as an evaporating black hole, perhaps the Iyer-Wald definition of dynamical entropy \cite{Iyer:1994ys} can be used.

Lastly, let us briefly comment on the prospect of whether our work is applicable to higher-dimensional models. Indeed, JT gravity is known to describe the physics of nearly extremal, charged four dimensional black holes, however, the question remains whether the techniques developed here and elsewhere are achievable for higher-dimensional systems. In the context of this article, properly addressing this question would involve solving the full backreacted Einstein's equations in higher dimensions, a notoriously difficult and open problem. Some progress can be made, however, as it is straightforward to compute the Wald entropy of the leading 1-loop quantum effective action induced by the trace anomaly in four dimensions \cite{Aros:2013taa}. It may be interesting to see how much of the generalized entropy the Wald functional encodes in this context, and this may shed new light on the black hole  information puzzle in higher dimensions.

\noindent\section*{Acknowledgments}

It is a pleasure to thank Erik Curiel, Daniel Harlow,  Ted Jacobson and Aron Wall for helpful and illuminating discussions. JP and AS are supported by the Simons Foundation through \emph{It from Qubit: Simons collaboration on Quantum fields, gravity, and information}. WS is supported by the Icelandic Research Fund (IRF) via a Personal Postdoctoral Fellowship Grant (185371-051). MV is supported by the Republic and canton of Geneva and the Swiss National Science Foundation, through Project Grants No. 200020- 182513 and No. 51NF40-141869 The Mathematics of Physics (SwissMAP).

 \appendix

 
  \section{Coordinate systems for $\text{AdS}_{2}$} \label{app:ebhcoord}

 Here we list various coordinate systems for the eternal $\text{AdS}_{2}$ black hole   and the transformations between them. \\

\noindent \textbf{Static coordinates:} A general solution to the \emph{vacuum} equations of motion  (\ref{eq:graveom}) and (\ref{eq:dileom})  in static  \emph{Schwarzschild   coordinates}  is (see, \emph{e.g.}, \cite{Witten:2020ert})
\beq ds^{2}=-N^{2}dt^{2}+N^{-2}dr^{2}\;,\quad N^{2}=\frac{r^{2}}{L^{2}}-\mu\;,\quad \phi(r)=\frac{r}{L}\phi_{r}\;,\label{eq:ads2bhstatapp}\eeq
where we emphasize the dilaton $\phi$ is purely classical. 
Introducing the tortoise coordinate $r_{\ast}$, 
\beq
\begin{split}
&r_{\ast}=-\int_{r}^{\infty}\frac{dr'}{r'^{2}/L^{2}-\mu}=\frac{L}{2\sqrt{\mu}}\log\left(\frac{r-L\sqrt{\mu}}{r+L\sqrt{\mu}}\right)\;,\\
&r=L\sqrt{\mu}\text{coth}(-\sqrt{\mu}r_{\ast}/L)\;, \label{eq:tortoiseschw}
\end{split}
\eeq
the line element in static coordinates becomes \cite{Moitra:2019xoj}
\beq ds^{2}=e^{2\rho}(-dt^{2}+dr_{\ast}^{2})\;,\quad e^{2\rho(r_{\ast})}=\frac{\mu}{\sinh^{2}(\sqrt{\mu}r_{\ast}/L)}\;.\label{eq:ads2conftor}\eeq
 Here $r_{\ast}$ ranges from $r_{\ast}\in[ -\infty,0]$, where  $r_{\ast}=-\infty$ is the location of the horizon and $r_{\ast}=0$ is the asymptotic boundary. In these coordinates the dilaton $\phi$ becomes
\beq \phi(r_{\ast})=\phi_{r}\sqrt{\mu}\text{coth}(-\sqrt{\mu}r_{\ast}/L)\;.\eeq
In terms of advanced and retarded time coordinates $(v,u)$
\beq 
\begin{split}
 v=t+r_{\ast}\;,  \quad u=t-r_{\ast}\;,
\end{split}
\label{eq:lccoorduv}\eeq
the static metric (\ref{eq:ads2conftor}) is cast in conformal gauge
\beq ds^{2}=-e^{2\rho(v,u)}dudv\;,\quad e^{2\rho}=\frac{\mu}{\sinh^{2}[\frac{\sqrt{\mu}}{2L}(v-u)]}\;,\label{eq:ads2confnull}\eeq
where the dilaton is now
\beq \phi(v,u)=\phi_{r}\sqrt{\mu}\text{coth}\left(-\frac{\sqrt{\mu}}{2L}(v-u)\right)\;.\eeq
The future horizon is at $u=\infty$, while the past horizon is at $v=-\infty$. As discussed in \cite{Spradlin:1999bn}, this system is characterized by the Boulware vacuum state, \emph{i.e.}, the state annihilated by modes which are positive frequency with respect to the static coordinates $(v,u)$. As a final comment, note that the surface gravity $\kappa_{H}=\frac{\sqrt{\mu}}{L}$ defining the Hawking   temperature $T_{\text H}=\frac{\kappa_{H}}{2\pi}$ appears in the above coordinate transformations, as well as those below.

\vspace{2mm}

\noindent \textbf{Poincar\'e coordinates:} Introducing a set of lightcone coordinates $(V,U)$ which are related to coordinates $(u,v)$  via \cite{Spradlin:1999bn}
\beq V=\frac{L}{\sqrt{\mu}}e^{\frac{\sqrt{\mu}}{L}v}\;,\quad  U=\frac{L}{\sqrt{\mu}}e^{\frac{\sqrt{\mu}}{L}u} \;,\label{eq:poincconfcoord}\eeq
the metric (\ref{eq:ads2confnull}) now takes the  \emph{Poincar\'e} form
\beq ds^{2}=-\frac{4L^{2}dUdV}{(V-U)^{2}}\;.\label{eq:ads2confPoinc}\eeq
The asymptotic boundary is located at $V=U$, while the future horizon is located at $U=\infty$ and the past horizon at $V=0$. The classical dilaton becomes\footnote{Alternatively, using a different definition $V = \frac{L}{\sqrt{\mu}}\tanh \left ( \frac{\sqrt{\mu}v}{2 L}\right),U = \frac{L}{\sqrt{\mu}}\tanh \left ( \frac{\sqrt{\mu}u}{2 L}\right)$ for the Poincar\'e coordinates, the line element of the Poincar\'{e} metric is unchanged, however, the dilaton now takes the form  $ \phi = \phi_r L \frac{1 - \mu U V/L^2}{U - V}\, $ \cite{Moitra:2019xoj,Almheiri:2019yqk}.}
\beq \phi(V,U)= \phi_{r}\sqrt{\mu}\text{coth}\left[\frac{1}{2}\log\left(\frac{U}{V}\right)\right]= \phi_{r}\sqrt{\mu}  \frac{U+V}{U-V} \;.\eeq

\vspace{2mm}

\noindent \textbf{Kruskal coordinates:} The line element in Schwarzschild form (\ref{eq:ads2bhstatapp}) only covers one of the exterior regions of the black hole at once, however, one can perform a Kruskal extension to analytically continue to the other exterior, as well as future and past regions. \emph{Kruskal coordinates} $\{\tilde V_{\text K}, \tilde U_{\text K}\}$ in terms of $(t,r)$ coordinates  are
\beq \tilde V_{\text K}= \sqrt{\frac{r-L\sqrt{\mu}}{r+L\sqrt{\mu}}}e^{\frac{\sqrt{\mu}}{L}t}\;,\quad \tilde U_{\text K}=-\sqrt{\frac{r-L\sqrt{\mu}}{r+L\sqrt{\mu}}}e^{-\frac{\sqrt{\mu}}{L}t}\;,\eeq
then
\beq ds^{2}=-\frac{4L^{2}d \tilde U_{\text K}d \tilde V_{\text K}}{(1+ \tilde U_{\text K} \tilde V_{\text K})^{2}}\;.\label{eq:ads2kruskal}\eeq
The inverse coordinate transformation is
\beq t=\frac{L}{2\sqrt{\mu}}\log\left(-\frac{\tilde V_{\text K}}{\tilde U_{\text K}}\right)\;, \quad r=L\sqrt{\mu}\frac{ 1-\tilde U_{\text K} \tilde V_{\text K} }{ 1+\tilde U_{\text K}\tilde V_{\text K} }\;.\eeq
The past horizon is located at $\tilde V_{\text K}=0$ and the future horizon at $ \tilde U_{\text K}=0$. In Kruskal coordinates the classical dilaton is
\beq \phi(\tilde V_{\text K}, \tilde U_{\text K})=\phi_{r}\sqrt{\mu}\frac{ 1-\tilde U_{\text K} \tilde V_{\text K} }{ 1+\tilde U_{\text K}\tilde V_{\text K} }\;.\eeq
Since Kruskal coordinates cover all of $\text{AdS}_{2}$, the metric (\ref{eq:ads2kruskal}) characterizes the global structure of $\text{AdS}_{2}$ in conformal coordinates. As pointed out in \cite{Spradlin:1999bn}, the vacuum state with respect to this set of coordinates is equivalent to the Hartle-Hawking state.

We can relate (dimensionless) Kruskal coordinates to static coordinates $(v,u)$ by
\beq \tilde V_{\text K}=e^{\frac{\sqrt{\mu}}{L}v}\;,\quad \tilde U_{\text K} =-e^{-\frac{\sqrt{\mu}}{L}u}\;,\eeq
 and to conformal Poincar\'e coordinates (\ref{eq:poincconfcoord}) as
\beq \tilde V_{\text K}=\frac{\sqrt{\mu}}{L}V\;,\quad \tilde U_{\text K}=- \frac{L}{\sqrt{\mu}}\frac{1}{U}\;.\eeq
Note we will often work with scaled (dimensionful) Kruskal coordinates as in \cite{Spradlin:1999bn}, 
\beq  {V}_{K}=\frac{L}{\sqrt{\mu}}e^{\frac{\sqrt{\mu}}{L}v}\;,\quad  {U}_{K}=-\frac{L}{\sqrt{\mu}}e^{-\frac{\sqrt{\mu}}{L}u}\;,\label{eq:scalekruskalcoord}\eeq
in which case the Kruskal line element becomes
\beq ds^{2}=-\frac{4 \mu d   U_{\text K}d   V_{\text K}}{(1+  \frac{\mu}{L^2} U_{\text K}  V_{\text K})^{2}}\;. \eeq

 \vspace{2mm}

\noindent \textbf{AdS-Rindler coordinates:} A  line element for \emph{AdS$_2$-Rindler space} is
\begin{equation}
    ds^2 = -  \kappa^2  \varrho^2   d \sigma^2 + \left ( \frac{\varrho^2}{L^2} +1 \right)^{-1} d \varrho^2.
\label{eq:ads2rindv2}\end{equation}
Surfaces of constant $\sigma$ parametrize the worldlines of   accelerating observers with  $a > 1/L$. Thus, $\sigma$ is the   proper Rindler time.  The AdS-Rindler horizon is located at $\varrho = 0$ and the asymptotic boundary at $\varrho = \infty.$ Further, $\kappa$ is the surface gravity of the future Rindler horizon, which is kept arbitrary here. In the flat space limit $L\to\infty$   one recovers flat $2D$ Rindler space.

AdS-Rindler coordinates   $(\sigma,\varrho)$ may be directly related to Schwarzschild   coordinates $(t,r)$ via
\beq \kappa \sigma = \sqrt{\mu}t/L \;, \quad \varrho=\frac{L}{\sqrt{\mu}}\sqrt{\frac{r^{2}}{L^{2}}-\mu}\;,\label{eq:adsrindlercoordtr}\eeq
so that Schwarzschild time $t$ is a rescaled proper Rindler time, and hence the eternal AdS$_2$ black hole is equivalent to AdS$_2$-Rindler space. If we set   the surface gravity equal to $\kappa  = \sqrt{\mu}/L$ then the two times $\sigma$ and $t$ coincide. Further, since conversely we have $r=L\sqrt{\mu}\sqrt{\frac{\varrho^{2}}{L^{2}}+1}$,   the static dilaton is cast as
\beq \phi(\varrho)=\phi_{r}\sqrt{\mu}\sqrt{\frac{\varrho^{2}}{L^{2}}+1}\;.\eeq
Defining the tortoise coordinate as  $\varrho_{\ast} = \frac{1}{\kappa} \text{arcsinh} [L/ \varrho ]$, AdS-Rindler space (\ref{eq:ads2rindv2}) can be written in conformal gauge, 
\beq ds^{2}=e^{2\rho(\varrho_{\ast})}(-d\sigma^{2}+d\varrho_{\ast}^{2})\;,\quad e^{2\rho(\varrho_{\ast})}=\frac{\kappa^2 L^2  }{\sinh^{2} ( \kappa \varrho_{\ast}  )}= \kappa^2 \varrho^{2}\;.\eeq
The horizon $\varrho=0$ is now located at $\varrho_{\ast}=-\infty$, while the conformal boundary is   at $\varrho_{\ast}=0$. Note that this tortoise coordinate is just a rescaling of the Schwarzschild tortoise coordinate~\eqref{eq:tortoiseschw}, namely  $\kappa \varrho_*  = \sqrt{\mu} r_*  / L.$


\subsection{Conformal gauge identities and the stress-energy tensor} \label{subsecapp:stress}

In conformal gauge $ds^2 = - e^{2 \rho} dy^+ dy^-$
the following identities are useful for working out the equations of motion
\begin{align} \label{eq:usefulid}
&g_{+-}=-\frac{1}{2}e^{2\rho}\;,\quad g^{+-}=-2e^{-2\rho}\;,\quad \Gamma^{+}_{\;++}=2\partial_{+}\rho\;,\quad \Gamma^{-}_{\;--}=2\partial_{-}\rho\;, \\
&R=-2\Box\rho=8e^{-2\rho}\partial_{+}\partial_{-}\rho\;,\quad \Box\Phi=-4e^{-2\rho}\partial_{+}\partial_{-}\Phi\;,\quad (\nabla\Phi)^{2}=-4e^{-2\rho}(\partial_{+}\Phi)(\partial_{-}\Phi)\;.  \nonumber
\end{align}
for a scalar field $\Phi$ (for us $\Phi$ denotes either $\phi$ or $\chi$).

In conformal static coordinates  \eqref{eq:lccoorduv} $(v,u)$,
the expectation value of the stress-energy tensor is
\beq
\begin{split}
& \langle T^{\chi}_{uv}\rangle=\langle T^{\chi}_{vu}\rangle=\frac{c}{48\pi L^{2}}e^{2\rho}(1-\lambda)\;,\\
&\langle T^{\chi}_{uu}\rangle=-\frac{c}{48\pi L^{2}}\mu-\frac{c}{12\pi}t_{u}(u)\;,\\
&\langle T^{\chi}_{vv}\rangle=-\frac{c}{48\pi L^{2}}\mu-\frac{c}{12\pi}t_{v}(v)\;,
\end{split}
\label{eq:tchivvs}\eeq
where $\langle T_{\mu\nu}^\chi \rangle \equiv  \langle \Psi | T_{\mu \nu}^\chi |\Psi \rangle$ for some unspecified quantum state $| \Psi \rangle$. Meanwhile, in null Kruskal coordinates (\ref{eq:scalekruskalcoord}) 
$(V_{\text K},U_{\text K})$, 
it is straightforward to verify
\beq
\begin{split}
& \langle T^{\chi}_{V_{\text K}U_{\text K}}\rangle=\langle T^{\chi}_{U_{\text K}V_{\text K}}\rangle=\frac{c}{48\pi L^{2}}e^{2\rho}(1-\lambda)\;,\\
&\langle T^{\chi}_{U_{\text K}U_{\text K}}\rangle=-\frac{c}{12\pi}t_{U_{\text K}}(U_{\text K})\;,\\
&\langle T^{\chi}_{V_{\text K}V_{\text K}}\rangle=-\frac{c}{12\pi}t_{V_{\text K}}(V_{\text K})\;.
\end{split}
\label{eq:tchiVVs}\eeq
We observe in either static or Kruskal coordinates for $\lambda=1$ the off-diagonal components $\langle T^{\chi}_{\pm \mp}\rangle$ vanish. 

Using the transformation
\beq \rho(v,u)=\rho(V_{\text K},U_{\text K})+ \frac{\sqrt{\mu}}{2 L}(v-u)=\rho(V_{\text K},U_{\text K})+\frac{1}{2}\log\left(-\frac{\mu}{L^{2}}U_{\text K}V_{\text K}\right)\;,\eeq
we find
\beq \xi_{u}=\xi_{U_{\text K}}-\frac{\sqrt{\mu}}{2 L}u=\xi_{U_{\text K}}+\frac{1}{2}\log\left(-\frac{\sqrt{\mu}}{L}U_{\text K}\right)\;,\eeq
\beq \xi_{v}=\xi_{V_{\text K}}+\frac{\sqrt{\mu}}{2L}v=\xi_{V_{\text K}}+\frac{1}{2}\log\left(\frac{\sqrt{\mu}}{L}V_{\text K}\right)\;,\eeq
such that 
\beq \xi(v,u)=\xi(V_{\text K},U_{\text K})+\frac{1}{2}\log\left(-\frac{\mu}{L^{2}}U_{\text K}V_{\text K}\right)\;.\eeq
Then with the following Schwarzian derivatives:
\beq
\begin{split}
& \frac{c}{24\pi}\{V_{\text K},v\}=\frac{c}{24\pi}\{U_{\text K},u\}=-\frac{c\mu}{48\pi L^{2}}\;,\\
&\frac{c}{24\pi}\{v,V_{\text K}\}=\frac{c}{48\pi V_{\text K}^{2}}\;,\quad \frac{c}{24\pi}\{u,U_{\text K}\}=\frac{c}{48\pi U_{\text K}^{2}}\;,
\end{split}
\eeq
the transformation rule for the normal-ordered stress tensors (\ref{eq:transnormord}) yields
\beq 
\begin{split}
:T^{\chi}_{vv}:&=\left(\frac{dV_{\text K}}{dv}\right)^{2}:T^{\chi}_{V_{\text K}V_{\text K}}:-\frac{c}{24\pi}\{V_{\text K},v\}=\left(\frac{\sqrt{\mu}}{L}V_{\text K}\right)^{2}:T^{\chi}_{V_{\text K}V_{\text K}}:+\frac{N\mu}{48\pi L^{2}}\\
&=4\pi^{2}T^{2}_{H} V_{\text K}^{2}:T^{\chi}_{V_{\text K}V_{\text K}}:+\frac{c\pi}{12}T^{2}_{H}\;,
\end{split}
\eeq
\beq 
\begin{split}
:T^{\chi}_{uu}:&=\left(\frac{dU_{\text K}}{du}\right)^{2}:T^{\chi}_{U_{\text K}U_{\text K}}:-\frac{c}{24\pi}\{U_{\text K},u\}=\left(\frac{\sqrt{\mu}}{L}U_{\text K}\right)^{2}:T^{\chi}_{U_{\text K}U_{\text K}}:+\frac{c\mu}{48\pi L^{2}}\\
&=4\pi^{2}T^{2}_{H}U_{\text K}^{2}:T^{\chi}_{U_{\text K}U_{\text K}}:+\frac{c\pi}{12}T^{2}_{H}\;.
\end{split}
\eeq
Lastly, the normal ordered relation (\ref{eq:normordgen}) gives
\beq \langle :T^{\chi}_{vv}:\rangle=-\frac{c}{12\pi}t_{v}(v)\;,\quad \langle :T^{\chi}_{V_{\text K}V_{\text K}}:\rangle=-\frac{c}{12\pi}t_{V_{\text K}}(V_{\text K})\;,\eeq
and similarly for $uu$ and $U_{\text K}U_{\text K}$ components, such that $t_{V_{\text K}}$ and $t_{v}$, and $t_{U_{\text K}}$ and $t_{u}$ are related by 
\beq t_{v}=\left(\frac{\sqrt{\mu}}{L}V_{\text K}\right)^{2}t_{V_{\text K}}-\frac{\mu}{4 L^{2}}\;,\quad t_{u}=\left(\frac{\sqrt{\mu}}{L}U_{\text K}\right)^{2}t_{U_{\text K}}-\frac{\mu}{4 L^{2}}\;.\eeq
Consequently, the diagonal stress-energy tensors in (\ref{eq:tchivvs}) and (\ref{eq:tchiVVs}) are, respectively,
\beq \langle T^{\chi}_{uu}\rangle=-\frac{c\mu}{12\pi L^{2}}U_{\text K}^{2}t_{U_{\text K}}\;,\quad \langle T^{\chi}_{vv}\rangle=-\frac{c\mu}{12\pi L^{2}}V_{\text K}^{2}t_{V_{\text K}}\;,\eeq
\beq \langle T^{\chi}_{U_{\text K}U_{\text K}}\rangle=-\frac{c}{48\pi}\frac{1}{U_{\text K}^{2}}-\frac{c L^{2}}{12\pi\mu}\frac{t_{u}}{U_{\text K}^{2}}\;,\quad \langle T^{\chi}_{V_{\text K}V_{\text K}}\rangle=-\frac{c}{48\pi}\frac{1}{V_{\text K}^{2}}-\frac{c L^{2}}{12\pi\mu}\frac{t_{v}}{V_{\text K}^{2}}\;.\eeq
It is worth comparing to the relation between stress tensors in static and conformal Poincar\'e coordinates (\ref{eq:poincconfcoord})
\beq
\begin{split}
:T_{vv}^{\chi}:&=\left(\frac{dV}{dv}\right)^{2}:T_{VV}^{\chi}:-\frac{c}{24\pi}\{V,v\}=\left(\frac{\sqrt{\mu}}{L}V\right)^{2}:T_{VV}^{\chi}:+\frac{c\mu}{48\pi L^{2}}\\
&=4\pi^{2}T_{\text H}^{2}V^{2}:T_{VV}^{\chi}:+\frac{c\pi}{12} T_{\text H}^{2}\;,
\end{split}
\label{eq:normordtransuvUV}\eeq
\beq
\begin{split}
:T_{uu}^{\chi}:&=\left(\frac{dU}{du}\right)^{2}:T_{UU}^{\chi}:-\frac{c}{24\pi}\{U,u\}=\left(\frac{\sqrt{\mu}}{L}U\right)^{2}:T_{UU}^{\chi}:+\frac{c\mu}{48\pi L^{2}}\\
&=4\pi^{2}T_{\text H}^{2}U^{2}:T_{VV}^{\chi}:+\frac{c\pi}{12} T_{\text H}^{2}\;.
\end{split}
\label{eq:normordtransuvUV2}\eeq
Using the relations 
\beq \langle :T_{vv}^{\chi}:\rangle = -\frac{c}{12\pi}t_{v}\;,\quad \langle :T_{VV}^{\chi}:\rangle=-\frac{c}{12\pi}t_{V}\;,\eeq
equation (\ref{eq:normordtransuvUV}) leads to a relation between $t_{v}$ and $t_{V}$, 
\beq t_{v}=\left(\frac{\sqrt{\mu}}{L}V\right)^{2}t_{V}-\frac{\mu}{4L^{2}}\;,\quad  t_{u}=\left(\frac{\sqrt{\mu}}{L}U\right)^{2}t_{U}-\frac{\mu}{4L^{2}}\;,\eeq
we find
\beq 
\begin{split}
&\langle T_{uu}^{\chi}\rangle=-\frac{c\mu}{12\pi L^{2}}U^{2}t_{U}\;,\quad \langle T_{vv}^{\chi}\rangle=-\frac{c\mu}{12\pi L^{2}}V^{2}t_{V}\;,\\
&\langle T_{UU}^{\chi}\rangle=-\frac{c}{48\pi}\frac{1}{U^{2}}-\frac{c L^{2}}{12\pi\mu}\frac{t_{u}}{U^{2}}\;,\quad \langle T_{VV}^{\chi}\rangle=-\frac{c}{48\pi}\frac{1}{V^{2}}-\frac{c L^{2}}{12\pi\mu}\frac{t_{v}}{V^{2}}\;.
\end{split}
\eeq
Thus, the relations between stress-energy tensors in static and Poincar\'e coordinates and static and Kruskal coordinates are identical. This is indicative of the fact the Kruskal and Poincar\'e vacuum states coincide \cite{Spradlin:1999bn}.


 \section{Boost Killing vector from the embedding formalism for AdS$_2$} \label{app:embeddingformalism}
 
 A quick method to compute the Killing vectors of AdS, particularly the boost Killing vector of AdS-Rindler space, follows from the embedding formalism. Locally two-dimensional AdS space  can be embedded in   three-dimensional flat space $\mathbb R^{2,1}$ with signature $(--+)$ and metric
 \begin{equation}
  ds^2 = - (dT^1)^2 - (dT^2)^2 +  dX^2, 
 \end{equation}
where AdS$_2$ is realized as a hyperboloid in the embedding space
\begin{equation}
    - (T^1)^2 - (T^2)^2 +X^2=- L^2.
\end{equation}
The embedding space induces a metric on this hyperboloid, corresponding to the metric of AdS$_2$. 
Furthermore, the symmetry group $SO(2,1)$ that preserves this hyperboloid  is the isometry group of AdS$_2$. In embedding coordinates, the rotation generator $J$ and the two boost generators $B_{1,2}$ of $SO(2,1)$ are given by 
 \begin{equation} \label{generatorsembedding}
J = T^1 \partial_{T^2} - T^2 \partial_{T^1} ,    \qquad B_1 = X \partial_{T^1} + T^1 \partial_X, \qquad B_2 = X \partial_{T^2} + T^2 \partial_X,
 \end{equation}
and satisfy the commutation relations
 \begin{equation}
 [J,B_1] = - B_2, \qquad [J,B_2]=   B_1, \qquad [B_1, B_2]=  J.
 \end{equation} 
 We can now compute the Killing vectors of AdS$_2$ from \eqref{generatorsembedding} by inserting the embeddding coordinates for specific AdS$_2$ coordinate systems. In particular, here we restrict ourselves to the AdS$_2$ metric in Schwarzschild coordinates \eqref{eq:ads2bhstatapp}, for which the embedding coordinates are given by 
  \begin{equation}
     T^1= \frac{r}{\sqrt{\mu}}, \quad
     T^2  = \frac{L}{\sqrt{\mu}} \sqrt{\frac{r^{2}}{L^{2}}- \mu}  \sinh \left(\frac{\sqrt{\mu} t}{L}\right),  \quad
     X  = \frac{L}{\sqrt{\mu}} \sqrt{\frac{r^2}{L^2} - \mu} \cosh \left(\frac{\sqrt{\mu}t}{L}\right),
 \end{equation}
 and also to the metric in retarded and advanced time coordinates $(v,u)$ \eqref{eq:ads2confnull}, for which the embedding coordinates are
   \begin{equation}
     T^1=L \frac{\cosh \left [    \frac{\sqrt{\mu}}{2L} (u-v)\right]}{\sinh \left [    \frac{\sqrt{\mu}}{2L} (u-v)\right]}, \quad
     T^2  = L \frac{\sinh \left [ \frac{\sqrt{\mu}}{2L}(u+v) \right]}{\sinh\left[\frac{\sqrt{\mu}}{2L} (u-v) \right]},   \quad
     X  =L \frac{\cosh \left [ \frac{\sqrt{\mu}}{2L}(u+v) \right]}{\sinh\left[\frac{\sqrt{\mu}}{2L} (u-v) \right]} .
 \label{eq:embeddnulluv}\end{equation}
  In Schwarzschild  coordinates the isometry generators  are
 \begin{align}
     J &= \frac{  r / \sqrt{\mu}}{\sqrt{r^2 / L^2 - \mu}} \cosh(\sqrt{\mu}t /L) \partial_t - L \sqrt{r^2 / L^2 - \mu} \sinh(\sqrt{\mu}t /L) \partial_r\;, \qquad B_2 = \frac{L}{\sqrt{\mu}} \partial_t\;, \nonumber \\
     B_1 &=-\frac{ r / \sqrt{\mu}}{\sqrt{r^2 / L^2 - \mu}} \sinh(\sqrt{\mu}t /L) \partial_t + L \sqrt{r^2 / L^2 - \mu} \cosh(\sqrt{\mu}t /L) \partial_r \;.
 \label{eq:gensSchw}\end{align}
Thus, $B_{2}$ is the time-translation Killing vector of the eternal AdS black hole. Meanwhile, in retarded and advanced null coordinates $(v,u)$, the isometry generators take the form\footnote{It is useful to know the inverse relationship $(u-v)=\frac{2L}{\sqrt{\mu}}\text{arccoth}(T^{1}/L)$ and $(u+v)=\frac{2L}{\sqrt{\mu}}\text{arctanh}(T^{2}/X)$.}
\begin{align}
 J &= \frac{L}{\sqrt{\mu}} \left [ \cosh (\sqrt{\mu} v / L)\partial_v + \cosh(\sqrt{\mu} u /L) \partial_u \right]\;, \qquad \quad B_2 = \frac{L}{\sqrt{\mu}} ( \partial_v + \partial_u)\;,
 \nonumber\\
  B_1  &= -\frac{L}{\sqrt{\mu}} \left [ \sinh (\sqrt{\mu} v / L)\partial_v + \sinh(\sqrt{\mu} u /L) \partial_u \right]\; .
\end{align}
Let us now compute the boost Killing vector field of an $\text{AdS}_{2}$-Rindler wedge nested inside an eternal $\text{AdS}_{2}$ black hole, \emph{i.e.}, a Rindler wedge inside a Rindler wedge. Denote $(\bar{t},\bar{r})$ as the Schwarzschild coordinates adapted to the nested AdS-Rindler wedge. Correspondingly, the adapted     advanced and retarded time coordinates are denoted as $(\bar{v},\bar{u})$, and are related to Schwarzschild coordinates in the same way as the eternal black hole, (\ref{eq:lccoorduv}). Thus, barred (adapted) coordinates, \emph{e.g.}, $(\bar{t},\bar{r})$, describe the generally smaller Rindler wedge nested inside the larger Rindler wedge, described by, say, $(t,r)$ coordinates. Succinctly,
\begin{align}
    \text{Enveloping eternal black hole} &:  \quad  (t,r) \quad \text{and} \quad (v,u) \quad  \text{coordinates}\,, \\
    \text{Nested AdS-Rindler wedge} &: \quad   (\bar t, \bar r) \quad  \text{and} \quad (\bar v,\bar u) \quad \text{coordinates}\,.
\end{align}
Note the future and past Killing horizons of the smaller Rindler wedge are located at $\bar{u}\to\infty$ and $\bar{v}\to-\infty$, respectively, which generally do not coincide with the future and past horizon of the $\text{AdS}_{2}$ black hole. Moreover, the timeslice $t=0$ is equal to the $\bar{t}=0$ slice. 

In adapted coordinates, the boost Killing vector is proportional to $\partial_{\bar{t}}$, \emph{i.e.}, the generator of proper Rindler time translations. Ultimately, however, we want to express this boost Killing vector in terms of the coordinates of the enveloping black hole, $(t,r)$ or $(v,u)$. To do so, let us relate $(\bar{v},\bar{u})$ coordinates to $(v,u)$ coordinates, which may be conveniently addressed in the   embedding space formalism. As proven in \cite{Casini:2011kv} and explained below, the small and large AdS-Rindler wedges can be transformed into each other by an isometry, corresponding to a boost in the ($T^1, X$) plane in embedding space. Below we mainly follow the derivation of the boost Killing vector in Appendix C.2 of \cite{Sarkar:2020yjs}, but we use different coordinates and we extend the analysis to nested Rindler wedges centered at arbitrary time slices. 
 
Consider the following boost in embedding space:
 \begin{equation}
     (T^1)' = \cosh (\beta) T^1 - \sinh (\beta) X, \quad  \qquad (X)'  = \cosh (\beta) X - \sinh (\beta) T^1,
 \label{eq:boostprime}\end{equation}
 where $\beta\in(-\infty,\infty)$ is the rapidity. Notice when $\beta=0$ we recover $(T^{1})'=T^{1}$ and $X'=X$.  The boost Killing vector of the transformed AdS-Rindler wedge  is   given by the boost generator in the $(T^2, X')$ plane,  which we   denote as $\zeta = X' \partial_{T^2} + T^2 \partial_{X'}$. We can now   express   $\zeta$   in terms of the umprimed embedding coordinates. Indeed, using \eqref{eq:boostprime}  we see   $\zeta$ becomes the following linear combination of isometry generators
 \begin{equation}
     \zeta = \cosh (\beta) B_2 - \sinh (\beta) J.
 \label{eq:boostzetaunp}\end{equation}
 Observe if $\beta=0$ we recover $\zeta=B_{2}$. 
 
 Since we are interested in the size of the nested Rindler wedge, we want to know how $\beta$ is related to the size of the Rindler wedge at the boundary, characterized by a dimensionful  parameter $\alpha$ we define momentarily. We begin by considering the following coordinate transformation between the enveloping AdS-Rindler wedge and the nested AdS-Rindler wedge:
 \beq
 \begin{split}
&T^1=  L\cosh (\beta) \frac{\cosh \left [\frac{\kappa}{2} (\bar u-\bar v)\right]}{\sinh \left [\frac{\kappa}{2} (\bar u-\bar v)\right]} + L\sinh (\beta)\frac{\cosh \left [ \frac{\kappa}{2}(\bar u+\bar v) \right]}{\sinh\left[\frac{\kappa}{2} (\bar u-\bar v) \right]}\;,\\
&X =  L\sinh (\beta)\frac{\cosh \left [\frac{\kappa}{2} (\bar u-\bar v)\right]}{\sinh \left [\frac{\kappa}{2} (\bar u-\bar v)\right]} + L\cosh (\beta)   \frac{\cosh \left [ \frac{\kappa}{2}(\bar u+\bar v) \right]}{\sinh\left[\frac{\kappa}{2} (\bar u-\bar v) \right]}\;,\\
&T^2  = L\frac{\sinh \left [ \frac{\kappa}{2}(\bar u+\bar v) \right]}{\sinh\left[\frac{\kappa}{2} (\bar u-\bar v) \right]}\;,
\end{split}
\eeq
where $\kappa$ is the (arbitrary) surface gravity of the Killing horizon of the nested wedge, associated to the boost Killing vector $\partial_{\bar t}$, which is analogous to the (fixed) surface gravity $\sqrt{\mu}/L$  of the Killing horizon of the enveloping   black hole, associated to the time-translation Killing vector~$\partial_t$.  We arrived to the above transformation by first inverting the boost (\ref{eq:boostprime}), such that $T^{1}=\cosh(\beta)(T^{1})'+\sinh(\beta)X'$ and $X=\sinh(\beta)(T^{1})'+\cosh(\beta)X'$. We then substituted the primed embedding coordinates for  $(\bar{v},\bar{u})$ of the smaller wedge, given by (\ref{eq:embeddnulluv}) with the replacements $(T^{1},T^{2},X)\to((T^{1})',(T^{2})',X')$,   $(v,u)\to(\bar{v},\bar{u})$ and $\sqrt{\mu}/L \to \kappa$. Substituting the unprimed embedding coordinates (\ref{eq:embeddnulluv}) for the enveloping eternal black hole into the left-hand side of the above transformation, with the help of Mathematica, we uncover the following simple transformation between the retarded and advanced time in the two wedges 
\begin{equation}
    \coth \left[ \frac{\sqrt{\mu}}{2L} u\right] = e^\beta \coth \left[ \frac{\kappa}{2} \bar u\right]\;, \quad   \coth \left[ \frac{\sqrt{\mu}}{2L} v\right]  = e^\beta \coth \left[ \frac{\kappa}{2} \bar v\right]\;.
\label{eq:retadvwedges}\end{equation}
Next, consider the limit to the bifurcation point $\mathcal B$  of the nested $\text{AdS}_{2}$-Rindler horizon, $\bar{u}\to\infty$, $\bar{v}\to-\infty$, and define a parameter $\alpha$ to be  the values of the retarded and advanced times at~$\mathcal B$,  \emph{i.e.},  $u = \alpha   $ and $v =- \alpha  $, 
\begin{equation}
    \bar u = \infty,\; \bar v = - \infty \quad \to \quad \sinh \beta = \frac{1}{\sinh     (\sqrt{\mu} \alpha / L)} \quad \cosh \beta = \frac{\cosh(\sqrt{\mu} \alpha / L)}{\sinh(\sqrt{\mu} \alpha / L)}\,. \label{eq:relationalphabeta}
\end{equation}
 Notice the limit $\alpha\to\infty$ corresponds to the rapidity $\beta=0$, while $\alpha=0$ corresponds to $\beta\to\infty$. The null boundaries of the nested Rindler wedge intersect  the AdS boundary at $t = \pm \alpha $ in static coordinates. The parameter~$\alpha$ therefore represents a (boundary) time interval, such that in the $\alpha\to\infty$ limit the time interval is the entire time boundary and the nested Rindler wedge coincides with the Rindler wedge of the enveloping black hole. 

Replacing rapidity $\beta$ with $\alpha$ in (\ref{eq:retadvwedges}) we now have
\begin{equation}
    \coth \left[ \frac{\sqrt{\mu}}{2L} u\right] =  \coth \left[ \frac{\sqrt{\mu}}{2L}\alpha\right] \coth \left[ \frac{\kappa}{2} \bar u\right], \quad   \coth \left[ \frac{\sqrt{\mu}}{2L} v\right]  =  \coth \left[ \frac{\sqrt{\mu}}{2L}\alpha\right] \coth \left[ \frac{\kappa}{2} \bar v\right]\;, 
\end{equation}
or equivalently,
\begin{equation}
     \frac{\sqrt{\mu}}{L}   u = \log \left [ \frac{1 + e^{\kappa \bar u + \sqrt{\mu} \alpha / L}}{ e^{\sqrt{\mu} \alpha / L}+e^{\kappa\bar u}} \right]\;,\quad  \frac{\sqrt{\mu}}{L}v = \log \left [ \frac{1 + e^{\kappa \bar v + \sqrt{\mu} \alpha / L}}{ e^{\sqrt{\mu} \alpha / L}+e^{\kappa \bar v}} \right]\;. \label{eq:coordtransflargesmallRindler}
\end{equation}
 Consequently, the boost Killing  vector $\zeta$ (\ref{eq:boostzetaunp}) in terms of $\alpha$ is
 \begin{equation}
     \zeta = \frac{1}{\sinh (\sqrt{\mu} \alpha / L)} \left ( \cosh (\sqrt{\mu} \alpha / L) B_2 - J \right)\;.
 \end{equation}
 In null coordinates $(v,u)$,
 \begin{equation}
   \zeta = \frac{1}{\sinh (\sqrt{\mu} \alpha / L)} \frac{L}{\sqrt{\mu}}\Big[ (  \cosh (\sqrt{\mu} \alpha / L)  -  \cosh (\sqrt{\mu} v / L))\partial_v+   ( \cosh (\sqrt{\mu} \alpha / L) -  \cosh(\sqrt{\mu} u /L)) \partial_u     \Big]\;,
 \end{equation}
 from which we observe $\zeta$ becomes null on the future and past Killing horizons  $u= \alpha  $ and $v=- \alpha  $, respectively. In Schwarzschild coordinates $(t,r)$, $\zeta$ takes the form
 \begin{equation}
     \zeta = \frac{1}{\sinh (\sqrt{\mu} \alpha / L)} \left [ \frac{L}{\sqrt{\mu}} \left ( \cosh (\sqrt{\mu} \alpha / L) - \frac{r / L}{\sqrt{\frac{r^2}{ L^2} - \mu}} \cosh (\sqrt{\mu} t / L) \right) \partial_t + L \sqrt{\frac{r^2}{L^2} - \mu} \sinh (\sqrt{\mu} t/ L) \partial_r  \right]\;.
 \end{equation}
 We see in the limit $\alpha \to \infty$  the boost Killing vector reduces to the isometry generator
  $B_2=(L / \sqrt{\mu}) \partial_t$, as expected. Moreover, the boundary vertices $\{t = \pm \alpha  , r= \infty \}$ and the bifurcation point $\{t=0, r = L \sqrt{\mu} \coth (\sqrt{\mu} \alpha / L)  \}$  are fixed points of the flow of $\zeta.$ We emphasize that here   $\zeta$ is normalized such that the surface gravity $\kappa=1$ at the future horizon. We can restore a generic surface gravity $\kappa \neq 1$ by replacing $\zeta \to \kappa \zeta.$

\subsection*{Nested wedge on arbitrary time slice} 

 Thus far we have considered a nested AdS-Rindler wedge centered at the $t=0$ slice. More generally one could consider an AdS-Rindler wedge where the bifurcation point is located at some other slice $t_{0}\neq0$. To see this, first note that $\zeta$ becomes null on the Killing horizon when the square of its norm,
 \beq
     \zeta^{2} 
     = - \frac{L^2/\mu}{\sinh^2 (\sqrt{\mu} \alpha / L)} \left ( \frac{r^2}{L^2} - \mu\right) \left [ \cosh \left(\frac{\sqrt{\mu}}{L} (t + \alpha) \right) - \frac{r/L}{\sqrt{\frac{r^2}{ L^2} - \mu}} \right] \left[ \cosh \left(\frac{\sqrt{\mu}}{L} (t -  \alpha) \right) - \frac{r/L}{\sqrt{\frac{r^2}{L^2} - \mu}}\right]\;,
 \label{eq:normzetaapp}\eeq
 vanishes. This occurs at 
 \begin{equation}
     \cosh \left(\frac{\sqrt{\mu}}{L} (t \pm \alpha) \right) = \frac{r/L}{\sqrt{r^2/ L^2 - \mu}}\;,
 \end{equation}
 where the plus sign corresponds to the past Killing horizon, and the minus sign to the future Killing horizon.
 This observation suggests the following generalization to an AdS-Rindler wedge which is centered at some arbitrary time $t_0$ slice, instead of just $t=0$,
 \begin{equation}
     \cosh \left (\frac{ \sqrt{\mu} }{L} (t- t_0 \pm \alpha) \right) = \frac{r / L}{\sqrt{r^2 / L^2 - \mu}}\;.
 \end{equation}
 The associated $\text{AdS}_{2}$ Killing vector is then
 \begin{equation}
 \begin{split}
          \zeta = \frac{1}{\sinh (\sqrt{\mu} \alpha / L)} &\Bigg [ \frac{L}{\sqrt{\mu}} \left ( \cosh \left(\frac{\sqrt{\mu}}{L} \alpha\right) - \frac{r / L}{\sqrt{\frac{r^2 }{L^2} - \mu}} \cosh \left(\frac{\sqrt{\mu}}{ L}(t-t_0)\right)\right) \partial_t \\
          &+ L \sqrt{\frac{r^2}{L^2}- \mu} \sinh \left(\frac{\sqrt{\mu}}{L}(t-t_0)\right) \partial_r  \Bigg],
\end{split}
 \end{equation}
 or in terms of $(v,u)$ coordinates
 \beq
 \begin{split}
\zeta =  \frac{1}{\sinh (\sqrt{\mu} \alpha / L)} \frac{L}{\sqrt{\mu}}&\Bigg[ \left (  \cosh \left(\frac{\sqrt{\mu}}{L} \alpha\right)  -  \cosh\left ( \frac{\sqrt{\mu}}{L} (  v - t_0)  \right) \right)\partial_v\\
&+  \left ( \cosh \left(\frac{\sqrt{\mu}}{L} \alpha\right) -  \cosh\left ( \frac{\sqrt{\mu}}{L} (  u - t_0)  \right) \right) \partial_u     \Bigg]    \,.
\end{split}
 \eeq
In fact, this expression for $\zeta$ corresponds in the embedding space to the following linear combination of   isometry generators 
 \begin{align}
     \zeta 
     &= \frac{1}{\sinh (\sqrt{\mu} \alpha / L)} \left [  \cosh (\sqrt{\mu} \alpha / L) B_2 - \cosh (\sqrt{\mu}t_0/L)J - \sinh (\sqrt{\mu}t_0/L) B_1\right]\,.
\label{eq:newboostzeta} \end{align}
This new $\zeta$ is associated to a shifted AdS-Rindler wedge centered at $t_0 \neq 0$, and can be obtained through another isometry transformation of the AdS-Rindler wedge centered at $t=0$. Specifically, if we perform a subsequent boost in the $(T^2, X)$ plane then the extremal slice of the nested Rindler wedge shifts from $t=0$ to some arbitrary $t=t_0\neq 0$. 

To see this explicitly, consider a boost in the  $(T^2, X)$ plane for a different rapidity $\gamma$ which we will identify momentarily,
  \begin{equation}
     (T^2)' = \cosh (\gamma) \, T^2 - \sinh (\gamma) X, \qquad  (X)'  = \cosh (\gamma) X - \sinh (\gamma) T^2.
 \end{equation}
 The boost Killing vector of the transformed AdS-Rindler wedge is now, analogous to (\ref{eq:boostzetaunp}),
 \begin{equation}
     \zeta = \cosh( \beta) B_2' - \sinh (\beta) J'\;,
 \end{equation}
 where $B_2' = X' \partial_{(T^2)'} + (T^2)'\partial_{X'} = X \partial_{T^2} + T^2 \partial_X=B_2$ and $J'= T^1 \partial_{(T^2)'}- (T^2)'\partial_{T^1}=\cosh( \gamma) J +  \sinh( \gamma) B_1.$ Thus, the transformed boost Killing vector can be written as
 \begin{equation}
    \zeta = \cosh (\beta) B_2 - \sinh (\beta) \cosh (\gamma) J - \sinh (\beta) \sinh (\gamma) B_1\;, \label{eq:zetageneralt0}
\end{equation}
so, upon comparing to (\ref{eq:newboostzeta}), we   identity $\gamma =\sqrt{\mu} t_0/L$. Thus, the AdS$_{2}$ Killing vector $\zeta$ for $t_{0}\neq0$   is a twice boosted Killing vector in the embedding space, where rapidity $\beta$ is related to a boundary time interval $\alpha$, and $\gamma$ to $t_{0}$.

Let us now derive the   transformation between the ($\bar v, \bar u$) coordinates of the nested wedge centered at an arbitrary time slice and the $(v,u)$ coordinates of eternal black hole, thereby extending the transformation \eqref{eq:coordtransflargesmallRindler} to $t_0 \neq 0$. The two successive boosts in the $(T^2, X)$ and $(T^1, X)$ planes lead to the following transformation in embedding space
\beq
\begin{split}
(T^{1})'&=\cosh(\beta) T^{1}+ \sinh (\beta) \sinh(\gamma) T^2 -\sinh(\beta) \cosh (\gamma) X  \;,\\
(T^{2})'&= \cosh( \gamma) T^2 - \sinh (\gamma) X\;,\\
X'&=-\sinh(\beta) T^1 -\cosh(\beta) \sinh(\gamma)T^2 +\cosh (\beta) \cosh (\gamma) X  \;,
\end{split}
\label{eq:embdtransf1}
\eeq
and the inverse transformation  is
\beq
\begin{split}
 T^{1} &=\cosh( \beta) (T^1)' + \sinh(\beta) X'  \;,\\
 T^{2} &=  \sinh(\beta) \sinh(\gamma)(T^1)'+\cosh( \gamma) (T^2)' + \cosh (\beta) \sinh(\gamma) X'  \;,\\
X &=  \sinh(\beta) \cosh(\gamma) (T^1)'+ \sinh(\gamma)(T^2)'  + \cosh(\beta) \cosh(\gamma) X' \;.
\end{split}
\label{eq:embdtransf2}
\eeq
Then, following the analogous steps which led us to (\ref{eq:retadvwedges}), we find the    relation:
\beq
 \coth \left[ \frac{\sqrt{\mu}}{2L} u\right] = \frac{e^{\kappa \bar u} \left(e^{\beta+\gamma}+e^\beta+e^\gamma-1\right) +e^{\beta+\gamma}+e^\beta-e^\gamma+1}{e^{\kappa \bar u }\left(e^{\beta+\gamma}-e^\beta+e^\gamma+1\right) +e^{\beta+\gamma}-e^\beta-e^\gamma-1}\;,  \label{eq:relationuvblabla} 
\eeq
and similarly for $v$ and $\bar v$. Next we take the limit $(\bar u\to  \infty, v\to- \infty)$ to the bifurcation point $\mathcal B$ of the AdS-Rindler Killing horizon to express the rapidities in terms of the  static coordinates of $\mathcal B$ $,(t_{\mathcal B} = t_0,  r_{*,\mathcal B} = \alpha )$,  
\beq 
 \gamma =\sqrt{\mu}t_0/L,  \quad e^\beta = \coth (\sqrt{\mu}\alpha/2L) \,.
\eeq 
The relation between $\alpha$ and $\beta$ is the same as in \eqref{eq:relationalphabeta}, so by replacing the rapidities in \eqref{eq:relationuvblabla} with variables $t_0$ and $\alpha$ we find
\begin{equation}
     \frac{\sqrt{\mu}}{L}   u = \log \left [ \frac{e^{\sqrt{\mu}t_0/L} + e^{\kappa \bar u +   \sqrt{\mu}(t_0 + \alpha)/L}}{ e^{\sqrt{\mu} \alpha / L}+e^{\kappa\bar u}} \right]\;,\quad  \frac{\sqrt{\mu}}{L}v = \log \left [ \frac{e^{\sqrt{\mu}t_0 /L} + e^{\kappa \bar v    +\sqrt{\mu}(t_0+ \alpha) /L}}{ e^{\sqrt{\mu} \alpha / L}+e^{\kappa \bar v}} \right]\;,
\end{equation}
which is desired extension of \eqref{eq:coordtransflargesmallRindler} to $t_0 \neq 0$.

As a final consistency check, let us   verify $\zeta$ in the embedding formalism reduces to the usual boost Killing vector in Poincar\'e coordinates. The embedding coordinates for the Poincar\'e patch are
\beq T^{1}=\frac{1}{2z}(L^{2}-\tau^{2}+z^{2})\;,\quad T^{2}=\frac{L\tau}{z}\;,\quad X=\frac{1}{2z}(L^{2}+\tau^{2}-z^{2})\;,\eeq
for which the following inverse relations are useful $z=\frac{L^{2}}{T^{1}+X}$ and $\tau=\frac{LT^{2}}{T^{1}+X}$. In these coordinates it is straightforward to show the generators (\ref{generatorsembedding}) become 
\beq
\begin{split}
&J=\frac{1}{2L}\left[(L^{2}+\tau^{2}+z^{2})\partial_{\tau}+2\tau z\partial_{z}\right]\;,\\
&B_{1}=-\tau\partial_{\tau}-z\partial_{z}\;,\\
&B_{2}=\frac{1}{2L}\left[(L^{2}-\tau^{2}-z^{2})\partial_{\tau}-2\tau z\partial_{z}\right]\;.
\end{split} \label{eq:generatorspoinc}
\eeq
Further, we need the relations between the rapidities $\beta$ and $\gamma$ and the Poincar\'e coordinates ($z_{\mathcal B}=R, \tau_{\mathcal B}=\tau_0$) of the bifurcation point of the nested Rindler wedge,
\beq
R = \frac{L}{\cosh (\beta) + \cosh(\gamma) \sinh(\beta)}\;, \qquad \tau_0 = L\frac{\sinh(\gamma)}{\cosh(\gamma)+ \coth(\beta)}\;,
\eeq 
or
\beq
\cosh (\beta)= \frac{L^2 + R^2 - \tau_0^2}{2 L R}, \quad \! \sinh(\beta) \cosh (\gamma) =\frac{L^2 - R^2 +\tau_0^2}{2 L R}, \quad  \! \sinh (\beta) \sinh(\gamma) = \frac{\tau_0}{R}\,. \label{eq:relationspoincare}
\eeq
Note for the special case $\gamma=0$ the relations reduce to  $R=e^{-\beta}L$  and  $\tau_0 =0$, derived in  \cite{Casini:2011kv} (just below their Eq. (3.18)).
The generic case $\gamma \neq 0$ follows from inserting the primed embedding coordinates for $(\bar v, \bar u)$ of the nested wedge and the unprimed embedding coordinates for ($z,\tau$) of the Poincar\'e patch into the dubble boost transformation  \eqref{eq:embdtransf2}, and inserting $\bar v = - \bar u$ (restricting to the $\tau =0$ slice) and $\bar u \to \infty$ (restricting to the bifurcation point). Next we substitute the generators \eqref{eq:generatorspoinc} into our boosted expression (\ref{eq:zetageneralt0}) for $\zeta$ and use the relations \eqref{eq:relationspoincare}.  The final result is
\beq \zeta=\frac{1}{2R}\left[(R^{2}-(\tau-\tau_0)^{2}-z^{2})\partial_{\tau}-2(\tau- \tau_0) z\partial_{z}\right]\;.\label{eq:zetapoincare}\eeq
This is the boost  Killing vector in $\text{AdS}_{2}$ in Poincar\'e coordinates, which is well known in the literature,  \emph{e.g.}, in Eq. (3.1) of \cite{Faulkner:2013ica} (their normalization is different since they  set  $\kappa=2\pi$). 

 \section{Covariant phase space formalism  for general $2D$ dilaton gravity} \label{app:Waldformalism}
 
 
In Section \ref{sec:semithermorindler},  where we derived the Smarr relation and   associated first law of Rindler-AdS$_2$, we made extensive use of the  covariant phase space formalism \cite{Lee:1990nz,Wald:1993nt,Iyer:1994ys,Wald:1999wa,Iyer:1995kg}. Thus, to keep this article self-contained, here we work out the covariant phase space formalism for generic two-dimensional dilaton-gravity theories, including the Polyakov action. For previous applications of the covariant phase formalism for specific two-dimensional dilaton-gravity models, namely, the classical CGHS and JT models, see, \emph{e.g.}, \cite{NavarroSalas:1992vy,NavarroSalas:1994gs,Iyer:1994ys,Kummer:1995qv,Harlow:2019yfa}.

\subsection*{Lagrangian formalism}
~Let $\psi = (g_{\mu\nu}, \Phi)$ denote a collection of dynamical fields, where $g_{\mu\nu}$ is an arbitrary background metric of a $(1+1)$-dimensional Lorenztian spacetime $M$ and $\Phi$ represents any   scalar field on~$M$, specifically, the dilaton $\phi$ or the local auxiliary field $\chi$ appearing in the Polyakov action. Consider the following covariant\footnote{By covariant we mean Lagrangian $L$ transforms as a 2-form under diffeomorphisms acting only on dynamical fields,  as opposed to local Lagrangians where the $L$ transforms under diffeomorphisms acting on both dynamical and (non-dynamical) background fields.} Lagrangian 2-form depending on $\psi$:
\begin{equation} \label{eq:genericdilatonlagr}
	L = L_{0}\epsilon\left [  R Z(\Phi) +   U(\Phi)( \nabla \Phi)^2 - V(\Phi)\right ]\;,
\end{equation}
where $\epsilon$ is the spacetime volume form,  $L_{0}$ is some coupling constant, $R$ is the Ricci scalar in $1+1$ dimensions, and $\Phi$ is a scalar field. This is the most general Lagrangian for two-dimensional Einstein gravity coupled non-minimally to a dynamical scalar field  (c.f. \cite{Grumiller:2002nm}). Here we keep functions $Z(\Phi), U(\Phi)$ and $V(\Phi)$  generic, however, it is worth highlighting three specific models characterized by the Lagrangian\footnote{If the function $Z(\Phi)$ is invertible, then under a suitable transformation, $Z(\Phi)$ may be replaced by~$\Phi$. Subsequently, under the field redefinition $ \Phi=e^{-2\varphi} $,  the Lagrangian becomes $L =   L_0 \epsilon    e^{-2 \varphi} \left [ R +  \tilde U(\varphi) ( \nabla \varphi)^2 - \tilde V(\varphi) \right]$, where now $\tilde U (\varphi) = 4 e^{-2\varphi}U(e^{-2 \varphi})$ and $\tilde V (\varphi) =   e^{2\varphi} V(e^{-2 \varphi})$, a more conventional form of the CGHS model.} (\ref{eq:genericdilatonlagr}):
\begin{enumerate}
\item  \emph{JT model}: $L_{0}=\frac{1}{16\pi G}$, $Z(\phi) = \phi + \phi_0$,  $U(\phi)=0$ and $V(\phi)=2 \phi \Lambda$\,, 
\item  \emph{CGHS model}: $L_{0}=\frac{1}{16\pi G}$, $Z(\phi) = \phi^2 $,  $U(\phi)= 4 $ and $V(\phi) =  - 4 \lambda^2 \phi^2$\,,
\item \emph{Polyakov term}: $L_{0}=-\frac{c}{24\pi}$\;, $Z(\chi)=\chi$, $U(\chi)=1$ and $V(\chi)=\frac{\tilde{\lambda}^{2}}{L^{2}}$\,,
\end{enumerate}
where $G$ is the gravitational Newton's constant,   $\Lambda$, $\lambda$ and $\tilde{\lambda}$ are cosmological constants, and $c$ is the central charge of massless scalar fields (quantifying the semi-classical   correction).

The variation of the Lagrangian \eqref{eq:genericdilatonlagr} yields
\begin{equation} \label{eq:applagrangianvar}
	\delta L = \epsilon E_\Phi \delta \Phi + \epsilon E_{\mu \nu} \delta g^{\mu \nu} + d \theta (  \psi, \delta \psi),
\end{equation}
with $d$ as the spacetime exterior derivative and $\delta\psi$ representative of field variations, and where the equations of motion $E_{\mu\nu}$ and $E_{\Phi}$ are, respectively,
\begin{align}
	E_{\mu \nu} &=L_{0}\left [ \frac{1}{2}V(\Phi) g_{\mu \nu} +\Box Z(\Phi) g_{\mu \nu} - \nabla_\mu \nabla_\nu Z(\Phi) + U(\Phi) \left ( \nabla_\mu \Phi \nabla_\nu \Phi - \frac{1}{2} (\nabla \Phi)^2 g_{\mu \nu}   \right) \right],\nonumber  \\
	E_\Phi &=L_{0}\left [  R Z'(\Phi) - V'(\Phi)-2 \nabla_\mu \left [ U(\Phi) \nabla^\mu \Phi \right] + U'(\Phi) \nabla_\mu \Phi \nabla^\mu\Phi  \right],
\end{align}
where we used $\delta\epsilon=\left(\frac{1}{2}g^{\mu\nu}\delta g_{\mu\nu}\right)\epsilon$. The symplectic potential 1-form $\theta$ for a theory of the type (\ref{eq:genericdilatonlagr}) is generally given by   \cite{Iyer:1994ys} 
\beq \theta=\epsilon_{\mu}\left[2P^{\mu\alpha\beta\nu}\nabla_{\nu}\delta g_{\alpha\beta}-2(\nabla_{\nu}P^{\mu\alpha\beta\nu})\delta g_{\alpha\beta}+2L_0 U(\Phi)(\nabla^{\mu}\Phi)\delta\Phi\right]\;,\eeq
where $P^{\mu\alpha\beta\nu}\equiv\frac{\partial L}{\partial R_{\mu\alpha\beta\nu}}$ such that for $L=\epsilon Z(\Phi)R$, one has $P^{\mu\alpha\beta\nu}=\frac{Z}{2}(g^{\mu\beta}g^{\alpha\nu}-g^{\alpha\beta}g^{\mu\nu})\epsilon$. Consequently, 
\begin{equation}
	 \theta =L_{0}\epsilon_\mu\left [ Z(\Phi)  (g^{\mu \beta} g^{\alpha \nu} - g^{\mu \nu} g^{\alpha \beta})\nabla_\nu \delta g_{\alpha \beta} + (g^{\alpha \beta} \nabla^\mu Z(\Phi) - g^{\beta \mu} \nabla^\alpha  Z(\Phi))\delta g_{\alpha \beta} + 2 U(\Phi) \nabla^\mu \Phi \delta \Phi \right].
\label{eq:sympotgen}\end{equation}
Here we used that in two dimensions, $R_{\mu \nu} = \frac{1}{2} R g_{\mu \nu}$, and $Z'(\Phi)=\frac{dZ}{d\Phi}$, etc.



For spacetimes $M$ with boundary $\partial M$  the pullback of the potential $\theta$ to the boundary   can be written as \cite{Iyer:1995kg,Harlow:2019yfa} 
\begin{equation}
\theta\big|_{\partial M}=\big  [p(\Phi) \delta \Phi + \frac{1}{2}\tau^{\mu \nu} \delta g_{\mu \nu} \big]  \epsilon_{\partial  M}+ \delta b  + d C\;,  \label{eq:pullbackoftheta}
\end{equation}
with
\begin{equation}
p(\Phi)\equiv 2L_{0}[Z'(\Phi)K+ U(\Phi) n^\mu  \nabla_\mu \Phi], \quad 	\tau^{\mu \nu}\equiv2L_{0}\gamma^{\mu \nu} n^\alpha \nabla_\alpha  Z(\Phi)\;, 
\label{eq:BYstresstensandP}\end{equation}
and
\begin{equation}
b =-2L_{0}\epsilon_{\partial  M} Z(\Phi)  K\, , \quad  C = c \cdot \epsilon_{\partial M}\;,\quad  c^\mu=-L_{0}Z(\Phi)  \gamma^{\mu \lambda}n^\nu \delta g_{\lambda \nu}\;.
\label{eq:Cdefn}\end{equation}
Here $\epsilon_{\partial M}$ is the volume form on $\partial M$, $K$ is the trace of the extrinsic curvature $K_{\mu \nu} = \frac{1}{2}\mathcal L_n \gamma_{\mu \nu}$  of the timelike boundary, and $\gamma_{\mu \nu} = - n_\mu n_\nu + g_{\mu \nu}$ is the induced metric on $\partial M$, with $n_\mu$ the unit normal to $\partial M$. 
Moreover, we used $\epsilon_{\mu}|_{\partial M}=n_{\mu}\epsilon_{\partial M}$,  $\delta\epsilon_{\partial M}=\left(\frac{1}{2}\gamma^{\mu\nu}\delta g_{\mu\nu}\right)\epsilon_{\partial M}$, and that in two dimensions the extrinsic curvature obeys $K_{\mu \nu} = K \gamma_{\mu \nu}$, such that\footnote{It is also useful to know $\delta n_{\mu}=\frac{1}{2} n_\mu n^\alpha n^\beta \delta g_{\alpha\beta}$, and
$$\delta K=-\frac{1}{2} K^{\mu\nu}\delta g_{\mu\nu}+\frac{1}{2}g^{\mu\nu}n^{\lambda}\nabla_{\lambda}\delta g_{\mu\nu}-\frac{1}{2}n^{\alpha}\nabla^{\beta}\delta g_{\alpha\beta}-\frac{1}{2}D_{\mu}(\gamma^{\mu\nu}n^{\alpha}\delta g_{\nu\alpha})\;,$$
where $D_{\mu}$ is the hypersurface covariant derivative, obeying $\gamma^{\mu\nu}D_{\mu}=D^{\nu}$.} 
\beq
\begin{split}
\delta b&=L_{0}\epsilon_{\partial M}\biggr[-2Z'(\Phi)K\delta\Phi-Z(\Phi)(g^{\mu\nu}n^{\lambda}\nabla_{\lambda}\delta g_{\mu\nu}-n^{\alpha}\nabla^{\beta}\delta g_{\alpha\beta})\\
&-n^{\alpha}(\nabla^{\beta}Z(\Phi))\delta g_{\alpha\beta}+D_{\mu}[Z(\Phi)\gamma^{\mu\nu}n^{\alpha}\delta g_{\alpha\nu}]\biggr]\;.
\end{split}
\eeq
  The quantity $b$ is a local 1-form and  $C$ is a local  $0$-form on the   boundary $\partial M$, which are both   covariant under diffeomorphisms which preserve the location of the (spatial) boundary \cite{Harlow:2019yfa}.
Further, $\tau^{\mu\nu}\equiv\frac{2}{\sqrt{-\gamma}}\frac{\delta I}{\delta \gamma_{\mu\nu}}$  with action given by  \eqref{eq:totalaction1}  is the (generalized) Brown-York stress tensor \cite{Brown:1992br}.  We will use the (renormalized)  Brown-York stress tensor below to derive the quasi-local and asymptotic energy for the semi-classical JT model.

The symplectic current 1-form, defined as $\omega (\psi, \delta_1 \psi, \delta_2 \psi) \equiv \delta_1 \theta (\psi, \delta_2 \psi) - \delta_2\theta (\psi, \delta_1\psi)$,  is explicitly given by
\beq
\begin{split}
	\omega  &= L_{0} \epsilon_\mu \Big [ Z(\Phi)S^{\mu \alpha \beta \nu \rho \sigma}\delta_1 g_{\rho \sigma} \nabla_\nu \delta_2 g_{\alpha \beta} + \frac{1}{2} g^{\mu \beta} g^{\alpha \nu} g^{\rho \sigma} \delta_1 g_{\rho \sigma} \delta_2 g_{\alpha\beta} \nabla_\nu Z(\Phi) \\
	&+ (g^{\mu \beta}g^{\alpha\nu} - g^{\mu \nu} g^{\alpha \beta}) ( \delta_1( Z(\Phi))  \nabla_\nu \delta_2 g_{\alpha \beta} - \delta_1 (\nabla_\nu Z(\Phi) )\delta_2 g_{\alpha \beta}) \\
	& + U(\Phi)\nabla^\mu \Phi g^{\alpha\beta}\delta_1 g_{\alpha \beta}\delta_2 \Phi  + 2 \delta_1 (U(\Phi) \nabla^\mu \Phi) \delta_2 \Phi - [1 \leftrightarrow 2] \Big ]\;,
\end{split}
\label{eq:sympcurrentgen}\eeq
where $S^{\mu\alpha\beta\nu\rho\sigma}$ is given by 
\beq S^{\mu\alpha\beta\nu\rho\sigma}=g^{\mu\rho}g^{\alpha\sigma}g^{\beta\nu}-\frac{1}{2}g^{\mu\nu}g^{\alpha\rho}g^{\beta\sigma}-\frac{1}{2}g^{\mu\alpha}g^{\beta\nu}g^{\rho\sigma}-\frac{1}{2}g^{\alpha\beta}g^{\mu\rho}g^{\sigma\nu}+\frac{1}{2}g^{\alpha\beta}g^{\mu\nu}g^{\rho\sigma}\;.\label{eq:tensorS}\eeq
To derive (\ref{eq:sympcurrentgen}), one can directly work with the potential $\theta$ in (\ref{eq:sympotgen}) and the definition of $\omega$, or, alternatively, one can utilize that a non-minimally coupled theory of gravity of the Brans-Dicke type is equivalent $f(R)$ gravity.  Following the latter approach, the non-minimally coupled contribution in $\omega$ follows from, \emph{e.g.}, Eq. (E.17) of \cite{Bueno:2016ypa} with $f'(R)$ replaced by $Z(\Phi)$.

Moving on, let $\zeta$ be an arbitrary smooth vector field on $M$, \emph{i.e.}, the infinitesimal generator of a diffeomorphism. The Noether current 1-form $j_{\zeta}$ associated with $\zeta$ and arbitrary field configuration $\psi$ is defined as $j_{\zeta}\equiv \theta (\psi, \mathcal{L}_{\zeta} \psi) - \zeta\cdot L$, with $\mathcal{L}_{\zeta}$ being the Lie derivative along $\zeta$. For the theory (\ref{eq:genericdilatonlagr}), explicitly we have
\begin{equation}
	j_\zeta = \epsilon_\mu \left [2L_{0}\nabla_\nu (Z(\Phi) \nabla^{[\nu} \zeta^{\mu ]} + 2 \zeta^{[\nu } \nabla^{\mu]} Z(\Phi)) + 2 {E^\mu}_\nu \zeta^\nu\right]\;,
\label{eq:Noethercurrent}\end{equation}
 where we can again use that non-minimally coupled scalar-Einstein theory is equivalent to $f(R)$ theory. From the current, we   easily read off the  Noether charge $0$-form, defined as $j_\zeta= d Q_\zeta$ on-shell,
\begin{equation}
	Q_\zeta = -L_{0}\epsilon_{\mu \nu} \left [Z(\Phi) \nabla^\mu \zeta^\nu + 2 \zeta^\mu \nabla^\nu Z(\Phi) \right]\;,
\label{eq:Noethercharge}\end{equation}
where $\epsilon_{\mu\nu}$ is the volume form for the codimension-0 surface $\partial\Sigma$, which   is a   cross section of the spatial part of $\partial M$.

Note that  the definitions of the aforementioned quantities   $L$, $\theta$ and $Q_\zeta$  are ambiguous up to the addition of exact forms. Specifically, adding a total derivative $d\mu$ to $L$ does not change the equations of motion, and the symplectic potential and Noether charge are defined up to a closed and hence locally exact form, denoted by $dY$ and $dZ$ respectively. These ambiguities   potentially alter other quantities as well, and can be summarized  as follows \cite{Iyer:1994ys}
\beq
\begin{split}
 &L\to L+d\mu \;, \\
 &\theta \to \theta + \delta \mu + d Y(\psi, \delta \psi) \;, \\
&Q_\zeta \to Q_\zeta + \zeta \cdot \mu + Y (\psi, \mathcal L_\zeta \psi) + d Z \;.  \\
\end{split}
\eeq
Moreover, the symplectic current and the form $\delta Q_\zeta - \zeta \cdot \theta (\psi, \delta \psi) $ are also ambiguous   \cite{Bueno:2016ypa}
  \begin{align}
    &\omega \to \omega + d[\delta_1 Y(\psi, \delta_2 \psi)- \delta_2 Y (\psi, \delta_1 \psi)]  \label{eq:ambiguitysymcurrent}\\
     &\delta Q_\zeta - \zeta \cdot \theta (\psi, \delta \psi) \to  \delta Q_\zeta - \zeta \cdot \theta (\psi, \delta \psi) + \delta Y (\psi, \mathcal L_\zeta \psi) - \mathcal L_\zeta Y (\psi, \delta \psi ) + d [\delta Z + \zeta \cdot Y(\psi, \delta \psi)] \nonumber
  \end{align} 
    Nonetheless, these ambiguities all cancel between the left- and right-hand side in the following on-shell fundamental variational identity  
\begin{equation}
	\omega (\psi, \delta \psi, \mathcal L_\zeta \psi) =d \left[ \delta Q_\zeta - \zeta \cdot \theta (\psi, \delta \psi) \right]\;.
\label{eq:variationalid}\end{equation}
Next, we briefly discuss a recent refinement of the covariant phase formalism for spacetimes with boundaries \cite{Harlow:2019yfa}.   Let us decompose the boundary as $\partial M=B\cup\Sigma_{-}\cup\Sigma_{+}$, where $\Sigma_{-}$ and $\Sigma_{+}$ are past and future boundaries, respectively, and $B$ (denoted by $\Gamma$ in \cite{Harlow:2019yfa}) is the spatial boundary.\footnote{In the main text we neglect the past and future boundaries, and only treat   the spatial boundary $B$  at infinity.} We require that the action is stationary under arbitrary variations of the dynamical fields up to  terms at the future and past boundary of the manifold. 
In order for the variational principle to be well posed  we require Dirichlet boundary conditions only on $B$, since boundary conditions at $\Sigma_\pm$ would also fix the initial or final state, which is too strong. From \eqref{eq:pullbackoftheta} it follows that  fixing $\Phi$ and the pullback of $g_{\mu \nu}$ to the spatial boundary~$B$, \emph{i.e.}, $\delta \Phi |_B = 0 = {\gamma_\alpha}^\mu {\gamma_\beta}^\nu \delta g_{\mu \nu} |_B$,    is sufficient to restrict the symplectic potential to the  form\footnote{We could have imposed a stronger boundary condition, $\delta \gamma_{\mu \nu} |_B =0$, by which   $C(\psi, \delta \psi)=0$ \eqref{eq:Cdefn} on $B$, since $\gamma^{\mu \lambda}n^\nu \delta g_{\lambda \nu}  = - {\gamma^{\mu}}_\nu \delta n^\nu = \gamma^{\mu \lambda} n^\nu \delta \gamma_{\lambda \nu}=0$. This amounts to a partial gauge fixing of the theory with the weaker  boundary condition ${\gamma_\alpha}^\mu {\gamma_\beta}^\nu \delta \gamma_{\mu \nu}|_B=0$ (see footnote 25 in \cite{Harlow:2019yfa}). We do this in the main text, but in the appendices we allow for a nonzero $C$ (see footnote \ref{fn:bdrycond}). \label{footnote:Cterm2}}
\beq
\theta \big |_{B} = \delta b + d C\;. \label{eq:importantrestrictiontheta}
\eeq
 Therefore, we    supplement the action   by the integral of (minus)~$b$ at $\partial M$, which thus serves as a Gibbons-Hawking-York (GHY) boundary term,\footnote{Our notation can be matched with that of \cite{Harlow:2019yfa} by replacing  $b  \to -\ell$.}
\beq I_{\text{tot}} = \int_M L - \int_{\partial M} b\;. \label{eq:totalaction1}\eeq
  Then, using \eqref{eq:applagrangianvar} and  \eqref{eq:importantrestrictiontheta}, 
  the variation of the total action becomes     \cite{Harlow:2019yfa}
\beq \delta I_{\text{tot}}=\int_{M}E_{\psi}\delta\psi+\int_{\Sigma_{+}-\Sigma_{-}}\hspace{-2mm}(\theta-\delta b-dC)\;. \label{eq:variationtotalactionapp}\eeq
This shows  the total action \eqref{eq:totalaction1} is stationary up to boundary terms at $\Sigma_\pm$. Since the combination $\Psi   \equiv  \theta - \delta b- dC$ appears a boundary term in the variation of the action, it is more natural to define the symplectic current instead as 
\beq
\tilde \omega (\psi, \delta_1 \psi, \delta_2 \psi) \equiv \delta_1 \Psi (\psi, \delta_2 \psi) - \delta_2 \Psi (\psi, \delta_1 \psi)\;.
\eeq
 The $b$ term drops out due to the antisymmetrization, hence the  new symplectic current   is  
\beq  \tilde{\omega}(\psi,\delta_{1}\psi,\delta_{2}\psi) =   \delta_{1}\theta(\psi,\delta_{2}\psi)-\delta_{2}\theta(\psi,\delta_{1}\psi)-d[\delta_{1}C(\psi,\delta_{2}\psi)-\delta_{2}C(\psi, \delta_{1}\psi)]\;, \label{eq:newsymplecticcurrentharlow}
\eeq
where the first two terms combine into the old symplectic current $\omega(\psi,\delta_{1}\psi,\delta_{2}\psi)$. By construction the symplectic current vanishes on the spatial boundary, $\tilde \omega |_B =0$. We note the new definition for the symplectic current can   be obtained by (partly) fixing the ambiguity in \eqref{eq:ambiguitysymcurrent}  as $Y= - C$ (there is a remaining ambiguity that involves a simultaneous shift in $\theta$ and~$C$). Now, the   on-shell variational identity (\ref{eq:variationalid}) for the new  symplectic current \eqref{eq:newsymplecticcurrentharlow} becomes
\beq
  \int_\Sigma \tilde \omega (\psi, \delta \psi, \mathcal L_\zeta \psi ) =   \oint_{\partial \Sigma} \left[ \delta Q_\zeta - \zeta \cdot \theta (\psi, \delta \psi)-  \delta C(\psi,\mathcal L_\zeta \psi) + \mathcal L_\zeta C(\psi,\delta \psi) \right]\;. \label{eq:identityharlowsymplecticform}
\eeq
One should be careful, however, \emph{not} to add the $C$ term at an inner boundary of $\Sigma$, for example at the bifurcation surface of a  Killing horizon, since $C$ does not extend covariantly into $M$. This issue would not have arisen if we had not included $C$ in the definition of the symplectic current \eqref{eq:newsymplecticcurrentharlow}, and instead just added a new boundary term at the intersection of $\Sigma$ and $B$  in the integral of the symplectic current, i.e., $\int_\Sigma \tilde \omega = \int_\Sigma \omega - \oint_{\Sigma \cap B} [\delta_1 C(\psi, \delta_2\psi) - \delta_2 C (\psi, \delta_1 \psi)]$.\footnote{We thank Daniel Harlow for   correspondence on this point.}  For the most part we will ignore this subtlety below.

\subsection*{Hamiltonian formalism}

Thus far we have only discussed the Lagrangian formalism. However, in our study of thermodynamics we need a notion of energy and hence we must make contact with the Hamiltonian formalism. We very briefly review this here and in the following subsection describe the quasi-local and asymptotic energy for the semi-classical JT model in both Boulware and Hartle-Hawking vacuum states.

The power of the covariant phase space formalism is that it provides a  way to understand the Hamiltonian dynamics of covariant Lagrangian field theories. The phase space $\mathcal{P}$, defined as the set of solutions to the equations of motion,\footnote{This definition differs from the standard, non-covariant, definition of phase space of a dynamical system, which labels the set of distinct initial conditions on a given timeslice to solve the systems equations of motion.} is   a symplectic manifold supplied with a closed and non-degenerate  2-form $\tilde{\Omega}$, called the symplectic form.\footnote{More accurately, in the context of theories with local symmetries, where the initial value problem is generally ill defined, \emph{e.g.}, general relativity and Maxwell theory, one instead works with the pre-phase space. Physical phase space $\mathcal{P}$ is given by the quotient of pre-phase space under the action of the group of continuous transformations whose generators are the zero modes of $\tilde{\Omega}$ \cite{Lee:1990nz}. Thus, more rigorously, $\tilde{\Omega}$ is really the pre-symplectic 2-form, and $\tilde{\omega}$ the pre-symplectic current.} In this context, the infinitesimal variation $\delta$ is viewed as the exterior derivative of differential forms on the configuration space -- the set of off-shell field configurations on spacetime obeying boundary conditions on $B$ -- such that the symplectic potential $\theta$ and   $C$ are interpreted as one-forms on the configuration space. Moreover, the symplectic current $\tilde{\omega}$ is related to $\tilde{\Omega}$ via
\beq \tilde{\Omega} \equiv \int_{\Sigma}\tilde{\omega}\;,\eeq
where $\Sigma$ is any Cauchy slice of $M$.  Notice the symplectic form $\tilde \Omega$ is independent of the choice of Cauchy slice, because $\tilde \omega |_B =0$ and on-shell $d \tilde \omega =0$.\footnote{The Cauchy slice independence also holds for the alternative definition of the symplectic form, see below equation \eqref{eq:identityharlowsymplecticform}, $\tilde \Omega \equiv \int_\Sigma \omega - \oint_{\Sigma \cap B} [\delta_1 C(\psi, \delta_2\psi) - \delta_2 C (\psi, \delta_1 \psi)]$, because the old symplectic current $\omega$ is closed on-shell and $\int_B \omega = \int_{\partial B} [\delta_1 C(\psi, \delta_2 \psi) - \delta_2 C (\psi, \delta_1, \psi)]$ due to \eqref{eq:importantrestrictiontheta}, where $\partial B$ consists of the two intersections $\Sigma \cap B$ and $\Sigma' \cap B$ (where $\Sigma'$ is another Cauchy slice on which the symplectic form can be defined). } We denote the induced metric on   $\Sigma$ by $h_{\mu \nu} = u_\mu u_\nu + g_{\mu \nu}$, with $u_\mu$ the unit normal to $\Sigma$.  

Of interest is the Hamiltonian $\tilde{H}_{\zeta}$ which generates time evolution  along the flow of the diffeomorphism generating  vector field $\zeta$.  Hamilton's equations read $\delta\tilde{H}_{\zeta}=\tilde{\Omega}(\psi,\delta\psi,\mathcal{L}_{\zeta}\psi)$. 
Therefore,
\beq \delta \tilde{H}_{\zeta}=\int_\Sigma \tilde{\omega} (\psi, \delta \psi, \mathcal L_\zeta \psi)=\oint_{\partial \Sigma} [\delta Q_\zeta - \zeta \cdot \theta (\psi, \delta \psi)- \delta C(\psi,\mathcal{L}_{\zeta}\psi)+\mathcal{L}_{\zeta}C(\psi,\delta\psi) ]\;,\label{eq:varH}\eeq
where the identity (\ref{eq:identityharlowsymplecticform}) was used. If we impose  $\theta|_{B}= \delta b+dC $ and make use of Cartan's ``magic" formula,\footnote{For a form $\sigma$, and vector field $X$ Cartan's magic formula says: $\mathcal{L}_{X}\sigma=X\cdot d\sigma+d(X\cdot \sigma)$. Note, when $\sigma$ is a zero form, then $X\cdot \sigma=0$.} the Hamiltonian variation simplifies to
\beq \delta\tilde{H}_{\zeta}=\oint_{\partial\Sigma}[\delta Q_{\zeta}-\zeta\cdot\delta b-\delta C(\psi,\mathcal{L}_{\zeta}\psi)]\;.\eeq
In Appendix \ref{app:extfirstlawdeets} we include variations of coupling constants in this equation. Thus, since $\delta\zeta=0$, the Hamiltonian $\tilde{H}_{\zeta}$ itself is
\beq \tilde{H}_{\zeta}=\int_{\partial\Sigma}[Q_{\zeta}-\zeta\cdot b- C(\psi,\mathcal{L}_{\zeta}\psi)]+ \text{cst} \;.\eeq
The constant  represents a standard ambiguity to the energy of any Hamiltonian system. Below we will set this arbitrary constant to zero. Notice, the ``integrability" of the Hamiltonian, allowing us to move from $\delta\tilde{H}_{\zeta}$ to $\tilde{H}_{\zeta}$, arises as a consequence of the boundary condition \eqref{eq:importantrestrictiontheta}, which appeared from demanding the    action $I_{\text{tot}}$ is stationary up to terms at the future and past boundaries.  

For the $2D$ dilaton-gravity model in question, the term on the right-hand side of $\tilde H_{\zeta}$ pulled back to $\partial\Sigma$ are given by 
\beq
Q_\zeta \big |_{\partial \Sigma}=+L_{0}\epsilon_{\partial \Sigma} (u_\mu n_\nu - u_\nu n_\mu) \left [  Z(\Phi) \nabla^\mu \zeta^\nu + 2 \zeta^\mu \nabla^\nu Z(\Phi) \right]\;,\eeq
\beq \zeta\cdot b |_{\partial \Sigma}= -2L_{0}\epsilon_{\partial \Sigma} \zeta^\mu u_\mu Z(\Phi)  K\;,\eeq
\beq C(\psi,\mathcal L_\zeta \psi )|_{\partial \Sigma}= -L_{0}\epsilon_{\partial \Sigma}(u^\mu n^\nu + u^\nu n^\mu) Z(\Phi)   \nabla_\mu \zeta_\nu\;,\label{eq:CLiepsi}\eeq
where we used $\epsilon_{\mu\nu}|_{\partial\Sigma}=n_{\mu}u_{\nu}-n_{\nu}u_{\mu} $ in (\ref{eq:Noethercharge}), $\delta g_{\lambda\nu}\to\nabla_{\lambda}\zeta_{\nu}$ together with $\gamma^{\mu\nu}=g^{\mu\nu}-n^{\mu}n^{\nu}$ in $C$ (\ref{eq:Cdefn}), and $(\epsilon_{\partial M})_{\mu}=+u_{\mu}\epsilon_{\partial\Sigma}$.\footnote{Here we take $u^{\mu}$ to be future-pointing, in contrast to \cite{Harlow:2019yfa}, where $u^{\mu}$ (denoted by $\tau^{\mu}$) is past-pointing.} Thus, the Hamiltonian for $2D$ dilaton gravity is\footnote{Notice if $\zeta$ is a Killing field, then   $C(\psi,\mathcal{L}_{\zeta}\psi) |_{\partial\Sigma}=0$  by Killing's equation. We do not use this though in deriving \eqref{eq:Hzetaexp}.}
\beq
\begin{split}
\tilde{H}_\zeta&=2L_{0}\int_{\partial\Sigma}\epsilon_{\partial\Sigma}\left[\gamma_{\mu\nu}u^{\mu}\zeta^{\nu}n_{\alpha}\nabla^{\alpha}Z(\Phi)+Z(\Phi)u^{\mu}n^{\nu}\nabla_{\mu}\zeta_{\nu}+\gamma_{\mu\nu}u^{\mu}\zeta^{\nu}Z(\Phi)K\right]\;,\\
&=\oint_{\partial \Sigma} \epsilon_{\partial \Sigma} \zeta^\mu u^\nu \tau_{\mu \nu}\;, 
\label{eq:Hzetaexp}
\end{split}
\eeq
with  $\tau_{\mu\nu}$   the   Brown-York stress-energy tensor (\ref{eq:BYstresstensandP}). To get to the second line we used $n^{\nu}\nabla_{\mu}\zeta_{\nu}\big |_{\partial \Sigma}=-\zeta^{\nu}\nabla_{\mu}n_{\nu}$, which follows from $n \cdot \zeta=0$ on $\partial M$, and we inserted the extrinsic curvature $K_{\mu\nu}=2\nabla_{(\mu}n_{\nu)}$ and the identity $K_{\mu\nu}=K \gamma_{\mu\nu}$ in $2D$. 



From here we note the quasi-local  momentum and spatial stress vanish, since there is no extra dimension, while the quasi-local energy density $\varepsilon$ is
\beq \varepsilon \equiv u_\mu u_\nu \tau^{\mu \nu} =-2L_{0} n^\alpha \nabla_\alpha Z(\Phi)\;.\label{eq:quasilocalen}\eeq
Thus, another way to express the Hamiltonian (\ref{eq:Hzetaexp}) is
\begin{equation}
\tilde{H}_\zeta= \oint_{\partial \Sigma} \epsilon_{\partial\Sigma} N \varepsilon\;, 
\label{eq:quasilocaHam}\end{equation}
where $N = - \zeta^\mu u_\mu$ is the lapse function. This expression for $\tilde{H}_{\zeta}$ is consistent with the definition given in \cite{Brown:1992br} (see their Eq. (4.13)).

\subsection{Quasi-local and asymptotic energies with semi-classical corrections}
\label{app:energies}

We now have all of the ingredients to compute the quasi-local and asymptotic energies for the nested Rindler wedge and the eternal black hole in the semi-classical JT model. 
Following \cite{Brown:1992br}, 
the quasi-local energy density is defined as in (\ref{eq:quasilocalen}), while
the asymptotic energy is equal to the Hamiltonian \eqref{eq:quasilocaHam} evaluated at spatial infinity. Since $C(\psi,\mathcal{L}_{\zeta}\psi)|_{\partial \Sigma}=0$ for the boost Killing vector $\zeta$, in what follows and in the main text we denote $\tilde{H}_{\zeta}$ as $H_{\zeta}$.

For this computation we need the unit normals $u$ and $n$, respectively, to $\Sigma=\{ t=t_0\}$ and  to  $B$ in the black hole background.  Away from   $\Sigma$ the future-pointing timelike unit vector   $u^\mu$ can be defined as the velocity vector of the flow of $\zeta$,
\beq
u^\mu = \frac{\zeta^\mu}{\sqrt{- \zeta \cdot \zeta}}\, , \quad \text{and at $\Sigma$} \quad u^\mu  =\frac{1}{\sqrt{r^2/L^2 -\mu}} \partial_t^\mu  \;.
\label{eq:normaluvec}
\eeq
By definition $u^\mu$ is   tangent to the spatial boundary $B$ of $M$.  The outward-pointing spacelike unit vector~$n^\mu$ is defined as the unit normal to $B$ 
\beq
n^\mu = \frac{\tilde n^\mu}{\sqrt{  \tilde n\cdot \tilde n}}\, , \quad\, \text{and at $\Sigma$} \quad n^\mu  = \sqrt{r_B^2/L^2 -\mu}  \partial_r^\mu  \;.
\label{eq:normalnvec}\eeq
where the non-normalized normal $\tilde n$ satisfies $\tilde n \cdot \zeta =0$ at $B$, so for convenience we take $\tilde n =    \zeta_r \partial_t -  \zeta_t \partial_r$.  Thus both $u$ and $n$ are defined in terms of the vector field $\zeta.$

\subsection*{Classical energy}

Specializing to JT gravity requires us to eliminate unwanted divergences arising in the energy upon evaluating the Hamiltonian at the conformal boundary. We accomplish this by supplementing the classical and semi-classical JT actions with appropriate local counterterms. Specifically, the boundary counterterm Lagrangian 1-form for classical JT gravity is
\beq b^{\text{ct}}_{\text{JT}}=\frac{\phi}{8\pi GL}\epsilon_{B}\;,\eeq
such that the total boundary Lagrangian 1-form $b$ for JT gravity is 
\beq b_{\text{JT}}=-\frac{\epsilon_{B}}{8\pi G}(\phi+\phi_{0})K+\frac{\epsilon_{B}}{8\pi G}\frac{\phi}{L}\;.\label{eq:btotclassJT}\eeq
Feeding this through the derivation of $H_{\zeta}$ we are led to the same expression (\ref{eq:Hzetaexp}), however with the BY stress tensor (\ref{eq:BYstresstensandP}) now given by (matching  Eq. (3.87) of \cite{Harlow:2019yfa})
\begin{equation}
	\tau^{\mu \nu}_{\phi_r}  =\frac{1}{8\pi G} \left ( n^\alpha \nabla_\alpha \phi - \frac{\phi}{L} \right) \gamma^{\mu \nu}\overset{\partial \Sigma}{=}\frac{\phi_{r}}{8\pi G L} \left (\sqrt{\frac{r_B^2}{L^2} - \mu}   -\frac{r_B}{L} \right) \gamma^{\mu \nu}\;.
\end{equation}
where we inserted the unit normal $n^\mu$ \eqref{eq:normalnvec} at $\partial \Sigma$ in the second equality. 
The corresponding classical quasi-local energy $\varepsilon$ is
\begin{equation}
	\varepsilon_{\phi_r} \big |_{\partial \Sigma}=\frac{\phi_{r}}{8\pi G L} \left ( \frac{r_B}{L} - \sqrt{\frac{r_B^2}{L^2} - \mu}\right) \;.
\end{equation}
Let us compute the asymptotic energy on the extremal slice $\Sigma$ associated with the boost Killing vector \eqref{eq:boostzetaschwarrz} of a nested Rindler wedge.    In the limit we approach   spatial infinity  $r_{B}\to\infty$ at $t=t_0$, where the boost Killing vector is $\zeta^{\mu}\to\frac{\kappa L}{\sqrt{\mu}}\tanh(\sqrt{\mu}\alpha/2L)\partial^{\mu}_{t}\equiv \mathcal{A} \partial^{\mu}_{t}$, we can compute the asymptotic energy from (\ref{eq:quasilocaHam})
\beq E_{\zeta}^{\phi_r} \equiv \lim_{r_{B}\to\infty}H_{\zeta}^{\phi_r}=\frac{\kappa\sqrt{\mu}\phi_{r}}{16\pi G} \tanh\left(\frac{\sqrt{\mu}\alpha}{2L} \right) =\mathcal{A}M_{\phi_r}\;,\label{eq:asympenergyclassJTapp}\eeq
where 
\beq \mathcal{A}\equiv \frac{\kappa L}{\sqrt{\mu}} \tanh\left(\frac{\sqrt{\mu}\alpha}{2L} \right) \;,\eeq
and $M_{\phi_{r}}$ is the classical ADM mass  of the black hole, given by  \eqref{eq:ADMmass}.  Further, in the $\alpha\to\infty$ limit,   upon identifying $\kappa=\sqrt{\mu}/L$, we recover $\lim_{\alpha\to\infty} E_{\zeta}^{\phi_r}\to M_{\phi_r}$ since $ \mathcal A \to    1$ in this limit.

\subsection*{Semi-classical energy}
Let's now see how the energy is modified under semi-classical corrections. This requires us to find the BY tensor solely associated with $\chi$, denoted $\tau^{\mu\nu}_{c,\chi}$, which will likewise be modified due to a local counterterm (Eq. (3.25) of \cite{Moitra:2019xoj}), such that the 1-form $b$ for the Polyakov action is
\beq b_{\chi}=\frac{c}{12\pi}\chi K\epsilon_{B}+\frac{c}{24\pi L}\epsilon_{B}\;.\label{eq:bchict}\eeq
The appropriate BY tensor (\ref{eq:BYstresstensandP}) with $L_{0}=-\frac{c}{24\pi}$, and $Z(\Phi)=\chi$, is found to be
\begin{equation}
\begin{split}
	\tau^{\mu \nu}_{c, \chi}&= -\frac{c}{12\pi} \left (   n^\alpha \nabla_\alpha \chi + \frac{1}{2L}  \right) \gamma^{\mu \nu}\;,\\
& \overset{\partial \Sigma}{=} -\frac{c}{12 \pi}   \left[\frac{1 }{ \sqrt{\mu}} \sinh \left(-\frac{\sqrt{\mu} r_{\ast B}}{L}\right) \left ( \frac{\sqrt{\mu}}{L}  \coth \left(\frac{\sqrt{\mu}r_{\ast B}}{L}\right)  + \partial_{r_\ast} \xi \right) +\frac{1}{2L}  \right] \gamma^{\mu \nu} \;,
\end{split}
\label{eq:BYstresschi}\end{equation}
where $r_{\ast,B}$ is the tortoise coordinate at the surface $B$, $\xi$ is the function defining $t_{\pm}$ in the quantum stress-energy tensor, and   the Polyakov cosmological constant is set to $\lambda=0$.

 We must also include in the BY tensor a contribution  coming from semi-classical corrections to the dilaton, denoted by $\phi_c$, whose BY tensor $\tau^{\mu\nu}_{c,\phi}$ is
\begin{equation}
	\tau^{\mu \nu}_{c, \phi} = \frac{1}{8\pi G} \left ( n^\alpha \nabla_\alpha \phi_c -\frac{\phi_{c}}{L} \right) \gamma^{\mu \nu}\;.
\end{equation}
Since the backreacted solutions $\phi_{c},\chi$ and $\xi$ depend on the choice of vacuum state, let us determine the energy in both the Boulware and Hartle-Hawking vacuum states.

\vspace{3mm}

\noindent \textbf{Boulware vacuum:} Specifically consider the static solution $\xi^{(1)} = c$, for which the quasi-local energy $\varepsilon_{c,\chi}$ associated to $\chi$ is easily worked out to be
\begin{equation}
	\varepsilon_{c,\chi} \big |_{\partial \Sigma} = \frac{c}{24 \pi L} \left[ 1 - 2 \cosh \left(\frac{\sqrt{\mu} r_{\ast B}}{L}\right) \right]\;,
\end{equation}
while the quasi-local energy $\varepsilon_{c,\phi}$ associated to $\phi_c =\frac{ G c }{3 } ( 1+  \frac{\sqrt{\mu} }{L} r_{\ast} )\coth (\sqrt{\mu} r_\ast / L)  $ is
\beq
\begin{split}
\varepsilon_{c, \phi}\big |_{\partial \Sigma}&= \frac{c}{24\pi L} \Biggr[\left(1+ \frac{\sqrt{\mu} r_{\ast B}}{L}\right) \coth \left(\frac{\sqrt{\mu} r_{\ast B}}{L}\right) \\
&+ \frac{L}{ \sqrt{\mu}} \sinh \left(\frac{\sqrt{\mu} r_{\ast B}}{L}\right)\partial_{r_\ast}   \Bigg (  \bigg( 1+  \frac{\sqrt{\mu} r_{\ast B}}{L} \bigg)  \coth \left(\frac{\sqrt{\mu} r_{\ast B}}{L}\right)\Bigg) \Biggr] \;,\\
&= \frac{c}{24\pi L} \left [ \left(1+  \frac{\sqrt{\mu} r_{\ast B}}{L}\right)  \tanh \left(\frac{\sqrt{\mu} r_\ast}{2L}\right) +  \cosh \left(\frac{\sqrt{\mu}r_{\ast B}}{L}\right)\right]\;.
\end{split}
\eeq
Thus, the total semi-classical quasi-local energy is
\begin{equation}
\varepsilon_{c}\equiv(\varepsilon_{c,\chi}+\varepsilon_{c,\phi})\big |_{\partial \Sigma}=\frac{c}{24\pi L} \left [1+ \left(1+  \frac{\sqrt{\mu} r_{\ast B}}{L}\right)  \tanh \left(\frac{\sqrt{\mu} r_{\ast B}}{2L}\right) -\cosh \left(\frac{\sqrt{\mu}r_{\ast B}}{L}\right)\right]\;.
\end{equation}
Consequently, 
\beq E_\zeta^c \equiv \lim_{r_{\ast B}\to0}H_{\zeta}^{c}=-\frac{c\kappa}{48\pi} \tanh\left(\frac{\sqrt{\mu}\alpha}{2L} \right) = \mathcal A M_c\;,\eeq
where 
\beq 
M_c = -\frac{c\sqrt{\mu}}{48\pi L}\;.
\eeq
In the $\alpha\to\infty$ limit we identify the above as the semi-classical correction to the asymptotic energy, \emph{i.e.}, $ \lim_{\alpha \to \infty} E_\zeta^c \to M_{c}$ (upon setting $\kappa=\sqrt{\mu}/L$), which is proportional to the expectation value of the stress-tensor (\ref{eq:expvalTuuB}).

\vspace{3mm}

\noindent \textbf{Hartle-Hawking vacuum:} First consider the static solution $\xi^{(4)} =  \sqrt{\mu} r_\ast /L+c$ , for which the quasi-local energy associated to $\chi$ is
\begin{equation}
	\varepsilon_{c,\chi}\big |_{\partial \Sigma} = \frac{c}{12 \pi L}   \left[1/2 + \sinh (-\sqrt{\mu} r_{\ast B} /L) -  \cosh (\sqrt{\mu} r_{\ast B} /L)  \right]\;,
\end{equation}
while the quasi-local energy associated to   $\phi_c =   Gc/3$ is
\begin{equation}
\varepsilon_{c,\phi}\big |_{\partial \Sigma}= \frac{c}{24 \pi L} \;.
\end{equation}
Hence, the total semi-classical quasi-local energy is
\beq
\varepsilon_{c} = ( \varepsilon_{c,\chi}  + \varepsilon_{c,\phi}) \big |_{\partial \Sigma} = \frac{c}{12 \pi L}   \left[1 + \sinh (-\sqrt{\mu} r_{\ast B} /L) -  \cosh (\sqrt{\mu} r_{\ast B} /L)  \right]\;,
\eeq
such that 
 \beq E^c_\zeta \equiv\lim_{r_{\ast B}\to0}H^{c}_{\zeta}=\frac{c\kappa}{12\pi}\tanh\left(\frac{\sqrt{\mu}\alpha}{2L} \right)\;.\label{eq:asymptotenSJT}\eeq
 In the $\alpha\to\infty$ limit we attain the semi-classical correction to the ADM energy of the black hole, upon setting $\kappa = \sqrt{\mu}/L $:
\beq 
M_{c}= \frac{c\sqrt{\mu}}{12\pi L}\;,
\label{eq:quasilocasymenHH}\eeq
matching what was found using  holographic renormalization methods \cite{Almheiri:2014cka,Moitra:2019xoj}.

Next consider the time-dependent solution for $\chi$:
\begin{equation}
	\xi^{(5)}_{u,v} = k - \log \left [  \left ( e^{- \sqrt{\mu} u/2L} + K_{U_{\text K}} e^{\sqrt{\mu} u/ 2L}\right)  \left ( e^{  \sqrt{\mu} v/2L} + K_{V_{\text K}} e^{-\sqrt{\mu} v/ 2L}\right) \right]\;.
\end{equation}
The derivative with respect to the tortoise coordinate $r_{\ast}$ is
\begin{equation}
	\partial_{r_\ast} \xi^{(5)}_{u,v}
=\frac{ \sqrt{\mu}}{L} \left [ \frac{1}{1 - V_{\text K} / V_{\text K}^B   }- \frac{1}{1 -  U_{\text K}^B / U_{\text K}  }\right]\;.
\end{equation}
Plugging this into the BY stress tensor (\ref{eq:BYstresschi}), the total quasi-local energy is
\begin{equation}
	\varepsilon_{c } \big |_{\partial \Sigma} = \frac{c}{12 \pi L}   \left[1  + \sinh \left(-\frac{\sqrt{\mu} r_{\ast B}}{L}\right)\left (\frac{1}{1 + \frac{1}{K_{V_{\text K}} }e^{\sqrt{\mu} v / L}} - \frac{1}{1 + K_{U_{\text K}}e^{ \sqrt{\mu}u/L}}  \right)    -  \cosh \left(\frac{\sqrt{\mu} r_{\ast B}}{L}\right) \right]\;.
\end{equation}
Meanwhile, the semi-classical asymptotic energy vanishes as $r_{\ast,B}\to0$ for any $\alpha$
\beq E_\zeta^c \equiv \lim_{r_{\ast, B}\to0}H^{c}_{\zeta}=0\;.\eeq

\section{Derivation of the  extended first law  with variations of couplings} \label{app:extfirstlawdeets}

Here we provide a detailed derivation of the extended first law in JT gravity for the nested AdS-Rindler wedge with boost Killing vector $\zeta$. This amounts to  evaluating the   fundamental variational identity in the covariant phase space formalism extended to include coupling variations \cite{Urano:2009xn,Caceres:2016xjz}
 \beq
\int_\Sigma \omega (\psi, \delta \psi, \mathcal L_\zeta \psi) = \oint_{\partial \Sigma} [\delta Q_\zeta - \zeta \cdot \theta (\psi, \delta \psi)]+ \int_{\Sigma}\zeta\cdot E^{\lambda_{i}}_L\delta\lambda_{i}\;. 
\label{eq:variationalidJTextapp} \eeq
  Here the `$\delta$' denotes a variation with respect to the coupling constants $\lambda_i$ and dynamical fields $\psi=(\phi, g_{\mu\nu})$. The two-form $E^{\lambda_i}_L \equiv\frac{\partial \mathcal L_{\text{bulk}}}{\partial \lambda_i} \epsilon $ is the derivative of the bulk off-shell Lagrangian density, defined via $L = \mathcal L_{\text{bulk}} (\psi, \lambda_i) \epsilon$,  with respect to the coupling constant $\lambda_i$, and summation over $i$ is implied in \eqref{eq:variationalidJTextapp}.  In the context of semi-classical JT gravity  the couplings are $\lambda_{i}=\{\phi_{0},G,L,c\}$. The relation (\ref{eq:variationalidJTextapp}) was   previously invoked in \cite{Svesko:2020yxo} to derive an extended first law for classical JT gravity, where background subtraction was used to regulate divergences.
  
 In this appendix we consider a manifold with boundary $\partial M$, with the same decomposition   $\partial M=B\cup\Sigma_{-}\cup\Sigma_{+}$  as in Appendix \ref{app:Waldformalism}. In the covariant phase space formalism with boundaries the relevant fundamental variational identity is slightly different compared to   (\ref{eq:variationalidJTextapp}). Below we derive this new variational identity, thereby simultaneously generalizing the work of \cite{Harlow:2019yfa} to allow for coupling variations and \cite{Urano:2009xn,Caceres:2016xjz} to the case of spacetimes with boundaries.

 Recall the total action (\ref{eq:totalaction1}), where $L $ is the covariant bulk Lagrangian defining a theory of dynamical fields $\psi$ and couplings $\lambda_{i}$. The   variation of the bulk Lagrangian  is 
 \beq \delta L =E_{\psi}\delta\psi+E^{\lambda_{i}}_L\delta_{\lambda_{i}}+d\theta(\psi, \delta \psi )\;.\label{eq:Lagrangianvarcoup}\eeq
 Thus, the variation of the total action is
 \beq \delta I_{\text{tot}}=\int_{M}(E_{\psi}\delta\psi+E^{\lambda_{i}}_L\delta\lambda_{i})+\int_{\partial M}(\theta-\delta b)\;.\eeq
 Here $\delta b$ refers to the total variation of boundary Lagrangian 1-form $b$, \emph{i.e.}, $\delta b=\delta_{\psi}b+\delta_{\lambda_{i}}b$. Following \cite{Harlow:2019yfa}, we invoke the boundary condition \eqref{eq:importantrestrictiontheta} on  the symplectic potential, given by $\theta |_B = \delta_\psi b + dC$. Implementing this boundary condition and expanding the total variation $\delta b$ on $B$, we find
 \beq
 \begin{split}
  \delta I_{\text{tot}}&=
 \int_{M}(E_{\psi}\delta\psi+E^{\lambda_{i}}_L\delta\lambda_{i})+\int_{\Sigma_{+}-\Sigma_{-}}(\theta-\delta b-dC)-\int_{B}\delta_{\lambda_{i}}b\;.
 \end{split}
 \eeq
Observe that  we recover  \eqref{eq:variationtotalactionapp} when we drop the coupling variations. 
 
Now consider the variation of  the Noether current $j_{\zeta}=\theta(\psi,\mathcal{L}_{\zeta}\psi)-\zeta\cdot L$. Using $\delta\zeta=0$, we   have the  following on-shell identity 
 \beq
 \begin{split}
  \delta j_{\zeta} 
  &=\delta\theta(\psi,\mathcal{L}_{\zeta}\psi)-\mathcal{L}_{\zeta}\theta(\psi,\mathcal{L}_{\zeta}\psi)+d[\zeta\cdot\theta(\psi , \delta \psi )]-\zeta\cdot E^{\lambda_{i}}_L\delta\lambda_{i}\\
  &=\omega(\psi,\delta\psi,\mathcal{L}_{\zeta}\psi)+d[\zeta\cdot\theta(\psi , \delta \psi )]-\zeta\cdot E^{\lambda_{i}}_L\delta\lambda_{i}\;,
 \end{split}
 \eeq
 where to get to the first line we used the Lagrangian variation (\ref{eq:Lagrangianvarcoup})  and Cartan's magic formula, and in the second line we identified the symplectic current $\omega$ evaluated on the Lie derivative of $\psi$ with respect to $\zeta$. Since, moreover, the Noether current is closed  on-shell and hence locally exact, $j_{\zeta}=dQ_{\zeta}$, such that $\delta j_{\zeta}=d\delta Q_{\zeta}$, we arrive at the   on-shell variational identity including coupling variations 
 \beq \omega(\psi,\delta\psi,\mathcal{L}_{\zeta}\psi)=d[\delta Q_{\zeta}-\zeta\cdot\theta(\psi, \delta \psi)]+\zeta\cdot E^{\lambda_{i}}_L \delta\lambda_{i}\;,\label{eq:variationalidcoupoff}\eeq
 which   agrees with the integral   identity (\ref{eq:variationalidJTextapp}). This is a true identity, but it is not the relevant fundamental identity in the covariant phase space formalism with boundaries. 
 
 Indeed, the appropriate symplectic current $\tilde \omega$ for spacetimes with boundaries differs by an exact form from $\omega$ and was defined in  \eqref{eq:newsymplecticcurrentharlow}   \cite{Harlow:2019yfa}. Inserting the identity \eqref{eq:variationalidcoupoff} into this definition and integrating over a Cauchy slice $\Sigma$ yields
 \beq
  \int_{\Sigma}\tilde{\omega}(\psi,\delta\psi,\mathcal{L}_{\zeta}\psi)= \oint_{\partial \Sigma}[\delta Q_{\zeta}-\zeta\cdot\theta(\psi, \delta \psi )-\delta C(\psi,\mathcal{L}_{\zeta}\psi)+\mathcal{L}_{\zeta}C(\psi,\delta\psi)]+\int_{\Sigma}\zeta\cdot E^{\lambda_{i}}_L\delta\lambda_{i}\;.
  \eeq
  This replaces the variational identity \eqref{eq:variationalidJTextapp}, and it is the generalization of (\ref{eq:varH}) to nonzero coupling variations.

Next, we apply this identity to the case where the Cauchy slice $\Sigma$ has an inner boundary at the bifurcation surface $\mathcal B$ of a Killing horizon and an outer boundary at spatial infinity.  We should be careful not to include the $C$ terms at $\mathcal B$, since the boundary condition \eqref{eq:importantrestrictiontheta} for the symplectic potential only applies to the spatial boundary $B\to \infty$   and not to the bifurcation surface $\mathcal B$, and moreover $C$ does not extend into the bulk in a covariant way.  Consequently, by imposing the bounday condition $\theta|_\infty=\delta_{\psi}b+dC$ and using Cartan's magic formula and $\delta_\psi b = \delta b - \delta_{\lambda_i} b$, we can rewrite the above as
  \beq \int_{\Sigma}\tilde{\omega}(\psi,\delta\psi,\mathcal{L}_{\zeta}\psi)=\delta\oint_{\infty}[Q_{\zeta}-\zeta\cdot b-C(\psi,\mathcal{L}_{\zeta}\psi)]- \delta \oint_{\mathcal B} Q_\zeta+\int_{\Sigma}\zeta\cdot E^{\lambda_{i}}_L\delta\lambda_{i}+\oint_{\infty}\zeta\cdot E_b^{\lambda_i}\delta {\lambda_{i}} \;.\label{eq:variationalidJTextappv3}\eeq
Here we introduced the notation
\beq
\delta_{\lambda_{i}}b = E_b^{\lambda_i} \delta \lambda_i \qquad \text{with} \qquad E_b^{\lambda_i} \equiv \frac{\partial   \mathcal L_{\text{bdy}}}{\partial \lambda_i}  \epsilon_{B}\;,   
\eeq
where $ \mathcal L_{\text{bdy}}$ is the boundary of-shell Lagrangian density. We will  include local counterterms in $b$ to regulate any divergences, so $b = b_{\text{GHY}} + b_{\text{ct}}$.     Note in the context of $2D$ dilaton gravity and in the case $\zeta$ is a Killing vector, we see from \eqref{eq:CLiepsi} that $C(\psi, \mathcal L_\zeta \psi)=0$ and $\delta C(\psi, \mathcal L_\zeta \psi)=0$, if the perturbed  metric is also stationary, \emph{i.e.}, $\delta [\nabla_{(\mu} \zeta_{\nu)}]=0$. Thus the $C$ term in \eqref{eq:variationalidJTextappv3} can be ignored in this context, which is consistent with the stronger boundary condition $C(\psi, \delta \psi )=0$ that we could have   imposed (see footnotes \ref{fn:bdrycond} and \ref{footnote:Cterm2}) 
  \beq \int_{\Sigma} \omega (\psi,\delta\psi,\mathcal{L}_{\zeta}\psi)=\delta\oint_{\infty} ( Q_{\zeta}-\zeta\cdot b )- \delta \oint_{\mathcal B} Q_\zeta+\int_{\Sigma}\zeta\cdot E^{\lambda_{i}}_L\delta\lambda_{i}+\oint_{\infty}\zeta\cdot E_b^{\lambda_i}\delta {\lambda_{i}} \;.\label{eq:variationalidJTextappv2}\eeq
  Below we will apply this variational integral identity   to classical and semi-classical JT gravity in order to derive the extended first laws of nested AdS-Rindler wedges    including variations of the couplings $\lambda_{i}=\{\phi_{0},G,L,c\}$. 


  \subsection*{Classical extended first law}
  
Let us first evaluate the left-hand side of the integral identity (\ref{eq:variationalidJTextappv2}). From (\ref{eq:sympcurrentJT}), where $\delta_{1}\psi=\delta_{\lambda_{i}}\psi$ and $\delta_{2}\psi=\mathcal{L}_{\zeta}\psi$, we find\footnote{More accurate would be to use the definition $\omega(\psi,\delta_{\lambda_i}\psi,\mathcal{L}_{\zeta}\psi)=\delta_{\lambda_i}\theta(\psi,\mathcal{L}_{\zeta}\psi)-\mathcal{L}_{\zeta}\theta(\psi,\delta_{\lambda_i}\psi)$ which considers all possible coupling variations. Using $\theta_{\text{JT}}(\psi,\mathcal{L}_{\zeta}\psi)|_{\Sigma}=0$ and $\theta_{\text{JT}}(\psi,\delta_{\lambda_i}\psi) =0$ for $\lambda_i =(\phi_{0},G)$ variations, we find the only non-zero coupling variation in the symplectic current of JT gravity is with respect to the AdS length~$L$, and so the result is the same in \eqref{eq:omegaappendixellclassical}.}
\beq \label{eq:omegaappendixellclassical}
\omega(\psi,\delta_{\lambda_{i}}\psi,\mathcal{L}_{\zeta}\psi)\Big|_{\Sigma}=\frac{\epsilon_{\mu}}{16\pi G}(g^{\mu\beta}g^{\alpha\nu}-g^{\mu\nu}g^{\alpha\beta})(\nabla_{\nu}\mathcal{L}_{\zeta}\phi) \delta_{\lambda_{i}}g_{\alpha\beta}  =-\frac{\kappa\phi'_{\mathcal{B}}}{16\pi G}h^{\alpha\beta}\delta_{L}g_{\alpha\beta}d\ell \;,
\eeq
where we used the metric is a function of $L$ and not of the other JT couplings. Thus the only nonzero coupling variation contribution to the left-hand side of (\ref{eq:variationalidJTextappv2}) is 
\beq \int_{\Sigma}\omega(\psi,\delta_{L}\psi,\mathcal{L}_{\zeta}\psi)=
-\frac{\kappa\phi'_{\mathcal{B}}}{8\pi G}\delta_{L}\ell 
\label{eq:coupomegavar}\;.\eeq
 Explicitly we have
\beq \int_{\Sigma}\omega(\psi,\delta_{L}\psi,\mathcal{L}_{\zeta}\psi)=\frac{\kappa\phi_{r}\sqrt{\mu}\delta L}{8\pi GL\sinh(\frac{\sqrt{\mu}\alpha}{L})}\left[\cosh \left(\frac{\sqrt{\mu}\alpha}{L}\right)-1+\log\left(\tanh
\left(\frac{\sqrt{\mu}\alpha}{2L}\right)\right)+\frac{1}{2}\log\left(\frac{4r_B^{2}}{L^{2}\mu}\right)\right]\;,\label{eq:omegaLvar}\eeq
which  is logarithmically divergent as the cutoff $r_B$ goes to infinity.

Next, we evaluate the coupling variations of the Noether charge on the right-hand side of (\ref{eq:variationalidJTextappv2}). Implicitly summing over each of the coupling variations, where we note the $\delta L$ contribution entirely drops out, we have
\beq \oint_{\infty}\delta_{\lambda_{i}}(Q_{\zeta}^{\text{JT}}-\zeta\cdot b)=-\frac{\kappa\sqrt{\mu}\phi_{r}}{16\pi G}\tanh(\sqrt{\mu}\alpha/2L)\frac{\delta G}{G}\;,\eeq
\beq \oint_{\mathcal{B}}\delta_{\lambda_{i}}Q_{\zeta}^{\text{JT}}=\frac{\kappa}{8\pi G}\delta\phi_{0}-\frac{\kappa}{8\pi G}(\phi_{0}+\phi_{\mathcal{B}})\frac{\delta G}{G}\;, \eeq
such that combined, with the field variations, 
\beq \delta\oint_{\infty}(Q_{\zeta}^{\text{JT}}-\zeta\cdot b)-\delta\oint_{\mathcal{B}}Q_{\zeta}^{\text{JT}}=\delta E_{\zeta}-\frac{\kappa}{8\pi}\delta\left(\frac{\phi_{0}+\phi_{\mathcal{B}}}{G}\right)\;,\label{eq:intmedcalcdeltaQ}\eeq
where we used $\delta_{G}E_{\zeta}=-E_{\zeta}\frac{\delta G}{G}$.

Lastly, consider the coupling variations of the off-shell Lagrangian and boundary counterterm on the right-hand side of the integral relation (\ref{eq:variationalidJTextappv2}). We have \cite{Svesko:2020yxo}
\beq \int_{\Sigma}\zeta\cdot E^{\lambda_{i}}_{L}\delta\lambda_{i}=\int_{\Sigma}\left(\frac{R}{16\pi G}\delta\phi_{0}-\frac{4\phi}{16\pi GL^{2}}\frac{\delta L}{L}-\mathcal{L}_{\text{JT}}\frac{\delta G}{G}\right)\zeta\cdot\epsilon\;,\label{eq:ELvarJT}\eeq
where $E^{\lambda_{i}}_{L}=\frac{\partial \mathcal{L}_{\text{JT}}}{\partial\lambda_{i}}\epsilon$, with {$\mathcal{L}_{\text{JT}}=\frac{1}{16\pi G}[(\phi+\phi_{0})R+\frac{2}{L^{2}}\phi]$. 
Similarly, the coupling variation  of the local boundary   Lagrangian (\ref{eq:btotclassJT}) yields 
\beq \oint_{\infty}\zeta\cdot E^{\lambda_{i}}_{b}\delta\lambda_{i}=\oint_{\infty}\left[\frac{K}{8\pi G}\delta\phi_{0}+\frac{\phi}{8\pi G L^{2}}\delta L-\frac{1}{8\pi G}\left((\phi_{0}+\phi)K-\frac{\phi}{L}\right)\frac{\delta G}{G}\right]\zeta\cdot\epsilon_{B}\;.\label{eq:EbvarJT}\eeq
Having taken the variations, on-shell the sum of (\ref{eq:ELvarJT}) and (\ref{eq:EbvarJT}) yields
\begin{align}
&\int_{\Sigma}\zeta\cdot E^{\lambda_{i}}_L\delta\lambda_{i} + \oint_\infty \zeta \cdot E^{\lambda_i}_b \delta \lambda_{i}= \left(  \frac{\phi_0}{8 \pi G L^2} \int_\Sigma \zeta \cdot \epsilon - \oint_\infty \zeta \cdot b\right) \frac{\delta G}{G}\\
&+\left( - \frac{1}{8 \pi G L^2}\int_\Sigma \zeta \cdot \epsilon + \frac{1}{8\pi G}\oint_\infty K \zeta \cdot \epsilon_{B} \right) \delta \phi_0 +\left( - \frac{1}{4 \pi G L^2} \int_\Sigma \phi \zeta \cdot \epsilon + \frac{1}{8\pi GL} \oint_\infty \phi \zeta \cdot \epsilon_{B} \right) \frac{\delta L}{L}\;. \nonumber
\end{align}
We recognize the first term on the right-hand side as being proportional to the  ``$G$-Killing volume'' $\Theta^{G}_{\zeta}$ (\ref{eq:thermovoldef}), and we introduce a  counterterm subtracted ``$\phi_0$-Killing volume'' 
\beq
\Theta^{\phi_0}_\zeta \equiv    \int_\Sigma \zeta \cdot \epsilon -L^2\oint_\infty K \zeta \cdot \epsilon_{B} \; ,
\eeq
and a counterterm subtracted ``$\Lambda$-Killing volume"  
\beq
\Theta_{\zeta}^{\Lambda}\equiv\int_\Sigma \phi \zeta \cdot \epsilon - \frac{L}{2} \oint_\infty \phi \zeta \cdot \epsilon_{B}\;.\label{eq:dilatonTV}\eeq
Notice we have normalized each Killing volume such that the coefficient in front of the volume integral is equal to unity.  
Then, the sum  simplifies to 
\beq 
\int_{\Sigma}\zeta\cdot E^{\lambda_{i}}_L\delta\lambda_{i} + \oint_\infty \zeta \cdot E^{\lambda_i}_b \delta \lambda_{i}=\frac{\phi_{0}\Theta^{G}_{\zeta}}{8\pi GL^{2}} \frac{\delta G}{G}-\frac{\Theta^{\phi_0}_\zeta }{8\pi GL^2}\delta\phi_{0} +\frac{\Theta^{\Lambda}_{\zeta}}{8\pi GL^{2}}\frac{\delta \Lambda}{\Lambda}\;.
\eeq
Here we replaced $\delta L / L$ for $- \delta\Lambda/(2\Lambda)$. Explicitly, the $\Lambda$-Killing volume $\Theta^{\Lambda}_{\zeta}$ is 
\beq \Theta^{\Lambda}_{\zeta}=-\frac{\kappa L^{2}\sqrt{\mu}\phi_{r}}{4\sinh(\sqrt{\mu}\alpha/L)}\left(-1+\cosh\left(\frac{\sqrt{\mu}\alpha}{L}\right)+\log\left(\frac{4r_B^{2}}{L^{2}\mu}\right)-2\log\left[\cosh\left(\frac{\sqrt{\mu}\alpha}{L}\right)+\text{csch}\left(\frac{\sqrt{\mu}\alpha}{L}\right)\right]\right)\;,\label{eq:dilatonKillvolexp}\eeq
where $r_B$ is a cutoff which  should be taken to infinity. So $\Theta^\Lambda_\zeta  $ also diverges logarithmically as $r_B \to \infty$. 
Furthermore, the $\phi_0$-Killing volume can be computed to be 
\beq
\Theta^{\phi_0}_\zeta = - \kappa L^2 \;.
\eeq
Putting everything together, we find the classical extended first law of the nested Rindler wedge is 
\beq
\delta E^{\phi_{r}}_\zeta= \frac{\kappa}{8\pi} \delta \left( \frac{\phi_0 + \phi_{\mathcal B}}{G} \right)-\frac{\kappa \phi'_{\mathcal B}}{8\pi G} \delta \ell +  \frac{\Theta_\zeta^{\phi_0}}{8\pi G L^2} \delta \phi_0  - \frac{\phi_0 \Theta^{G}_{\zeta}}{8 \pi G L^2}\frac{\delta G}{G} - \frac{\Theta_{\zeta}^{\Lambda}}{8 \pi G L^2}\frac{\delta \Lambda}{\Lambda}\;,\label{eq:classextfirstlawapp}\eeq
where here $\delta$ includes both field and coupling variations. Notice the variations with respect to $\phi_{0}$ cancel on the right-hand side, and upon invoking the classical Smarr relation \eqref{eq:smarrJTv2}, the $\delta G$ variations cancel between the left- and right-hand sides. The $\delta L$ variation on the right-hand side is divergent as $r_B\to\infty$, however, this divergence is precisely cancelled by the $r_B\to\infty$ divergence appearing in the $\delta_{L}\ell$ term (\ref{eq:omegaLvar}). In fact, combining these two $\delta L$ variations leads to the remarkable simplification:
\beq -\frac{\kappa\phi'_{\mathcal{B}}}{8\pi G}\delta_L\ell+\frac{\Theta_{\zeta}^{\Lambda}}{4\pi GL^{2}}\frac{\delta L}{L}=-\frac{\kappa\sqrt{\mu}\phi_{r}}{16\pi G L} \tanh(\sqrt{\mu}\alpha/2L)\delta L=-E^{\phi_r}_{\zeta}\frac{\delta L}{L}\;. \label{eq:checkofdeltaLvariation}
\eeq
This is consistent with the $L$ variation of the left-hand side of the first law, $\delta_{L} E^{\phi_r}_{\zeta}=-E_{\zeta}^{\phi_r}\frac{\delta L}{L}$, which follows from $\kappa \sim 1/L$. 

\subsection*{Semi-classical extension}

Including semi-classical effects introduces a new coupling, the central charge $c$, and the auxiliary field $\chi$.  The auxiliary field $\chi$ generally depends on $L$ and the dilaton receives a $c$ correction,   $\phi_c = \frac{Gc}{3}$ (working in the Hartle-Hawking state).

It is straightforward to show the combined coupling and field variation  of the Noether charge yields
\beq \delta\oint_{\infty}(Q^{c}_{\zeta}-\zeta\cdot b_{c})-\delta\oint_{\mathcal{B}}Q_{\zeta}^{c}=\delta E^{c}_{\zeta}-\frac{\kappa}{2\pi}\delta\left(\frac{\phi_{c}}{4G}-\frac{c\chi_{\mathcal{B}}}{6}\right)\;,
\label{eq:varQctot}\eeq
where $Q^{c}_{\zeta}= Q^{\chi}_{\zeta}+Q^{\phi_{c}}_{\zeta}$ and $b_{c}=b_{\chi}+b_{\phi_{c}}$. 


We also have contributions from the coupling variations of the Lagrangian 2-forms and boundary counterterm 1-forms associated with the correction $\phi_{c}$ and $\chi$. The $\phi_{c}$ correction only provides   an additional $\delta L$ variation,
\beq \int_{\Sigma}\zeta\cdot E^{\lambda_{i}}_{L_{\phi_{c}}}\delta\lambda_{i} + \oint_\infty \zeta \cdot E^{\lambda_i}_{b_{\phi_{c}}} \delta \lambda_{i}=-\frac{\Theta_{\zeta}^{\Lambda,c}}{4 \pi G L^2}  \frac{\delta L}{L}\;,\label{eq:EbphicLvar}\eeq
where we have defined the  counterterm subtracted ``semi-classical $\Lambda$-Killing volume"  
\beq \Theta_{\zeta}^{\Lambda,c} \equiv \phi_{c}\int_\Sigma  \zeta \cdot \epsilon -\frac{L}{2}\phi_{c}\oint_\infty \zeta \cdot \epsilon_{B}  \;.\eeq
This term diverges as $r_B$, 
more specifically it is given by  
\beq \Theta^{\Lambda,c}_{\zeta}=\frac{cG\kappa L}{6}\left(-2L+\frac{r_B}{\sqrt{\mu}}\tanh(\sqrt{\mu}\alpha/2L)\right)\;.\label{eq:semidilatonKV}\eeq
Moreover, the non-vanishing coupling variations of the Lagrangian 2-form and boundary counterterm associated with $\chi$ include
\beq 
\begin{split}
\int_{\Sigma}\zeta\cdot E^{c}_{L_{\chi}}\delta c+\oint_{\infty}\zeta\cdot E_{b_{\chi}}^{c}\delta c=\left(\int_{\Sigma}\zeta \cdot L_{\chi}+\oint_{\infty}\zeta\cdot b_{\chi}\right)\frac{\delta c}{c}=\frac{c}{12\pi L^{2}}\Theta_{\zeta}^{c}\frac{\delta c}{c}\;,
\end{split}
\label{eq:varEctot}\eeq
and
\beq \oint_{\infty}\zeta \cdot E^{L}_{b_{\chi}}\delta L=\frac{c}{24\pi L^{2}}\Theta^{B}_\zeta\delta L=\lim_{r_B \to \infty }\frac{c\kappa\tanh(\sqrt{\mu}\alpha/2L)}{24\pi \sqrt{\mu}L^{2}}r_B\delta L\;,
\label{eq:LvarEbchi}\eeq
where $\Theta^{c}_{\zeta}$ is the ``semi-classical $c$-Killing volume'' (\ref{semiclassicalvolumeccc}), and 
\beq
\Theta^{B}_\zeta\equiv\oint_{\infty}\zeta\cdot \epsilon_{B}
\eeq
 is the Killing volume of the asymptotic boundary $B\to \infty$. Combining (\ref{eq:LvarEbchi}) with (\ref{eq:EbphicLvar}) cancels a common divergence resulting in,
\beq -\frac{\Theta^{\Lambda,c}_{\zeta}}{4\pi GL^{2}}\frac{\delta L}{L}+\frac{c\Theta^{B}_\zeta}{24\pi L }\frac{\delta L}{L}=\frac{c\kappa}{12\pi L}\delta L\;,\label{eq:LvarEphiEc}\eeq
however, we will not make use of this simplification in the extended first law below.

As in the classical case, only variations with respect to $L$ appear in the symplectic current associated with $\chi$ \eqref{eq:omegabeginningchi}. 
Explicitly, 
\beq \omega_\chi(\psi,\delta_{L}\psi,\mathcal{L}_{\zeta}\psi)\Big|_{\Sigma}=-\frac{c}{24\pi}f(r,\alpha)\left[h^{\alpha \beta}\delta_{L}g_{\alpha\beta}-2\delta_{L}\chi\right](u\cdot \epsilon)\;, \label{eq:additionalsymplecticLblabla}\eeq
where we used   $\nabla_{\mu}(\mathcal{L}_{\zeta}\chi)|_{\Sigma}=u_{\mu}f(r,\alpha)$ (\ref{eq:covdliechiu}). Using $h^{\alpha\beta}\delta_{L}g_{\alpha\beta}=\frac{2}{L}\delta L$, and considering the time-\emph{independent} solution $\chi^{(4)}$  \eqref{eq:chi4sol}  for illustrative purposes,\footnote{For  the time-independent solution for $\chi$ \eqref{eq:chi4sol} we have $\nabla_{\nu}(\mathcal{L}_{\zeta}\chi)\big|_{\Sigma}=\frac{\kappa\sqrt{\mu}}{L}\frac{1}{\sinh(\sqrt{\mu}\alpha/L)}\frac{u_{\nu}}{\left(\frac{r}{L}+\sqrt{\mu}\right)}$. Also, note that when performing the variation of $\chi$ and $h^{\alpha\beta}\delta_{L}g_{\alpha\beta}$, one should use the unscaled Kruskal coordinates (\ref{eq:ads2kruskal}) as it is in these coordinates the $L$ variations of $\chi$ and $g_{\mu\nu}$ are covariant.} we find this additional term \eqref{eq:additionalsymplecticLblabla} vanishes in the $\alpha\to\infty$ limit, however, below and in the main text we simply absorb this contribution into $\delta H^{\chi}_{\zeta}$. 

Putting together (\ref{eq:varQctot}), (\ref{eq:EbphicLvar}), (\ref{eq:varEctot}), and (\ref{eq:LvarEbchi}), we add to the classical extended first law the following combination of terms
\beq \delta E^{c}_{\zeta}=\frac{\kappa}{2\pi}\delta\left(\frac{\phi_{c}}{4G}-\frac{c\chi_{\mathcal{B}}}{6}\right)-\frac{\Theta^{\Lambda,c}}{8\pi GL^{2}}\frac{\delta \Lambda}{\Lambda}-\frac{c\Theta^{c}_{\zeta}}{12\pi L^{2}}\frac{\delta c}{c}+\frac{c\Theta^{B}_\zeta}{48\pi L}\frac{\delta \Lambda}{\Lambda}+\delta H^{\chi}_{\zeta}\;,\eeq
where $\delta H^{\chi}_{\zeta}=\int_{\Sigma}\omega_{\chi}(\psi,\delta \psi,\mathcal{L}_{\zeta}\psi)$ and here again $\delta$ include both field and coupling variations.  If the variation is induced by only changing the central charge $c$ then it reduces to
\beq
\delta_c E^c_\zeta = E^c_\zeta \frac{\delta c}{c}= \frac{\kappa}{2 \pi} \left ( \frac{\phi_c}{4G}- \frac{c \chi_{\mathcal B}}{6} \right) \frac{\delta c}{c} -\frac{c\Theta^{c}_{\zeta}}{12\pi L^{2}}\frac{\delta c}{c}\;,
\eeq
which holds because of the semi-classical part of the Smarr relation \eqref{eq:semiJTsmarr}. Further, for variations that only change the cosmological constant the extended first law becomes
\beq
\delta_L E^c_\zeta= -E_\zeta^c \frac{\delta L}{L}= -\frac{\Theta^{\Lambda,c}}{8\pi GL^{2}}\frac{\delta \Lambda}{\Lambda} +\frac{c\Theta^{B}_\zeta}{48\pi L}\frac{\delta \Lambda}{\Lambda}+\delta_L H^{\chi}_{\zeta}.
\eeq
In the limit $\alpha \to \infty$ both the left- and right-hand side reduce  to $- M_c \frac{\delta L}{L}= - \frac{c\sqrt{\mu}}{12 \pi L} \frac{\delta L}{L}$, due to   relation \eqref{eq:LvarEphiEc} between the Killing volumes and the vanishing of the Hamiltonian variation $\lim_{\alpha \to \infty}\delta_L H_\zeta^\chi = 0$. 

\section{A heuristic argument for the generalized second law} \label{app:secondlaw}

Under some particular assumptions about the final state of the background, here we provide a heuristic derivation of the generalized second law. That is, upon throwing some matter into the system, the semi-classical generalized entropy monotonically increases along the future Killing horizon of the nested Rindler wedge. Our approach is  similar in spirit to the presentation given in \cite{Moitra:2019xoj} (see also \cite{Myers:1994sg} for a flat space analog), and so we only present the broad details.

 \subsection*{Classical JT and the second law}
 
 Consider including infalling matter, characterized by a stress-energy tensor $T^{\text{m}}_{\mu\nu}$. The dilaton equation (\ref{eq:dileom}) remains unchanged such that the background is still $\text{AdS}_{2}$, however, the gravitational field equation  (\ref{eq:graveom}) becomes
 \beq (g_{\mu\nu}\Box\phi-\nabla_{\mu}\nabla_{\nu}\phi-\frac{1}{L^{2}}g_{\mu\nu}\phi)= 8\pi G T^{\text{m}}_{\mu\nu}\;.\label{eq:mateom}\eeq
 Following \cite{Moitra:2019xoj}, assume conformally invariant matter  $g^{\mu\nu}T^{\text{m}}_{\mu\nu}=0$, such that in conformal gauge $T^{\text{m}}_{+-}=0$, which is infalling, \emph{i.e.}, $T^{\text{m}}_{++}=0$. The infalling matter, encoded in $T^{\text{m}}_{--}$ is assumed to obey the classical null energy condition, $T^{\text{m}}_{\mu\nu}k^{\mu}k^{\nu}\geq0$ for any future directed vector $k^{\mu}=\frac{dx^{\mu}}{d\lambda}$ tangent to a null geodesic $x^{\mu}$ parametrized by an affine parameter $\lambda$; equivalently, $T^{\text{m}}_{--}\geq0$.
 
 Solving the equation of motion  (\ref{eq:mateom}) under these conditions leads to a positive increase to the mass of the black hole $\Delta M=\frac{L}{2\phi_{r}}\int dy^{-}h(y^{-})T^{\text{m}}_{--}$, where $h(y^{-})$ is a ``constant" of integration arising from solving the field equations, $h(y^{-})=h_{0}(y^{-})-16\pi G\int_{y^{-}=0}T^{\text{m}}_{--}$, with $h_{0}$ being its value when there is no infalling matter. 
 
 Accompanied by the increase in mass is an increase in the black hole entropy. That is, the classical ``area'' of the black hole, given by the dilaton $\phi$, increases along the (future) event horizon $y^{+}_{H}$ of the black hole. This is seen as follows. Allow the matter to fall in for some finite duration in $y^{-}$, such that after the matter has fallen in and the system reaches equilibrium it remains an eternal black hole. For an affine parameter   $\lambda$ along the event horizon, the null generator satisfies $\frac{dy^{-}}{d\lambda}=(y^{+}_{H}-y^{-})^{2}$, such that the first derivative of the Bekenstein-Hawking entropy $S_{\text{BH}}=\frac{1}{4G}(\phi_{0}+\phi)$ with respect to $\lambda$ is 
 \beq \frac{dS_{\text{BH}}}{d\lambda}\Big|_{H}=(y^{+}_{H}-y^{-})^{2}\frac{dS_{\text{BH}}}{dy^{-}}\Big|_{H}\;.\eeq
 Using the $--$ component of the gravitational field equations (\ref{eq:mateom}) and the null energy condition $T^{\text{m}}_{--}\geq0$, the second derivative of $S_{\text{BH}}$ with respect to $\lambda$ satisfies \cite{Moitra:2019xoj}
 \beq \frac{d^{2}S_{\text{BH}}}{d\lambda^{2}}\Big|_{H}=\frac{1}{4G}(y^{+}_{H}-y^{-})^{4}\nabla_{-}\nabla_{-}\phi\Big|_{H}=-2\pi (y^{+}_{H}-y^{-})^{4}T^{\text{m}}_{--}\leq0\;.\label{eq:secondDSsign}\eeq 
 At late times, after the system has reached equilibrium, where $\frac{d\phi}{d y^{-}}\Big|_{H}=0$,\footnote{This is nothing more than the extremization of the classical entropy. More explicitly, in a generic conformal gauge $ds^{2}=-e^{2\rho(y^{+},y^{-})}dy^{+}dy^{-}$, the generic solution for the dilaton $\phi$ is \cite{Moitra:2019xoj}
 $$\phi(y^{+},y^{-})=\frac{a+b(y^{+}+y^{-})+c y^{+}y^{-}}{(y^{+}-y^{-})}\frac{\phi_{r}}{L},$$
 with black hole mass parameter $\mu/L^2=b^{2}-ac$ for real constants $a,b,c$. The future horizon $y^{+}_{H}=-\frac{1}{c}(b+\sqrt{\mu}L)$ is found by computing $\partial_{y^{-}}\phi=0$, and similarly for the past horizon.} it follows from (\ref{eq:secondDSsign}) that $\frac{dS_{\text{BH}}}{d\lambda}\geq0$ at all times. Hence, the classical entropy obeys the second law, \emph{i.e.}, it monotonically increases along the future event horizon. Mechanically, where $\phi$ is interpreted as the area of the horizon, the second law is completely analogous to the area increase theorem of black hole horizons. While we won't review this here, the classical entropy also monotonically increases along the apparent horizon \cite{Moitra:2019xoj}. Note that the second law also holds from the viewpoint of the nested Rindler wedge. This follows trivially since the above derivation can be worked out with respect to the nested coordinates identically, differing only by constant factors of $\alpha$ which do not influence the sign of the first or second derivative of the entropy. 
 
 \subsection*{Semi-classical JT and the generalized second law}
 
 At equilibrium, we have $\frac{dS_{\text{gen}}}{d\lambda}=0$ along the QES, by definition of the QES. Let us thus consider the semi-classical JT model where, as above, we include classical infalling conformal matter obeying the null energy condition, $T^{\text{m}}_{--}\geq0$. Including infalling matter will alter the solutions for the dilaton $\phi$.
 The semi-classical gravitational equations of motion (\ref{eq:semiclassgraveom}) now become
 \beq T^{\phi}_{\mu\nu}+\langle T^{\chi}_{\mu\nu}\rangle+T^{\text{m}}_{\mu\nu}=0\;.\label{eq:semiclasseominfallmat}\eeq
 such that in conformal gauge the equation of motion (\ref{eq:graveomsjt2}) changes into
 \beq e^{2\rho}\partial_{\pm}(e^{-2\rho}\partial_{\pm}\phi)=-8\pi G\langle T^{\chi}_{\pm\pm}\rangle-8\pi G T^{\text{m}}_{\pm\pm}\;,\label{eq:graveomsjtmat}\eeq
 while (\ref{eq:graveomsjt1}) remains unchanged since $T^{\text{m}}_{\pm\mp}=0$. 
 
 For convenience, below we will work in Poincar\'e coordinates $\{y^{+},y^{-}\}=\{V,U\}$,
\beq ds^{2}=-e^{2\rho}dVdU\;,\quad e^{2\rho}=\frac{4L^{2}}{(V-U)^{2}}\;.\eeq
 We will also consider, when we are near equilibrium,  that the system is in the Hartle-Hawking vacuum state, for which   the expectation values of the $\chi$ stress-energy tensor take the following form in Poincar\'e coordinates (see Appendix \ref{app:ebhcoord})
 \beq \langle T^{\chi}_{\pm\pm}\rangle=0\;,\quad \langle T^{\chi}_{\pm\mp}\rangle=\frac{c}{12\pi}\frac{1}{(y^{+}-y^{-})^{2}}\;.\eeq
To study more realistic evolution as matter is dropped in during some short time duration, one should not restrict   to the Hartle-Hawking state. 
One way this assumption shows up is that we do not take into account backreaction effects on $\chi$ which directly enter into the generalized entropy. 
Namely, while the black hole is settling towards its new equilibrium it is reasonable to expect a mismatch between left- and right-moving modes, which is reflected in $\chi$.
We explicitly assume that $h(y^{-})$ is mild enough to keep the error under control.
A more careful   analysis was presented in \cite{Moitra:2019xoj}. Here we will only focus on the system when it is near equilibrium where our   assumption about being in the Hartle-Hawking state holds. 

The $++$ component of (\ref{eq:graveomsjtmat}) together with (\ref{eq:graveomsjt1}) yield
\beq\phi=\frac{4L^{2}}{y^{-}-y^{+}}h(y^{-})+\frac{cG}{3}-2L^{2}h'(y^{-})\;,\eeq
here $h(y^{-})$ is some ``constant'' of integration and $h'(y^{-})=\frac{dh}{dy^{-}}$. Substituting this into the $--$ component of (\ref{eq:graveomsjtmat}), where $T^{\text{m}}_{--}\neq0$, we find
\beq  2L^{2}h'''(y^{-})=8\pi G T_{--}^{\text{m}}\;.\eeq
 We may solve this differential equation explicitly for $T^{\text{m}}_{--}(y^{+},y^{-})$, however, we won't need to know the exact form of the solution for the derivation of the generalized second law. Importantly, carrying out an identical quasinormal mode analysis as in \cite{Moitra:2019xoj}, at late times, where $y^{-}\to y^{+}_{\mathcal{B}}$, one has $h(y^{-})\sim (y^{+}_{\mathcal{B}}-y^{-})$ where $y^{+}_{\mathcal{B}}$ denotes the   null coordinate along the future  Killing horizon   associated to the quantum extremal surface. 
 
 As in the classical case, we have $\frac{dy^{-}}{d\lambda}=(y^{+}_{\mathcal{B}}-y^{-})^{2}$. Using that the auxiliary field $\chi^{(5)}$, for example, in Poincar\'e coordinates is simply\footnote{This solution follows straightforwardly from a nearly identical analysis presented in Section \ref{sec:JTandsemiclass}, where now we use the transformation  between static coordinates $(v,u)$ and Poincar\'e coordinates $(V,U)$.}
 \beq \chi^{(5)}=-\frac{1}{2}\log\left[\frac{4L^{2}}{(y^{-}-y^{+})^{2}}\right]-\log\left[\left(-\frac{\sqrt{\mu}}{L}y^{-}+k_{-}\right)\left(\frac{\sqrt{\mu}}{L}y^{+}+k_{+}\right)\right]+\tilde{c}\;,\eeq
 where $\tilde{c},k_{\pm}$ are real constants, we find that the $\frac{d S_{\text{gen}}}{d\lambda}$ is simply
 \begin{align}
\frac{dS_{\text{gen}}}{d\lambda}&=(y^{+}_{\mathcal{B}}-y^{-})^{2}(\partial_{-}S_{\text{gen}})=(y^{+}_{\mathcal{B}}-y^{-})^{2}\biggr\{\frac{1}{4G}\left(-\frac{4L^{2}}{(y^{-}-y^{+})^{2}}h+\frac{4L^{2}}{(y^{-}+y^{+})}h'-2L^{2}h''\right) \nonumber \\
&-\frac{c}{6}\left[\frac{1}{y^{-}-y^{+}}+\frac{\sqrt{\mu}}{L}\frac{1}{\left(k_{-}-\frac{\sqrt{\mu}}{L}y^{-}\right)}\right]\biggr\}\;.
 \end{align}
 Upon using $h(y^{-})\sim (y^{+}_{\mathcal{B}}-y^{-})$, we find at late times $\frac{dS_{\text{gen}}}{d\lambda}$ vanishes.
 
 Consider next the second derivative of the generalized entropy. We have
 \beq
 \begin{split}
\frac{d^{2}S_{\text{gen}}}{d\lambda^{2}}&=16 e^{-2\rho}\partial_{-}\left(e^{-2\rho}\partial_{-}S_{\text{gen}}\right)\\
&=\frac{4e^{-4\rho}}{G}\left([\partial^{2}_{-}\phi-2(\partial_{-}\phi)(\partial_{-}\rho)]+\frac{2Gc}{3}[-\partial_{-}^{2}\chi+2(\partial_{-}\rho)(\partial_{-}\chi)]\right)\;.
\end{split}
\eeq
Invoking the gravitational equations of motion (\ref{eq:graveomsjtmat}) and (\ref{eq:graveomsjt1}), with $\langle T_{--}^{\chi}\rangle=0$ in the Hartle-Hawking vacuum, yields
\beq \frac{d^{2}S_{\text{gen}}}{d\lambda^{2}}=-\frac{4e^{-4\rho}}{G}\left(8\pi GT^{\text{m}}_{--}+\frac{2Gc}{3}(\partial_{-}\chi)^{2}\right)\leq0\;,\label{eq:2ndderivSgen}\eeq
where the inequality follows from the classical null energy condition. Note that the second derivative calculation did not make use of the explicit form of $\chi$ or $\phi$. We also see that had one considered the Boulware state, where $\langle T_{\pm\pm}^{\chi}\rangle\neq0$, the second derivative of $S_{\text{gen}}$ is not guaranteed to satisfy (\ref{eq:2ndderivSgen}). 

Lastly, using that $\frac{dS_{\text{gen}}}{d\lambda}\to0$ at equilibrium and (\ref{eq:2ndderivSgen}), it follows 
\beq \frac{dS_{\text{gen}}}{d\lambda}\geq0\;,\eeq
such that the generalized entropy increases monotonically along the future   Killing horizon associated with the quantum extremal surface, as in the case of the black hole Killing horizon \cite{Moitra:2019xoj}. This completes our heuristic derivation. A more careful analysis would solve the entire problem without assuming we are in the Hartle-Hawking state, such that we study the entire evolution of the system. This would be interesting and we save it for future work.

\bibliography{nowormrefs}

\end{document}